\documentclass[twoside,twocolumn,english,prc,amsmath,a4paper,myheadings]{article}
\usepackage{mathpazo}
\usepackage[T1]{fontenc}
\usepackage[latin9]{inputenc}
\usepackage{geometry}
\usepackage{color}
\usepackage{verbatim}
\usepackage{float}
\usepackage{calc}
\usepackage{amsmath}
\usepackage{amssymb}
\usepackage{graphicx}

\makeatletter

\providecommand{\tabularnewline}{\\}

\newcommand{\lyxaddress}[1]{
	\par {\raggedright #1
	\vspace{1.4em}
	\noindent\par}
}

\def\@oddhead{\rightmark \hfill Parallel scattering, saturation, and generalized 
Abramovskii-Gribov-Kancheli (AGK) theorem in EPOS4 \hfill \thepage}
\def\@evenhead{\thepage \hfill Klaus WERNER\hfill}
\def\fnum@table{\tablename~{\bf\thetable}}
\def\fnum@figure{\figurename~{\bf\thefigure}}
\def\tablename{\footnotesize{\bf Table}}
\def\figurename{\footnotesize{\bf Figure}}

\usepackage{dcolumn}
\usepackage[font=small,labelfont=bf]{caption}

\def\citet{\cite}

\AtBeginDocument{
  
}

\makeatother

\usepackage{babel}
\begin{document}
\twocolumn[   \begin{@twocolumnfalse}  
\title{Parallel scattering, saturation, and generalized Abramovskii-Gribov-Kancheli (AGK) theorem in the
EPOS4 framework, with applications for heavy-ion collisions at $\sqrt{s_{NN}}$ of 5.02
TeV and 200 GeV}
\date{}
\author{K.$\,$Werner}
\maketitle

\lyxaddress{\begin{center}
SUBATECH, Nantes University \textendash{} IN2P3/CNRS \textendash{}
IMT Atlantique, Nantes, France
\par\end{center}}
\begin{abstract}
Ultrarelativistic heavy-ion collisions will first realize many nucleon-nucleon
scatterings, happening instantaneously and therefore necessarily in parallel, due to the short
collision time. An appropriate quantum mechanical tool to treat that
problem is S-matrix theory, and it has been known for a long time how
to derive a simple geometric probabilistic picture, still widely used,
and here the Abramovskii-Gribov-Kancheli (AGK) theorem plays a crucial role. All this is done
in a scenario where energy conservation is not taken care of, but
this is needed, in particular for Monte Carlo simulations. When introducing
energy-momentum sharing properly, the AGK theorem does not apply anymore, 
nor do simple geometric concepts such as binary scaling. I will discuss this
(very serious)
problem, and how it can be solved, in the EPOS4 framework. 
When connecting the multiple Pomeron approach (for parallel scatterings) and perturbative QCD, one
is actually forced to implement in a very particular way
saturation scales, in order to get an approach
free of contradictions. One recovers a 
generalized AGK theorem
(gAGK), valid at large $p_{t}$ (larger than the relevant saturation
scales). I discuss how gAGK is related to factorization (in proton-proton scatterings) and
binary scaling (in heavy-ion collisions). I will show some applications, using this new
approach as an initial condition for hydrodynamical evolutions, for 
heavy-ion collisions at $\sqrt{s_{NN}}$ of 5.02 TeV and 200 GeV, to get some idea about
the energy dependence.

~~
\end{abstract}
\end{@twocolumnfalse}]

\section{Introduction \label{======= Introduction =======}}

It is of fundamental importance to realize that multiple nucleon-nucleon
scatterings in heavy-ion collisions must happen in parallel, and not
sequentially, based on very elementary considerations concerning time scales.
The appropriate tool to take this into account has been known for a long
time: an S-matrix approach, referred to as Gribov-Regge (GR) theory \cite{Gribov:1967vfb,Gribov:1968jf,GribovLipatov:1972}.
A key property is the so-called 
Abramovskii-Gribov-Kancheli (AGK) theorem, 
also referred to
as ``AGK cancellations'' \cite{Abramovsky:1973fm}, which allows
to make the link between the multiple scattering approach and geometric
properties such as binary scaling. 

However, for a realistic scenario \textendash{}
in particular as a basis for event-by-event Monte Carlo procedures \textendash{}
it is mandatory to include energy-momentum conservation, which is not considered
in the approach mentioned above. This sounds trivial, but actually
implementing it is a highly complex operation, in a scenario
where all $NN$ collisions happen in parallel, avoiding any ordering
of these collisions which would make no sense. If a projectile nucleon interacts
with several target nucleons, there is nothing like a first and subsequent collisions, 
they are all equal. Despite many technical difficulties,
an unbiased energy-momentum sharing can be dealt with \cite{Drescher:2000ha},
but unfortunately it destroys the ``AGK cancellations'', and as
a consequence, all the nice and simple geometric properties which follow,
in particular binary scaling. 

The aim of this paper is to first discuss and finally understand the serious problems which arise (unavoidably)
in a scenario with appropriate energy-momentum sharing and at the same time unbiased parallel scattering, 
which at the end leads to an amazingly simple solution of all problems based on a particular implementation of saturation. 

The present paper is the fourth in a series of publications 
\cite{werner:2023-epos4-overview,werner:2023-epos4-heavy,werner:2023-epos4-micro} 
about the EPOS4 framework. One distinguishes between primary and secondary interactions. 
The former refer to the multiple parallel scatterings
happening (at high energies) instantaneously and which result in complex configurations composed of many strings,
whereas the latter refer to subsequent interactions of the string decay products, 
which amounts to first a core-corona separation based on the string segments,
and then the fluid formation, evolution, and decay of the core part.
It should also be mentioned that the discussion of primary interactions covers two topics, 
namely the parallel scattering formalism, developed in terms of abstract objects for single scatterings called ``Pomerons'',
and the internal structure of the Pomerons, showing how the Pomeron is related to parton-parton cross sections expressed 
in terms of QCD diagrams.  
Reference \cite{werner:2023-epos4-overview} represents an overview of EPOS4, in Ref. \cite{werner:2023-epos4-heavy}, 
the internal Pomeron structure is treated, with very detailed discussions of the parton-partons scatterings based on 
perturbative QCD (which makes the link between the multiple scattering formalism and QCD). 
Reference \cite{werner:2023-epos4-micro} focuses on the secondary interactions, in particular on the core-corona procedure 
and microcanonical hadronization of the fluid. 

The present paper refers to the primary interactions, and it focuses on a very detailed and rigorous 
treatment of the multiple scattering aspect based on Pomerons, referring when necessary to the internal structure 
discussed in Ref. \cite{werner:2023-epos4-heavy}. 
The aim is the presentation of a consistent multiple scattering formalism, which takes into account energy-momentum sharing 
(which is mandatory) and rigorous unbiased parallel scattering (which is mandatory as well), 
and which finally solves the problem of violating AGK (and simple geometrical pictures). 
The key element is an appropriate implementation of dynamical saturation scales.

In the following sections, I will first discuss the goal and the philosophy of the formalism developed in this paper, 
with precisions concerning the assumptions and the limitations.
Then I will discuss time scales
and the corresponding applicability of parallel scattering as a function
of the collision energy.
Since ``AGK cancellations'' are the key element in the parallel
scattering scheme, I will briefly review this concept
in the classical approach \cite{Gribov:1967vfb,Gribov:1968jf,GribovLipatov:1972}
and \cite{Abramovsky:1973fm}, before generalizing towards a scenario
including energy-momentum conservation.  I will discuss all
the difficulties that show up, and eventually the solution, based
on saturation scales, leading to the generalized AGK theorem (gAGK),
valid at large $p_{t}$, which allows to restore the above-mentioned
geometrical properties like binary scaling.
Finally, I  show results for PbPb at $\sqrt{s_{NN}}$ of 5.02 TeV and AuAu at 200 GeV.

\section{The philosophy, basic assumptions, and objectives}

Let me make some remarks about the philosophy behind the present work,
the main goals, and the assumptions employed. The aim is not
to construct an event generator, but to provide a theoretical framework
for ``multiple scatterings'', where one considers multiple partonic
scatterings in proton-proton and multiple nucleon-nucleon scatterings
in nucleus-nucleus collisions. 

A crucial point is the fact that one is aiming at a formalism which
allows access to event classes. Most theoretical work based
on QCD is devoted to inclusive cross sections, and here factorization
plays a dominant role. The approach in this paper does not contribute
to these admittedly very important achievements, being very useful
to study rare processes. But one of the highlights at the LHC
and a major activity for (at least) the next decade are studies concerning
particle production as a function of the event activity (measured
in terms of multiplicity or transverse energy). Models are requested
to compute particle yields or two-particle correlations, but with
a very important additional condition: for given event classes. And
this makes life complicated. Minimum bias results based on factorization
are not able to provide answers or provide wrong answers, citing,
for example, the (experimentally well-known) increase of the mean transverse
momentum with multiplicity in proton-proton scattering. 

From the huge amount of experimental data and phenomenological attempts
to interpret these, there is strong evidence that event activity
is very strongly correlated with multiple parton interactions (MPI).
Reference \cite{MPI:2018} discusses the role of multiple parton
scatterings at the LHC: one distinguishes between ``hard MPI'' and
``soft MPI'', where the former treats the scattering of two hard partons,
based on multiple parton distribution functions, generalizing the
factorization approach. But this does not address the questions related
to studies of properties for given event classes, defined via ``event
activity''. This is discussed in Ref. \cite{MPI:2018} in the part ``soft
MPI'', and here the situation is not satisfactory; the ``phenomenological
approaches'' have often no solid theoretical basis, and here I am
not even talking about a first-principles basis. 

The discussion of ``event classes'' is clearly also very relevant
for scattering of nuclei with mass numbers $A$ and $B$ ($A\!+\!B$ scattering).
Nobody considers minimum bias results; everything is presented again
for event classes, again defined via event activity. Also here many
things are known from phenomenological studies, like the (measurable)
event activity being very strongly correlated with the (non measurable)
impact parameter and also with the (non measurable) number of nucleons
being involved. There are also formulas concerning factorization in
nucleus-nucleus scattering, but also here one needs a framework where
one has access to information for given event classes, and one needs
to identify the theoretical counterpart of these event activities.

It is completely out of reach to construct such a framework (with
access to event classes) from first principles (QCD), and this is
not at all what will be attempted in this paper. What can be done
(and this will be developed in this paper) is to construct a scheme
in the spirit of S-matrix theory, which was the main theoretical tool
to treat scatterings before the QCD era. One assumes certain fundamental
properties, like Lorentz invariance and analyticity of the T-matrix,
although it is not possible to derive the particular T-matrix from
first principles. But at least the assumptions are very clear and
transparent. Here the same strategy is adopted. One starts with
a certain assumption about the form of the T-matrix, which is only
based on a qualitative (or phenomenological) understanding of experimental
data. It cannot be proven from first principles, but it is well defined,
and even very simple, and one respects perfectly the framework of
quantum mechanics (this is worth noticing, because many ``models''
for heavy-ion scatterings are based purely classical considerations).

Before discussing more in detail the assumptions of the present approach,
some historical remarks are in order. In his famous lecture \cite{Glauber},
Glauber discusses the scattering of high energy protons with nuclei.
There are several assumptions used, like a factorization of the wave
function into a plane wave times a slowly (in space) changing function.
The assumptions are clear and plausible, and one gets finally a simple
result with a nice interpretation, of a particle moving on a straight
line, accumulating phases (the so-called eikonal phase is an integral
over the potential along the longitudinal axis). Another early (pre-QCD)
approach is GR theory \cite{Gribov:1967vfb,Gribov:1968jf,GribovLipatov:1972}.
Also here very strong assumptions are made about the structure of
the T-matrix, and the form of the sub-T-matrices. But also here, the
assumptions cannot be proven from some fundamental theory, but they
are plausible, clear, and transparent. Interesting enough, the Glauber
and the GR approaches (the latter being equal to the EPOS
approach without energy sharing and without saturation) give
similar results.

The scheme to be developed in this paper has to be understood in the
spirit of these two (historical) approaches. It is not attempted
to ``derive'' the formalism from a fundamental theory (QCD), but
to construct a model, in the framework of quantum mechanics (with
the possibility to treat elastic and inelastic scattering, related
by the optical theorem), and which is finally compatible
with QCD. There are assumptions, which are purely based on a phenomenological
(and partly qualitative) understanding of experimental data. But,
the assumptions are very clear and transparent (for this purpose I
will try to summarize all the important points in the following). 

\begin{figure}[h]
\centering{}\includegraphics[scale=0.22]
{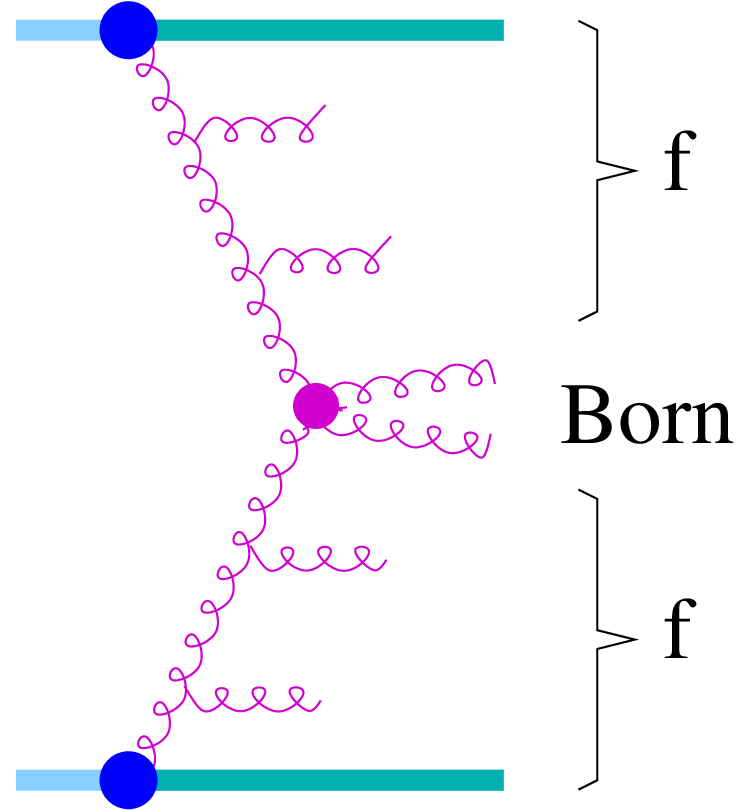}
\caption{Factorization in $pp$ scattering. \label{factorization}}
\end{figure}

\begin{figure}[h]
\centering{}\includegraphics[scale=0.22]
{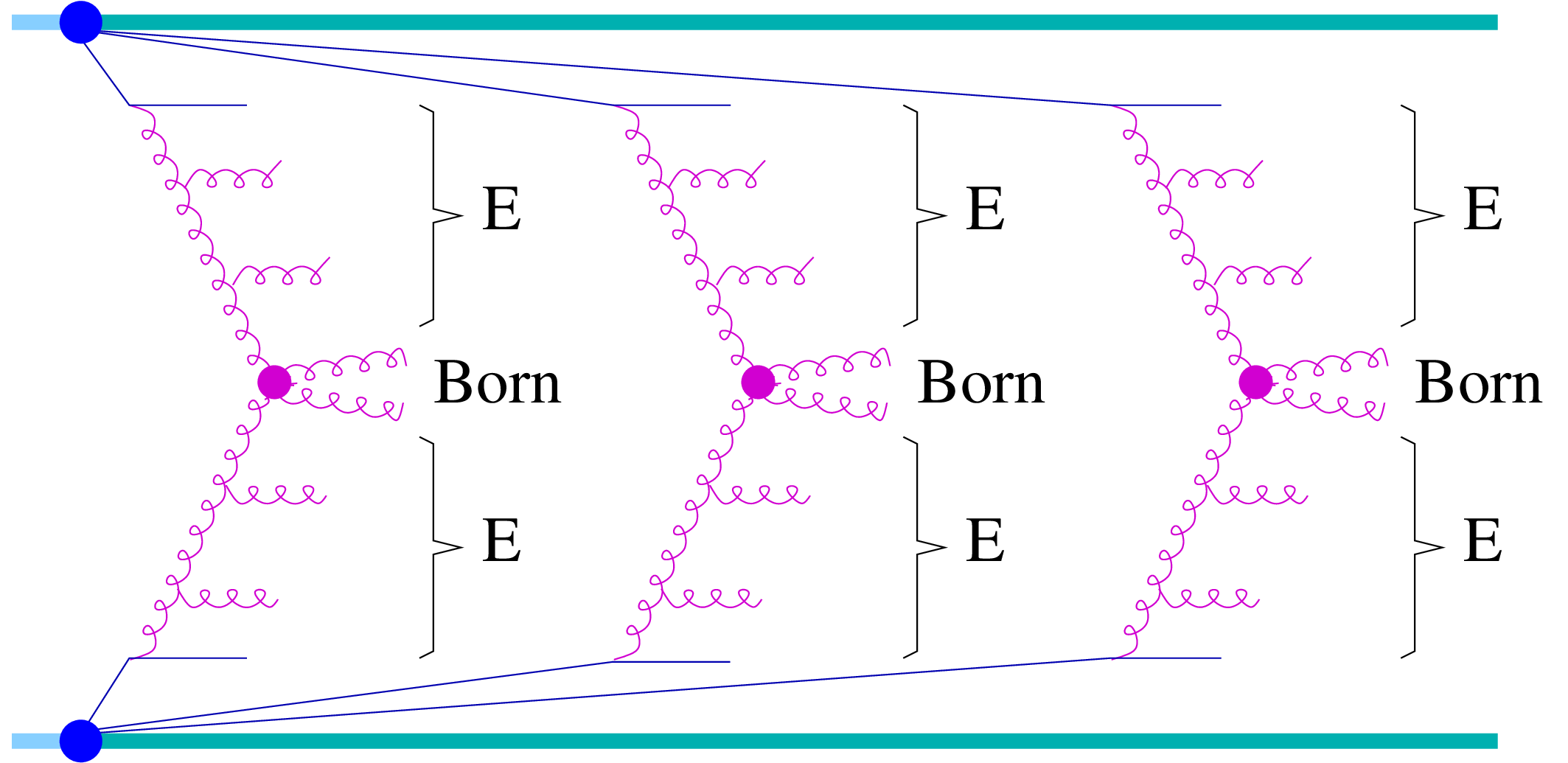}
\caption{Multiple scattering. \label{multiple scattering}}
\end{figure}

\begin{figure}[h]
\centering{}\includegraphics[scale=0.22]
{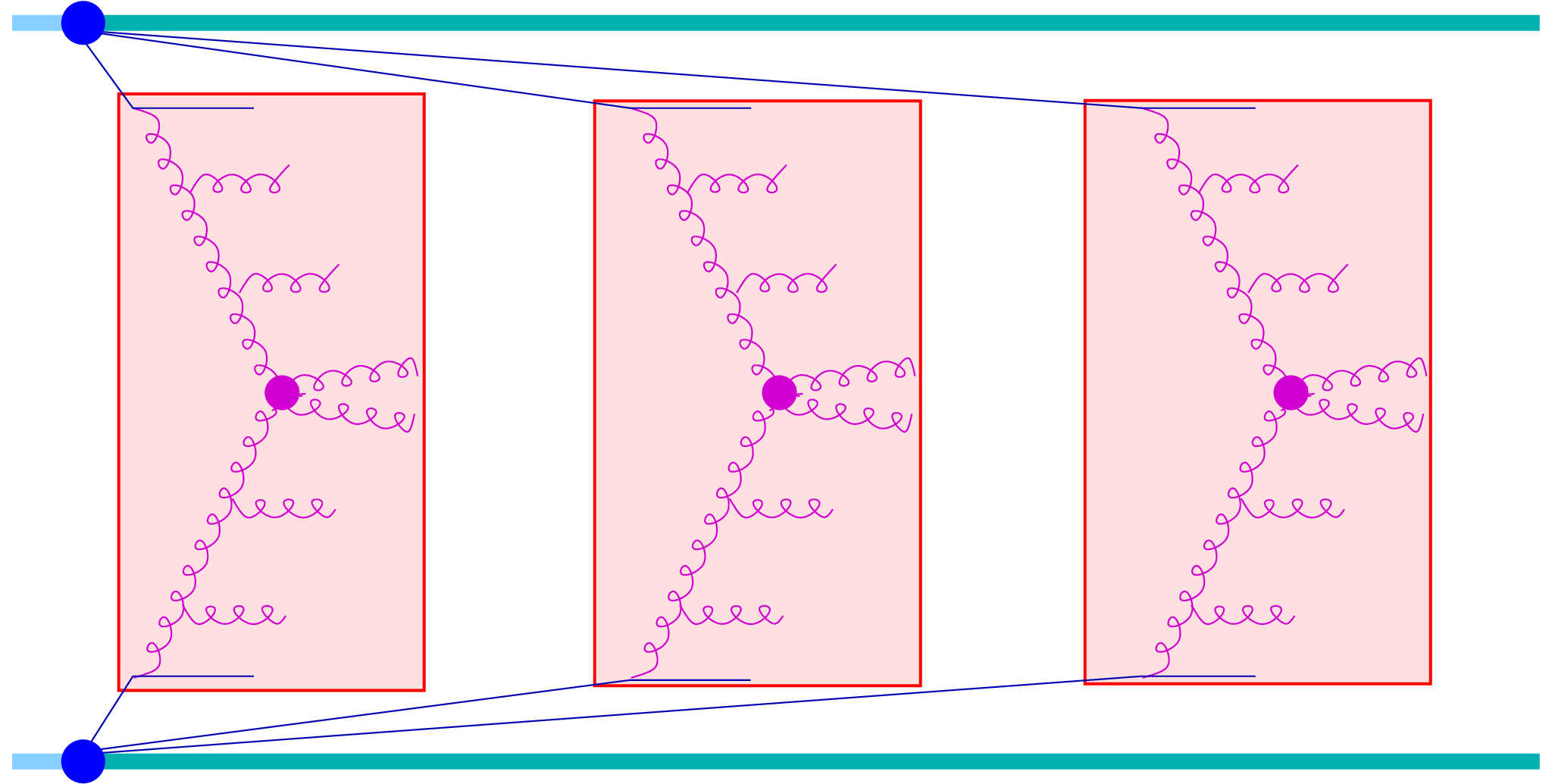}
\caption{Modular structure of multiple scattering. \label{modular structure}}
\end{figure}

But let me first have some qualitative discussion. The usual factorization
picture where one attempts to separate \textquotedblleft long range
(soft)\textquotedblright{} and \textquotedblleft short range (hard)\textquotedblright{}
parts of the interaction, may be graphically presented as shown in
Fig. \ref{factorization}. 
In this plot and all the following ones, I show for simplicity only
gluons, in reality all kinds of partons are present. The two light
blue thick lines represent the projectile and the target protons.
The proton structure and the so-called spacelike parton cascade are
taken care of by using parton distribution functions (PDFs) $f$, which
allow writing the jet cross section as a convolution of these PDFs
and an elementary QCD cross section for the Born process in the middle.
This approach provides excellent results concerning inclusive particle
production. From studies of particle production for given event classes,
for example charm production versus charged particle multiplicity,
it seems that the ``real events'' look more like that sketched in Fig.
\ref{multiple scattering},
where one has not only several (three in the figure) scatterings,
but even several parallel ``objects'' (multiple scattering). The fact
of having several of these objects not only increases (for example)
the rate of heavy flavor production, it also increases the charged
particle multiplicity, and each object produces particles over the
whole range in rapidity. This is why in the sketch of Fig. \ref{multiple scattering}
the three objects are similar to the one shown in Fig. \ref{factorization},
with parton emissions, but per object. Some ``evolution
function'' $E$ is indicated which should obey the same evolution equations
as the $f$ in Fig. \ref{factorization}. The picture sketched in
Fig. \ref{multiple scattering} suggests some modular structure, as
indicated in Fig. \ref{modular structure},
where one first simply considers ``objects'' indicated by the red boxes,
where one might separate the multiple scattering aspect (expressed
in terms of the objects) and the content of the objects, representing
parton-parton scattering based on QCD diagrams.\\

The main objective of this work (with this paper being an important
piece of) is to investigate (and understand) how a picture as sketched
in Fig. \ref{multiple scattering} breaks down to the one sketched
in Fig. \ref{factorization} when it comes to inclusive particle production
(in minimum bias scattering). Even more drastic, one has to show
how in $A\!+\!B$ scattering a picture of $A\times B$ possible interactions
each one of the form of Fig. \ref{multiple scattering} breaks down
as well to the one sketched in Fig. \ref{factorization}, simply multiplied
by a factor $A\times B$ (which seems to be an experimental fact).
These are crucial questions; any multiple scattering approach, which
claims to describe observables ``per event class'', must do this
exercise.\\

The ideas sketched in the previous paragraph will be realized based
on a model, generalizing the GR approach [to include energy-momentum
sharing (GR\textsuperscript{+}), and to make it QCD compatible]. This is compatible with the
picture sketched in Fig. \ref{modular structure} and the experimental
data which suggest such a picture. There are serious problems coming
up, for which solutions can finally be proposed. But they might be questioned.
This is why I will present in the following all the important assumptions
and hypotheses (marked \textbf{H1a}, \textbf{H1b}, etc.). 

\begin{description}
\item [{(H1a)}] The most important assumption says that the T-matrix (first
for $pp$ elastic scattering) has a modular structure: it is given
in terms of products of sub-T-matrices $T_{\mathrm{Pom}}$ referring
to elementary scatterings between ``proton constituents'' by the
exchange of a ``Pomeron'' (whatever this may be). I consider first
the case without energy-momentum sharing. This is precisely what is
done in Refs. \cite{Gribov:1967vfb,Gribov:1968jf,GribovLipatov:1972}. One
gets {[}see Eq. (\ref{multiple-tmatrix-pp-1-1}){]}
\begin{equation}
iT=\sum_{n=0}^{\infty}\,\frac{1}{n!}\,\left\{ iT_{\mathrm{Pom}}\right\} ^{n}\,.\label{multiple-tmatrix-pp-1-1-1}
\end{equation}
\item [{(H1b)}] A fundamental assumption is the complete separation of
the multiple scattering structure, expressed in terms of the sub-T-matrices
$T_{\mathrm{Pom}}$ and the underlying microscopic theory (QCD), which
allows to first develop the multiple scattering theory based on $T_{\mathrm{Pom}}$,
and postpone the precise ``internal structure'' of $T_{\mathrm{Pom}}$
in terms of QCD diagrams.
\end{description}
The optical theorem (in impact parameter representation,
see Sec. \ref{======= S-matrix-theory-basic-definitions =======})
allows to compute the inelastic cross section, as $\sigma_{\mathrm{in}}=\int d^{2}b\,\mathrm{cut}\,T(b)$,
with $b$ being the impact parameter. Cutting a complex diagram amounts
to summing all possible cuts. 
\begin{description}
\item [{(H1c)}] 
Here one assumes that a Pomeron is cut
completely, or not at all. Summing over all cuts then means summing over all
possibilities of cut and uncut Pomerons. Assuming in addition $T_{\mathrm{Pom}}$
to be purely imaginary, and defining the ``cut Pomeron'' $G=2\,\mathrm{Im}\,T_{\mathrm{Pom}}$,
one gets 
\begin{equation}
\sigma_{\mathrm{in}}=\int d^{2}b\,\mathrm{\sum_{n=1}^{\infty}\,\frac{1}{n!}\,\sum_{m=0}^{n}\left(\begin{array}{c}
n\\
m
\end{array}\right)}G^{m}(-G)^{n-m}.\label{sigma-inel-2-2}
\end{equation}
\end{description}
In principle the T-matrices  depend on the collision energy
(expressed in terms of $s$) but for simplicity I will not write this
dependence explicitly here (but it will be done properly later).
In order to generalize towards $A\!+\!B$ scattering, one defines $b_{i}^{A}$
and $b_{j}^{B}$ being the transverse coordinates of the nucleons
in the nuclei $A$ and $B$, and an integration
\begin{equation}
\int db\!_{A\!B}=\int d^{2}b\int\prod_{i=1}^{A}d^{2}b_{i}^{A}\,T_{A}(b_{i}^{A})\int\prod_{j=1}^{B}d^{2}b_{j}^{B}\,T_{B}(b_{j}^{B}),\label{b-AB-integration-1}
\end{equation}
with the nuclear thickness functions $T_{A}$ and $T_{B}$ {[}see Eq.
(\ref{thickness-function}){]} representing the nuclear geometry.
\begin{description}
\item [{(H2a)}] It is assumed that the T-matrix expression for $A\!+\!B$ scattering
is given as an integral $\int db\!_{A\!B}...$ .
\item [{(H2b)}] The integrand is assumed to be a product of $A\times B$
nucleon-nucleon expressions as Eq. (\ref{multiple-tmatrix-pp-1-1-1}),
where the impact parameter argument of the T-matrices corresponding
to the $k$-th nucleon-nucleon pair is $b_{k}=|b+b_{\pi(k)}^{A}-b_{\tau(k)}^{B}|$.
The functions $\pi(k)$ and $\tau(k)$ refer to the nucleons associated
to the pair $k$. Defining $G_{k}=G(b_{k})$, one gets
\begin{equation}
\sigma_{\mathrm{in}}^{AB}=\int db\!_{A\!B}\sum_{m_{1}l_{1}}\ldots\sum_{m_{AB}l_{AB}}\prod_{k=1}^{AB}\mathrm{\frac{1}{m_{k}!l_{k}!}}(G_{k})^{m_{k}}(-G_{k})^{l_{k}}.\label{sigma-inel-AB-6}
\end{equation}
\end{description}
What is discussed sofar is essentially the GR approach,
which does not address energy-momentum sharing. To incorporate that,
one introduces the light-cone momentum fractions $x^{\pm}$ with respect
to the initial light-cone momenta of the nucleons, with $0\le x^{+},x^{-}\le1$.
One defines $x_{k\mu}^{\pm}$ to be the light-cone momentum fractions
of the external legs of the $\mu$-th Pomeron of pair $k$, all of
them connected to projectile nucleon $i=\pi(k)$ and target nucleon
$j=\tau(k)$. The light-cone momenta of the projectile and target
remnants are named $x_{\mathrm{remn},i}^{+}$ and $x_{\mathrm{remn},j}^{-}$.
Energy-momentum conservation amounts to 
\begin{equation}
x_{\mathrm{remn},i}^{+}=1-\!\!\sum_{\underset{\pi(k)=i}{k=1}}^{AB}\sum_{\mu=1}^{n_{k}}\!x_{k\mu}^{+}\:,\;x_{\mathrm{remn},j}^{-}=1-\!\!\sum_{\underset{\tau(k)=j}{k=1}}^{AB}\sum_{\mu=1}^{n_{k}}\!x_{k\mu}^{-},\label{energy-conservation-1}
\end{equation}
which means that the initial values ($x^{\pm}=1$) are shared among
Pomerons and remnants. 
\begin{description}
\item [{(H3a)}] One assumes, that energy-momentum conservation can be incorporated
by simply using $x_{k\mu}^{\pm}$ as arguments of the sub-T-matrices and
\item [{(H3b)}] by adding Pomeron-nucleon vertices $V(x_{\mathrm{remn},i}^{+})$
and $V(x_{\mathrm{remn},j}^{-})$ representing the coupling of the
Pomerons to the projectile and target remnants, which ensures energy-momentum
conservation.
\end{description}
As a consequence, defining $G_{k\mu}\!=G\!\left(\!x_{k\mu}^{+},x_{k\mu}^{-},b_{k}\!\right)$,
and an integration $\int dX\!_{A\!B}$ over all light-cone momentum
fractions {[}see Eq. (\ref{x-AB-integration}){]}, one gets (see Sec.
\ref{-----The-sign-problem-----}) $\sigma_{\mathrm{in}}^{AB}=\int\!\!db\!_{A\!B}G^{AB}(\{b_{A\!B}\})$,
with $\{b_{A\!B}\}$ being the multidimensional variable $\{b,\{b_{i}^{A}\},\{b_{j}^{B}\}\}$,
and with
\begin{align}
G^{AB}(\{b_{A\!B}\}) & =\sum_{n_{1}=0}^{\infty}\!\!\ldots\!\!\sum_{n_{AB}=0}^{\infty}\quad\sum_{m_{1}\le n_{1}}\!\!\ldots\!\!\sum_{m_{AB}\le n_{AB}}\int dX\!_{A\!B}\nonumber \\
 & \qquad\prod_{k=1}^{AB}\frac{1}{n_{k}!}\mathrm{\Big(\!\!\begin{array}{c}
n_{k}\\
m_{k}
\end{array}\!\!\Big)}\prod_{\mu=1}^{m_{k}}G_{k\mu}\prod_{\mu=m_{k}+1}^{n_{k}}-G_{k\mu}\nonumber \\
 & \qquad\prod_{i=1}^{A}V(x_{\mathrm{remn},i}^{+})\prod_{j=1}^{B}V(x_{\mathrm{remn},j}^{-}),\label{sigma-inel-AB-2-1}
\end{align}
with at least one $n_{k}$ being nonzero. 
This extension of the GR approach (by including energy-momentum sharing) will be referred to as GR\textsuperscript{+}.

\begin{figure}[h]
\centering{}\includegraphics[scale=0.22]
{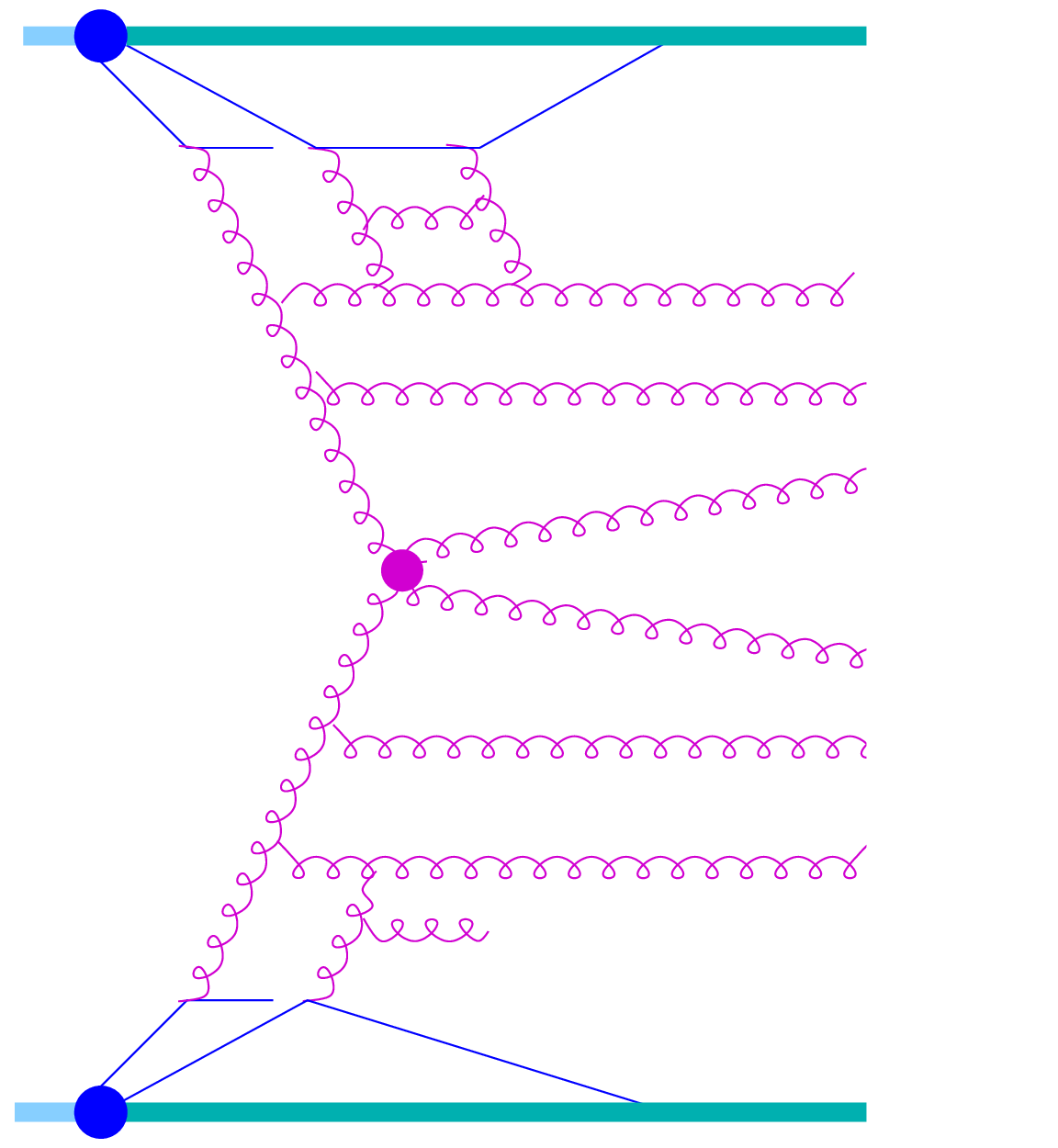}
\caption{Nonlinear effects: ladders which evolve first independently and in
parallel, finally fuse. \label{nonlinear-effects}}
\end{figure}

One may define an expression $\bar{G}^{AB}(\{b_{A\!B}\})$ as in Eq.
(\ref{sigma-inel-AB-2-1}), but for all $n_{k}=0$, which represents
the case excluded in the summation of Eq. (\ref{sigma-inel-AB-2-1}).
One can prove {[}see Eq. (\ref{G-plus-Gbar}){]} the following relation:
\begin{equation}
G^{AB}(\{b_{A\!B}\})+\bar{G}^{AB}(\{b_{A\!B}\})=1.\label{G-plus-Gbar-1}
\end{equation}
In case of $A=B=1$ ($pp$ scattering) one then gets 
\begin{equation}
\sigma_{\mathrm{in}}^{pp}=\int\!\!db\big\{1-\bar{G}^{11}(b)\big\}.
\end{equation}
It is very tempting to interpret the expression $\big\{1-\bar{G}^{11}(b)\big\}$
as the probability of an interaction at given impact parameter $b$.
In that case, $\bar{G}^{11}(b)$ must be non-negative, otherwise one
exceeds the ``black disk limit''. It is also very plausible that this
condition should hold for $A>1$ and/or $B>1$. So one requires the following:
\begin{description}
\item [{(H4)}] Whatever approximation might be employed, it is mandatory
that $\bar{G}^{AB}(\{b_{A\!B}\})$ is non-negative, for any value of $A$ and $B$.
 This is a fundamental expectation which will guide the following hypotheses.
 It can be proven in the case without energy sharing, and is here expected to be true as well.
\end{description}
This is a crucial requirement, it has enormous consequences, as will
be seen later. In order to investigate $G^{AB}$ and$\bar{G}^{AB}$
(and in particular the sign of the latter), one has to be more specific.
At this point the $G$ functions are just abstract objects, but eventually
they should correspond to ``real'' expressions based on QCD diagrams
representing parton-parton scattering. Studying such expressions,
it could be shown \cite{Drescher:2000ha} that one may obtain an almost
perfect fit of the numerically computed functions $G_{\mathrm{QCD}}$,
with a parametrization being a sum of expressions of
the form $\alpha_{N}(x^{+}x^{-})^{\beta_{N}}$. This parametric form
has been inspired by the asymptotic expressions for T-matrices (see
Appendix \ref{======= asymptotic-behavior =======}).
So one postulates the following: 
\begin{description}
\item [{(H5a)}] With $\alpha_{N}$ and $\beta_{N}$ being coefficients
depending on $b$ in terms of a few parameters, the functions $G(\!x^{+},x^{-},b\!)$
can be written as (see Appendix \ref{======= W-AB =======})
\begin{equation}
G_{\mathrm{QCDpar}}(\!x^{+},x^{-},b\!)=\sum_{N=1}^{4}\alpha_{N}(x^{+}x^{-})^{\beta_{N}}.\label{QCDpar-2}
\end{equation}
\item [{(H5b)}] Furthermore, the vertices can be parametrized as $V(x)=x^{\alpha_{\mathrm{remn}}},$
again motivated by the asymptotic expressions for T-matrices. 
\end{description}
These postulates are crucial, they allow one to do the integrals, which
could not be done numerically. Defining $\{m_{k}\}$ to be the set
of the $m_{k}$ variables, and $\{x_{k\mu}^{\pm}\}$ the set of the
$x_{k\mu}^{\pm}$ variables, one finds (see Sec. \ref{-----The-sign-problem-----})
\begin{equation}
G^{AB}(\{b_{A\!B}\})=\sum_{\{m_{k}\}}\int\prod_{k=1}^{AB}\prod_{\mu=1}^{m_{k}}dx_{k\mu}^{+}dx_{k\mu}^{-}\;P(\{m_{k}\},\{x_{k\mu}^{\pm}\}),\label{proba-law-1-1}
\end{equation}
and $\bar{G}^{AB}=P(\{m_{k}\}=0),$ with 
\begin{equation}
P(\{m_{k}\},\{x_{k\mu}^{\pm}\})=\prod_{k=1}^{AB}\left[\frac{1}{m_{k}!}\prod_{\mu=1}^{m_{k}}G_{k\mu}\right]\times W_{AB},\label{proba-law-2-1}
\end{equation}
where the expression $W_{AB}$ is a function of the variables 
\begin{equation}
x_{i}^{+}=1-\!\!\sum_{\underset{\pi(k)=i}{k=1}}^{AB}\sum_{\mu=1}^{m_{k}}\!x_{k\mu}^{+},\quad x_{j}^{-}=1-\!\!\sum_{\underset{\tau(k)=j}{k=1}}^{AB}\sum_{\mu=1}^{m_{k}}\!x_{k\mu}^{-},\label{W-2-2}
\end{equation}
 with $1\le i\le A$ and $1\le j\le B$. One finds {[}see Eq. (\ref{W-AB-1-1}){]}
\begin{align}
 & W_{AB}(\{x_{i}^{+}\},\!\{x_{j}^{-}\})=\label{W-AB-1-1-1}\\
 & \prod_{i=1}^{A}(x_{i}^{+})^{\alpha_{\mathrm{remn}}}\prod_{j=1}^{B}(x_{j}^{-})^{\alpha_{\mathrm{remn}}}\sum_{\{r_{Nk}\}}\Bigg\{\prod_{k=1}^{AB}\prod_{N=1}^{4}\frac{(-\alpha_{N})^{r_{Nk}}}{r_{Nk}!}\nonumber \\
 & \prod_{i=1}^{A}\Bigg[\prod_{\underset{\pi(k)=i}{k=1}}^{AB}\prod_{N=1}^{4}\left(\Gamma(\tilde{\beta}_{N})(x_{i}^{+})^{\tilde{\beta}_{N}}\right)^{r_{Nk}}g(\sum_{\underset{\pi(k)=i}{k=1}}^{AB}\sum_{N=1}^{4}r_{Nk}\tilde{\beta}_{N})\Bigg]\nonumber \\
 & \prod_{j=1}^{B}\Bigg[\prod_{\underset{\tau(k)=j}{k=1}}^{AB}\prod_{N=1}^{4}\left(\Gamma(\tilde{\beta}_{N})(x_{j}^{-})^{\tilde{\beta}_{N}}\right)^{r_{Nk}}g(\sum_{\underset{\tau(k)=j}{k=1}}^{AB}\sum_{N=1}^{4}r_{Nk}\tilde{\beta}_{N})\Bigg]\,\Bigg\},\nonumber 
\end{align}
where $\sum_{\{r_{Nk}\}}$ means summing all the indices $r_{Nk}$,
with $1\le N\le4$ and with $1\le k\le AB$, from zero to infinity, where
$r_{Nk}$ refers to the number of uncut ``Pomerons of type $N$''
of nucleon-nucleon pair $k$. It is useful for the discussion to consider
``Pomeron types'' $N$, although they are not physical objects,
just coming from the parametrization in Eq. (\ref{QCDpar}). I use $\tilde{\beta}_{N}=\beta_{N}+1$,
and a function $g$ defined as 
\begin{equation}
g(z)=\frac{\Gamma(1+\alpha_{\mathrm{remn}})}{\Gamma(1+\alpha_{\mathrm{remn}}+z)}.\label{g-definition-1}
\end{equation}
All the integrals could be done (see Appendix \ref{======= W-AB =======}).
One gets $\bar{G}^{AB}=W_{AB}(\{x_{i}^{+}=1\},\!\{x_{j}^{-}=1\}),$
and since $\bar{G}^{AB}(\{b_{A\!B}\})$ has to be non-negative, one
has the requirement
\begin{equation}
W_{AB}(\{x_{i}^{+}=1\},\!\{x_{j}^{-}=1\})\ge0.\label{positivity}
\end{equation}
At least for $A=B=1$ the numerical calculation of $W_{AB}$ can be
done. Although not written explicitly, this quantity depends on $b$.
For a large impact parameter $b$, one finds indeed $W_{11}(1,1)>0$,
but below $b=1\,$fm, one gets negative values, up to roughly $-0.4$
at $b=0$ (see Sec. \ref{-----The-sign-problem-----}). 
\begin{itemize}
\item
So here the
requirement Eq. (\ref{positivity}) is violated.
\item
The result one finds
is ``unreasonable'', it contradicts common sense. Let me refer to
this as the ``sign problem''. 
\end{itemize}
What does this mean? It seems that something
is missing. Here one needs again some input from a qualitative understanding
of high energy scattering. It is known that with increasing energy,
partons with very small momentum fractions $x\ll1$ become increasingly
important, the parton density becomes very large, and therefore the
linear DGLAP evolution scheme is not valid anymore. Nonlinear evolution
takes over, considering explicitly gluon-gluon fusion. These phenomena
are known as ``saturation'' \cite{Gribov:1983ivg,McLerran:1993ni,McLerran:1993ka,kov95,kov96,kov97,kov97a,jal97,jal97a,kov98,kra98,jal99a,jal99b,jal99}.
The diagrams for each scattering actually look more like the one shown
in Fig. \ref{nonlinear-effects}. 
One expect ``nonlinear effects'', which means that two ladders which
evolve first independently and in parallel, finally fuse. And only
after that, the (linear) DGLAP evolution might be realized. So ideally
one should generalize the framework, in addition to simple exchanges
of Pomerons, on should include ``Pomeron fusions'', i.e. triple Pomeron
graphs, in an iterative fashion up to infinite order. This ``more realistic
scenario'' should solve the sign problem. But to do that in a framework
with energy-momentum sharing (considered to be crucial) seems impossible.

\begin{figure}[h]
\centering{}\includegraphics[scale=0.22]
{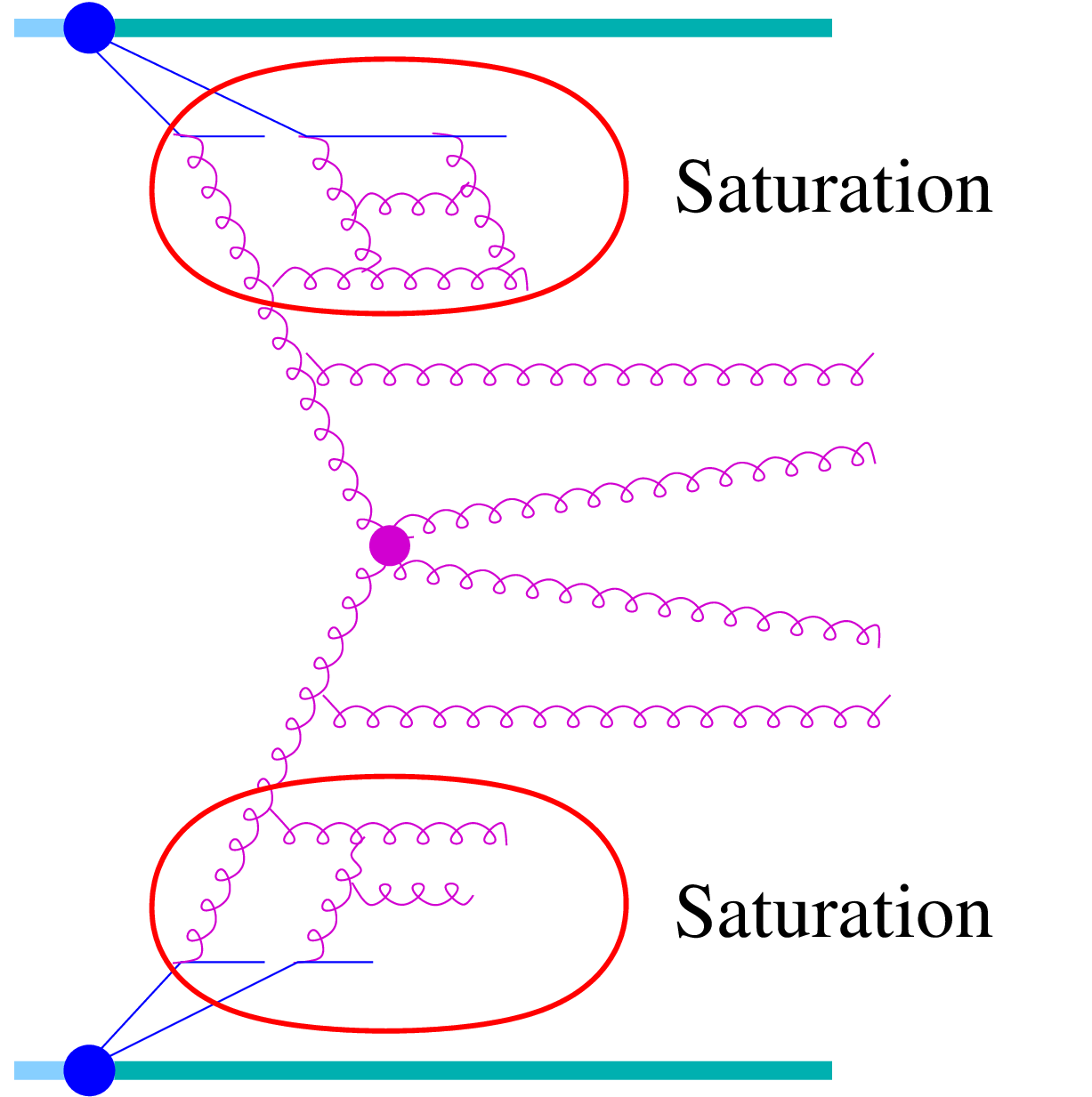}
\caption{Nonlinear effects (inside the red ellipses) are ``summarized'' in
the form of saturation scales. \label{saturation-one-pom}}
\end{figure}

\begin{figure}[h]
\centering{}\includegraphics[scale=0.22]
{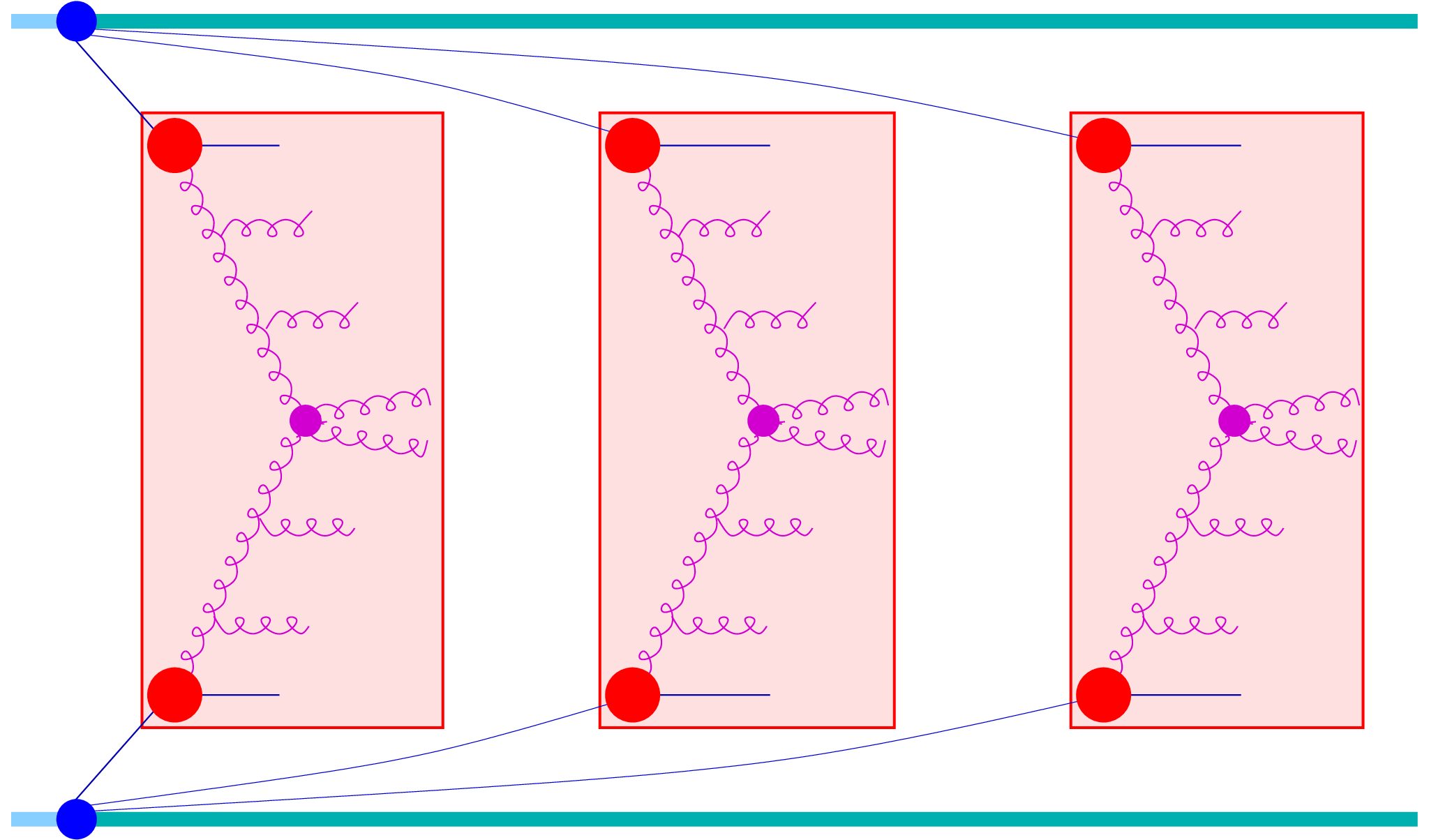}
\caption{Modular structure of multiple scattering, including saturation scales,
indicated by the big red dots. \label{modular structure-1}}
\end{figure}

Here comes another important fact: nonlinear effects lead to strong
destructive interference, which may be summarized in terms of a saturation
scale \cite{McLerran:1993ni,McLerran:1993ka}. So one might think
of treating these saturation phenomena not explicitly, but by
introducing saturation scales as the lower limit of the virtualities
for the DGLAP evolutions, as sketched in Fig. \ref{saturation-one-pom}.
So the diagrams inside the red ellipses are replaced by two scales
$Q_{\mathrm{sat,proj}}^{2}$ and $Q_{\mathrm{sat,targ}}^{2}$. Based
on these considerations, one might think of keeping the modular structure
as presented above, but introduce saturation via saturation scales,
as sketched in Fig. \ref{modular structure-1},
where one still uses objects called Pomerons as basic modules, but
the intrinsic structure of these modules takes into account saturation
scales. 

So far I have discussed the idea, in the following I will discuss how to develop the
corresponding formalism. But before doing so, there is another important
element to be discussed, and this is related to binary scaling, which
states that the inclusive cross section concerning the production of hard
probes (high $p_{t}$ or heavy flavor particles) in $A\!+\!B$ scattering
is $AB$ times the corresponding $pp$ cross section. This is shown
experimentally, and anyway it is difficult to imagine how it could be different.
Based on the formalism discussed so far (the one which leads to the
sign problem), one can prove that binary scaling holds {[}see Eq. (\ref{binary-scaling-1}){]}.
One can as well prove the AGK theorem, saying that the inclusive $A\!+\!B$
cross section is $AB$ times the single Pomeron contribution, being
an even stronger statement {[}see Eq. (\ref{AGK cancellations}){]}.
These proofs are based on a probability interpretation (like $\big\{1-W_{11}(1,1)\big\}$
being the probability of an interaction at given impact parameter
$b$). But this makes no sense if $W_{11}(1,1)$ has negative values. 

One postulates that the qualitative picture discussed above (and therefore
a ``realistic'' description) can be achieved by the following:
\begin{description}
\item [{(H6a)}] Solve the sign problem by modifying $g$ in Eq. (\ref{W-AB-1-1-1})
such that one gets rigorously non-negative results for whatever choice
of parameters, without changing the large-$b$ behavior (this will
be called ``regularization'').
\item [{(H7a)}] Introduce saturation scales when connecting $G$
(used in the multiple scattering formalism) with $G_{\mathrm{QCD}}$
(the corresponding QCD expression) such that for hard processes, binary
scaling and the AGK theorem are recovered {[}their very significant
violation is an unavoidable consequence of H6a{]}. 
The validity of AGK (proven for the case without energy sharing) is a strong expectation
which significantly guides the proposed hypotheses.
\end{description}
Concerning H6a, one finds that the infinite sums in Eq. (\ref{W-AB-1-1-1})
are finally simply a product of exponentials {[}see Eq. (\ref{W-AB-7}){]}
and therefore definitely always non-negative, leaving the large-$b$
behavior unchanged as required in H6a, if one postulates the following:
\begin{description}
\item [{(H6b)}] For given coefficients $\tilde{\beta}_{\lambda}$ (arbitrary,
but in practice of order unity) and three parameters $c_{\mu}$, the
$g$ functions have to be modified as (referred to as ``$g$-factorization'')
\begin{equation}
g\left(\sum_{\lambda}\tilde{\beta}_{\lambda}\right)\mapsto c_{1}\prod_{\lambda}c_{2}\,g(c_{3}\,\tilde{\beta}_{\lambda}).\label{g-property-1-1}
\end{equation}
\end{description}
As shown in Figs. \ref{gfunction-1} and \ref{gfunction-1-1},
this $g$-factorization looks like a reasonable approximation, and for
large $b$ it does not change results (concerning $W_{AB}$ for example).
But for small $b$ this modification is crucial; it changes results
from ``nonphysical'' to ``physical'' ($W_{AB}$ always non-negative).
It is a fundamental change, forcing the expression to become ``physical''
(to mimic the elements not treated explicitly.

\begin{figure}[h]
\centering{}\includegraphics[bb=50bp 100bp 700bp 505bp,clip,scale=0.3]
{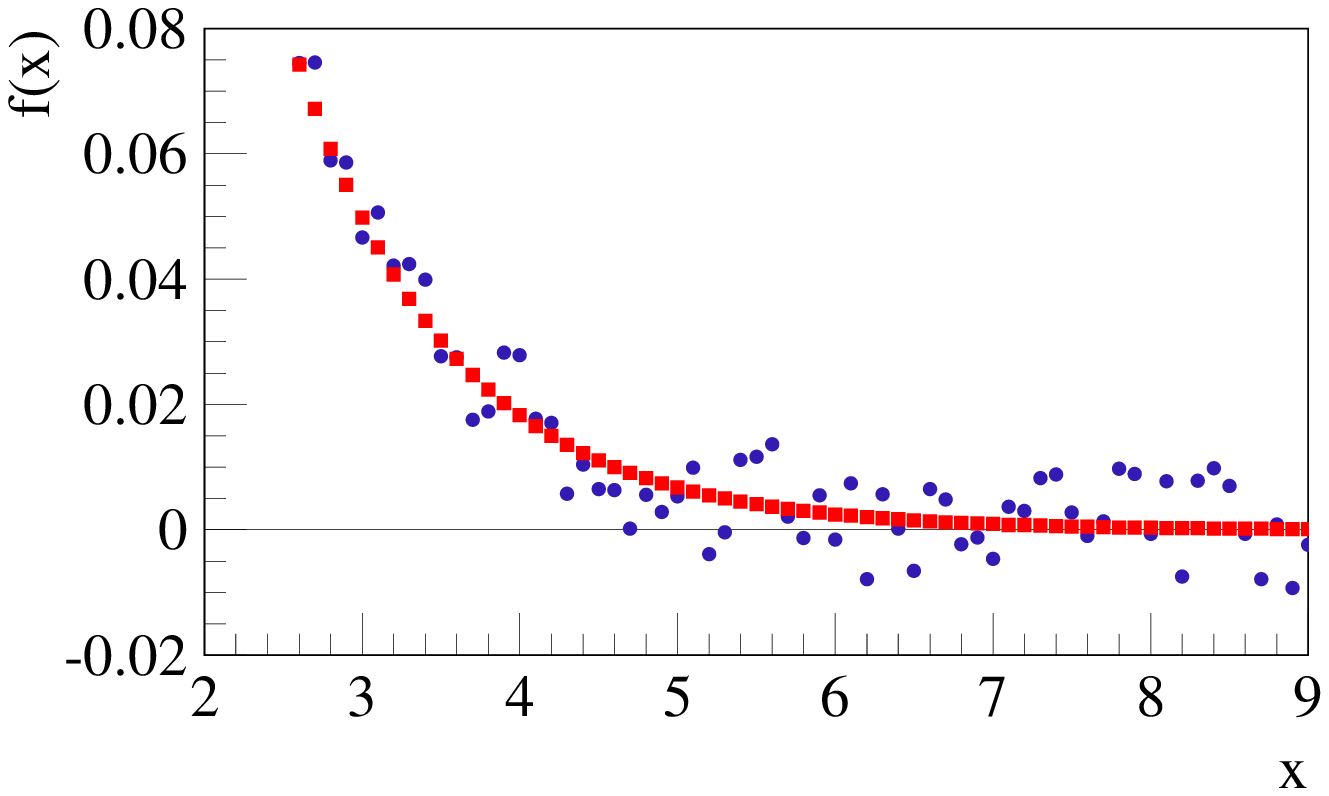}
\caption{Exponential function (red squares) and ``almost exponential'' function
(blue circles), see text.\label{exponential}}
\end{figure}

This is clearly an ad hoc solution. But let me discuss some analog
to better understand what is done: consider the power series $f(x)=\sum_{i=0}^{\infty}a_{i}\frac{1}{i!}(-x)^{i}$.
Choosing $a_{i}=1$, one gets the well-behaved exponential function,
shown in Fig. \ref{exponential}
as red squares. If one changes the values $a_{i}=1$ randomly by 1\%
(let me call the corresponding function $f^{(1\%)}$, one gets the result
shown as blue circles in Fig. \ref{exponential}: the values strongly
fluctuate, and beyond $x=5$ one gets also negative ones. So even small
deviations of the coefficients from the ``ideal ones'' (namely, $\frac{1}{i!}$)
lead to conceptual changes, namely, negative values where the ideal
case gives rigorously positive values. Or, the other way around, if
one takes the strongly fluctuating function $f^{(1\%)}$, and makes
a small (1\%) change of the coefficients (back to $a_{i}=1$), then
one gets the perfectly smooth and positive exponential function. One
might see Eq. (\ref{W-AB-1-1-1}) similarly: one has an ``almost exponential''
function (with negative arguments as well), and -- not surprisingly
-- one gets negative results. And a small correction (H6a) makes
it well behaved. This is why I use the term ``regularization''.

But anyway, this regularization is not unique, and this should
be further investigated. Already in Eq. (\ref{g-property-1-1}) one
has three parameters. The present choice is one ``which works'', but
one might consider an optimized set of parameters which minimizes
the difference between the factorized expression and the original
$g$. One might also simply linearize $g$ as $g(z)\mapsto g(0)+g'(0)\times z$
and then use $g(0)\exp(g'(0)/g(0)\times z)$ which ``optimizes'' $g$ at small
$z$. And then one should see to what extent the final results are
affected by theses different choices.

The regularization saves the probabilistic interpretation, or, to be
more precise, expressions, which must represent probabilities (as,
for example, \{$1-W_{11}(1,1)\}$) behave properly (values between
0 and 1, correct normalization). One has probability laws allowing
to generate all possible event configurations and compute all kinds
of distributions, all based on the assumption ``$G$ is equal to $G_{\mathrm{QCD}}$'',
the latter being the QCD expression, discussed in detail in Ref. \cite{werner:2023-epos4-heavy}
(this is a preliminary assumption, to be changed later). One finds
quickly that binary scaling and more generally the AGK theorem are
badly violated, and this can be traced back to the so-called deformation
function $R_{\mathrm{deform}}(x^{+},x^{-})$ being different from
unity in the case of event classes of high event activity. Deformation
functions are given as the inclusive distribution of Pomeron light-cone momentum
fractions considering particular event classes (EC), $\Big\{\frac{1}{\sigma_{\mathrm{incl}}^{EC}}\,\frac{d\sigma_{\mathrm{incl}}^{EC}}{dx^{+}dx^{-}}\Big\}$,
divided by the corresponding expression for an event class with only isolated Pomerons. Having understood the origin of the problem, one parametrizes
it; i.e., one parametrizes $R_{\mathrm{deform}}(x^{+},x^{-})$ as
a function of a variable called the ``connection number'', $N_{\mathrm{conn}}=\frac{N_{\mathrm{P}}+N_{\mathrm{T}}}{2},$
with $N_{\mathrm{P}}$ being the number of Pomerons connected to \emph{i,
}and with $N_{\mathrm{T}}$ being the number of Pomerons connected
to $j$, where $i$ and $j$ are the projectile and target nucleons
a given Pomeron is connected to (see Sec. \pageref{======= AGK-violation-deformation =======}). 

One understands that because of energy-momentum sharing, $R_{\mathrm{deform}}(x^{+},x^{-})$
is unavoidably different from unity (actually $<1$ at large $x^{+}x^{-}$),
and therefore the AGK theorem and binary scaling are unavoidably violated.
There is only one way out: giving up $G$$=G_{\mathrm{QCD}}$, which
was thought to be the natural choice. In addition, one needs to
take care of H7a, namely, incorporation of saturation scales. Putting
all the pieces together, in particular the quantitative understanding of
why AGK and binary scaling fail, and how this is related to the deformation
function, it is almost mandatory to postulate the following:
\begin{description}
\item [{(H7b)}] The $G$ of the multiple scattering formalism and $G_{\mathrm{QCD}}$
are related as 
\begin{equation}
G(x^{+},x^{-})=\frac{n}{R_{\mathrm{deform}}(x^{+},x^{-})}G_{\mathrm{QCD}}(Q_{\mathrm{sat}}^{2},x^{+},x^{-}),\label{fundamental-epos4-equation-2}
\end{equation}
such that \textbf{$G$ }itself does not depend on the environment
(the latter represented via the connection number).
\end{description}
This means that the deformation does not affect $G$ but it changes (very
much) the saturation scale $Q_{\mathrm{sat}}^{2},$ and since the saturation
scale affects eventually low-$p_{t}$ particle production, the high
$p_{t}$ behavior will not be affected, and one recovers binary scaling
at high $p_{t}$, as it should be. Going through the proofs in Sec.
\ref{======= the solution =======},
the choice of (H7b) seems mandatory,
\begin{description}
\item [{(H7c)}] based on the expectation that AGK must hold [leading to factorization ($pp$) and binary scaling ($A\!+\!B$) at large $p_t$].
\end{description}
But still, H7b  and H7c are  assumptions. \\

Let me close this section with an important remark. One should clearly
distinguish between ``multiple scattering formalism'' and ``Monte
Carlo implementation''. A very important aspect of this paper is
the attempt to provide a formalism which is 100\% compatible with
its Monte Carlo implementation. This sounds obvious, but this is in
general not the case. If energy conservation is enforced in the code
(via if statements), then this should have a correspondence in the
theoretical formulas. It is easy to add an ``if'' in the code, but
very difficult to have the corresponding feature in the theoretical
formalism. Concerning all aspects of the current work, the model is
completely defined via formulas (finally cross sections expressed
in terms of multidimensional integrals), and the Monte Carlo implementation
is only a numerical procedure to solve mathematical equations.\\

To summarize this section: The present work tries to put together
theoretical knowledge on S-matrix theory, perturbative QCD, and saturation,
together with a large amount of phenomenological (or even qualitative)
understanding of experimental data (in particular at the RHIC and
LHC), to construct a multiple scattering formalism, which
goes beyond factorization, and which allows to address particle production
per event classes (a major activity at the LHC for the next decade).
Multiple scatterings must happen in parallel, there is no question
about that, but the big assumption is the hypothesis that saturation
can be incorporated via saturation scales, without any need to split
Pomerons, so one can keep the picture of multiple parallel Pomerons.
But the fact that saturation is not treated explicitly clearly requires
some corrections at some stage. Doing cross section calculations, one
eventually finds expressions, usually interpreted as probabilities,
which are negative at small impact parameters (one kind of exceeds
the black disk limit). Being convinced that these ``probability-like''
expressions MUST be positive, one forces them to be so. One provides a
method which sounds reasonable and which works, but the solution 
is not unique; one can even propose other possible solutions, and this should be explored further.
The final step is the redefinition of the relation between $G$ and
$G_{\mathrm{QCD}}$ by incorporating saturation scales. One proposes
Eq. (\ref{fundamental-epos4-equation-2}), and from very detailed
discussions about the origin of the violation of binary scaling, and
the discussion of why and how this proposal solves the problem, it looks
like this choice is mandatory. At the end one has constructed
a formalism that incorporates perturbative QCD (pQCD) and saturation, 
does parallel
scattering in $pp$ and $AA$ scattering appropriately, can do analysis
of particle production per event class, and is perfectly compatible
with the so-called QCD Monte Carlo generators (with fewer hard processes
implemented though).
\vspace{1cm}

\section{Parallel scattering \label{======= parallel-scattering =======}}

\begin{figure}[h]
\centering{}{\Large{}\includegraphics[scale=0.27]
{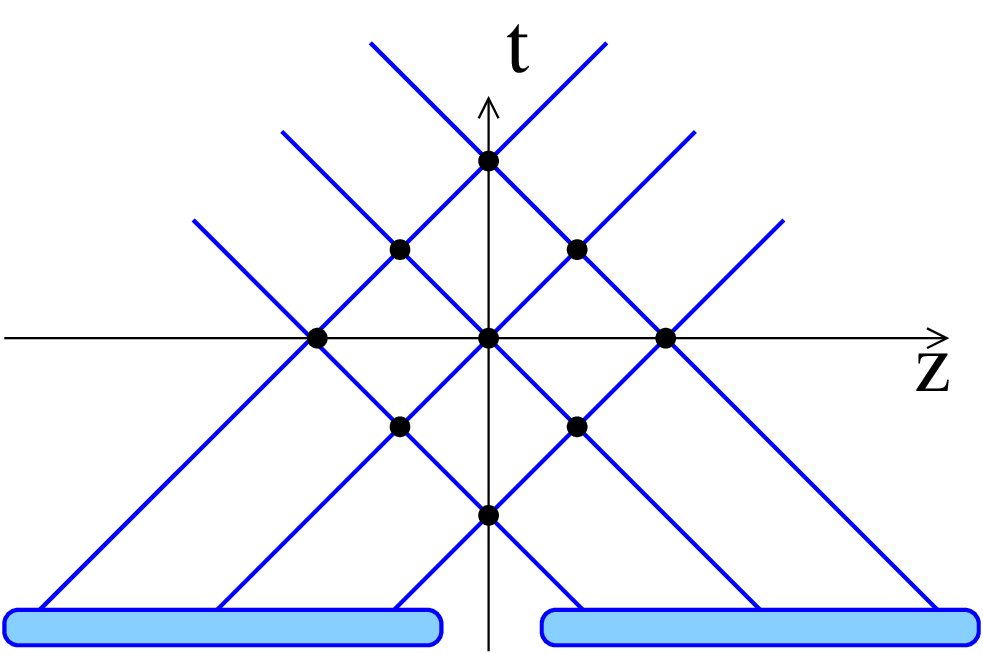}}
\caption{Sketch of a nucleus-nucleus collision in space-time. Here, $t$ is
the time and $z$ the longitudinal coordinate, and the blue lines represent
the nucleons. \label{space time}}
\end{figure}

\begin{figure}[h]
\centering{}{\Large{}\includegraphics[scale=0.27]
{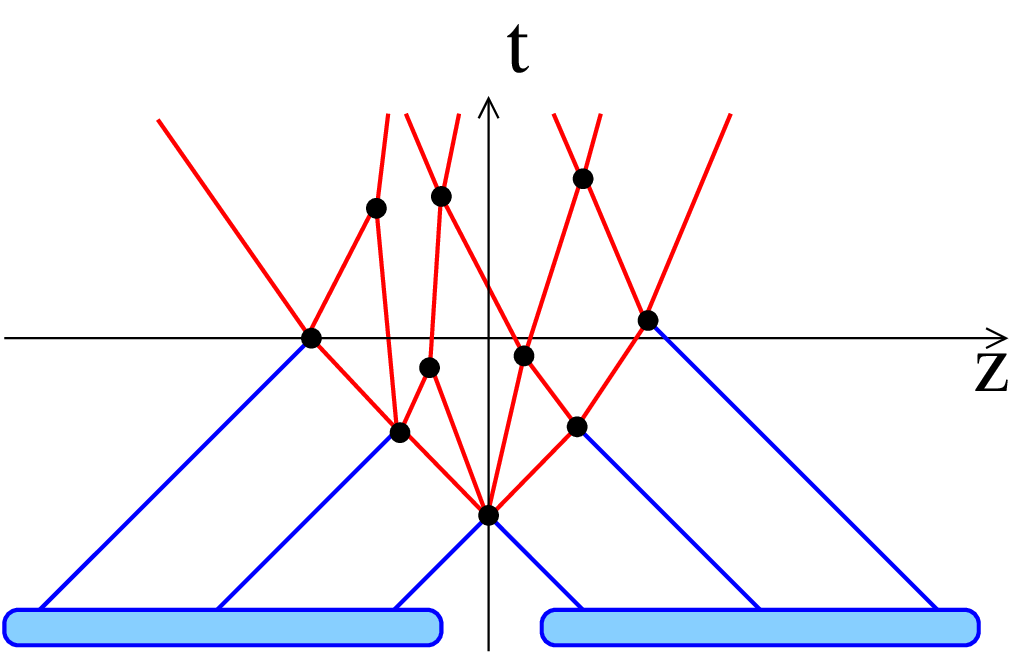}}
\caption{Sketch of a nucleus-nucleus collision at very low energies, with sequential
scatterings. \label{space time 2-1}}
\end{figure}

\begin{figure}[h]
\centering{}{\Large{}\includegraphics[scale=0.27]
{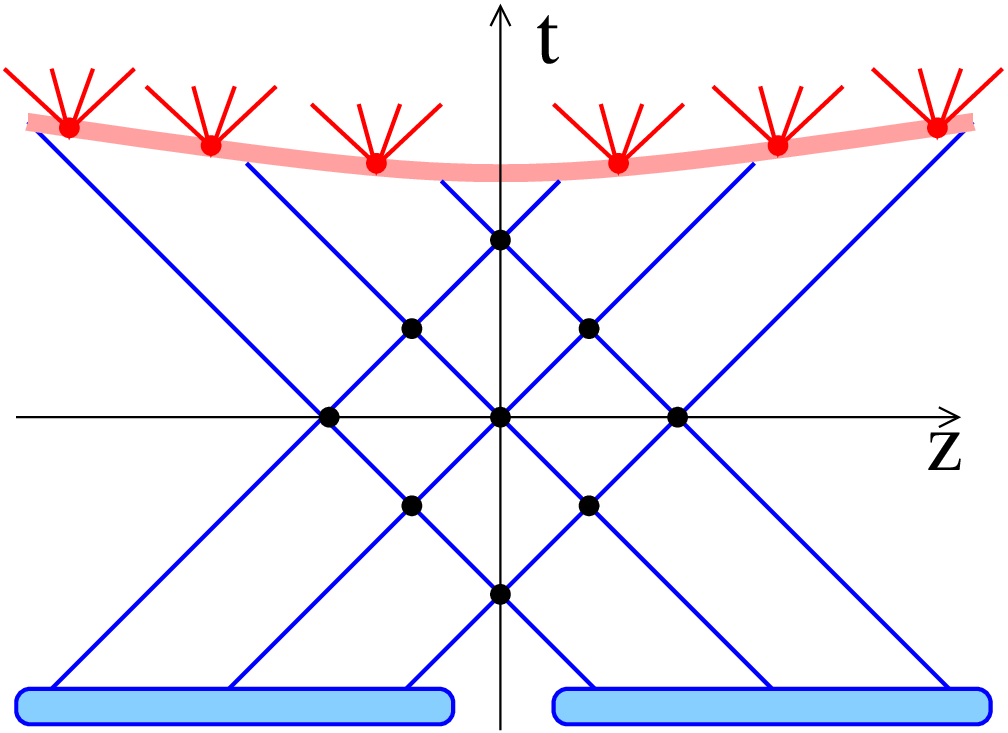}}
\caption{Sketch of a nucleus-nucleus collision at very high energies, with
all nucleon-nucleon scatterings happening in parallel, before
particle production starts. \label{space time 2}}
\end{figure}

\begin{figure}[h]
\centering{}{\Large{}\includegraphics[scale=0.27]
{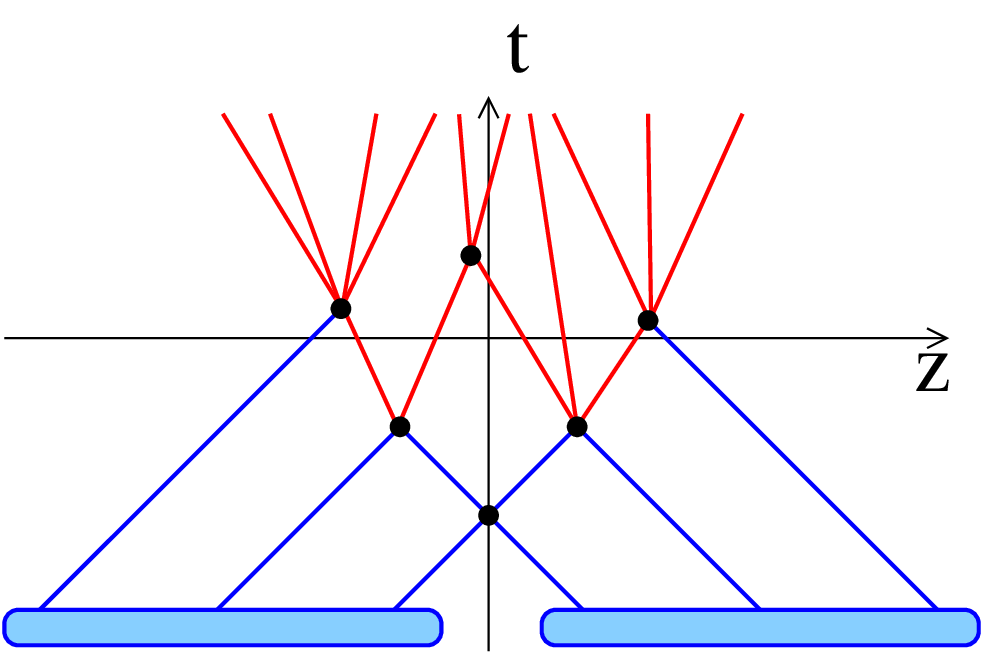}}
\caption{Sketch of a nucleus-nucleus collision at intermediate energies, with
only some (but not all) $NN$ scatterings being realized, before the particle
production starts. \label{space time 3}}
\end{figure}

One of the most important aspects of the formalism developed in this paper is the fact that collisions must happen in parallel -- at high energies. In the following the energy dependence of this aspect will be studied. 
Let me consider a central nucleus-nucleus collision in space-time
in its center-of-mass system, as sketched in Fig. \ref{space time}.
The points (intersections between projectile and target nucleon trajectories,
assuming that they are close enough in the transverse direction) represent
possible interactions. But what precisely happens depends on the relations
between three crucial time scales: \textbf{
\begin{equation}
\tau_{\mathrm{collision}}=\frac{2R}{\gamma v}\,,
\end{equation}
}which is the duration of the $AA$ collision\textbf{,
\begin{equation}
\tau_{\mathrm{interaction}}=\frac{2R}{n\gamma v}\,,
\end{equation}
}which is the time between two $NN$ interactions, and\textbf{
\begin{equation}
\tau_{\mathrm{form}}=\gamma_{\mathrm{hadron}}\tau_{\mathrm{form,\,CMS}}
\end{equation}
}which is the particle (hadron) formation time, after an interaction
of two nucleons, with $R$ being the nuclear radius, $v$ the (average)
velocity of the nucleons, $\gamma$ the corresponding Lorentz gamma factor, $n$
the (average) number of nucleons in a row, 
$\tau_{\mathrm{form,\,CMS}}$ the (average) formation time of produced particles in their rest frames, 
and $\gamma_{\mathrm{hadron}}$ their (average) gamma factor.
The collision energy
enters essentially because of the gamma factors in the denominators
of $\tau_{\mathrm{collision}}$ and $\tau_{\mathrm{interaction}}$,
which leads to strong Lorentz contractions at high energies. 
At very low energies, defined by 
\begin{equation}
\tau_{\mathrm{form}}<\tau_{\mathrm{interaction}}\:,
\end{equation}
after each collision, the particles are formed before the next interaction
happens, so one has sequential scatterings (a cascade), as sketched
in Fig. \ref{space time 2-1}.
The blue lines are the trajectories of projectile and target nucleons,
and the red lines are produced hadrons. 
However, at very high energies, defined by 
\begin{equation}
\tau_{\mathrm{collision}}<\tau_{\mathrm{form}},
\end{equation}
one has completely ``parallel scatterings''; all nucleon-nucleon
scatterings are realized before particle production starts, as sketched
in Fig. \ref{space time 2}.
The blue lines are the trajectories of projectile and target nucleons,
the red lines are produced hadrons. 
At TeV energies, the longitudinal dimensions of the nuclei are smaller
than $0.01\,$fm/c, so the overlap area is essentially pointlike. 
In the intermediate energy range, defined by 
\begin{equation}
\tau_{\mathrm{interaction}}<\tau_{\mathrm{form}}<\tau_{\mathrm{collision}},
\end{equation}
one needs a \textquotedblleft partially parallel approach\textquotedblright :
several (but not all) $NN$ scatterings are realized, before the particle
production starts, as sketched in Fig. \ref{space time 3}.
The blue lines are the trajectories of projectile and target nucleons;
the red lines are produced hadrons. 

Let me define the low and high energy (per nucleon-nucleon pair) thresholds
$E_{\mathrm{HE}}$ by the identity $\tau_{\mathrm{collision}}=\tau_{\mathrm{form}}$
and $E_{\mathrm{LE}}$ by the identity $\tau_{\mathrm{form}}=\tau_{\mathrm{interaction}}$.
The high energy value $E_{\mathrm{HE}}$ is then obtained from $\frac{2R}{\gamma v}=\tau_{\mathrm{form}}$.
For a rough numerical estimate, I consider central rapidity
hadrons ($\gamma_{\mathrm{hadron}}=1$) and $\tau_{\mathrm{form}}=1\,$fm/c,
and a big nucleus with $R=6.5\,$fm, which gives $\gamma\frac{v}{c}=13$,
so one gets 
\begin{equation}
E_{\mathrm{HE}}\approx24\,\mathrm{GeV}.
\end{equation}
 The low energy value $E_{\mathrm{LE}}$ is given as $\tau_{\mathrm{form}}=\frac{2R}{n\gamma v}$,
and using $n=7$, one gets $\gamma\frac{v}{c}=\frac{2R}{nc\tau_{\mathrm{form}}}\approx\frac{13}{7}$,
which gives\textbf{\textcolor{blue}{\Large{}}}%
\begin{comment}
$n=2R\sigma_{pp}\rho_{\mathrm{nucl}}$=13{*}0.14{*}4.0=7.28

a=13/7.; b\textasciicircum\{2\}/(1-b\textasciicircum\{2\})=a\textasciicircum\{2\};
b\textasciicircum\{2\}=a\textasciicircum\{2\}-a\textasciicircum\{2\}b\textasciicircum\{2\};

b=sqrt(a{*}{*}2/(1+a{*}{*}2)); g=1/sqrt(1-b{*}{*}2); E=g{*}0.94; E=1.98
\end{comment}
\begin{equation}
E_{\mathrm{LE}}\approx4\,\mathrm{GeV}.
\end{equation}
To conclude this part: For energies (in the sense of $E=\sqrt{s_{NN}}$ ) beyond $E_{\mathrm{HE}}\approx24\,\mathrm{GeV}$
one has to employ a ``parallel scattering approach'', for energies
below $E_{\mathrm{LE}}\approx4\,\mathrm{GeV}$ a pure hadron cascade
is appropriate, and in between one needs a ``partially parallel scattering
approach''. The estimates for $E_{\mathrm{LE}}$ and $E_{\mathrm{HE}}$
are conservative in the sense that considering $\gamma_{\mathrm{hadron}}>1$
will lead to even smaller values.\\

\begin{figure}[h]
\centering{}\includegraphics[scale=0.25]
{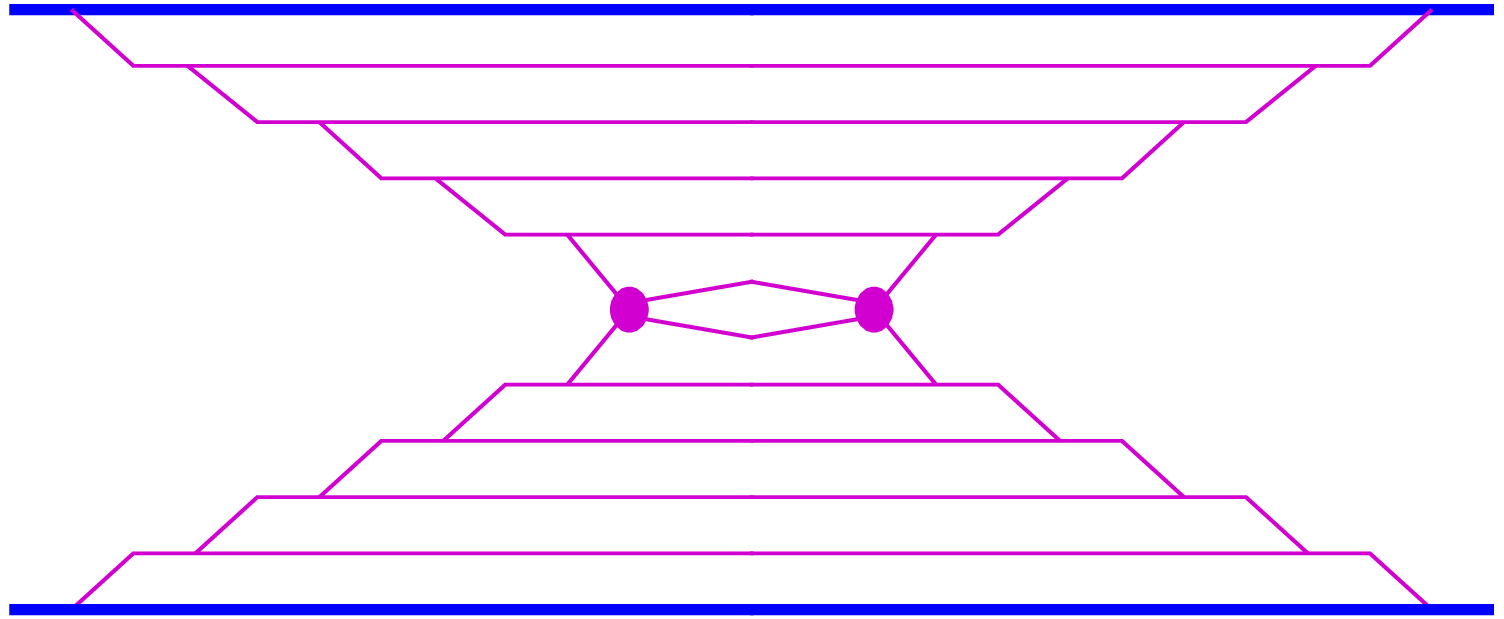}
\caption{Sketch of a nucleon-nucleon scattering, with successive parton emissions
before the hard process in the middle. \label{single pomeron}}
\end{figure}

\begin{figure}[h]
\centering{}\includegraphics[scale=0.22]
{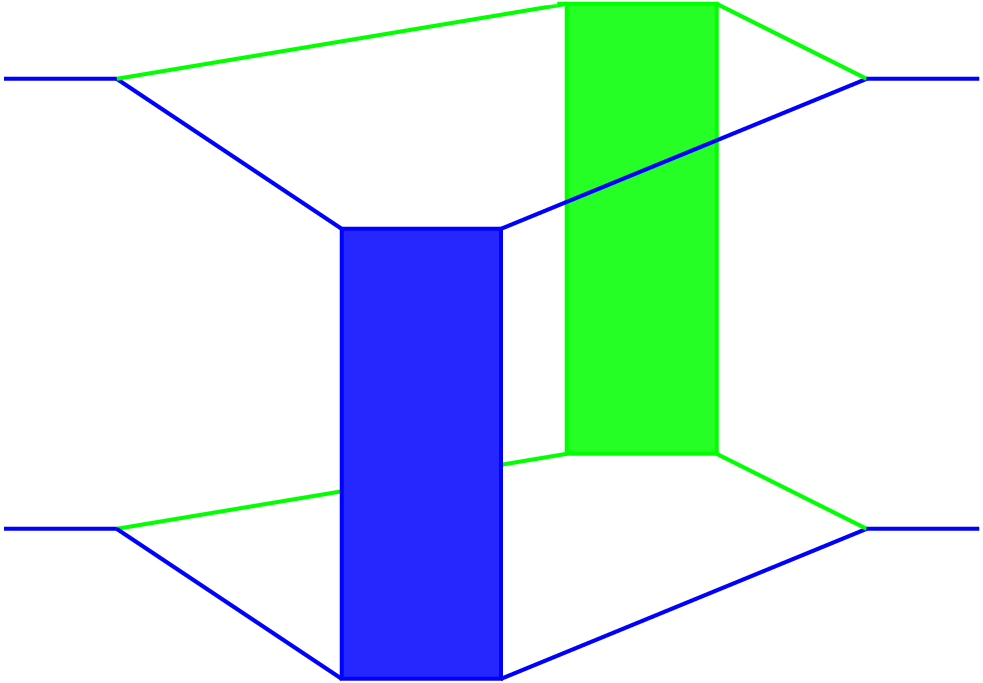}
\caption{Sketch of a parallel scattering in nucleon-nucleon scattering. \label{parallel scattering}}
\end{figure}

In the case of nucleon-nucleon scattering, multiple parton scattering
occurs with the corresponding weight increasing with energy. Here
again, the $\gamma$ factor plays a crucial role, forcing multiple scatterings
to happen in parallel. Looking at a single scattering, I sketch
in Fig. \ref{single pomeron}
a typical example of a hard scattering in the middle of the diagram,
preceded by successive parton emission from both sides. In particular,
the first emissions lead to very fast partons, with large $\gamma$ factors,
and therefore long lifetimes $\gamma\tau$, with $\tau$ of the order
of $1\,\mathrm{fm}/c$. This means that at high energies with $\gamma\gg1$,
the duration of the scattering is very large, which makes it impossible
to have several such collisions one after the other. They must
occur in parallel; one needs a structure as shown in Fig. \ref{parallel scattering}.
In principle, this is relevant for energies beyond several tens of
GeV, but really important at TeV energies, with a multiple scattering
weight being high. \\

\begin{figure}[h]
\centering{}\includegraphics[scale=0.37]
{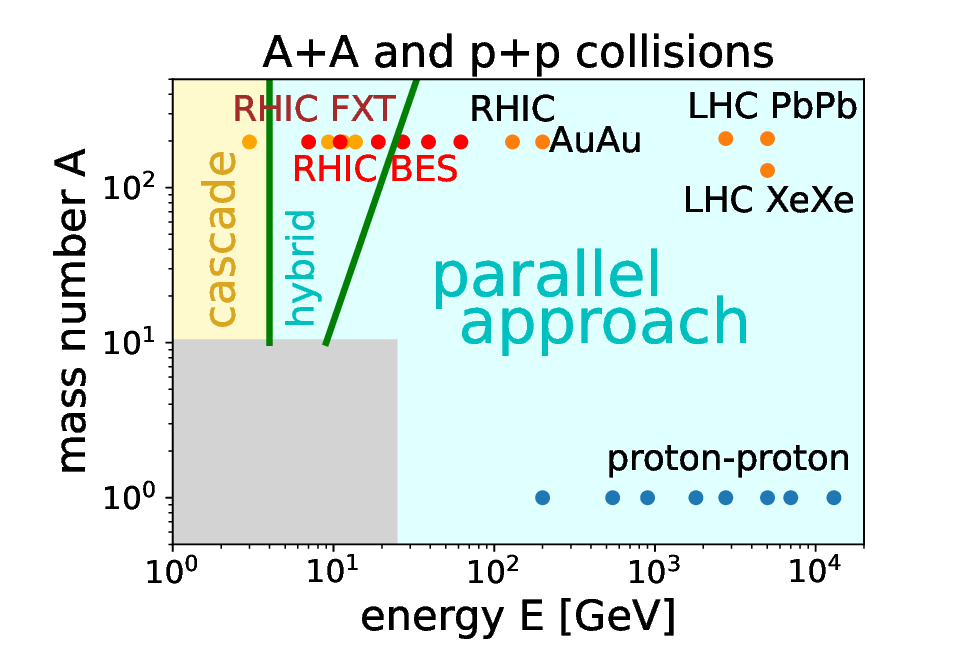}
\caption{Validity of the parallel scattering scenario. The "hybrid" area refers to a partially parallel scenario. \label{parallel scattering-1}}
\end{figure}

From the above discussion, only based on elementary arguments based
on time scales, I  conclude that a parallel scattering scenario
is mandatory, for $pp$ and $AA$ scattering, beyond center-of-mass energies
of several tens of GeV (see Fig. \ref{parallel scattering-1}).
So not only LHC energies are concerned, but even the energies of the RHIC Beam Energy Scan (BES) program. 

The appropriate framework to treat strictly parallel scatterings is
S-matrix theory, and before coming to the multiple scattering approach,
I will first recall some basic facts about the S-matrix.

\section{S-matrix theory: basic definitions, assumptions, facts \label{======= S-matrix-theory-basic-definitions =======}}

\begin{figure}[h]
\begin{centering}
(a)$\qquad\qquad\qquad\qquad\qquad$\\
\includegraphics[scale=0.33]
{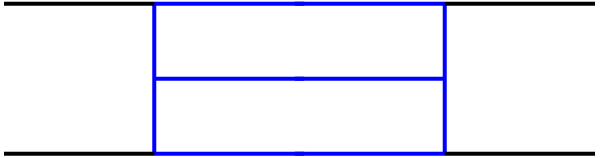}
\par\end{centering}
\begin{centering}
(b)$\qquad\qquad\qquad\qquad\qquad\qquad\qquad\qquad$\vspace{-0.4cm}
\\
\includegraphics[scale=0.33]
{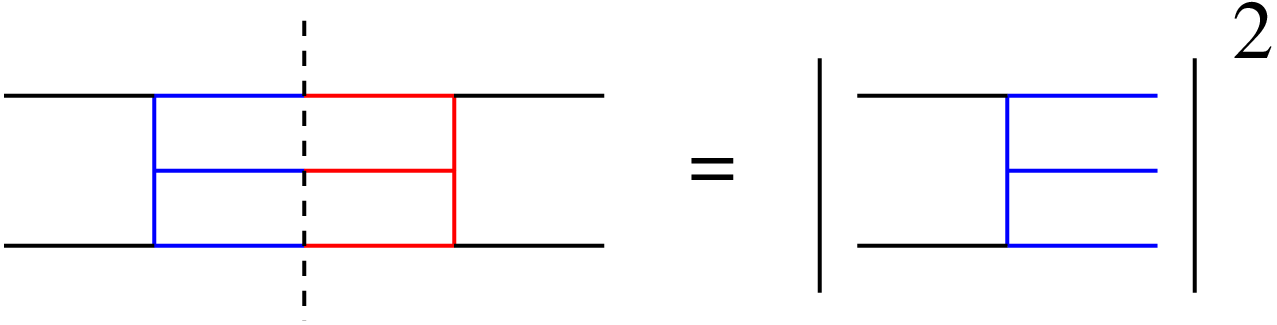}
\par\end{centering}
\centering{}\caption{(a) Elastic scattering diagram and (b) the corresponding cut diagram.
\label{cut-diagram}}
\end{figure}

Let me recall some basic facts about S-matric theory, needed to understand
the paper. %
\begin{comment}
The scattering operator $\hat{S}$ is defined via
\begin{equation}
\left|\psi(t=+\infty\right\rangle =\hat{S}\left|\psi(t=-\infty\right\rangle 
\end{equation}
\end{comment}
The S-matrix is the representation of the scattering operator $\hat{S}$,
i.e. $\boldsymbol{\mathrm{S}}_{ab}=\left\langle a\right|\hat{S}\left|b\right\rangle $
for basis states $\left|a\right\rangle $ and $\left|b\right\rangle .$
The T-matrix is defined as 
\begin{equation}
\boldsymbol{\mathrm{S}}_{fi}\!=\!\delta_{fi}+i(2\pi)^{4}\delta(p_{f}-p_{i})\boldsymbol{\mathrm{T}}_{fi}\,,
\end{equation}
where $i$ and $f$ refer to initial and final state, respectively, 
and $p_i$ and $p_f$ are the corresponding momentum four-vectors.    
The operator $\hat{S}$ must be unitary, $\hat{S}^{\dagger}\hat{S}=1$,
which means that the scattering does not change the normalization
of a state. One adds three ``very plausible'' hypotheses:
\begin{itemize}
\item $\boldsymbol{\mathrm{T}}_{ii}$ is Lorentz invariant \\
($\to$ use Mandelstam variables $s$, $t$)
\item $\boldsymbol{\mathrm{T}}_{ii}(s,t)$ is an analytic function of $s$,
with $s$ considered to be a complex variable (Hermitian analyticity)
\item $\boldsymbol{\mathrm{T}}_{ii}(s,t)$ is real on some part of the real
axis 
\end{itemize}
Using the Schwarz reflection principle (a theorem), $\boldsymbol{\mathrm{T}}_{ii}(s,t)$
first defined for $\mathrm{Im}s\geq0$ can be continued in a unique
fashion via $\boldsymbol{\mathrm{T}}_{ii}(s^{*},t)=\boldsymbol{\mathrm{T}}_{ii}(s,t)^{*}$.
The cross section is given as a sum over final states,%
\begin{comment}
{\scriptsize{}$1=\sum_{f}S_{fi}^{*}S_{fi}$}{\scriptsize\par}

{\scriptsize{}$1=\sum_{f}\left\{ \delta_{fi}+i(2\pi)^{4}\delta(p_{f}-p_{i})\mathrm{\boldsymbol{T}}_{fi}\right\} ^{*}\left\{ \delta_{fi}+i(2\pi)^{4}\delta(p_{f}-p_{i})\mathrm{\boldsymbol{T}}_{fi}\right\} $}{\scriptsize\par}

{\scriptsize{}$1=\sum_{f}\left\{ \;\delta_{fi}+\delta_{fi}\left\{ i(2\pi)^{4}\delta(p_{f}-p_{i})\mathrm{\boldsymbol{T}}_{fi}\right\} -\left\{ i(2\pi)^{4}\delta(p_{f}-p_{i})\mathrm{\boldsymbol{T}}_{fi}^{*}\right\} \delta_{fi}-\left\{ i(2\pi)^{4}\delta(p_{f}-p_{i})\mathrm{\boldsymbol{T}}_{fi}^{*}\right\} \left\{ i(2\pi)^{4}\delta(p_{f}-p_{i})\mathrm{\boldsymbol{T}}_{fi}\right\} \;\right\} $}{\scriptsize\par}

{\scriptsize{}$0=\left\{ i(2\pi)^{4}\delta(p_{i}-p_{i})\mathrm{\boldsymbol{T}}_{fi}\right\} -\left\{ i(2\pi)^{4}\delta(p_{i}-p_{i})\mathrm{\boldsymbol{T}}_{fi}^{*}\right\} -\sum_{f}\left\{ i(2\pi)^{4}\delta(p_{i}-p_{i})\mathrm{\boldsymbol{T}}_{fi}^{*}\right\} \left\{ i(2\pi)^{4}\delta(p_{f}-p_{i})\mathrm{\boldsymbol{T}}_{fi}\right\} $}{\scriptsize\par}

{\scriptsize{}divide by $i(2\pi)^{4}\delta(0)$ : $0=\left\{ \mathrm{\boldsymbol{T}}_{fi}\right\} -\left\{ \mathrm{\boldsymbol{T}}_{fi}^{*}\right\} -\sum_{f}\left\{ \mathrm{\boldsymbol{T}}_{fi}^{*}\right\} \left\{ i(2\pi)^{4}\delta(p_{f}-p_{i})\mathrm{\boldsymbol{T}}_{fi}\right\} $}{\scriptsize\par}

{\scriptsize{}$(\mathrm{\boldsymbol{T}}_{fi}-\mathrm{\boldsymbol{T}}_{fi}^{*})/i=\sum_{f}(2\pi)^{4}\delta(p_{f}-p_{i})\mathrm{\boldsymbol{T}}_{fi}\mathrm{\boldsymbol{T}}_{fi}^{*}$}
\end{comment}
\begin{equation}
\sigma_{\mathrm{tot}}=\frac{1}{w}\sum_{f}|\boldsymbol{\mathrm{T}}_{fi}|^{2}\,(2\pi)^{4}\delta^{4}(p_{f}-p_{i}),
\end{equation}
with %
\begin{comment}
$w=2\sqrt{\left(s-\left(m_{1}+m_{2}\right)^{2}\right)\left(s-\left(m_{1}-m_{2}\right)^{2}\right)}\,\to2s\;\mathrm{for}\,s\to\infty,$
\end{comment}
{} $w=2s$ for large $s$. Using unitarity and the Schwarz reflection
principle, and using here and in the following $\boldsymbol{\mathrm{T}}=\boldsymbol{\mathrm{T}}_{ii}$,
one gets 
\begin{equation}
2s\,\sigma_{\mathrm{tot}}=\sum_{f}(2\pi)^{4}\delta(p_{f}-p_{i})|\boldsymbol{\mathrm{T}}_{fi}|^{2}=\mathrm{\frac{1}{i}disc}\,\boldsymbol{\mathrm{T}},\label{sigma-tot}
\end{equation}
with $\mathrm{disc}\,\boldsymbol{\mathrm{T}}=\boldsymbol{\mathrm{T}}(s+i\epsilon,t)-\boldsymbol{\mathrm{T}}(s-i\epsilon,t)$.
Interpretation: $\mathrm{\frac{1}{i}disc}\,\boldsymbol{\mathrm{T}}$
can be seen as a so-called ``cut diagram'' 
(I will also use $\mathrm{cut}\,\boldsymbol{\mathrm{T}}=\mathrm{\frac{1}{i}disc}\,\boldsymbol{\mathrm{T}}$)
with modified Feynman rules [complex conjugate expressions on the
right side of the cut, and ``intermediate particles'' on mass shell
via $2\pi\theta(p^{0})\delta(p^{2}-m^{2})$ ]; see Fig. \ref{cut-diagram},
where I show an elastic scattering diagram and the corresponding
cut diagram, the latter being equal to an inelastic amplitude squared.

For applications such as relativistic proton-proton and heavy-ion scattering,
one usually assumes purely transverse momentum transfer, i.e., $t=-p_{\bot}^{2}$,
which allows to define a Fourier transform with respect to $p_{\bot}^{2}$.
In addition, it is useful to divide by $2s$, so I define
the so-called ``impact parameter representation'' as
\begin{equation}
T(s,b)=\frac{1}{8\pi^{2}s}\int d^{2}p_{\bot}e^{-i\vec{p}_{\bot}\vec{b}}\boldsymbol{\mathrm{T}}(s,t),\label{T-definition}
\end{equation}
with $b=|\vec{b}|$.
I also define 
\begin{equation}
G(s,b)=\frac{1}{i}\mathrm{disc}T(s,b)=2\mathrm{Im}\,T(s,b)=\mathrm{cut}\,T(s,b).\label{G-definition}
\end{equation}
Having $t=0$ as the argument of $\boldsymbol{\mathrm{T}}$ in Eq. (\ref{sigma-tot}),
and using $\boldsymbol{\mathrm{T}}(s,0)=\int d^{2}b\,2s\,T(s,b)$
obtained from Eq. (\ref{T-definition}), one finds
\begin{equation}
\sigma_{\mathrm{tot}}
=\int d^{2}b\,\mathrm{cut}\,T(s,b)
=\int d^{2}b\,G(s,b)
,\label{sigma-tot-1}
\end{equation}
which allows a simple geometrical interpretation:
\begin{itemize}
\item One may interpret $G$ as the probability of an interaction at impact
parameter $b$, provided $G\le1$. 
\end{itemize}
\begin{table}[h]
\begin{tabular*}{1\columnwidth}{@{\extracolsep{\fill}}@{\extracolsep{\fill}}|c|l|}
\hline 
$\boldsymbol{\mathrm{T}}$  & %
\begin{minipage}[t]{0.85\columnwidth}%
Diagonal element of the elastic scattering T-matrix as defined in
standard quantum mechanics textbooks, where the asymptotic state is
a system of two protons or two nuclei%
\end{minipage}\tabularnewline
\hline 
$T$  & %
\begin{minipage}[t]{0.85\columnwidth}%
Fourier transform with respect to the transverse momentum exchange
of the elastic scattering T-matrix $\boldsymbol{\mathrm{T}}$, divided
by $2s$ (formulas are simpler using this representation)%
\end{minipage}\tabularnewline
\hline 
$G$  & %
\begin{minipage}[t]{0.85\columnwidth}%
Defined as $G=\mathrm{cut}\,T=2\mathrm{Im}T=\frac{1}{i}\mathrm{disc}T$
(where ``disc'' refers to the variable $s$), referring to the inelastic
process associated with the cut of the elastic diagram corresponding
to $T$%
\end{minipage}\tabularnewline
\hline 
\end{tabular*}\caption{The symbols $\boldsymbol{\mathrm{T}}$, $T$, and $G$. \label{the-four-symbols}}
\end{table}
Throughout the paper, I will systematically use the symbols
$\boldsymbol{\mathrm{T}}$, $T$, and $G$, as defined above. For
clarity, I recall the definitions in Table \ref{the-four-symbols}. 

Let me finally recall that cutting a complex diagram [corresponding
to $\mathrm{cut}\,T(s,b)$] amounts to summing over all possible cuts
(cutting rules \cite{Cutkosky60}). I will illustrate what this means
in the next section.

\section{AGK cancellations, factorization, and binary scaling in a simple scenario
without energy sharing \label{======= AGK-without energy sharing =======}}

To better understand the importance of AGK cancellations, and its
relation with factorization and binary scaling, I first discuss
a simple scenario, without energy sharing.

\subsection{A simple scenario}

\begin{figure}[h]
\begin{centering}
\includegraphics[scale=0.33]
{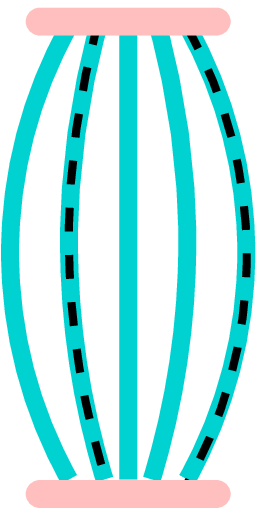}
\par\end{centering}
\centering{}\caption{Five Pomeron graph, with two Pomerons being cut.\label{cut-diagram-1}}
\end{figure}

Let me consider the original Gribov-Regge approach for multiple scattering,
following Refs. \cite{Gribov:1967vfb,Gribov:1968jf,GribovLipatov:1972}
and \cite{Abramovsky:1973fm}. For $pp$ scattering, the T-matrix is
given in terms of products of sub-T-matrices, 

\begin{equation}
iT=\sum_{n=0}^{\infty}\,\frac{1}{n!}\,\left\{ iT_{\mathrm{Pom}}\right\} ^{n}\,,\label{multiple-tmatrix-pp-1-1}
\end{equation}
with $T_{\mathrm{Pom}}$ referring to an elementary scattering between
``proton constituents'' by the exchange of a ``Pomeron'' (whatever
this may be). Originally, lacking an underlying microscopic theory,
one simply used asymptotic expressions (limit $s/t\to\infty$) of
the form $\boldsymbol{\mathrm{T}}_{\mathrm{Pom}}(s,t)=As^{B+Ct}$
with parameters $A$, $B$, and $C$ (see Appendix \ref{======= asymptotic-behavior =======}),
which allows to easily compute $T_{\mathrm{Pom}}(s,b)$ according
to Eq. (\ref{T-definition}). %
\begin{comment}
\[
T(s,b)=\frac{1}{2s}\,\frac{As^{B}}{4\pi^{2}}\int d^{2}p_{\bot}e^{-i\vec{p}_{\bot}\vec{b}}e^{-\ln s\,C\,p_{\bot}^{2}}=\frac{As^{B}}{2s}\,\frac{1}{4\pi\ln s\,C}\exp\left(-\frac{b^{2}}{4\ln s\,C}\right).
\]
\[
T(s,b)=\frac{As^{B}}{2s}\,\frac{1}{4\pi R^{2}}\exp\left(-\frac{b^{2}}{4R^{2}}\right).
\]
(using $t=-p_{\bot}^{2}$, Eq. (\ref{T-definition}), and $R^{2}=\ln s\,C$)
\end{comment}
The limit of infinite energy implies that energy sharing plays no
role, and the arguments $s,t$ of $\boldsymbol{\mathrm{T}}_{\mathrm{Pom}}$
refer to $pp$ scattering, and not to parton-parton scattering (I
will come back to this later). The inelastic cross section is given
as [similar to Eq. (\ref{sigma-tot-1})] 
\begin{equation}
\sigma_{\mathrm{in}}=\int d^{2}b\,\mathrm{cut}\,T(s,b)
\end{equation}
\begin{equation}
\qquad\qquad=\int d^{2}b\,\mathrm{\sum_{n=0}^{\infty}\,\frac{1}{n!}\,\sum_{cuts}}\left\{ iT_{\mathrm{Pom}}\right\} ^{n}.
\end{equation}
Summing all cuts simply means considering a graph
with $n$ Pomeron exchanges, seen as a three-dimensional (3D) plot, and considering all
possibilities to cut them, 
with at least one Pomeron being cut. The cuts are 
indicated by the dashed lines in Fig. \ref{cut-diagram-1}.
A cut should be imagined as an infinite plane! One usually assumes
the sub-T-matrix to be purely imaginary, i.e. $T_{\mathrm{Pom}}=i\frac{G}{2}$,
with some real function $G=G(b)$, and a factor $1\!/\!2$ for convenience.
Then one gets for the cut Pomeron $\mathrm{cut}\,T_{\mathrm{Pom}}=2\,\mathrm{Im}\,T_{\mathrm{Pom}}=G$.
Concerning the uncut Pomerons, one sums up the contributions where
the Pomeron is to the left or to the right of the cut, which gives
$\{iT_{\mathrm{Pom}}\}+\{iT_{\mathrm{Pom}}\}^{*}=-G$. So cut and
uncut Pomerons have opposite signs, and one gets 
\begin{equation}
\sigma_{\mathrm{in}}=\int d^{2}b\,\mathrm{\sum_{n=1}^{\infty}\,\frac{1}{n!}\,\sum_{m=0}^{n}\left(\begin{array}{c}
n\\
m
\end{array}\right)}G^{m}(-G)^{n-m},\label{sigma-inel-2}
\end{equation}
where $m$ refers to the number of cut Pomerons. One may also write
\begin{equation}
\sigma_{\mathrm{in}}=\int d^{2}b\,\mathrm{\sum_{m=1}^{\infty}\,\sum_{l=0}^{\infty}\frac{1}{m!\,l!}}G^{m}(-G)^{l}.\label{sigma-inel-2-1}
\end{equation}

\subsection{AGK cancellations in $pp$ scattering}

Let me consider inclusive cross sections, like jet cross sections,
where $m$-cut-Pomeron events contribute $m$ times more than single
Pomeron events, so one gets
\begin{equation}
\sigma_{\mathrm{incl}}=\int d^{2}b\,\sum_{m=1}^{\infty}\mathrm{\sum_{l=0}^{\infty}m\frac{1}{m!\,l!}}G^{m}(-G)^{l},\label{sigma-inclusive}
\end{equation}
which is equal to
\begin{equation}
\sigma_{\mathrm{incl}}=\int d^{2}b\,G\,\underbrace{\sum_{m=0}^{\infty}\mathrm{\sum_{l=0}^{\infty}\frac{1}{m!l!}}G^{m}(-G)^{l}}_{1}\,,
\end{equation}
\begin{comment}
\begin{equation}
\sigma_{\mathrm{incl}}=\int d^{2}b\,\sum_{n=1}^{\infty}\,\frac{1}{n!}\,\left\{ \mathrm{\sum_{m=0}^{n}m\left(\begin{array}{c}
n\\
m
\end{array}\right)}G^{m}(-G)^{n-m}\right\} ,\label{sigma-inclusive-2}
\end{equation}
where the term in curly brackets represents the sum over all cuts.

\[
\mathrm{\sum_{m=0}^{n}m\left(\begin{array}{c}
n\\
m
\end{array}\right)}(-1)^{n-m}=\mathrm{\sum_{m=0}^{n}m\left(\begin{array}{c}
n\\
m
\end{array}\right)}x^{m}y{}^{n-m}=x\frac{d}{dx}\mathrm{\sum_{m=0}^{n}\left(\begin{array}{c}
n\\
m
\end{array}\right)}x^{m}y{}^{n-m}=x\frac{d}{dx}(x+y)^{n}=xn(x+y)^{n-1}=\left\{ \begin{array}{c}
1\,\,\mathrm{if}\,\,n=1\\
0\,\,\mathrm{if}\,\,n>1
\end{array}\right.
\]
For a given number $n$ of Pomerons, an elementary calculation allows
to compute the sum over all possible cuts, and one finds
\begin{equation}
\mathrm{\sum_{m=0}^{n}m\left(\begin{array}{c}
n\\
m
\end{array}\right)}G^{m}(-G)^{n-m}=G^{n}\times\left\{ \begin{array}{c}
1\,\,\mathrm{if}\,\,n=1\\
0\,\,\mathrm{if}\,\,n>1
\end{array}\right.,\label{cancellations}
\end{equation}
\end{comment}
and so one finds
\begin{equation}
\sigma_{\mathrm{incl}}=\int d^{2}b\,G=\int d^{2}b\,\mathrm{cut}\,T_{\mathrm{Pom}}\,,
\end{equation}
which is an amazing result: \\

\begin{itemize}
\item
Only a single cut Pomeron ($\mathrm{cut}\,T_{\mathrm{Pom}}$) contributes
to the inclusive cross section, 
\item
All higher-order terms  cancel each
other (\textbf{AGK cancellations}). 
\item
This is the \textbf{AGK theorem}
\cite{Abramovsky:1973fm}.%
\end{itemize}

\begin{figure}[h]
\begin{centering}
\includegraphics[scale=0.3]
{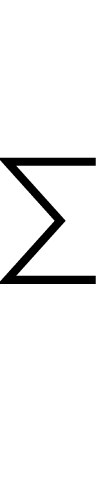}
$\quad$\includegraphics[scale=0.33]
{Home/2/komm/conf/99aktuell/epos/poms0}
$\quad$\includegraphics[scale=0.3]
{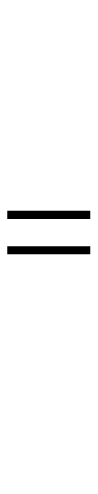}
\includegraphics[scale=0.33]
{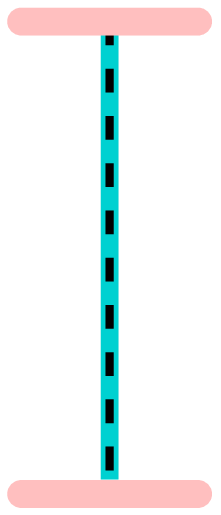}
\par\end{centering}
\centering{}\caption{The only diagram contributing to inclusive cross sections: a single
cut Pomeron. \label{cut-diagram-1-1}}
\end{figure}

So although there is an infinite number of multiple Pomeron exchange
diagrams, with an infinite number of possibilities to cut, they all
cancel each other, with the exception of one diagram (see Fig. \ref{cut-diagram-1-1}).

\subsection{AGK cancellations in \emph{A+B} scattering}

Let me come to the scattering of two nuclei of mass numbers $A$ and
$B$. The inelastic cross section is a straightforward generalization
of the $pp$ result of Eq. (\ref{sigma-inclusive}); one simply needs to
add the nuclear geometry in terms of an integration over the positions
(in the transverse plane) of the nuclei, in addition to the integration
over the nuclear impact parameter $b$, as
\begin{equation}
\int db\!_{A\!B}=\int d^{2}b\int\prod_{i=1}^{A}d^{2}b_{i}^{A}\,T_A(b_{i}^{A})\int\prod_{j=1}^{B}d^{2}b_{j}^{B}\,T_B(b_{j}^{B}),\label{b-AB-integration}
\end{equation}
with the nuclear thickness function 
\begin{equation}
T_A(b)=\int dz\,\rho_A\left(\sqrt{b^{2}+z{}^{2}}\right),\label{thickness-function}
\end{equation}
where $\rho_A$ is the nuclear density for nucleus $A$ (and correspondingly for $B$). Considering all $AB$ possible
nucleon-nucleon pairs $k$, and summing over all possible numbers
of cut ($m_{k}$) and uncut ($l_{k}$) Pomerons, one gets for the
cross section $\sigma_{\mathrm{in}}^{AB}$ $=\int\!\!db\!_{A\!B}\mathrm{cut}\,T^{AB}(s,b)$
the following expression:
\begin{equation}
\sigma_{\mathrm{in}}^{AB}=\int db\!_{A\!B}\sum_{m_{1}l_{1}}\ldots\sum_{m_{AB}l_{AB}}\prod_{k=1}^{AB}\mathrm{\frac{1}{m_{k}!l_{k}!}}(G_{k})^{m_{k}}(-G_{k})^{l_{k}},\label{sigma-inel-AB}
\end{equation}
with at least one $m_{k}$ being nonzero, and with $G_{k}=G(b_{k})$
and $b_{k}=|b+b_{\pi(k)}^{A}-b_{\tau(k)}^{B}|$ referring to the impact
parameter of the nucleon-nucleon pair number $k$. 

The inclusive cross section (again like jet cross sections), where
pairs $k$ with $m_{k}$ cut Pomerons count with a factor $m_{k}$,
is given as 
\begin{align}
\sigma_{\mathrm{incl}}^{AB} & =\sum_{m_{1}l_{1}}\ldots\sum_{m_{AB}l_{AB}}\Big\{\sum_{k'=1}^{AB}\,m_{k'}\Big\}\\
 & \qquad\int db\!_{A\!B}\prod_{k=1}^{AB}\mathrm{\frac{1}{m_{k}!l_{k}!}}(G_{k})^{m_{k}}(-G_{k})^{l_{k}},\nonumber 
\end{align}
where in $\{...\}$ one sums up the cut Pomeron numbers from the different
nucleon-nucleon collisions. Using $m_{k'}/(m_{k'})!$ $=1/(m_{k'}-1)!$,
and after renaming $m_{k'}-1$ to $m_{k'}$ in the product $\prod_{k=1}^{AB}$,
one gets 
\begin{align}
 & \sigma_{\mathrm{incl}}^{AB}=\sum_{k'=1}^{AB}\int db\!_{A\!B}\,G_{k'}\nonumber \\
 & \qquad\Big\{\sum_{m_{1}l_{1}}\ldots\sum_{m_{AB}l_{AB}}\prod_{k=1}^{AB}\mathrm{\frac{1}{m_{k}!l_{k}!}}G_{k}^{m_{k}}(-G_{k})^{l_{k}}\Big\},
\end{align}
without any constraint concerning the $m_k$ summations.
Therefore  the term $\{...\}$ is unity, which means that the different multiple Pomeron
terms cancel each other (AGK cancellations). The thickness functions
are normalized as $\int d^{2}b_{i}^{A}\,T_{A}(b_{i}^{A})=1,$ and
correspondingly for $B$, so all the integrations $\int d^{2}b_{i}^{A}\,T_{A}(b_{i}^{A})$
and $\int d^{2}b_{j}^{B}\,T_B(b_{j}^{B})$ in $\int db\!_{A\!B}$ give
unity, except for the indices $i'$ and $j'$ corresponding to $k'$.
So one finds
\begin{align}
\sigma_{\mathrm{incl}}^{AB} & =\sum_{k'=1}^{AB}\int d^{2}b_{i'}^{A}\,T_A(b_{i'}^{A})\int d^{2}b_{j'}^{B}\,T_B(b_{j'}^{B})\nonumber \\
 & \qquad\qquad\qquad\int d^{2}b\,G(|b+b_{i'}^{A}-b_{j'}^{B}|),
\end{align}
A variable change such as $\vec{b}'=\vec{b}+\vec{b}_{i'}^{A}-\vec{b}_{j'}^{B}$
allows one to write
\begin{equation}
\sigma_{\mathrm{incl}}^{AB}=\sum_{k'=1}^{AB}\int d^{2}b_{i'}^{A}\,T_A(b_{i'}^{A})\!\!\int\!\!d^{2}b_{j'}^{B}\,T_B(b_{j'}^{B})\!\!\int\!\!d^{2}b'\,G(b').
\end{equation}
After using the normalization $\int d^{2}b_{i'}^{A}\,T_A(b_{i'}^{A})=1$
and $\int d^{2}b_{j'}^{B}\,T_B(b_{j'}^{B})=1$, one replaces $\sum_{k'=1}^{AB}$
by simply $AB$, and one gets
\begin{equation}
\sigma_{\mathrm{incl}}^{AB}=AB\times\int d^{2}b\,G(b)=AB\times\int d^{2}b\,\mathrm{cut}\,T_{\mathrm{Pom}}\,,
\end{equation}
which is again an amazing result: \\

\begin{figure}[h]
\begin{centering}
\begin{minipage}[c]{0.2\columnwidth}%
\begin{center}
{\huge{}$\sum$}
\par\end{center}%
\end{minipage}%
\begin{minipage}[c]{0.35\columnwidth}%
\begin{center}
\includegraphics[scale=0.33]
{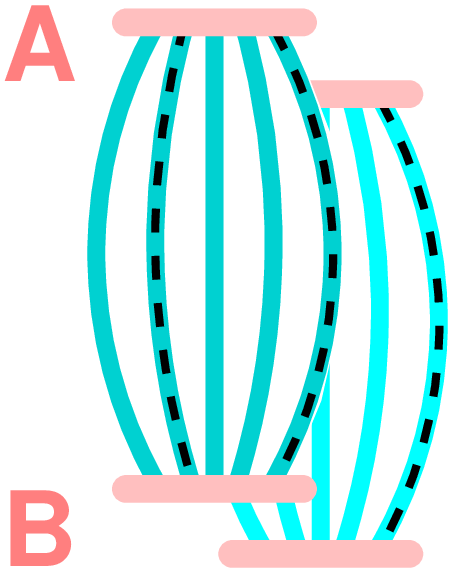}
\par\end{center}%
\end{minipage}%
\begin{minipage}[c]{0.25\columnwidth}%
\begin{center}
{\huge{}$\boldsymbol{\boldsymbol{=AB\,\times}}$}
\par\end{center}%
\end{minipage}%
\begin{minipage}[c]{0.2\columnwidth}%
\begin{center}
\includegraphics[scale=0.33]
{Home/2/komm/conf/99aktuell/epos/poms2}
\par\end{center}%
\end{minipage}
\par\end{centering}
\centering{}\caption{The only diagram contributing to inclusive cross sections in $A\!+\!B$ scattering:
a single cut Pomeron (multiplied by $AB$). \label{cut-diagram-1-1-1}}
\end{figure}

\begin{itemize}
\item
The inclusive $A\!+\!B$ cross section is $AB$ times the single cut-Pomeron
contribution ($\mathrm{cut}\,T_{\mathrm{Pom}}$), see Fig. \ref{cut-diagram-1-1-1}.
\item
This is the \textbf{AGK theorem} \cite{Abramovsky:1973fm} for (minimum
bias) scattering of nuclei with mass numbers $A$ and $B$. 
\item
It also
implies 
%%\begin{equation}
$\sigma_{\mathrm{incl}}^{AB}=AB\times\sigma_{\mathrm{incl}}^{pp}$,
%%\label{binary-scaling}
%%\end{equation}
also referred to as \textbf{binary scaling}.%
\end{itemize}

\noindent So although there is an infinite number of multiple Pomeron
exchange diagrams, with an infinite number of possibilities to cut,
they all cancel each other, with the exception of one diagram.\\

\subsection{Factorization}

So I have shown that for $pp$ scattering and also for nuclear collisions,
the inclusive cross sections are given in terms of one single cut
Pomeron ($G=\mathrm{cut}\,T_{\mathrm{Pom}}$). What does this mean
concerning factorization? To answer this question, one needs to specify
the internal structure of the Pomeron. At the time of Gribov et al.,
it was unknown, and one could not do more than using parametrized
expressions for the T-matrices, based on their asymptotic behavior.
But nowadays one knows more. It is shown in Ref. \cite{werner:2023-epos4-heavy}
how to compute T-matrices and then $G=\mathrm{cut}\,T$, corresponding
to (a single) parton-parton scattering, based on perturbative QCD.
The result is named $G_{\mathrm{QCD}}$.
\begin{figure}[h]
\centering{}%
\begin{minipage}[c]{0.59\columnwidth}%
(a)\\
 \includegraphics[bb=0bp 0bp 460bp 405bp,scale=0.3]
{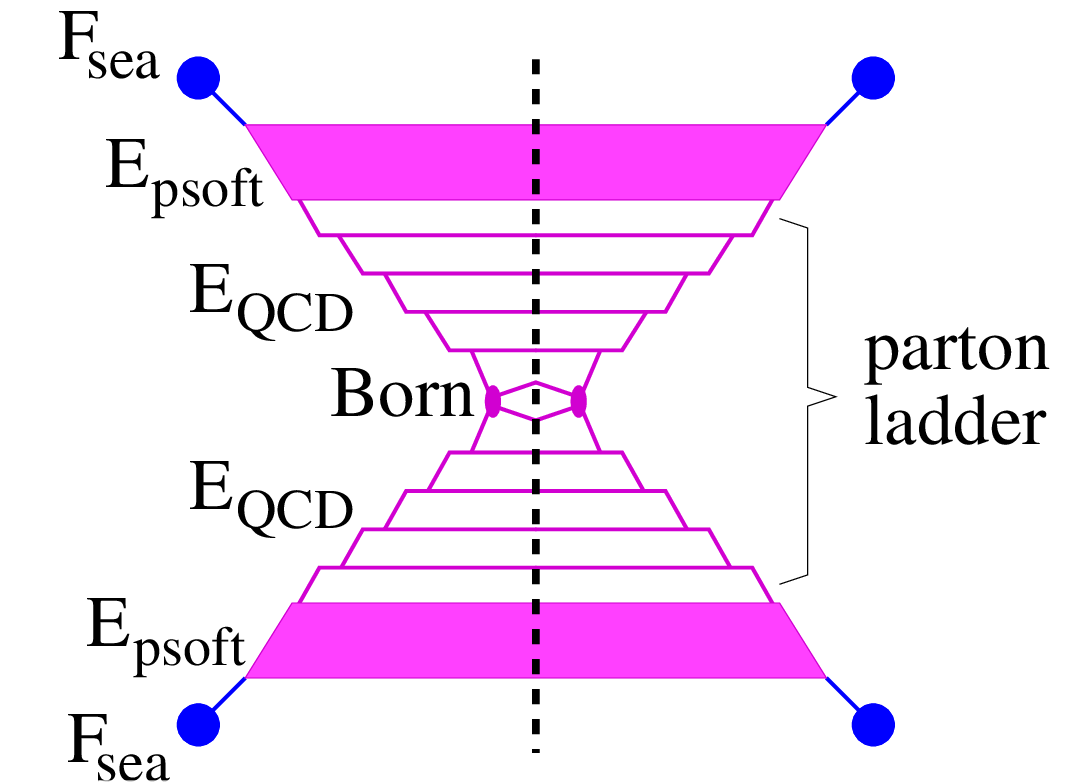}%
\end{minipage}%
\begin{minipage}[c]{0.3\columnwidth}%
(b)\\
\includegraphics[bb=0bp 0bp 235bp 405bp,clip,scale=0.3]
{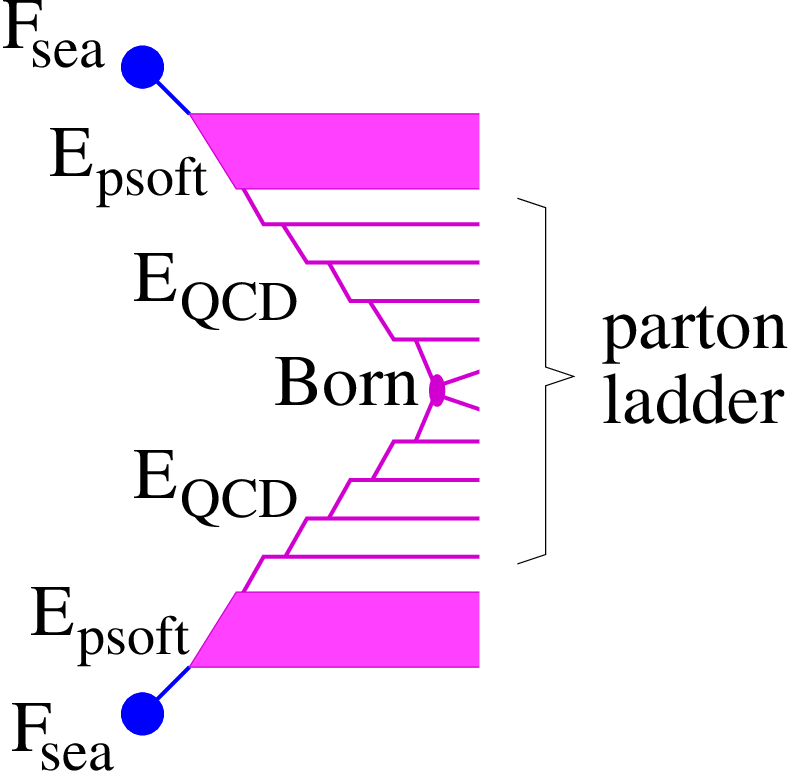}%
\end{minipage}%
\begin{minipage}[c]{0.15\columnwidth}%
~\bigskip{}

PDF\medskip{}

Born\medskip{}

PDF%
\end{minipage}\caption{(a) A single cut Pomeron based on pQCD and (b) the corresponding inelastic
process.\label{single-pomeron-graph-1}}
\end{figure}
It is a sum of several terms, the most important one shown in Fig.
\ref{single-pomeron-graph-1}, representing an integration expressed
in terms of several modules ($F_{\mathrm{sea}}$, $E_{\mathrm{psoft}}$, $E_{\mathrm{QCD}}$,
Born), 
%%$\sigma_{\mathrm{QCD}}$), 
with $E_{\mathrm{QCD}}$
representing a spacelike parton evolution, and 
``Born'' the QCD Born process.
%%$\sigma_{\mathrm{QCD}}$
%%an elementary QCD cross section. 
Assuming that the $G$ of the multiple
scattering formalism is equal to $G_{\mathrm{QCD}}$, i.e.
\begin{equation}
G=G_{\mathrm{QCD}},\label{G-equal-G-QCD}
\end{equation}
one gets for the inclusive $pp$ cross section (using the AGK theorem, and thererfore considering only one single Pomeron)
the following result (see Ref. \cite{werner:2023-epos4-heavy}):
\begin{align}
 & \sigma_{\mathrm{incl}}^{pp}=\int\frac{d^{3}p_{3}d^{3}p_{4}}{E_{3}E_{4}}\nonumber \\
 & \quad\Bigg\{\sum_{klmn}\int\!\!\int\!\!d\xi_{1}d\xi_{2}\,f_{\mathrm{PDF}}^{k}(\xi_{1},\mu_{\mathrm{F}}^{2})f_{\mathrm{PDF}}^{l}(\xi_{2},\mu_{\mathrm{F}}^{2})\nonumber \\
 & \qquad\frac{1}{32s\pi^{2}}\bar{\sum}|\mathcal{M}^{kl\to mn}|^{2}\delta^{4}(p_{\mathrm{in\,}}-p_{\mathrm{out}})\frac{1}{1+\delta_{mn}}\Bigg\},
\end{align}
with $f_{\mathrm{PDF}}^{k}$ being a convolution of modules like $F_{\mathrm{sea}}$,
$E_{\mathrm{psoft}}$, and $E_{\mathrm{QCD}}$,
with $\mathcal{M}^{kl\to mn}$ being the
QCD matrix element for elementary parton-parton $2\to2$ scatterings
(Born process), and with $\vec{p}_{3}$, $E_{3}$ and $\vec{p}_{4}$,
$E_{4}$ being the momenta and energies of the outgoing partons. This
amounts to factorization. \\

\subsection{Conclusion}

It is interesting that one can derive in a quantum mechanical multiple
scattering approach important features such as binary scaling and factorization,
as a consequence of AGK cancellations. It is very useful to understand
these phenomena in a simple approach where calculations can be easily
done. But the main problem is the fact that energy sharing is not
present. The cut-Pomeron expressions $G$ should depend explicitly
on the Pomeron energy $s_{\mathrm{Pom}}$, i.e., $G=G(s_{\mathrm{Pom}},b)$,
and when one has ten Pomerons in a $pp$ scattering, then the available
energy has to be shared among these ten Pomerons. It is also known
that binary scaling fails at low $p_{t}$, whereas here binary scaling
is always true. So the scenario discussed
in this section is a first step, but not the final solution.\\

But even in more realistic scenarios, to be discussed later, one feature
remains always correct: the fundamental property is the ``validity
of AGK cancellations'', for $pp$ scatterings, and for $AA$ scatterings,
whereas binary scaling and factorization are merely consequences.%
\bigskip{}

To summarize this section: 
\begin{itemize}
\item In a simplified parallel scattering S-matricx approach, without energy
sharing (and consequently simple formulas), one can prove the ``AGK
theorem'', in $pp$ and $AA$ scattering (first shown in Ref. \cite{Abramovsky:1973fm}). 
\item ``AGK theorem'' means for $pp$ scattering that inclusive cross sections
$\sigma_{\mathrm{incl}}^{pp}$ are given in terms of a single cut
Pomeron, although the real events represent multiple scatterings.
\item ``AGK theorem'' means for $A\!+\!B$ scattering that inclusive cross sections
$\sigma_{\mathrm{incl}}^{AB}$ are given as $AB$ times the single
cut Pomeron contribution.
\item As a consequence, one gets $\sigma_{\mathrm{incl}}^{AB}=AB\times\sigma_{\mathrm{incl}}^{pp}$,
which is called binary scaling. So it is a direct consequence
of the AGK theorem.
\item One can make statements about factorization, but only when one specifies
the internal structure of the Pomeron, more precisely the microscopic
picture underlying $T_{\mathrm{Pom}}$ and $G=\mathrm{cut}\,T_{\mathrm{Pom}}$.
Assuming $G=G_{\mathrm{QCD}}$, the latter being based on a QCD calculation
of parton-parton scattering, including a DGLAP evolution as defined
in Ref. \cite{werner:2023-epos4-heavy}, one can deduce ``factorization''.
\item The AGK theorem is the fundamental feature; others (like binary
scaling and factorization) are just consequences. 
\item The simple scenario of this section is useful for understanding the relation
between the AGK theorem, binary scaling, and factorization.
However, a major problem is the fact that energy sharing is not present,
but it should be.
\end{itemize}
In the following, I will discuss a scenario including energy sharing. 

\section{Multiple scattering S-matrix approach with energy sharing and ``sign
problem'' \label{======= S-matrix-with-energy-sharing =======}}

In this section, I take over the S-matrix approach discussed 
in Sec. \ref{======= AGK-without energy sharing =======} (GR approach), but I add energy-momentum sharing (GR\textsuperscript{+}). I will discuss
right away collisions of nuclei with mass numbers $A$ and $B$,
where $pp$ is just a special case ($A=B=1$). 

\subsection{Scenario with energy-momentum sharing}

\begin{figure}[h]
(a)\hspace*{3.5cm}

\begin{minipage}[c]{0.5\columnwidth}%
\begin{center}
\includegraphics[scale=0.3]
{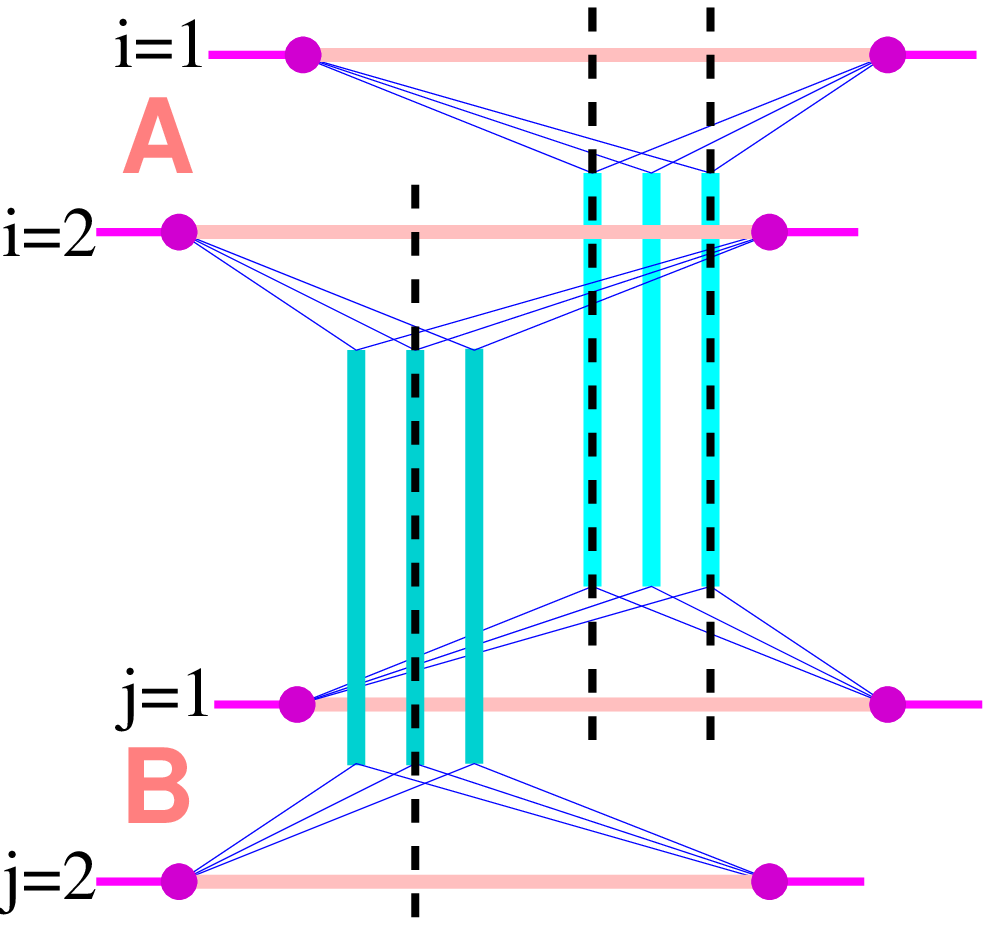}
\par\end{center}%
\end{minipage}%
\begin{minipage}[c]{0.4\columnwidth}%
~

~

~

(b)\vspace*{-0.1cm}

\includegraphics[scale=0.22]
{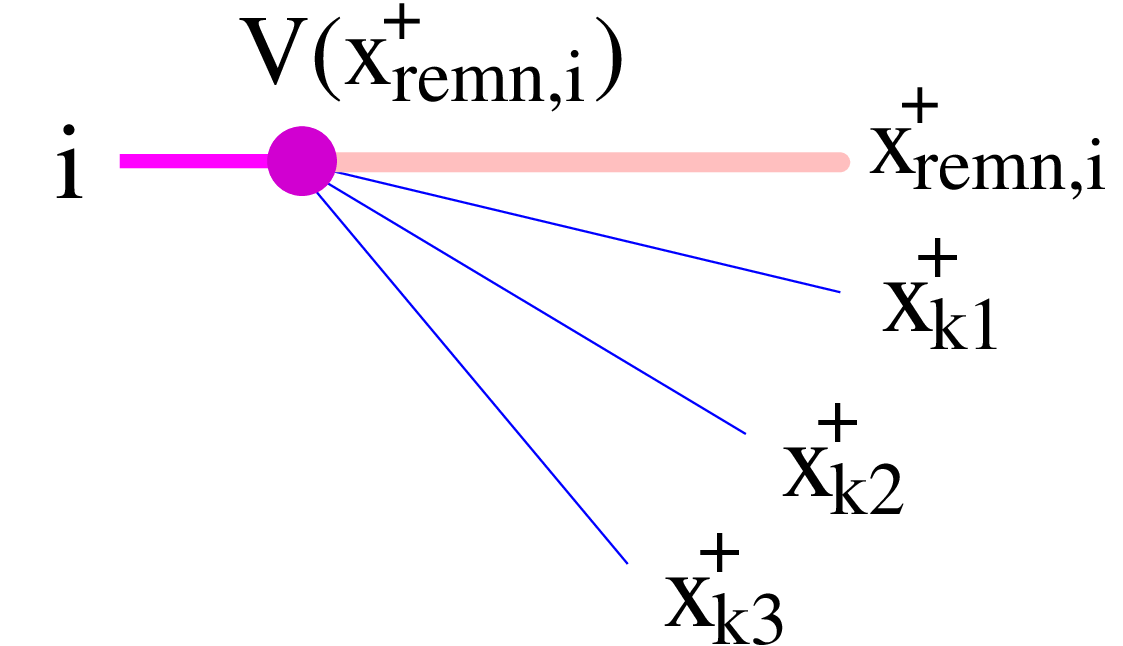}%
\end{minipage}\caption{(a) Multiple scattering with energy sharing, for two colliding nuclei
$A$ and $B$, with Pomerons (cyan vertical lines), remnants (pink
horizontal lines), vertices (magenta dots), and cuts (black, vertical).
(b) Detailed view of a vertex connected to projectile $i$, associated
to pair $k$, with remnants light-cone momentum fraction $x_{\mathrm{remn},i}^{+}$
and Pomeron leg momenta $x_{k\mu}^{+}$. \label{mutiple-scattering-EM-sharing}}
\end{figure}

I generalize Eq. (\ref{sigma-inel-AB}) by taking into account energy-momentum
sharing, as sketched in Fig. \ref{mutiple-scattering-EM-sharing}(a).
In general, one considers all possible interactions between the $A$
projectile and the $B$ target nucleons, and one therefore introduces
a ``pair index'' $k$, going from $1$ to $AB$. I define $i=\pi(k)$
and $j=\tau(k)$ to be the associated nucleon indices on the projectile
and target sides. In Fig. \ref{mutiple-scattering-EM-sharing}(a),
I consider $A=B=2$ (for simplicity), and I consider multiple (three)
Pomeron interactions between $i=1$ and $j=1$ and between $i=2$
and $j=2$, which means zero Pomerons exchanged between $i=1$ and
$j=2$ and between $i=2$ and $j=1$. In general, one sums over any
number of Pomerons for all possible pairs. In Fig. \ref{mutiple-scattering-EM-sharing}(b),
I show a detailed view of one of the projectile vertices connected
to projectile $i$, with the associated pair numbering $k$. For the
different components, one uses light-cone momentum fractions $x^{\pm}$
with respect to the initial light-cone momenta of the nucleons, with
$0\le x^{+},x^{-}\le1$. I define $x_{k\mu}^{\pm}$ to be the light-cone
momentum fractions of the external legs of the $\mu$-th Pomeron of
pair $k$, all of them connected to projectile nucleon $i=\pi(k)$
and target nucleon $j=\tau(k)$. For each cut Pomeron one has a term
$G$ (equal to $\mathrm{cut}\,T_{\mathrm{Pom}}$ $=2\mathrm{Im}\,T_{\mathrm{Pom}}$;
see Sec. \ref{======= S-matrix-theory-basic-definitions =======}
for basic definitions), and for each uncut Pomeron one has a term
$-G$, both with arguments $x_{k\mu}^{\pm}$. The light-cone momenta
of the projectile and target remnants are named $x_{\mathrm{remn},i}^{+}$
and $x_{\mathrm{remn},j}^{-}$. In Fig. \ref{mutiple-scattering-EM-sharing}(b),
I only show the projectile part. Energy-momentum conservation amounts
to 
\begin{equation}
x_{\mathrm{remn},i}^{+}=1-\!\!\sum_{\underset{\pi(k)=i}{k=1}}^{AB}\sum_{\mu=1}^{n_{k}}\!x_{k\mu}^{+}\:,\;x_{\mathrm{remn},j}^{-}=1-\!\!\sum_{\underset{\tau(k)=j}{k=1}}^{AB}\sum_{\mu=1}^{n_{k}}\!x_{k\mu}^{-},\label{energy-conservation}
\end{equation}
which means that the initial values ($x^{\pm}=1$) are shared among
Pomerons and remnants. The nuclear geometry will be taken care of
by using $\int db\!_{A\!B}$, defined in Eqs. (\ref{b-AB-integration}) and (\ref{thickness-function}),
which integrates over the impact parameter and the transverse coordinates
$b_{i}^{A}$ and $b_{j}^{B}$ of projectile and target nucleons. With
all these definitions, one may write the expression for the inelastic
cross section being a sum over all possible numbers of Pomerons per
pair $k$ and a sum over all possibilities to cut them, as
\begin{align}
\sigma_{\mathrm{in}}^{AB} & =\sum_{n_{1}=0}^{\infty}\!\!\ldots\!\!\sum_{n_{AB}=0}^{\infty}\quad\sum_{m_{1}\le n_{1}}\!\!\ldots\!\!\sum_{m_{AB}\le n_{AB}}\int db\!_{A\!B}\int dX\!_{A\!B}\nonumber \\
 & \qquad\prod_{k=1}^{AB}\frac{1}{n_{k}!}\mathrm{\Big(\!\!\begin{array}{c}
n_{k}\\
m_{k}
\end{array}\!\!\Big)}\prod_{\mu=1}^{m_{k}}G_{k\mu}\prod_{\mu=m_{k}+1}^{n_{k}}-G_{k\mu}\nonumber \\
 & \qquad\prod_{i=1}^{A}V(x_{\mathrm{remn},i}^{+})\prod_{j=1}^{B}V(x_{\mathrm{remn},j}^{-})\label{sigma-inel-AB-1}
\end{align}
[generalizing Eq. (\ref{sigma-inel-AB})], with at least one $n_{k}$
being nonzero, with $\prod_{\mu=1}^{m_{k}}G_{k\mu}=1$ for $m_{k}=0$,
and with 
\begin{equation}
G_{k\mu}\!=G\!\left(\!x_{k\mu}^{+},x_{k\mu}^{-},s,b_{k}\!\right),\label{G-k-mu}
\end{equation}
with $b_{k}=|b+b_{\pi(k)}^{A}-b_{\tau(k)}^{B}|$ referring to the
impact parameter of the nucleon-nucleon pair number $k$. The indices
$n_{k}$ refer to the numbers of Pomerons (cut and uncut) for pair
$k$, whereas the indices $m_{k}$ refer to the number of cut Pomerons.
The symbol $\int dX\!_{A\!B}$ represents the integration over light-cone
momentum fractions:
\begin{equation}
\int dX\!_{A\!B}=\int\,\prod_{k=1}^{AB}\prod_{\nu=1}^{n_{k}}dx_{k\nu}^{+}dx_{k\nu}^{-}.\label{x-AB-integration}
\end{equation}
Equation (\ref{energy-conservation}) ensures energy-momentum conservation.
If one would remove the vertex part (replace $V$ by unity), the $\int dX\!_{A\!B}$
could be done, and one recovers Eq. (\ref{sigma-inel-AB}), the case
without energy sharing (I just organize the summations somewhat differently).
With the vertex part, and full energy-momentum conservation, the up
to 10~000~000 dimensional integrations are not separable, but this
``technical problem'' can be handled: see Ref. \cite{Drescher:2000ha}
for a detailed discussion about how to compute expressions like Eq.
(\ref{sigma-inel-AB-1}). \\

\subsection{AGK theorem}

As discussed in Sec. \ref{======= AGK-without energy sharing =======}, a crucial feature of any multiple
scattering approach is the validity of the AGK theorem, which is needed
to get for inclusive cross sections binary scaling in $AA$ collisions
and factorization in $pp$. Therefore I am going to investigate this
in the following.\\

\begin{figure}[h]
\begin{centering}
\noindent\begin{minipage}[c]{0.1\columnwidth}%
\begin{center}
{\huge{}$\sum$}
\par\end{center}%
\end{minipage}\hspace*{-0.5cm}%
\begin{minipage}[c]{0.37\columnwidth}%
\begin{center}
\includegraphics[scale=0.3]
{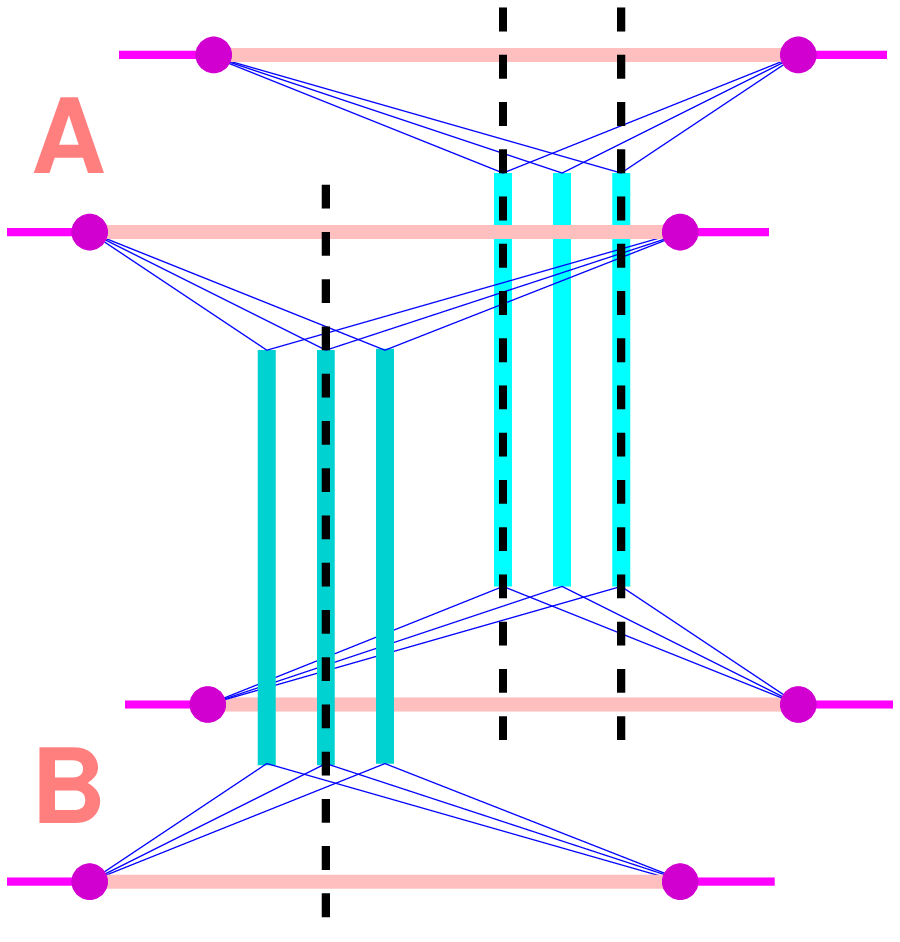}
\par\end{center}%
\end{minipage}$\quad\quad$%
\begin{minipage}[c]{0.22\columnwidth}%
\begin{center}
{\huge{}$\boldsymbol{\boldsymbol{=AB\,\times}}$}
\par\end{center}%
\end{minipage}%
\begin{minipage}[c]{0.22\columnwidth}%
\begin{center}
\includegraphics[scale=0.3]
{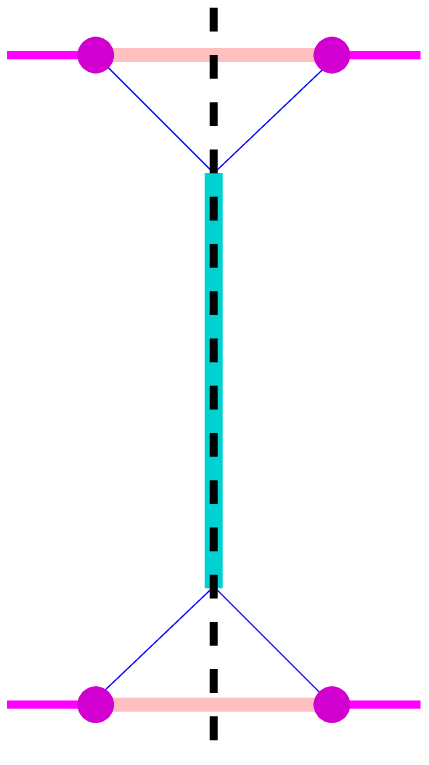}
\par\end{center}%
\end{minipage}
\par\end{centering}
\centering{}\caption{The only diagram contributing to inclusive cross sections in $A\!+\!B$ scattering:
a single cut Pomeron (multiplied by $AB$). \label{cut-diagram-2}}
\end{figure}

Again I am considering collisions of nuclei with mass numbers $A$
and $B$, where $pp$ is just a special case ($A=B=1$). The inclusive
cross section is a modification of the inelastic cross section in Eq.
(\ref{sigma-inel-AB-1}). In the case of ``no energy sharing'' as
discussed in Sec. \ref{======= AGK-without energy sharing =======},
I simply added a factor $\Big\{\sum_{k'=1}^{AB}\,m_{k'}\Big\}$, which
amounts to counting the number of cut Pomerons. Now, I introduce energy-momentum
sharing, and each Pomeron is characterized by the light-cone momentum
fractions $x^{+}$ and $x^{-}$, such that the squared (transverse)
mass of the Pomeron is given by $x^{+}x^{-}s$, with $s$ being the
Mandelstam $s$ of the nucleon-nucleon collisions. So one needs to
count the Pomerons for given $x^{+}$ and $x^{-}$ intervals, which
can be done by adding a factor 
\begin{equation}
\sum_{k'=1}^{AB}\sum_{\mu'=1}^{m_{k'}}\delta(x^{+}-x_{k'\mu'}^{+})\delta(x^{-}-x_{k'\mu'}^{-})dx^{+}dx^{-}\label{delta-times-delta}
\end{equation}
into Eq. (\ref{sigma-inel-AB-1}), and after dividing by $dx^{+}dx^{-}$,
one gets
\begin{align}
 & \frac{d^{2}\sigma_{\mathrm{incl}}^{AB}}{dx^{+}dx^{-}}=\sum_{k'=1}^{AB}\;\sum_{n_{1}=0}^{\infty}\!\!\ldots\!\!\sum_{n_{AB}=0}^{\infty}\quad\sum_{m_{1}\le n_{1}}\!\!\ldots\!\!\sum_{m_{AB}\le n_{AB}}\;\sum_{\mu'=1}^{m_{k'}}\label{sigma-incl-AB-1}\\
 & \int\!\!db\!_{A\!B}\!\!\int\!\!dX\!_{A\!B}\Bigg\{\prod_{k=1}^{AB}\!\left[\!\mathrm{\frac{1}{m_{k}!(n_{k}-m_{k})!}}\prod_{\mu=1}^{m_{k}}\!\!G_{k\mu}\!\!\prod_{\mu=m_{k}+1}^{n_{k}}\!\!\!\!-G_{k\mu}\!\right]\nonumber \\
 & \prod_{i=1}^{A}\!V(x_{\mathrm{remn},i}^{+})\prod_{j=1}^{B}\!V(x_{\mathrm{remn},j}^{-})\,\delta(x^{+}-x_{k'\mu'}^{+})\delta(x^{-}-x_{k'\mu'}^{-})\Bigg\},\nonumber 
\end{align}
with at least one $n_{k}$ being nonzero. The $\delta$ functions make
some of the integrations of $\int dX\!_{A\!B}$ trivial. Integrating
over $x_{k'\mu'}^{+}$ and $x_{k'\mu'}^{-}$ allows one to replace $G_{k',\mu'}$
by $G(x^{+},x^{-},s,b_{k'})$, which may be written in front of $\int dX\!_{A\!B}$.
Then one may rename the integration variables such that the variables
$x_{k'\mu'}^{\pm}$ disappear (to have cut Pomeron variables $x_{k'\mu}^{\pm}$
with $\mu\le m_{k'}-1$ and uncut Pomeron variables $x_{k'\mu}^{\pm}$
with $m_{k'}\le\mu\le n_{k'}-1$). In $x_{\mathrm{remn},i}^{+}$ for $i=\pi(k')$
and in $x_{\mathrm{remn},j}^{-}$ for $j=\tau(k')$ one replaces $1$ by $1-x^{+}$ and $1-x^{-}$,
respectively. The same procedure applies for all values of $\mu'$,
giving identical expressions, so the sum $\sum_{\mu'=1}^{m_{k'}}$
gives simply a factor $m_{k'}$, which one uses to replace $m_{k'}!$
by $(m_{k'}-1)!$. The expression $[...]$ in the second line of Eq.
(\ref{sigma-incl-AB-1}) for $k=k'$ may then be written as
\begin{equation}
\mathrm{\frac{1}{(m_{k}\!-\!1)!(n_{k}\!-\!1\!-\!(m_{k}\!-\!1))!}}\!\prod_{\mu=1}^{m_{k}-1}\!G_{k\mu}\!\prod_{\mu=m_{k}-1+1}^{n_{k}-1}\!-G_{k\mu}.
\end{equation}
I then rename $(m_{k}-1)$ to $m_{k}$ and $(n_{k}-1)$ to $n_{k}$,
which allows one to drop the condition ``at least one $n_{k}$ nonzero''.
In addition, one may now write 
\begin{equation}
\sum_{m_{1}\le n_{1}}\!\!\ldots\!\!\sum_{m_{AB}\le n_{AB}}\ldots\prod_{k=1}^{AB}=\prod_{k=1}^{AB}\sum_{m_{k}\le n_{k}},
\end{equation}
and so one gets 

\begin{align}
\frac{d^{2}\sigma_{\mathrm{incl}}^{AB}}{dx^{+}dx^{-}} & \!=\!\sum_{k'=1}^{AB}\;\sum_{n_{1}=0}^{\infty}\!\!\ldots\!\!\sum_{n_{AB}=0}^{\infty}\int\!db\!_{A\!B}\,G(x^{+},x^{-},s,b_{k'})\!\!\int\!\!dX\!_{A\!B}\,\nonumber \\
 & \qquad\prod_{k=1}^{AB}\mathrm{\frac{1}{n_{k}!}}\prod_{\mu=1}^{n_{k}}G_{k\mu}\sum_{m_{k}\le n_{k}}\Big(\!\!\begin{array}{c}
n_{k}\\
m_{k}
\end{array}\!\!\Big)\,(-1)^{n_{k}-m_{k}}\nonumber \\
 & \qquad\prod_{i=1}^{A}V(x_{\mathrm{remn},i}^{+})\prod_{j=1}^{B}V(x_{\mathrm{remn},j}^{-}).
\end{align}
without any constraints on the $n_{k}$. The relation
\begin{equation}
\sum_{m_{k}\le n_{k}}\Big(\!\!\begin{array}{c}
n_{k}\\
m_{k}
\end{array}\!\!\Big)\,(-1)^{n_{k}-m_{k}}=(1-1)^{n_{k}}=\delta_{0n_{k}}\,,
\end{equation}
for all values of $k$, means that only one term in the sum $\sum_{n_{1}}\ldots\sum_{n_{AB}}$contributes,
namely the one with $n_{1}=n_{2}=...=n_{AB}=0,$ and one gets%
\begin{comment}
\begin{align*}
\frac{d^{2}\sigma_{\mathrm{incl}}^{AB}}{dx^{+}dx^{-}} & =\sum_{k'=1}^{AB}\;\sum_{n_{1}}\ldots\sum_{n_{AB}}\;\int db\!_{A\!B}\,G(x^{+},x^{-},s,b_{k'})\,\int dX\!_{A\!B}\,\\
 & \qquad\prod_{k=1}^{AB}\mathrm{\frac{1}{n_{k}!}}\prod_{\mu=1}^{n_{k}}G_{k\mu}\delta_{0n_{k}}\\
 & \qquad\prod_{i=1}^{A}V(x_{\mathrm{remn},i}^{+})\prod_{j=1}^{B}V(x_{\mathrm{remn},j}^{-}).
\end{align*}
\end{comment}
\begin{align}
\frac{d^{2}\sigma_{\mathrm{incl}}^{AB}}{dx^{+}dx^{-}} & =\sum_{k'=1}^{AB}\;\int db\!_{A\!B}\,G(x^{+},x^{-},s,b_{k'})\,\nonumber \\
 & \qquad V(1-x^{+})V(1-x^{-}).
\end{align}
Note that only one pair of vertices $V$ contributes, namely that corresponding
to the projectile $i=\pi(k')$ and to the target $j=\tau(k'$); for
all others one has $V=V(1)$ which is per definition unity. All the
integrations $\int d^{2}b_{i}^{A}\,T_A(b_{i}^{A})$ and $\int d^{2}b_{j}^{B}\,T_B(b_{j}^{B})$
in $\int db\!_{A\!B}$ give unity (normalization of the thickness
functions), except for the indices $i'$ and $j'$ corresponding to
$k'$, so one gets
\begin{align}
\frac{d^{2}\sigma_{\mathrm{incl}}^{AB}}{dx^{+}dx^{-}} & =\sum_{k'=1}^{AB}\;\int d^{2}b_{i'}^{A}\,T_A(b_{i'}^{A})\int d^{2}b_{j'}^{B}\,T_B(b_{j'}^{B})\\
 & \int d^{2}b\,G(x^{+},x^{-},s,b_{k'})\,V(1-x^{+})V(1-x^{-}),\nonumber 
\end{align}
with $\vec{b}_{k'}\!\!=\!\vec{b}\!+\!\vec{b}_{i'}^{A}\!-\!\vec{b}_{j'}^{B}$.
A variable change such as $\vec{b}'\!\!=\!\!\vec{b}\!+\!\vec{b}_{i'}^{A}\!-\!\vec{b}_{j'}^{B}$
allows one to separate the three integrations, to use again the normalization
of the thickness functions, and to replace $\sum_{k'=1}^{AB}$ by
$AB$, to get 
\begin{equation}
\frac{d^{2}\sigma_{\mathrm{incl}}^{AB}}{dx^{+}dx^{-}}=AB\;\int d^{2}b\,G(x^{+},x^{-},s,b)\,V(1\!-\!x^{+})V(1\!-\!x^{-}).
\end{equation}
One recalls that $G(x^{+},x^{-},s,b)$ is a single cut Pomeron, so
one may write
\begin{equation}
\frac{d^{2}\sigma_{\mathrm{incl}}^{AB}}{dx^{+}dx^{-}}=AB\times\int d^{2}b\,\mathrm{cut}\,T_{\mathrm{Pom}}\times V(1\!-\!x^{+})V(1\!-\!x^{-}),
\end{equation}
and again one finds this important result: 

\begin{itemize}
\item
The inclusive $A\!+\!B$ cross
section is $AB$ times the single cut Pomeron contribution ($\mathrm{cut}\,T_{\mathrm{Pom}}$)
including a vertex part (i.e. the single Pomeron inclusive cross section).
\item
So one has \vspace{-0.5cm}
\begin{equation}
\frac{d\sigma_{\mathrm{incl}}^{AB}}{dx^{+}dx^{-}}=AB\times\frac{d\sigma_{\mathrm{incl}}^{\mathrm{single\,Pom}}}{dx^{+}dx^{-}}
\label{AGK cancellations}
\end{equation}\vspace{-0.6cm}

(see Fig. \ref{cut-diagram-2}), referred to as the \textbf{AGK theorem} 
(actually being an extension of the original AGK theorem, taking into account energy-momentum sharing).
\item
As a direct consequence, one gets \vspace{-0.3cm}
\begin{equation}
\frac{d\sigma_{\mathrm{incl}}^{AB}}{dx^{+}dx^{-}}=AB\times\frac{d\sigma_{\mathrm{incl}}^{pp}}{dx^{+}dx^{-}},
\label{binary-scaling-1}
\end{equation}\vspace{-0.6cm}

referred to as \textbf{binary scaling}.
\end{itemize}

\noindent \medskip{}

So I have proven in the \textbf{scenario with energy sharing}, for
(minimum bias) scattering of nuclei with mass numbers $A$ and $B$:
\begin{itemize}
\item the validity of the \textbf{AGK theorem,}
\item and \textbf{binary scaling}.
\end{itemize}
So although there is an infinite number of multiple Pomeron exchange
diagrams, with an infinite number of possibilities to cut, they all
cancel each other, with the exception of one diagram.\\

I have been discussing the inclusive cross section $d\sigma_{\mathrm{incl}}^{AB}/dx^{+}dx^{-}$,
which counts the number of Pomerons with given values of $x^{+}$
and $x^{-}.$ This is kind of a ``master distribution'', which allows one
to obtain the inclusive cross section for the production of measurable
quantities like transverse momentum, provided the nature of the Pomeron
is known, as discussed in Sec. \ref{======= AGK-without energy sharing =======}.

But before implementing the microscopic structure of the Pomeron,
one has  to confront a quite serious problem, related to probabilistic
interpretations, as discussed in the following.

\subsection{The sign problem}  \label{-----The-sign-problem-----}

\begin{figure}[h]
\begin{centering}
\includegraphics[scale=0.28]
{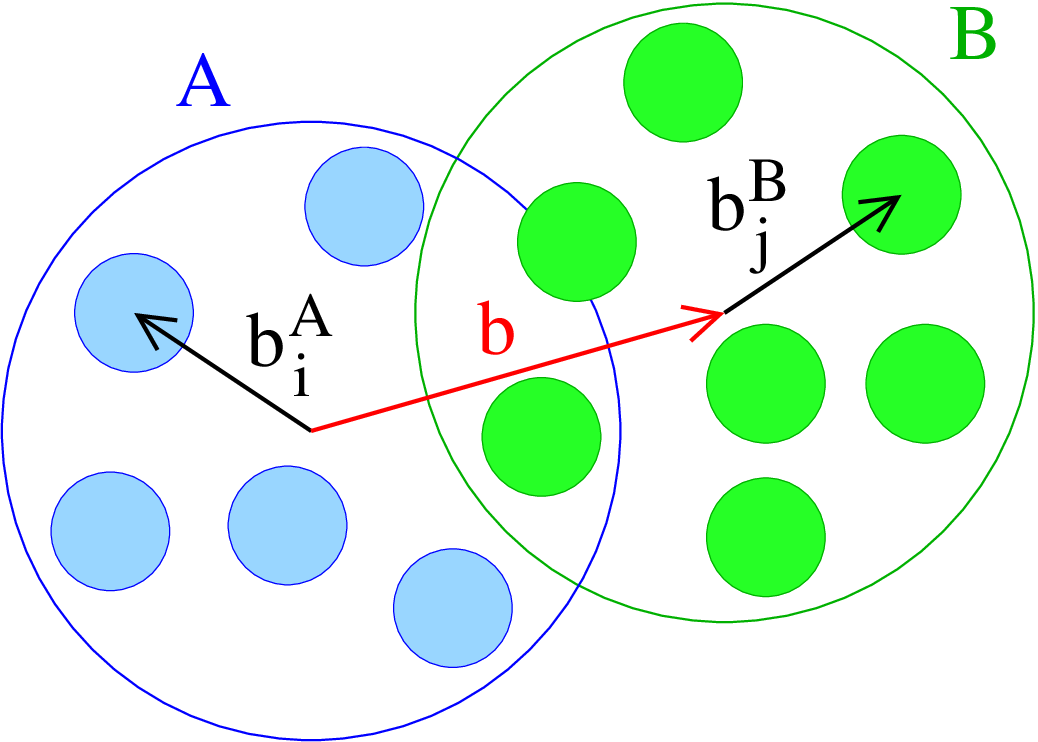}
\par\end{centering}
\centering{}\caption{The multidimensional variable $\{b,\{b_{i}^{A}\},\{b_{j}^{B}\}\}$,
with $b$ being the impact parameter, and with $b_{i}^{A}$ and $b_{j}^{B}$
being the transverse coordinates of the projectile and target nucleons,
respectively (they are meant to be two-dimensional vectors). \label{b-variables}}
\end{figure}

Let me consider again the inclusive cross section for the scattering of two nuclei, which may be written as [see Eq. (\ref{sigma-inel-AB-1})]
\begin{equation}
\sigma_{\mathrm{in}}^{AB}=\int\!\!db\!_{A\!B}\,\mathrm{cut}\,T^{AB}(s,\{b_{A\!B}\})=\int\!\!db\!_{A\!B}G^{AB}(s,\{b_{A\!B}\}),\label{G-AB}
\end{equation}
where one employs the usual
relation $G=\mathrm{cut}\,T$. Here, $\{b_{A\!B}\}$ is the multidimensional
variable $\{b,\{b_{i}^{A}\},\{b_{j}^{B}\}\}$ (see Fig. \ref{b-variables}),
and $\int\!\!db\!_{A\!B}$ the corresponding integration, defined
in Eqs. (\ref{b-AB-integration}) and (\ref{thickness-function}).
Equation (\ref{G-AB}) indicates that $G^{AB}$ may be interpreted
as the probability of having an interaction for a given impact parameter $b$
and given transverse coordinates $\{b_{i}^{A}\}$ and $\{b_{j}^{B}\}$ of
the nucleons. Comparing with Eq. (\ref{sigma-inel-AB-1}), one gets
\begin{align}
G^{AB}(s,\{b_{A\!B}\}) & =\sum_{n_{1}=0}^{\infty}\!\!\ldots\!\!\sum_{n_{AB}=0}^{\infty}\quad\sum_{m_{1}\le n_{1}}\!\!\ldots\!\!\sum_{m_{AB}\le n_{AB}}\int dX\!_{A\!B}\nonumber \\
 & \qquad\prod_{k=1}^{AB}\frac{1}{n_{k}!}\mathrm{\Big(\!\!\begin{array}{c}
n_{k}\\
m_{k}
\end{array}\!\!\Big)}\prod_{\mu=1}^{m_{k}}G_{k\mu}\prod_{\mu=m_{k}+1}^{n_{k}}-G_{k\mu}\nonumber \\
 & \qquad\prod_{i=1}^{A}V(x_{\mathrm{remn},i}^{+})\prod_{j=1}^{B}V(x_{\mathrm{remn},j}^{-}),\label{sigma-inel-AB-2}
\end{align}
with at least one $n_{k}$ being nonzero, and with the momentum fraction
integrations $\int dX\!_{A\!B}$ being defined in Eq. (\ref{x-AB-integration}).
One may define an expression $\bar{G}^{AB}(s,\{b_{A\!B}\})$ as Eq.
(\ref{sigma-inel-AB-2}), but for all $n_{k}=0$, which represents
the case excluded in the summation of Eq. (\ref{sigma-inel-AB-2}),
which might be interpreted as the probability of ``no interaction''.
One then gets 
\begin{align}
 & G^{AB}+\bar{G}^{AB}\!=\!\sum_{n_{1}=0}^{\infty}\!\!\ldots\!\!\sum_{n_{AB}=0}^{\infty}\quad\sum_{m_{1}\le n_{1}}\!\!\ldots\!\!\sum_{m_{AB}\le n_{AB}}\int dX\!_{A\!B}\nonumber \\
 & \qquad\qquad\prod_{k=1}^{AB}\frac{1}{n_{k}!}\mathrm{\Big(\!\!\begin{array}{c}
n_{k}\\
m_{k}
\end{array}\!\!\Big)}\prod_{\mu=1}^{m_{k}}G_{k\mu}\prod_{\mu=m_{k}+1}^{n_{k}}-G_{k\mu}\nonumber \\
 & \qquad\qquad\prod_{i=1}^{A}V(x_{\mathrm{remn},i}^{+})\prod_{j=1}^{B}V(x_{\mathrm{remn},j}^{-}),\label{sigma-inel-AB-3}
\end{align}
without any constraint for the $n_{k}$. Using similar manipulations
as for the $d\sigma_{\mathrm{incl}}^{AB}/dx^{+}dx^{-}$ calculation above,
one gets 
\begin{equation}
G^{AB}(s,\{b_{A\!B}\})+\bar{G}^{AB}(s,\{b_{A\!B}\})=\prod_{i=1}^{A}V(1)\prod_{j=1}^{B}V(1),
\end{equation}
and with (by definition) $V(1)=1$, one finds that the right-hand side (rhs) of this
equation is unity.\medskip{}

So I have proven
\begin{equation}
G^{AB}(s,\{b_{A\!B}\})+\bar{G}^{AB}(s,\{b_{A\!B}\})=1,\label{G-plus-Gbar}
\end{equation}
with $G^{AB}(s,\{b_{A\!B}\})$ and $\bar{G}^{AB}(s,\{b_{A\!B}\})$
being interpreted as probabilities to interact for a given impact parameter
$b$ and given transverse coordinates $\{b_{i}^{A}\}$ and $\{b_{j}^{B}\}$ of
the nucleons, and to not interact, respectively. %
This sounds promising. So far the interpretations seem to be consistent.\medskip{} 

From Eqs. (\ref{G-plus-Gbar}) and (\ref{sigma-inel-AB-3}), one gets
\begin{align}
 & 1=\sum_{n_{1}=0}^{\infty}\!\!\ldots\!\!\sum_{n_{AB}=0}^{\infty}\quad\sum_{m_{1}\le n_{1}}\!\!\ldots\!\!\sum_{m_{AB}\le n_{AB}}\int dX\!_{A\!B}\nonumber \\
 & \qquad\qquad\prod_{k=1}^{AB}\frac{1}{n_{k}!}\mathrm{\Big(\!\!\begin{array}{c}
n_{k}\\
m_{k}
\end{array}\!\!\Big)}\prod_{\mu=1}^{m_{k}}G_{k\mu}\prod_{\mu=m_{k}+1}^{n_{k}}-G_{k\mu}\nonumber \\
 & \qquad\qquad\prod_{i=1}^{A}V(x_{\mathrm{remn},i}^{+})\prod_{j=1}^{B}V(x_{\mathrm{remn},j}^{-}).\label{sigma-inel-AB-4}
\end{align}
Let me rename the light-cone momentum variables: let $x_{k\mu}^{\pm}$
be the light-cone momenta of the $\mu$th cut Pomeron for
pair $k$, and $\tilde{x}_{k\lambda}^{\pm}$ those of the $\lambda$th
uncut Pomeron for pair $k$. Instead of $\int dX\!_{A\!B}$, one has
 $\int dX\!_{A\!B}\:\int d\tilde{X}\!_{A\!B}$, where now 
\begin{equation}
\int dX\!_{A\!B}=\int\,\prod_{k=1}^{AB}\bigg(\prod_{\mu=1}^{m_{k}}dx_{k\mu}^{+}dx_{k\mu}^{-}\bigg)\label{x-AB-integration-1}
\end{equation}
refers to cut Pomerons and 
\begin{equation}
\int d\tilde{X}\!_{A\!B}=\int\,\prod_{k=1}^{AB}\bigg(\prod_{\lambda=1}^{l_{k}}d\tilde{x}_{k\lambda}^{+}d\tilde{x}_{k\lambda}^{-}\bigg)\label{x-AB-integration-2}
\end{equation}
to uncut Pomerons. I use $G'_{k\lambda}\!=G\!(\!\tilde{x}_{k\lambda}^{+},\tilde{x}_{k\lambda}^{-},s,b_{k}\!)$.
All this allows one to reorganize the sums in Eq. (\ref{sigma-inel-AB-4}),
and one may write 
\begin{align}
 & 1=\sum_{m_{1}=0}^{\infty}\!\!\ldots\!\!\sum_{m_{AB}=0}^{\infty}\quad\sum_{l_{1}=0}^{\infty}\!\!\ldots\!\!\sum_{l_{AB}=0}^{\infty}\int dX\!_{A\!B}\int d\tilde{X}\!_{A\!B}\nonumber \\
 & \qquad\qquad\prod_{k=1}^{AB}\left[\frac{1}{m_{k}!}\prod_{\mu=1}^{m_{k}}G_{k\mu}\right]\prod_{k=1}^{AB}\left[\frac{1}{l_{k}!}\prod_{\lambda=1}^{l_{k}}-G'_{k\lambda}\right]\nonumber \\
 & \qquad\qquad\prod_{i=1}^{A}V(x_{\mathrm{remn},i}^{+})\prod_{j=1}^{B}V(x_{\mathrm{remn},j}^{-}),\label{sigma-inel-AB-5}
\end{align}
where $m_{k}$ refers to the number of cut Pomerons and $l_{k}$ to
the number of uncut ones, for pair $k$. One recalls that $G_{k\mu}$
is defined as $G\!(\!x_{k\mu}^{+},x_{k\mu}^{-},s,b_{k}\!)$ with $b_{k}=|b+b_{\pi(k)}^{A}-b_{\tau(k)}^{B}|$.
I define ``cut Pomeron configurations''
\begin{equation}
K=\big\{\{m_{k}\},\{x_{k\mu}^{\pm}\}\big\},
\end{equation}
with $\{m_{k}\}$ being the set of the $m_{k}$ variables, and $\{x_{k\mu}^{\pm}\}$
the set of the $x_{k\mu}^{\pm}$ variables, for given values of the
nuclear impact parameter $b$ and the transverse coordinates $b_{i}^{A}$
and $b_{j}^{B}$ of all the nucleons in nuclei $A$ and $B$. The
sum $\sum_{m_{1}=0}^{\infty}\ldots\sum_{m_{AB}=0}^{\infty}$ is then
simply the sum over all possible values of $\{m_{k}\}$, and Eq. (\ref{sigma-inel-AB-5})
becomes 
\begin{equation}
1=\sum_{\{m_{k}\}}\int dX\!_{A\!B}P\,,
\end{equation}
or explicitly
\begin{equation}
1=\sum_{\{m_{k}\}}\int\prod_{k=1}^{AB}\prod_{\mu=1}^{m_{k}}dx_{k\mu}^{+}dx_{k\mu}^{-}\;P(\{m_{k}\},\{x_{k\mu}^{\pm}\}),\label{proba-law-1}
\end{equation}
with
\begin{equation}
P(\{m_{k}\},\{x_{k\mu}^{\pm}\})=\prod_{k=1}^{AB}\left[\frac{1}{m_{k}!}\prod_{\mu=1}^{m_{k}}G_{k\mu}\right]\times W_{AB},\label{proba-law-2}
\end{equation}
with
\begin{align}
W_{AB} & =\sum_{l_{1}=0}^{\infty}\!\!\ldots\!\!\sum_{l_{AB}=0}^{\infty}\int d\tilde{X}\!_{A\!B}\prod_{k=1}^{AB}\left[\frac{1}{l_{k}!}\prod_{\lambda=1}^{l_{k}}-G'_{k\lambda}\right]\label{W-1}\\
 & \prod_{i=1}^{A}V(x_{i}^{+}-\!\!\sum_{\underset{\pi(k)=i}{k=1}}^{AB}\sum_{\lambda=1}^{l_{k}}\!\tilde{x}_{k\lambda}^{+})\prod_{j=1}^{B}V(x_{j}^{-}-\!\!\sum_{\underset{\tau(k)=j}{k=1}}^{AB}\sum_{\lambda=1}^{l_{k}}\!\tilde{x}_{k\lambda}^{-}),\nonumber 
\end{align}
with
\begin{equation}
x_{i}^{+}=1-\!\!\sum_{\underset{\pi(k)=i}{k=1}}^{AB}\sum_{\mu=1}^{m_{k}}\!x_{k\mu}^{+},\quad x_{j}^{-}=1-\!\!\sum_{\underset{\tau(k)=j}{k=1}}^{AB}\sum_{\mu=1}^{m_{k}}\!x_{k\mu}^{-}.\label{W-2}
\end{equation}
The term $W_{AB}$ is a function of the variables $\{x_{i}^{+}\}$
with $1\le i\le A$ and $\{x_{j}^{-}\}$ with $1\le j\le B$, i.e.
$W_{AB}=W_{AB}(\{x_{i}^{+}\},\{x_{j}^{-}\})$. It also depends on
$s$ and $b$ and the transverse coordinates $b_{i}^{A}$ and $b_{j}^{B}$
of all the nucleons, not written explicitly. \medskip{}

One concludes:
\begin{itemize}
\item
The equation $1=\sum_{\{m_{k}\}}\int dX\!_{A\!B}P$ allows the interpretation
of 
\begin{equation}
P=\prod_{k=1}^{AB}\left[\frac{1}{m_{k}!}\prod_{\mu=1}^{m_{k}}G_{k\mu}\right]\times W_{AB}(\{x_{i}^{+}\},\{x_{j}^{-}\})\label{proba-interpretation}
\end{equation}
to be the probability of the ``configuration'' $K$,
\item
where $K=\big\{\{m_{k}\},\{x_{k\mu}^{\pm}\}\big\}$ represents $m_{k}$ cut Pomerons per pair $k$, with light-cone
momentum fractions $x_{k\mu}^{\pm}$ .%
\end{itemize}
\medskip{}

This would be a perfect basis for Monte Carlo applications: one ``simply''
needs to generate configurations according to that law. But what is
still missing: one needs to prove that all the $P(\{m_{k}\},\{x_{k\mu}^{\pm}\})$
are non-negative, and to do so, one has  to prove that $W_{AB}$ is
non-negative. \\

As a side remark (just to understand the meaning of the quantity $W_{AB}$),
one has  
\begin{equation}
W_{AB}(\{x_{i}^{+}=1\},\{x_{j}^{-}=1\})=P(\{m_{k}=0\}),
\end{equation}
corresponding to no cut Pomerons at all, which means ``no interaction''
in the sense of inelastic scattering. The expression $(1-W_{AB})$
with all momentum fractions $x_{i}^{+}$ and $x_{j}^{-}$ being unity, corresponds to inelastic
scattering, summed over all possible configurations, and the inelastic
cross section is obtained by integrating $\int\!\!db\!_{A\!B}...$,
as
\begin{equation}
\sigma_{\mathrm{in}}=\int\!\!db\!_{A\!B}\left\{ 1-W_{AB}(\{x_{i}^{+}=1\},\{x_{j}^{-}=1\})\right\} .
\end{equation}
For example, the $pp$ inelastic cross section is an integral over the impact
parameter of $1-W_{11}(1,1)$. This again confirms the probabilistic
interpretation, but still, under the condition $W_{AB}\ge0$.\\

Let me now discuss how to calculate $W_{AB}.$ To do so, one needs
to specify $G(\!x^{+},x^{-},s,b\!)$. Using $G=G_{\mathrm{QCD}}$
(see Eq. (\ref{G-equal-G-QCD}) and the discussion before), one can
show \cite{Drescher:2000ha} that one may obtain an almost perfect
fit of the numerically computed functions $G_{\mathrm{QCD}}$, with
a parametrization of the form
\begin{equation}
G_{\mathrm{QCDpar}}(\!x^{+},x^{-},s,b\!)=\sum_{N=1}^{N_{\mathrm{par}}}\alpha_{N}(x^{+}x^{-})^{\beta_{N}},\label{QCDpar}
\end{equation}
where $\alpha_{N}$ and $\beta_{N}$ depend on $s$ and $b$ given
in terms of a few parameters, as shown in Appendix \ref{======= W-AB =======}.
This parametric form has been inspired by the asymptotic expressions
for T-matrices (see Appendix \ref{======= asymptotic-behavior =======}).
So I will use in the following
\begin{equation}
G=G_{\mathrm{QCDpar}}.
\end{equation}
Furthermore, the vertices are parametrized as
\begin{equation}
V(x)=x^{\alpha_{\mathrm{remn}}},\label{fit-4-1}
\end{equation}
again motivated by the asymptotic expressions for T-matrices. \\

Now one has  all the ingredients to compute $W_{AB}$. All the formulas
will be given with the current choice of $N_{\mathrm{par}}=4$, it
is straightforward to change this value. From Eq. (\ref{W-1}), one
gets
\begin{align}
 & W_{AB}(\{x_{i}^{+}\},\!\{x_{j}^{-}\})=\sum_{\{l_{k}\}}\int\prod_{k=1}^{AB}\bigg(\prod_{\lambda=1}^{l_{k}}d\tilde{x}_{k\nu}^{+}d\tilde{x}_{k\nu}^{-}\bigg)\label{W-1-4}\\
 & \quad\Bigg\{\prod_{k=1}^{AB}\left[\frac{1}{l_{k}!}\prod_{\lambda=1}^{l_{k}}-G_{\mathrm{QCDpar}}(\!\tilde{x}_{k\lambda}^{+},\tilde{x}_{k\lambda}^{-},s,b_{k}\!)\right]\nonumber \\
 & \qquad\quad\prod_{i=1}^{A}\Big(x_{i}^{+}-\!\!\sum_{\underset{\pi(k)=i}{k=1}}^{AB}\sum_{\lambda=1}^{l_{k}}\!\tilde{x}_{k\lambda}^{+}\Big)^{\alpha_{\mathrm{remn}}}\nonumber \\
 & \qquad\quad\prod_{j=1}^{B}\Big(x_{j}^{-}-\!\!\sum_{\underset{\tau(k)=j}{k=1}}^{AB}\sum_{\lambda=1}^{l_{k}}\!\tilde{x}_{k\lambda}^{-}\Big)^{\alpha_{\mathrm{remn}}}\Bigg\},\nonumber
\end{align}
where $\sum_{\{l_{k}\}}$ means summing all the indices $l_{k}$,
with $1\le k\le AB$, from zero to infinity, where $l_{k}$ refers to
the number of uncut Pomerons of nucleon-nucleon pair $k$. Remember
that computing $W_{AB}$ amounts to summing over and integrating out
all uncut Pomerons. As proven in Appendix \ref{======= W-AB =======}
[see Eq. (\ref{W-AB-1})], one finds 
\begin{align}
 & W_{AB}(\{x_{i}^{+}\},\!\{x_{j}^{-}\})=\label{W-AB-1-1}\\
 & \prod_{i=1}^{A}(x_{i}^{+})^{\alpha_{\mathrm{remn}}}\prod_{j=1}^{B}(x_{j}^{-})^{\alpha_{\mathrm{remn}}}\sum_{\{r_{Nk}\}}\Bigg\{\prod_{k=1}^{AB}\prod_{N=1}^{4}\frac{(-\alpha_{N})^{r_{Nk}}}{r_{Nk}!}\nonumber \\
 & \prod_{i=1}^{A}\Bigg[\prod_{\underset{\pi(k)=i}{k=1}}^{AB}\prod_{N=1}^{4}\left(\Gamma(\tilde{\beta}_{N})(x_{i}^{+})^{\tilde{\beta}_{N}}\right)^{r_{Nk}}g(\sum_{\underset{\pi(k)=i}{k=1}}^{AB}\sum_{N=1}^{4}r_{Nk}\tilde{\beta}_{N})\Bigg]\nonumber \\
 & \prod_{j=1}^{B}\Bigg[\prod_{\underset{\tau(k)=j}{k=1}}^{AB}\prod_{N=1}^{4}\left(\Gamma(\tilde{\beta}_{N})(x_{j}^{-})^{\tilde{\beta}_{N}}\right)^{r_{Nk}}g(\sum_{\underset{\tau(k)=j}{k=1}}^{AB}\sum_{N=1}^{4}r_{Nk}\tilde{\beta}_{N})\Bigg]\,\Bigg\},\nonumber 
\end{align}
where $\sum_{\{r_{Nk}\}}$ means summing all the indices $r_{Nk}$,
with $1\le N\le4$ and with $1\le k\le AB$, from zero to infinity, where
$r_{Nk}$ refers to the number of uncut Pomerons of type $N$
of nucleon-nucleon pair $k$. It is useful for the discussion to consider
``Pomeron types'' $N$, although they are not physical objects,
just coming from the parametrization in Eq. (\ref{QCDpar}). I use $\tilde{\beta}_{N}=\beta_{N}+1$,
and a function $g$ defined as 
\begin{equation}
g(z)=\frac{\Gamma(1+\alpha_{\mathrm{remn}})}{\Gamma(1+\alpha_{\mathrm{remn}}+z)}.\label{g-definition}
\end{equation}
The result in Eq. (\ref{W-AB-1-1}) is remarkable in the sense that all
the integrations could be done analytically, expressed in terms of
gamma functions $\Gamma$. \\

One cannot simplify Eq. (\ref{W-AB-1-1}) any further, without approximations.
Looking at this expression, one sees that the two terms $g(...)$ ``make
the problem''. Without them, the sums could be done, and one would
get products of exponentials. But of course, one cannot simply drop
these terms. However, if the $g$ functions would have a particular
property, namely, a factorization such as
\begin{equation}
g\left(\sum_{\lambda}\tilde{\beta}_{\lambda}\right)=c_{1}\prod_{\lambda}c_{2}\,g(c_{3}\,\tilde{\beta}_{\lambda}),\label{g-property-1}
\end{equation}
for given coefficients $\tilde{\beta}_{\lambda}$ (arbitrary, but
in practice of order unity) and three parameters $c_{\mu}$, then the
expression in Eq. (\ref{W-AB-1-1}) can be factorized such that the infinite
sums are finally just a product of exponentials, and as proven in
Appendix \ref{======= W-AB =======}
[see Eq. (\ref{W-AB-7})], and one gets
\begin{align}
 & W_{AB}=\prod_{i=1}^{A}c_{1}(x_{i}^{+})^{\alpha_{\mathrm{remn}}}\prod_{j=1}^{B}c_{1}(x_{j}^{-})^{\alpha_{\mathrm{remn}}}\label{W-AB-7-1}\\
 & \qquad\qquad\prod_{k=1}^{AB}\exp\left(-\tilde{G}(x_{\pi(k)}^{+}x_{\tau(k)}^{-})\right),\nonumber 
\end{align}
for any choice of the parameters $c_{\mu}$, where $\tilde{G}$ is
given as 
\begin{equation}
\tilde{G}(x)=\sum_{N=1}^{4}\tilde{\alpha}_{N}x^{\tilde{\beta}_{N}},
\end{equation}
with
\begin{align}
\tilde{\alpha}_{N} & =\alpha_{N}\left(\frac{\Gamma(\tilde{\beta}_{N})\,c_{2}\,\Gamma(1+\alpha_{\mathrm{remn}})}{\Gamma(1+\alpha_{\mathrm{remn}}+c_{3}\,\tilde{\beta}_{N})}\right)^{2},\\
\tilde{\beta}_{N} & =\beta_{N}+1\;.
\end{align}
What is very nice: $W_{AB}$ as given in Eq. (\ref{W-AB-7-1}) is
strictly non-negative, and this was the missing piece which allows a
probabilistic interpretation of the formulas [see the discussion around
Eq. (\ref{proba-interpretation})], which is extremely important for
any Monte Carlo application.\\

But the question arises: Is the property in Eq. (\ref{g-property-1})
really true? To answer this question, I make some simulations, where
I take $\alpha_{\mathrm{remn}}=1$ (currently used), so one has  $g(z)=\Gamma(2)/\Gamma(2+z)$.
In principle, Eq. (\ref{g-property-1}) should be valid for any choice
of sequences $\tilde{\beta}_{\lambda}$. So I generate randomly integer
numbers $\lambda_{\mathrm{MAX}}$ between zero and some upper limit
(20), and then I generate a sequence $\tilde{\beta}_{1}$, $\tilde{\beta}_{2}$,
$...$, $\tilde{\beta}_{\lambda_{\mathrm{MAX}}}$ of $\lambda_{\mathrm{MAX}}$
uniformly
\begin{figure}[h]
\begin{centering}
\includegraphics[bb=10bp 20bp 595bp 420bp,clip,scale=0.39]
{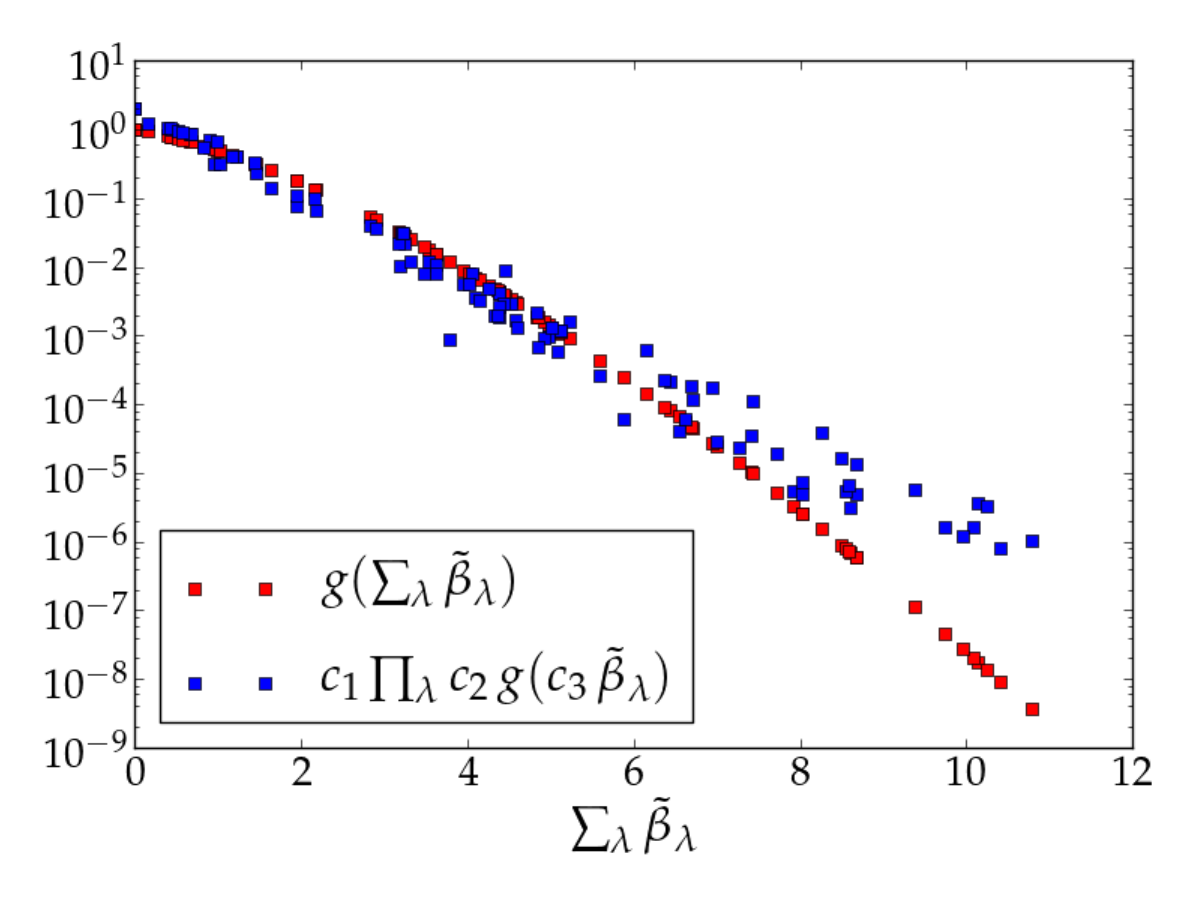}
\par\end{centering}
\centering{}\caption{Comparing $g\left(\sum_{\lambda}\tilde{\beta}_{\lambda}\right)$ and
$c_{1}\prod_{\lambda}c_{2}\,g(c_{3}\,\tilde{\beta}_{\lambda})$ for
randomly created sequences $\tilde{\beta}_{\lambda}$, for the parameter
choice (A) $c_{1}=2$, $c_{2}=0.65$, and $c_{3}=1$. \label{gfunction-1}}
\end{figure}
\begin{figure}[h]
\begin{centering}
\includegraphics[bb=10bp 20bp 595bp 420bp,clip,scale=0.39]
{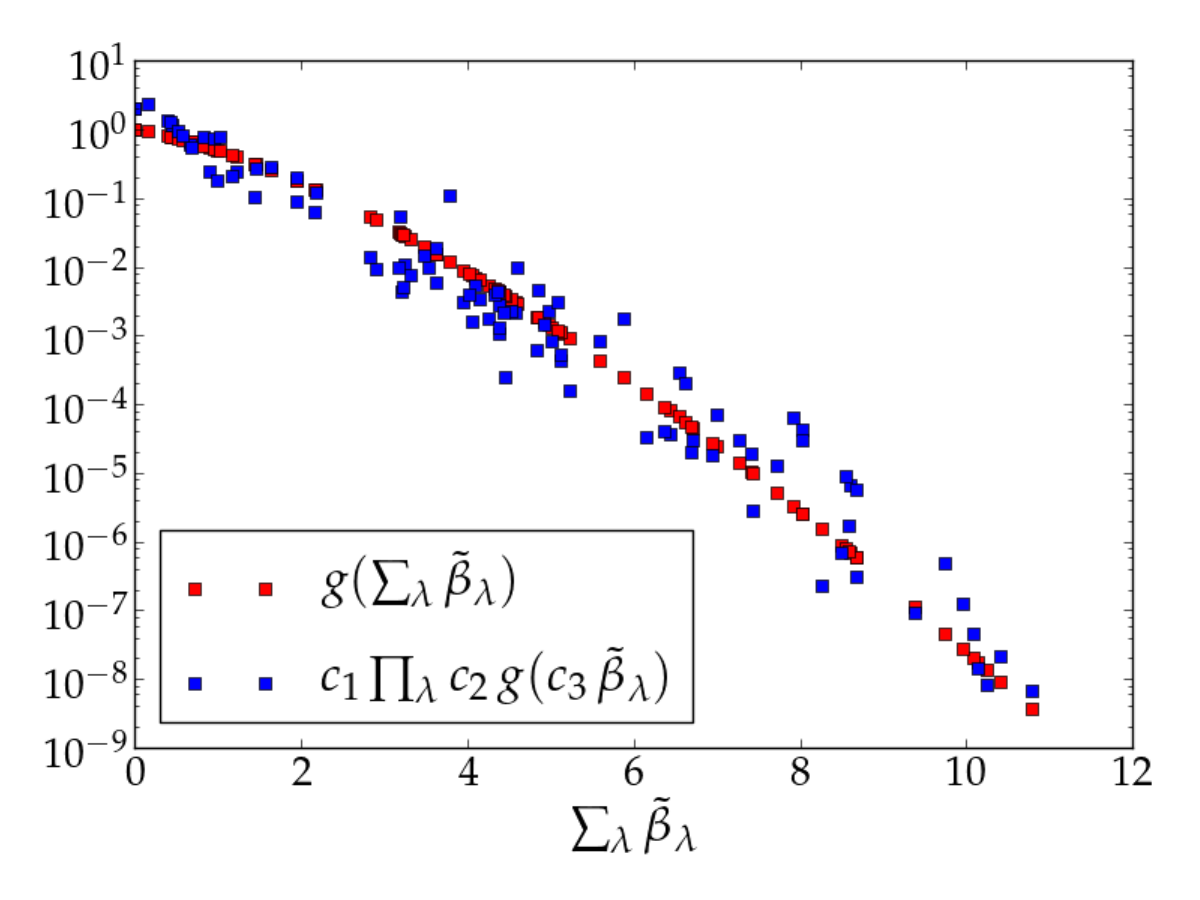}
\par\end{centering}
\centering{}\caption{Same as Fig. \ref{gfunction-1}, but for the parameter choice (B)
$c_{1}=2$, $c_{2}=1.5$, and $c_{3}=2.8$. \label{gfunction-1-1}}
\end{figure}
distributed random numbers. For each sequence, I compute $g\left(\sum_{\lambda}\tilde{\beta}_{\lambda}\right)$
and $c_{1}\prod_{\lambda}c_{2}\,g(c_{3}\,\tilde{\beta}_{\lambda})$,
and then I compare the two. In Fig. \ref{gfunction-1}, I plot
the two quantities as a function of $\sum_{\lambda}\tilde{\beta}_{\lambda}$,
for the parameter choice (A) $c_{1}=2$, $c_{2}=0.65$, and $c_{3}=1$.
In Fig. \ref{gfunction-1-1}, I show the comparison for a second parameter
choice (B) $c_{1}=2$, $c_{2}=1.5$, and $c_{3}=2.8$. Neither of
the two choices is perfect (one will understand later that a perfect
choice cannot exist). Choice A is somewhat better for small $\sum_{\lambda}\tilde{\beta}_{\lambda}$,
but then deviates for large values, whereas choice B gives an overall
good description, but with bigger fluctuations. In the following,
I will take choice A, mainly because B has only been discovered
recently, and all EPOS4 simulations are based on A. A $g$-factorization
similar to choice A was employed in Ref. \cite{pierog:2002}.\\

\begin{figure}[h]
\begin{centering}
\includegraphics[bb=40bp 50bp 595bp 490bp,clip,scale=0.33]
{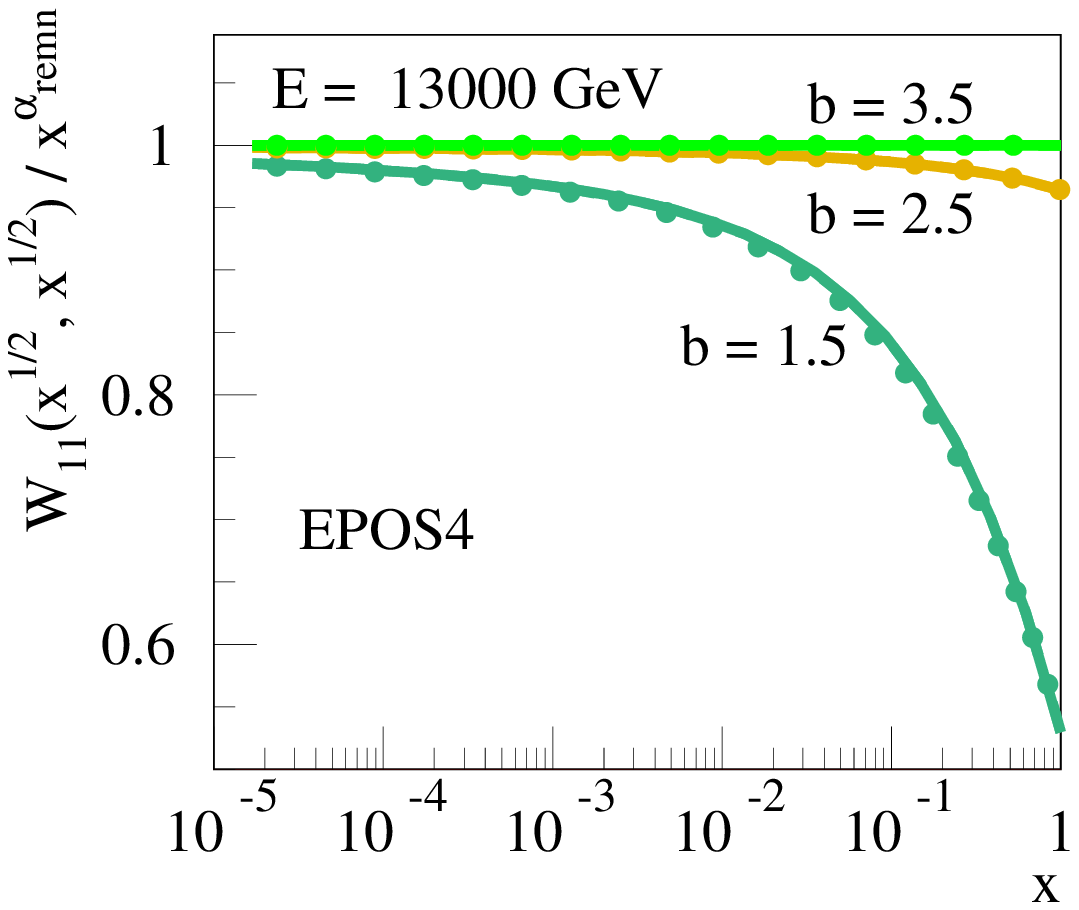}
\par\end{centering}
\centering{}\caption{$W_{11}(\sqrt{x},\sqrt{x})$ $/$ $x^{\alpha_{\mathrm{remn}}}$ as
a function of $x=x^{+}x^{-}$, for $pp$ scattering at $13\,\mathrm{TeV}$,
for the impact parameters $b=3.5$, $2.5$, and $1.5$ fm. The
solid curves refer to the exact results, and the dotted ones to the
calculations based on $g$-factorization [Eq. (\ref{g-property-1})].
\label{w11-versus-x}}
\end{figure}

The next obvious question is: How does the $g$-factorization [see Eq.
(\ref{g-property-1})] compare to the exact result, concerning $W_{AB}$?
By exact result I mean a numerical calculation of Eq. (\ref{W-AB-1-1}),
which at least for $pp$ scattering ($A=B=1$) can be easily done, since
the infinite sums converge fast. One recalls that $W_{11}$ depends
also on the impact parameter, and on the collision energy. In Fig.
\ref{w11-versus-x},
I show $W_{11}(\sqrt{x},\sqrt{x})$ divided by the trivial factor
$x^{\alpha_{\mathrm{remn}}}$, as a function of $x=x^{+}x^{-}$ (so
I use here $x^{+}=x^{-}$), for $pp$ scattering at $13\,\mathrm{TeV}$,
for the impact parameters $b=3.5$, $2.5$, and $1.5$ fm. The
solid curves refer to the exact results, and the dotted ones to the calculations
based on $g$-factorization [see Eq. (\ref{g-property-1})]. The two methods
give very similar results, and they look reasonable: at very large
$b$, one gets for $x=1$ unity, as it should be, since $W_{11}(1,1)$
is interpreted as the probability of no interaction. With decreasing $b$,
the probability of no interaction decreases.

\begin{figure}[h]
\begin{centering}
\includegraphics[bb=40bp 50bp 595bp 490bp,clip,scale=0.33]
{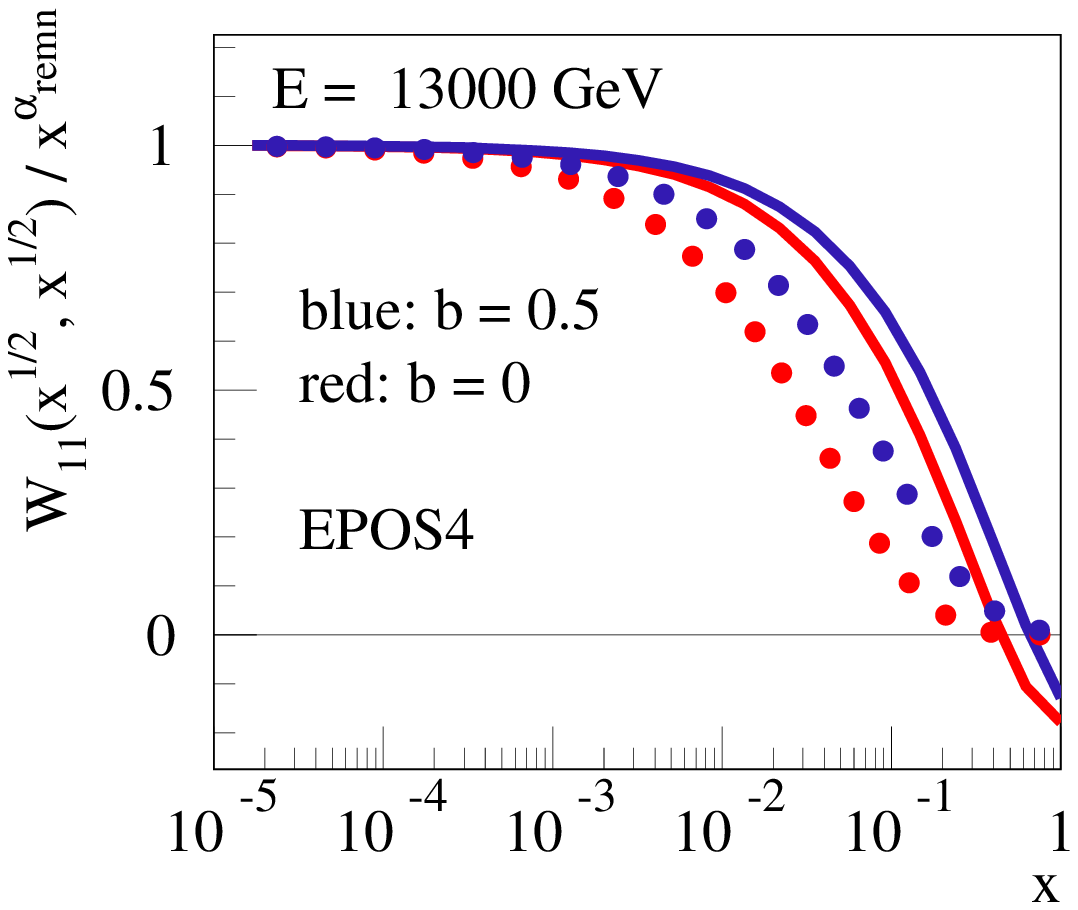}
\par\end{centering}
\centering{}\caption{As Fig. \ref{w11-versus-x}, but for impact parameters $b=0.5$ and
$0$ fm. \label{w11-versus-x-1}}
\end{figure}

The situation changes completely when one considers small values
of the impact parameter $b$. In Fig. \ref{w11-versus-x-1},
\begin{figure}[h]
\begin{centering}
\includegraphics[bb=40bp 50bp 595bp 490bp,clip,scale=0.33]
{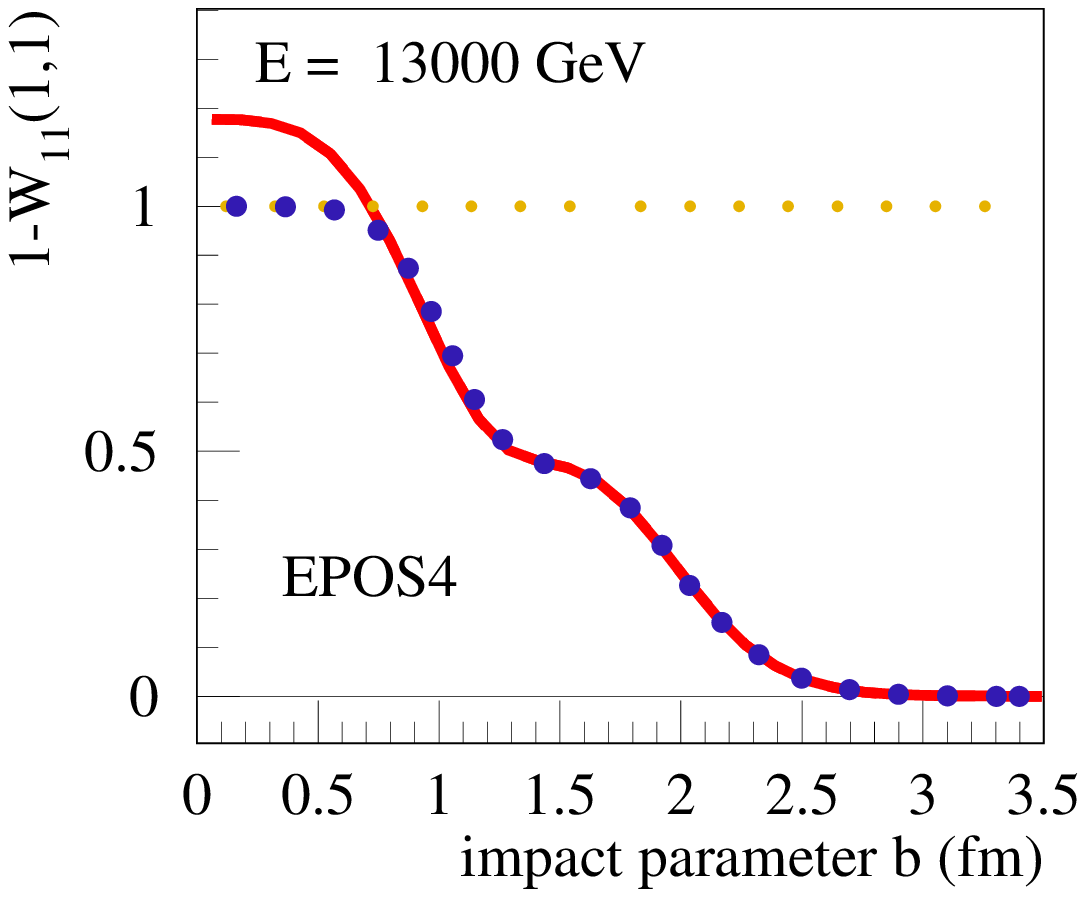}
\par\end{centering}
\centering{}\caption{$1-W_{11}(1,1)$ as a function of $b$, for $pp$ scattering at $13\,\mathrm{TeV}$.
The solid curve refers to the exact result, and the dotted one to
the calculations based on $g$-factorization [see Eq. (\ref{g-property-1})].
\label{w11-versus-b}}
\end{figure}
I show $W_{11}(\sqrt{x},\sqrt{x})$ divided by $x^{\alpha_{\mathrm{remn}}}$
as a function of $x=x^{+}x^{-}$, for impact parameters $b=0.5$ and
$0$ (in fm). Again the solid curves refer to the exact results, and the
dotted ones to the calculations based on $g$-factorization. The
results are amazing: Not only are the curves from the two methods
substantially different, but $W_{11}$ even becomes negative at large
$x$ for the exact calculations, whereas the g-factorized results
converge towards zero, as it should be. 

I will now focus on the large $x$ behavior, and consider $1-W_{11}$
for $x^{+}=x^{-}=1$. As discussed earlier, $1-W_{11}(1,1)$ is interpreted
as the probability of an inelastic interaction (summed over all possible
numbers of exchanged Pomerons). I plot $1-W_{11}(1,1)$ as a function
of the impact parameter $b$, as shown in Fig. \ref{w11-versus-b},
where the solid curve refers to the exact results, and the dotted one
to the calculations based on $g$-factorization [see Eq. (\ref{g-property-1})].
The two curves are very close for large impact parameters but differ
for small $b$ (compatible with Figs. \ref{w11-versus-x} and \ref{w11-versus-x-1}).
For small $b$, the curve based on $g$-factorization approaches unity,
as it should, but the exact calculation (solid curve) exceeds unity.

\subsection{Conclusion}

So one observes some ``unphysical behavior'' in the sense that $1-W_{11}(1,1)$,
interpreted as the ``probability of an inelastic scattering'', exceeds
unity at small impact parameter $b$, or the other way around, the
quantity $W_{11}(1,1)$, interpreted as ``probability of no interaction'',
becomes negative. This is what I refer to as the ``sign problem''.
Imposing $g$-factorization [see Eq. (\ref{g-property-1})] provides an
excellent approximation to the exact solution at large impact parameter,
but deviates at small $b$ by correcting the sign problem. Therefore
it is a ``regularization'' (which fixes a problem) rather than a
simple approximation.\\

Let me summarize this section (containing a large amount of information):
\begin{itemize}
\item One generalizes the Gribov-Regge (GR) S-matrix approach, to include
energy-momentum sharing (GR\textsuperscript{+}). 
\item One writes an expression for the inelastic nucleus-nucleus cross
section $\sigma_{\mathrm{in}}^{AB}$ summing over all possible numbers
of cut and uncut Pomerons for each nucleon-nucleon pair, integrating
over all Pomeron momenta, imposing strict energy-momentum sharing.
The elementary object representing a single Pomeron is $G$ (cut T-matrix
element).
\item One proves the validity of the AGK theorem, which is an extremely
important issue, since it allows one to deduce binary scaling in $AA$ scattering
and factorization in pp scattering.
\item From the expression for $\sigma_{\mathrm{in}}^{AB}$, one deduces
an equation $1=\sum_{\{m_{k}\}}\int dX\!_{A\!B}P$, which allows the
interpretation of 
\begin{equation}
P=\prod_{k=1}^{AB}\Big[\frac{1}{m_{k}!}\prod_{\mu=1}^{m_{k}}G_{k\mu}\Big]\times W_{AB}(\{x_{i}^{+}\},\{x_{j}^{-}\})\label{proba-interpretation-1}
\end{equation}
to be the probability of the ``configuration'' $K=\big\{\{m_{k}\},\{x_{k\mu}^{\pm}\}\big\}$,
representing $m_{k}$ cut Pomerons per pair $k$, with light-cone
momentum fractions $x_{k\mu}^{\pm}$ . This is very important since
it allows Monte Carlo applications perfectly compatible with the theoretical
S-matrix-based formulas. But one has  to prove: $W_{AB}\ge0$.
\item The computation of $W_{AB}$ is challenging due to a very high-dimensional
nonseparable integration. Introducing a particular (and well-justified)
parametrization of the $x^{\pm}$ dependence of $G$ as $\sum_{N=1}^{N_{\mathrm{par}}}\alpha_{N}(x^{+}x^{-})^{\beta_{N}},$
the integration can be done, and one deduces an analytic formula for $W_{AB}$. 
\item $W_{AB}$ is still a complex expression, given as $\sum_{\{r_{Nk}\}}...$,
which means summing indices $r_{Nk}$ (from 0 to $\infty$), with
$1\le N\le4$ and with $1\le k\le AB$. Doing a ``small'' manipulation
(let me call it ``regularization''), the sums can be done, one gets
a simple product of exponentials, and, most importantly, $W_{AB}\ge0$,
as it should be.
\item For $pp$ scattering, the regularized results and the exact results can be shown to be identical,
for small values of the remnant momentum fractions $x^{\pm}$ (1 minus
the sum of the Pomeron momentum fractions) and always for large impact
parameters.
\item But for small impact parameter, for large $x^{\pm}$, and in particular
for $x^{\pm}=1$, the value of $W_{11}$ becomes negative.
And $1-W_{11}$, to be interpreted as the probability of an inelastic
interaction, exceeds unity. Here, the regularized results deviate
from the exact ones, since for the former one gets strictly $1-W_{11}=1$ for small impact
parameters. 
\end{itemize}
Imposing this regularization seems to solve the problem of negative
probabilities, but also indicates that there still is something missing,
and the theoretical framework is not yet complete. This will be discussed
in the next section. 

\section{AGK violation and deformation functions in the regularized theory
\label{======= AGK-violation-deformation =======}}

In the last section, I developed the parallel scattering S-matrix
approach for heavy-ion scattering, \textbf{with energy-momentum sharing},
the latter being crucial for any realistic application. I could
prove the validity of the AGK theorem, which is an essential property,
the condition for binary scaling and factorization. 
An expression [see Eq. (\ref{proba-interpretation})]
\begin{equation}
P(K)=\prod_{k=1}^{AB}\left[\frac{1}{m_{k}!}\prod_{\mu=1}^{m_{k}}G_{k\mu}\right]\times W_{AB}(\{x_{i}^{+}\},\{x_{j}^{-}\})\label{proba-interpretation-2}
\end{equation}
could be derived, to be interpreted as a probability law for particular multi-Pomeron
configurations $K=\big\{\{m_{k}\},\{x_{k\mu}^{\pm}\}\big\}$.
The $W_{AB}$ term turned out to be
negative, for certain values of the arguments, which is unphysical.
A ``regularization'' allowed one to solve this problem, and all $W_{AB}$
(and the P's) are finally non-negative. In addition, this regularization
changes $\sigma_{\mathrm{in}}^{AB}$ only  little compared to
the exact calculation (at least for $A=B=1$). 

\subsection{AGK theorem in the regularized approach}

One has the impression that after the regularization the problem
is solved, but before coming to this conclusion, 
at least the (very important) AGK theorem has to be checked. 

Based on the (reestablished) probability interpretation
of the $P(K)$ expressions, one may generate corresponding configurations,
and assuming 
\begin{equation}
G=G_{\mathrm{QCD}}
\end{equation}
with $G_{\mathrm{QCD}}$ being discussed in detail in Ref. \cite{werner:2023-epos4-heavy}
[see also Fig. \ref{single-pomeron-graph-1} and Eq. (\ref{G-equal-G-QCD}],
one may generate partons, and employing the EPOS method to translate
partons into strings and then hadrons (as well discussed in Ref. \cite{werner:2023-epos4-heavy}),
one finally produces a final state of hadrons.
This allows the computation of inclusive cross sections with respect to 
the transverse momenta of partons or of hadrons. 
But I will first consider the inclusive cross  section with respect 
to the light-cone momentum fractions $x^{\pm}$ of the Pomerons, 
obtained from $P(K)$ defined in Eq. (\ref{proba-interpretation-2}) 
by summing over all $\{m_{k}\}$ and integrating over $b\!_{A\!B}$ and $\{x_{k\mu}^{\pm}\}$,
and by adding a factor 
\begin{equation}
\sum_{k'=1}^{AB}\sum_{\mu'=1}^{m_{k'}}\delta(x^{+}-x_{k'\mu'}^{+})\delta(x^{-}-x_{k'\mu'}^{-})dx^{+}dx^{-}\label{delta-times-delta-2}
\end{equation}
[see Eqs. (\ref{delta-times-delta}) and (\ref{sigma-incl-AB-1})]. 
One gets
\begin{comment}
\[
P(K)=\prod_{k=1}^{AB}\left[\frac{1}{m_{k}!}\prod_{\mu=1}^{m_{k}}G_{k\mu}\right]\times W_{AB}(\{x_{i}^{+}\},\{x_{j}^{-}\})
\]
see Eq. (\ref{proba-interpretation})

\begin{align*}
 & W_{AB}=\prod_{i=1}^{A}c_{1}(x_{i}^{+})^{\alpha_{\mathrm{remn}}}\prod_{j=1}^{B}c_{1}(x_{j}^{-})^{\alpha_{\mathrm{remn}}}\\
 & \qquad\qquad\prod_{k=1}^{AB}\exp\left(-\tilde{G}(x_{\pi(k)}^{+}x_{\tau(k)}^{-})\right),
\end{align*}
see Eq. (\ref{W-AB-7-1})
\[
x_{i}^{+}=1-\!\!\sum_{\underset{\pi(k)=i}{k=1}}^{AB}\sum_{\mu=1}^{m_{k}}\!x_{k\mu}^{+},\quad x_{j}^{-}=1-\!\!\sum_{\underset{\tau(k)=j}{k=1}}^{AB}\sum_{\mu=1}^{m_{k}}\!x_{k\mu}^{-}.
\]
\end{comment}
\begin{align}
 & \frac{d^{2}\sigma_{\mathrm{incl}}^{AB}}{dx^{+}dx^{-}}=\sum_{k'=1}^{AB}\;\sum_{\{m_{k}\}\ne0}\sum_{\mu'=1}^{m_{k'}}\int\!\!db\!_{A\!B}\!\!\int\!\!dX\!_{A\!B}\label{sigma-incl-AB-2}\\
 & \qquad\quad\Big\{ P(K)\delta(x^{+}-x_{k'\mu'}^{+})\delta(x^{-}-x_{k'\mu'}^{-})\Big\},\nonumber 
\end{align}
where $\int dX\!_{A\!B}$ represents the integration over $\{x_{k\mu}^{\pm}\}$,
see Eq. (\ref{x-AB-integration-1}), and $\int db\!_{A\!B}$, defined
in Eqs. (\ref{b-AB-integration}) and (\ref{thickness-function}), integrates
over the impact parameter and the transverse coordinates $b_{i}^{A}$
and $b_{j}^{B}$ of projectile and target nucleons. The weight $P$
is given in Eq. (\ref{proba-interpretation-2}), with $W_{AB}$ from
Eq. (\ref{W-AB-7-1}), and with the arguments $\{x_{i}^{+}\},\{x_{j}^{-}\}$
of $W_{AB}$ from Eq. (\ref{W-2}). Equation (\ref{sigma-incl-AB-2}) can
be evaluated numerically, based on Monte Carlo simulations using Markov
chains \cite{Drescher:2000ha}.

\begin{figure}[h]
\begin{centering}
\includegraphics[bb=40bp 50bp 595bp 490bp,clip,scale=0.3]
{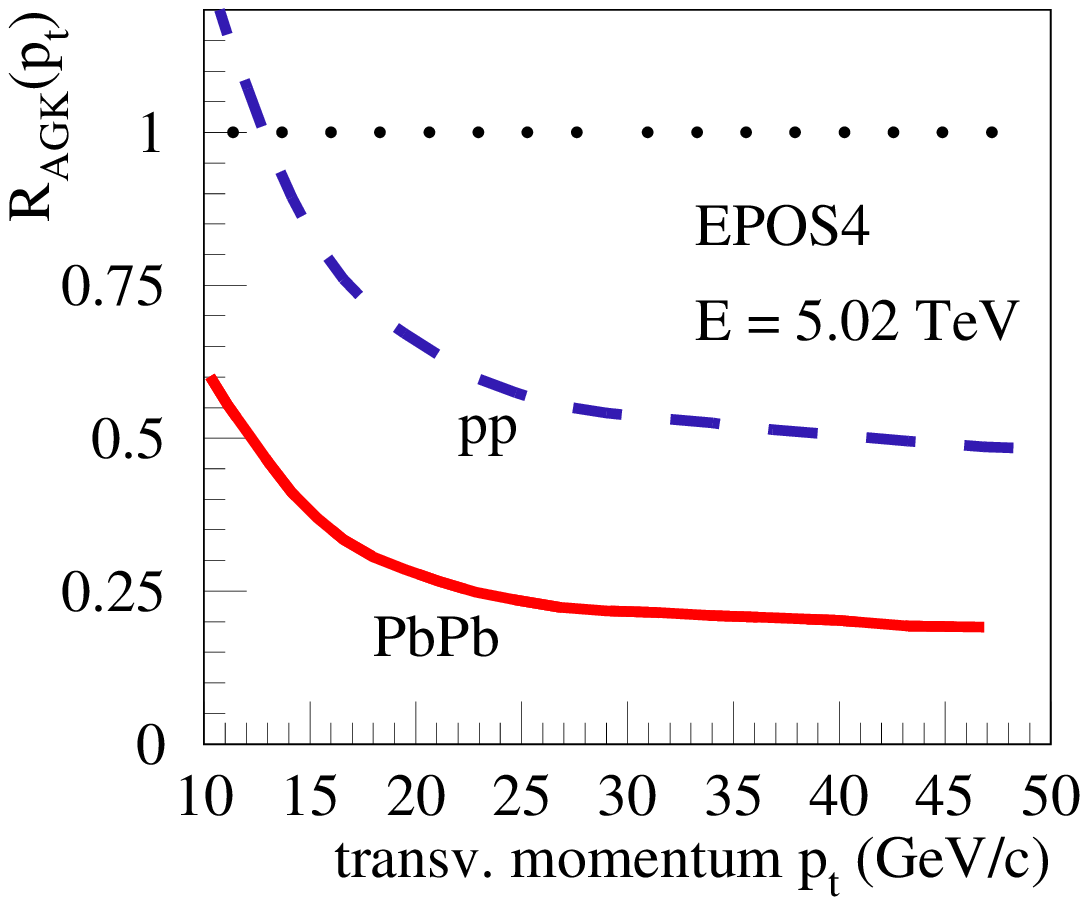}
\par\end{centering}
\centering{}\caption{$R_{\mathrm{AGK}}$ as a function of the $p_{t}$ of partons for minimum bias
PbPb and $pp$ scatterings at 5.02 TeV. \label{R-AGK}}
\end{figure}

\begin{figure}[h]
\begin{centering}
\includegraphics[bb=40bp 50bp 595bp 490bp,clip,scale=0.33]
{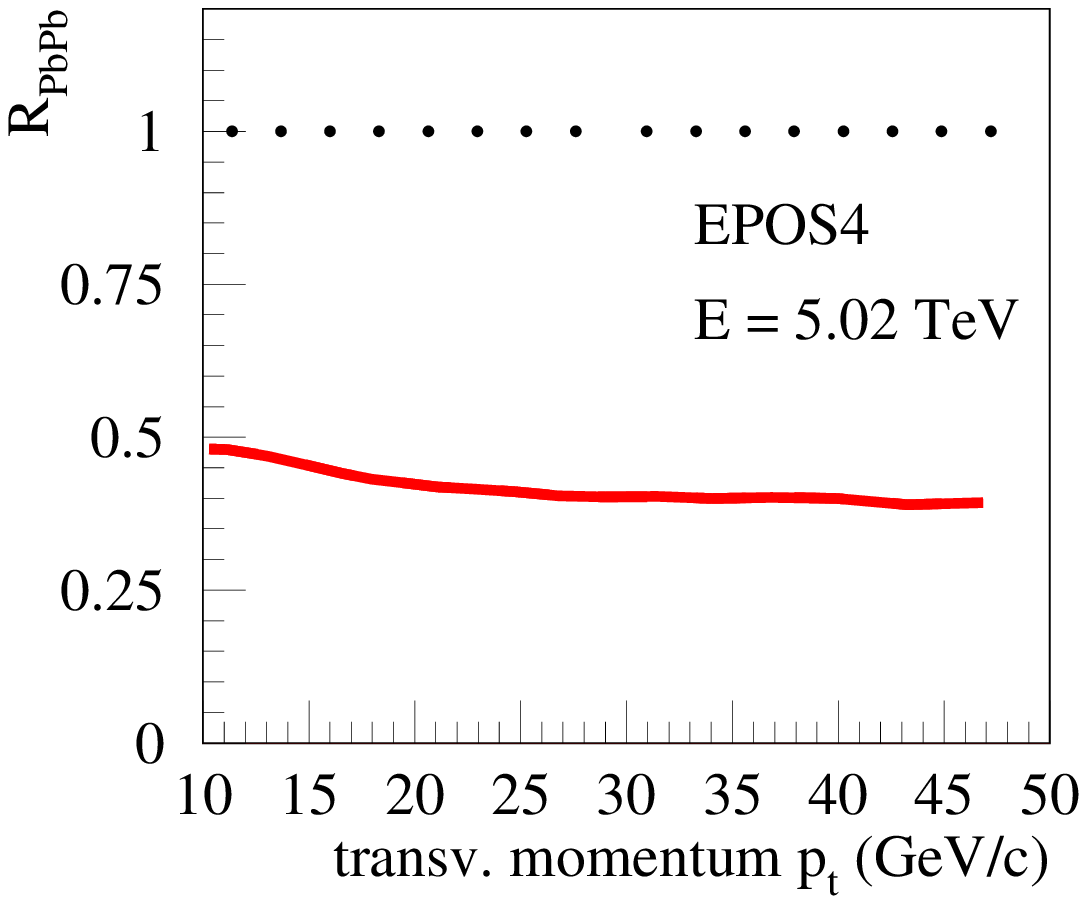}
\par\end{centering}
\centering{}\caption{$R_{\mathrm{PbPb}}$ as a function of the $p_{t}$ of partons for minimum bias PbPb
 scatterings at 5.02 TeV. \label{R-AGK-1}}
\end{figure}

Let me make some checks. The ``regularization'' affects $\sigma_{\mathrm{in}}^{AB}$
very little, but this is not necessarily true for other quantities,
and most important is the validity of AGK theorem. Does it still hold
in the regularized theory? Validity of the AGK theorem means that inclusive
cross sections in the full nuclear ($A\!+\!B$) scattering is $AB$ times
the result for a single Pomeron, 
\begin{equation}
\frac{d\sigma_{\mathrm{incl}}^{AB}}{dx^{+}dx^{-}}=AB\times\frac{d\sigma_{\mathrm{incl}}^{\mathrm{single\,Pom}}}{dx^{+}dx^{-}}\label{AGK cancellations-1}
\end{equation}
[see Eq. (\ref{AGK cancellations}) and Fig. \ref{cut-diagram-2}].
But since for a given $x^{\pm}$ distribution, and knowing the structure
of a Pomeron ($G=G_{\mathrm{QCD}}$), one obtains from $d\sigma_{\mathrm{incl}}/dx^{+}dx^{-}$
in a unique fashion $d\sigma_{\mathrm{incl}}/dq$ for any single particle
variable $q$, like for example the transverse momentum $p_{t}$ of
partons. So validity of the AGK theorem implies
\begin{equation}
\frac{d\sigma_{\mathrm{incl}}^{AB}}{dp_{t}}=AB\times\frac{d\sigma_{\mathrm{incl}}^{\mathrm{single\,Pom}}}{dp_{t}}.\label{AGK cancellations-2}
\end{equation}
And this is easy to check: One computes $d\sigma_{\mathrm{incl}}^{\mathrm{single\,Pom}}/dp_{t}$
based on $G=G_{\mathrm{QCD}}$ (see \cite{werner:2023-epos4-heavy}).
Then one runs the full Monte Carlo simulation and computes $d\sigma_{\mathrm{incl}}^{\mathrm{AB}}/dp_{t}$
for minimum bias scatterings. Finally, one computes the ratio
\begin{equation}
R_{\mathrm{AGK}}(p_{t})=\frac{d\sigma_{\mathrm{incl}}^{AB}}{dp_{t}}\Big/\Big\{ AB\times\frac{d\sigma_{\mathrm{incl}}^{\mathrm{single\,Pom}}}{dp_{t}}\Big\}.
\end{equation}
Validity of the AGK theorem then simply means $R_{\mathrm{AGK}}(p_{t})=1$;
everything else means violation of AGK. In Fig. \ref{R-AGK},
I plot $R_{\mathrm{AGK}}$ for PbPb and for $pp$ at 5.02 TeV, and one
sees that in both cases one is far from unity, so AGK is badly violated,
and the violation is much stronger for PbPb compared to $pp$, and it
increases with $p_{t}$. One should keep in mind that one has  for
the moment just primary scatterings, no final state effects, and one
expects therefore at least at large $p_{t}$ a value of $R_{\mathrm{AGK}}$
equal to unity. The commonly used ``nuclear modification factor'' $R_{\mathrm{PbPb}}$
is given as the ratio of the two $R_{\mathrm{AGK}}$ curves, 
\begin{equation}
R_{\mathrm{PbPb}}=R_{\mathrm{AGK}}^{\mathrm{PbPb}}/R_{\mathrm{AGK}}^{\mathrm{pp}},
\end{equation}
shown in Fig. \ref{R-AGK-1}.
The result is way below unity, around 0.4 at 50 GeV/c, and it decreases
with $p_{t}$. As a side remark: Usually $R_{\mathrm{PbPb}}$ is considered
for given centrality classes, and the ratio is computed as $d\sigma_{\mathrm{incl}}^{AB}/dp_{t}\Big/\{N_{\mathrm{coll}}\times d\sigma_{\mathrm{incl}}^{\mathrm{single\,Pom}}/dp_{t}\}\Big\}$,
but $N_{\mathrm{coll}}$ is not a measurable quantity, whereas rescaling
with $AB$ is well defined. The PbPb result at 5.02 TeV is just an
example, similar results can be found for many different systems at
many different energies. The effect increases with increasing nuclear
mass and with energy. 

\subsection{Deformation functions}

So one has  clearly a problem. $R_{\mathrm{PbPb}}$ is not as it should
be, and this is a consequence of the violation of the AGK theorem,
with $R_{\mathrm{AGK}}<1$, actually substantially less than unity.
As mentioned above, the $p_{t}$ distributions can be derived from
$x^{\pm}$ distributions $d\sigma_{\mathrm{incl}}^{AB}/dx^{+}dx^{-}$.
Validity of AGK  would mean
\begin{equation}
R_{\mathrm{AGK}}(x^{+},x^{-})=\frac{d\sigma_{\mathrm{incl}}^{AB}}{dx^{+}dx^{-}}\Big/\Big\{ AB\times\frac{d\sigma_{\mathrm{incl}}^{\mathrm{single\,Pom}}}{dx^{+}dx^{-}}\Big\}=1,\label{AGK cancellations-3}
\end{equation}
and this is what one needs, to have $R_{\mathrm{AGK}}(p_{t})=1$.
The latter is expected for large $p_{t}$, which corresponds to large
values of $x^{+}x^{-}$. In the following, I will investigate how
the $x^{\pm}$ dependence of $d\sigma_{\mathrm{incl}}^{AB}/dx^{+}dx^{-}$
changes (gets ``deformed'') with respect to the reference curve
for a single Pomeron, so I define the ratio of the normalized distributions,
\begin{align}
R_{\mathrm{deform}}(x^{+},x^{-})\! & =\!\Big\{\frac{1}{\sigma_{\mathrm{incl}}^{AB}}\,\frac{d\sigma_{\mathrm{incl}}^{AB}}{dx^{+}dx^{-}}\Big\}\!\label{R-deform-definition}\\
 & \Big/\!\Big\{\frac{1}{d\sigma_{\mathrm{incl}}^{\mathrm{single\,Pom}}}\,\frac{d\sigma_{\mathrm{incl}}^{\mathrm{single\,Pom}}}{dx^{+}dx^{-}}\!\Big\},\nonumber 
\end{align}
referred to as the ``deformation function''. It is useful to define
the ``Pomeron squared energy fraction'' $x_{\mathrm{PE}}$ and the
Pomeron rapidity $y_{\mathrm{PE}}$ as
\begin{equation}
x_{\mathrm{PE}}=x^{+}x^{-}=\frac{M_{\mathrm{Pom}}^{2}}{s},\quad y_{\mathrm{PE}}=0.5\ln\frac{x^{+}}{x^{-}}.\label{definition-p-PE}
\end{equation}
The quantity $M_{\mathrm{Pom}}$ is the transverse mass of the Pomeron.
The corresponding distributions (for $\sigma_{\mathrm{incl}}^{AB}$
and $\sigma_{\mathrm{incl}}^{\mathrm{single\,Pom}}$) are,
\begin{equation}
\frac{d\sigma_{\mathrm{incl}}}{dx_{\mathrm{PE}}dy_{\mathrm{PE}}}=J\times\frac{d\sigma_{\mathrm{incl}}}{dx^{+}dx^{-}},\label{x-PE-distri}
\end{equation}
with $J$ being the corresponding Jacobian determinant. This defines
$R_{\mathrm{deform}}(x_{\mathrm{PE}},y_{\mathrm{PE}})$ as the corresponding
ratio. The Pomeron rapidity $y_{\mathrm{PE}}$ turns out to be close
to zero, so one usually integrates over $y_{\mathrm{PE}}$, which
gives (again for $\sigma_{\mathrm{incl}}^{AB}$ and $\sigma_{\mathrm{incl}}^{\mathrm{single\,Pom}}$)
the distribution $d\sigma_{\mathrm{incl}}/dx_{\mathrm{PE}}$, and
the ratio of the $A\!+\!B$ cross section over the single Pomeron one provides
$R_{\mathrm{deform}}(x_{\mathrm{PE}})$.

\begin{figure}[h]
\begin{centering}
\includegraphics[bb=40bp 50bp 595bp 500bp,clip,scale=0.33]
{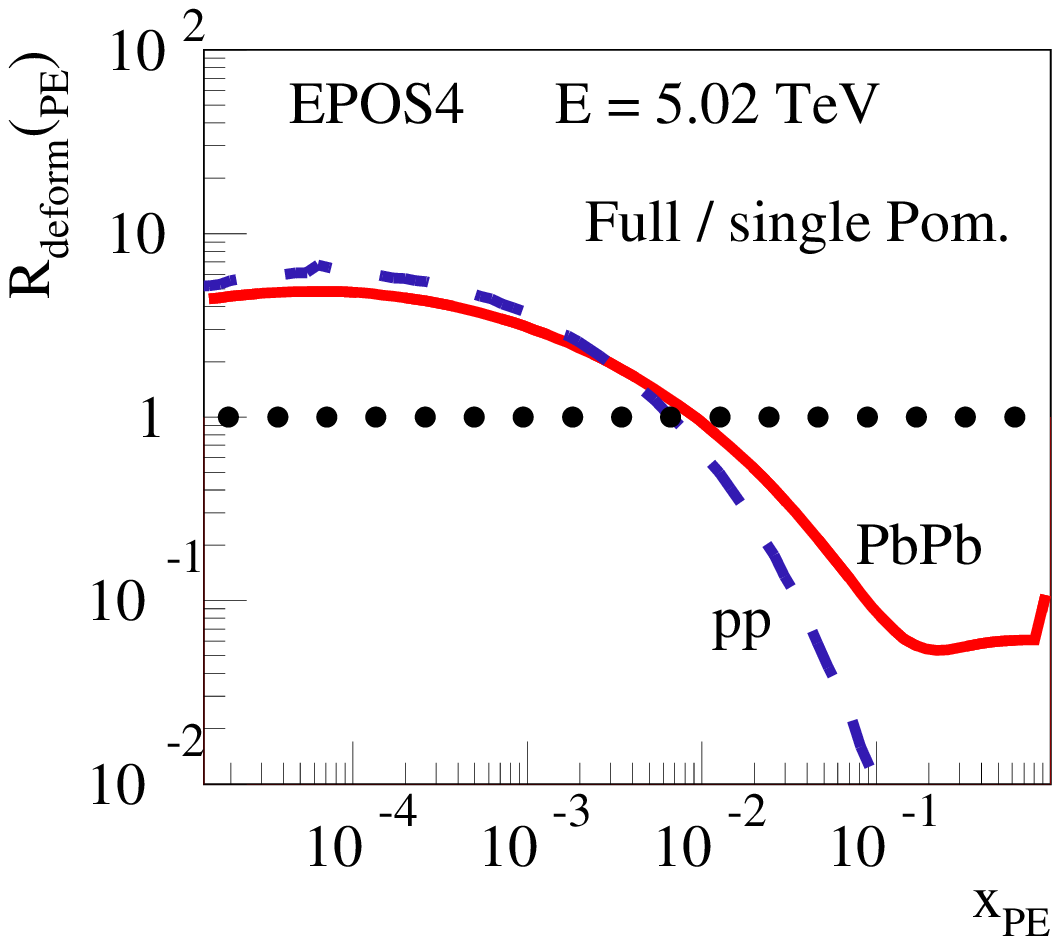}
\par\end{centering}
\centering{}\caption{$R_{\mathrm{deform}}(x_{\mathrm{PE}})$ as a function of $x_{\mathrm{PE}}$
for PbPb (central, 0-5\% ) and $pp$ (event class 16-20 Pomerons) scatterings
at 5.02 TeV. \label{R-AGK-1-1}}
\end{figure}

\begin{figure}[h]
\begin{centering}
\includegraphics[bb=28bp 30bp 595bp 770bp,clip,scale=0.45]
{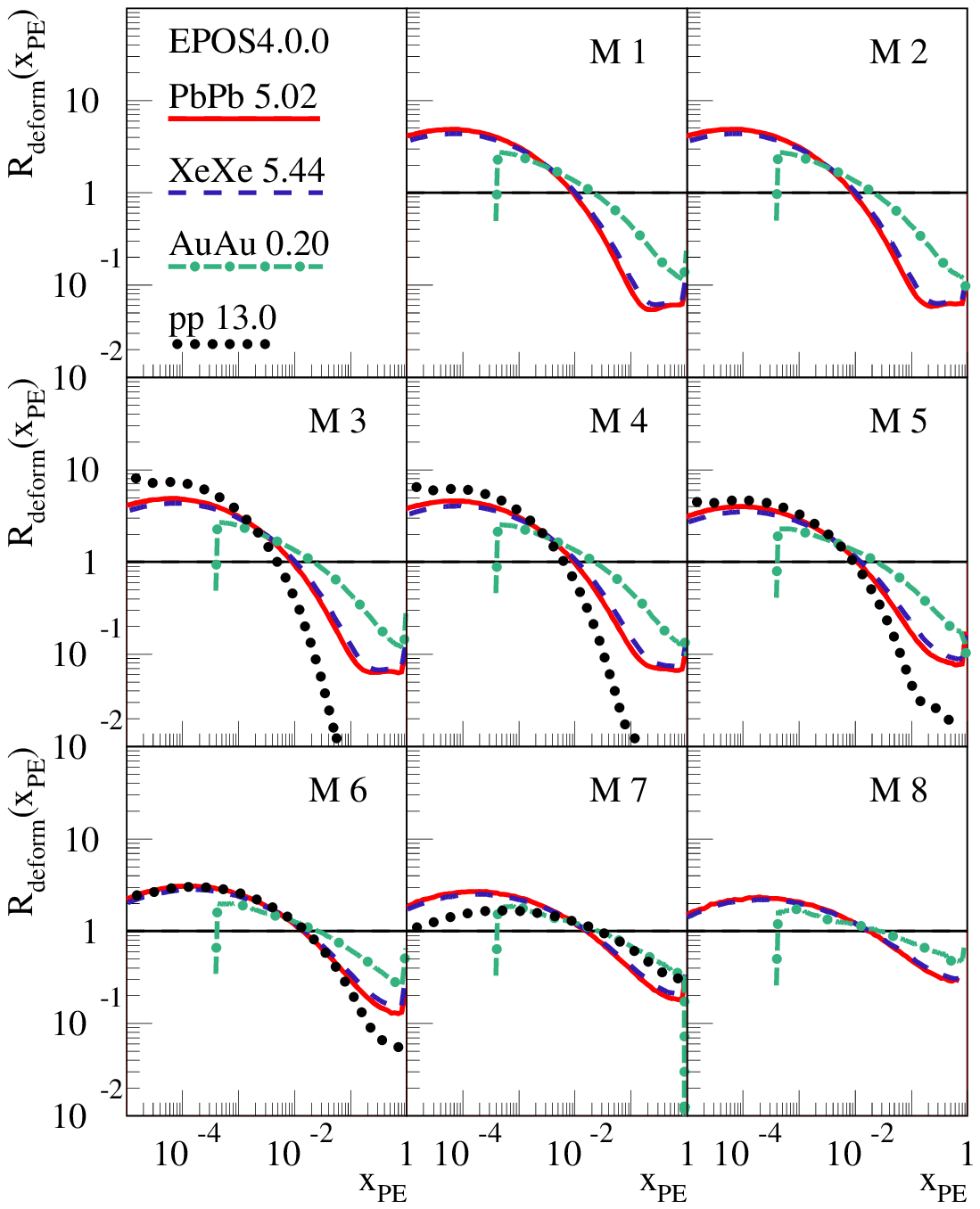}
\par\end{centering}
\centering{}\caption{$R_{\mathrm{deform}}(x_{\mathrm{PE}})$ as a function of $x_{\mathrm{PE}}$
for different systems, different energies (in TeV), different events
classes ``Mi'' (M1 = highest and M8 = lowest multiplicity). See
text. \label{R-deform}}
\end{figure}

In Fig. \ref{R-AGK-1-1},
I show results for $R_{\mathrm{deform}}(x_{\mathrm{PE}})$ in PbPb
(central, 0-5\% ) and $pp$ (event class 16-20 Pomerons) scatterings
at 5.02 TeV. These curves should be unity, but they are far from that.
At large $x_{\mathrm{PE}}$, the ratios are much smaller than unity,
whereas they exceed unity at small $x_{\mathrm{PE}}$. Many simulations
were done to compute $R_{\mathrm{deform}}(x_{\mathrm{PE}})$ for different
energies and systems, see Fig. \ref{R-deform},
where I show the results for PbPb at 5.02 TeV, XeXe at 5.44 TeV, AuAu
at 0.2 TeV (centralities 0-5\%, 5-10\%, 10-20\%, 30-40\%, 50-60\%, 70-80\%, 80-90\%,
90-100\%), and  $pp$ at 13 TeV (number of Pomerons: 51-60, 31-40, 21-25, 16-20,
10-15, 5-9, 2-4, 1). In all cases, one sees a suppression at large
$x_{\mathrm{PE}}$, and the effect gets bigger with increasing ``event
activity''. The effect is actually biggest in $pp$, for rare events
with very large Pomeron numbers (the average number is around two).\\

The current status is as follows: Whereas originally one could prove
the validity of AGK, one needed to introduce a ``regularization''
to avoid negative probabilities, which in turn ruined this very important
AGK property: for all systems, energies, and centralities, one gets
deformed $x_{\mathrm{PE}}$ distributions with respect to the reference
distribution of a single Pomeron, which leads to AGK violation with
respect to $x_{\mathrm{PE}}$, which as a consequence violates AGK
with respect to $p_{t}$, which finally leads to a violation of binary
scaling in $AA$ scattering. It also leads to a violation of factorization
in $pp$, but this requires in addition to consider the internal structure
of $G$. \\

\begin{figure}[h]
\centering{}\includegraphics[scale=0.5]
{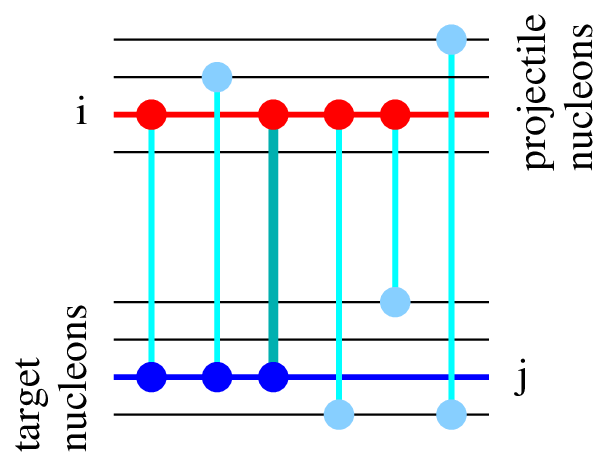}
\caption{A Pomeron connected to projectile nucleon \emph{i} and target nucleon
\emph{j,} together with other Pomerons connected to one (or both)
of these nucleons. \label{nconn-variable}}
\end{figure}

So at the heart of all these problems are the ``deformed'' $x^{\pm}$
or $x_{\mathrm{PE}}$ distributions. In the following, I will try
to better understand and eventually parametrize these deformations,
as a first step towards a solution. Let me consider in $AA$ collisions
(including $pp$ as a special case) a particular multi-Pomeron configurations
$K=\big\{\{m_{k}\},\{x_{k\mu}^{\pm}\}\big\}$, with $m_{k}$ cut Pomerons
per nucleon-nucleon pair $k$, with Pomeron light-cone momentum fractions
$x_{k\mu}^{\pm}$. And let me consider a particular Pomeron, connected
to projectile nucleon \emph{i} and target nucleon \emph{j} (see Fig.
\ref{nconn-variable}). 
In the configuration $K$, there might be other Pomerons, connected
to one (or both) of these nucleons. The corresponding Pomeron-nucleon
connections are marked asf red and blue dots. It is obvious that the
additional Pomerons connected to the same nucleons $i$ and $j$ compete
with each other; they have to share the initial energy-momentum of
the two nucleons. The more Pomerons are connected, the less energy
is available for one particular Pomeron. To quantify this statement,
I define the ``connection number'' 
\begin{equation}
N_{\mathrm{conn}}=\frac{N_{\mathrm{P}}+N_{\mathrm{T}}}{2},
\end{equation}
with $N_{\mathrm{P}}$ being the number of Pomerons connected to \emph{i,
}and with $N_{\mathrm{T}}$ being the number of Pomerons connected
to \emph{j }(the variable $N_{\mathrm{conn}}$ corresponds to half
of the number of red and blue points in Fig. \ref{nconn-variable}). 

In the following, I will discuss the effect of energy sharing related
to the connection number. One wants to understand the dependence of
the $x^{\pm}$ distributions on the connection number $N_{\mathrm{conn}}$,
so rather than Eq. (\ref{sigma-incl-AB-2}), I will compute 
\begin{align}
 & \frac{d^{2}\sigma_{\mathrm{incl}}^{AB\,(N_{\mathrm{conn}})}}{dx^{+}dx^{-}}=\!\!\!\!\sum_{k'=1}^{AB}\;\sum_{\{m_{k}\}\ne0}\sum_{\mu'=1}^{m_{k'}}\delta_{N_{\mathrm{conn}}(k',\mu')}^{N_{\mathrm{conn}}}\int\!\!db\!_{A\!B}\!\!\int\!\!dX\!_{A\!B}\nonumber \\
 & \qquad\qquad\quad\Big\{ P(K)\delta(x^{+}-x_{k'\mu'}^{+})\delta(x^{-}-x_{k'\mu'}^{-})\Big\},\label{sigma-incl-AB-2-1}
\end{align}
where $\delta_{a}^{b}$ is the Kronecker delta. So I only consider
Pomerons $k',\mu'$ with connection number $N_{\mathrm{conn}}(k',\mu')$
equal to $N_{\mathrm{conn}}$. Equation (\ref{sigma-incl-AB-2-1}) can
be easily evaluated numerically, based on Monte Carlo simulations. 

\begin{figure}[h]
\begin{centering}
\includegraphics[bb=28bp 45bp 595bp 700bp,clip,scale=0.45]
{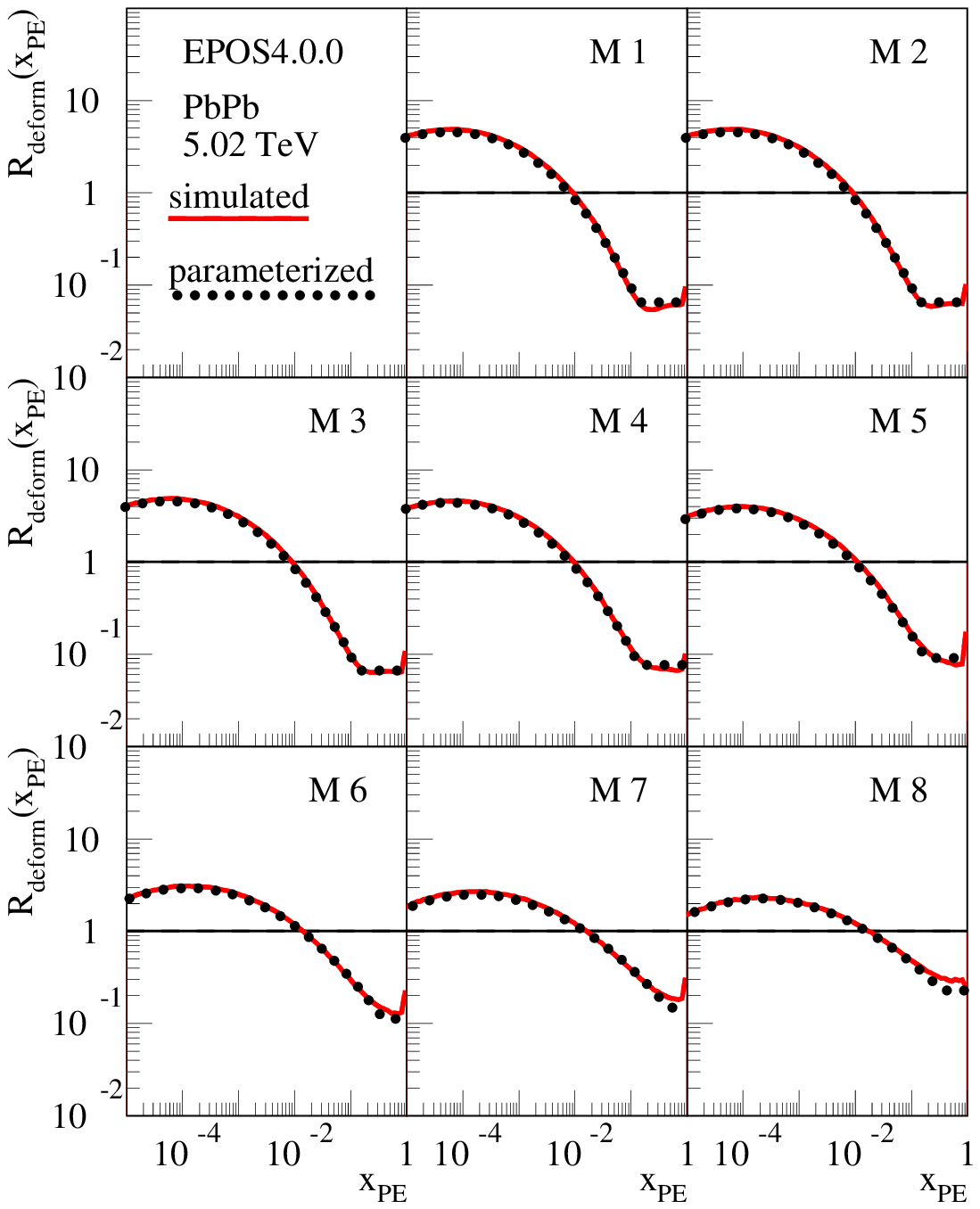}
\par\end{centering}
\centering{}\caption{$R_{\mathrm{deform}}(x_{\mathrm{PE}})$ as a function of $x_{\mathrm{PE}}$
for PbPb at 5.02 TeV, for the centralities M1 to M8.
See text. \label{R-deform-1}}
\end{figure}

Although the above method to compute the inclusive $x^{\pm}$ cross
sections for given $N_{\mathrm{conn}}$ is perfectly doable, I proceed
somewhat differently. I define event classes (centrality in AA
scattering, number of Pomerons in pp), and then  compute the average values 
of the cross sections 
$d^{2}\sigma_{\mathrm{incl}}^{AB} / dx_{\mathrm{PE}}$ 
and $N_{\mathrm{conn}}$ per event
class.
This allows to compute and then investigate $R_{\mathrm{deform}}(x_{\mathrm{PE}})$
for particular event classes,
with the associated mean $N_{\mathrm{conn}}$, 
in different collision systems at various energies. Studying
the corresponding plots, one realizes that all curves $R_{\mathrm{deform}}(x_{\mathrm{PE}})$
can be parametrized in the following way: one first defines
\begin{equation}
\left\{ \begin{array}{l}
w=\ln\big(R_{\mathrm{deform}}\big)/20\\
u=-\ln(x_{\mathrm{PE}})/20
\end{array}\right.,
\end{equation}
and then uses
\begin{equation}
\left\{ \begin{array}{l}
v=|u-a_{1}|/0.5\\
w=a_{2}-v^{a_{3}} a_{4}\\
\mathrm{if}(u>a_{1})w=\mathrm{max}(0.,w)\\
w=\max(a_{5},w)
\end{array}\right..
\end{equation}
In Fig. \ref{R-deform-1},
 I plot $R_{\mathrm{deform}}(x_{\mathrm{PE}})$ as a function of $x_{\mathrm{PE}}$,
for PbPb at 5.02 TeV, for the centralities M1 to M8, i.e., 0-5\%, 5-10\%,
10-20\%, 30-40\%, 50-60\%, 70-80\%, 80-90\%, 90-100\%. The red solid curves
refer to the simulated results, the black dotted ones to the parametrized
curves.
\begin{table}[h]
\begin{centering}
\begin{tabular}{|c|c|c|c|c|c|c|}
\hline 
Class &  $\left\langle N_{\mathrm{conn}}\right\rangle $ & $a_{1}$ & $a_{2}$ & $a_{3}$ & $a_{4}$ & $a_{5}$\tabularnewline
\hline 
\hline 
M1  & 7.69  & 0.49  & 0.076  & 2.28  & 0.38  & -0.15 \tabularnewline
\hline 
M2  & 7.68  & 0.49  & 0.076  & 2.28  & 0.38  & -0.15 \tabularnewline
\hline 
M3  & 7.63  & 0.49  & 0.076  & 2.28  & 0.38  & -0.14 \tabularnewline
\hline 
M4  & 7.25  & 0.49  & 0.075  & 2.26  & 0.36  & -0.13 \tabularnewline
\hline 
M5  & 6.12  & 0.47  & 0.067  & 2.12  & 0.33  & -0.12 \tabularnewline
\hline 
M6  & 4.18  & 0.45  & 0.054  & 2.08  & 0.26  & -0.11 \tabularnewline
\hline 
M7  & 3.23  & 0.44  & 0.046  & 2.00  & 0.22  & -0.10 \tabularnewline
\hline 
M8  & 2.21  & 0.42  & 0.040  & 1.90  & 0.20  & -0.07 \tabularnewline
\hline 
\end{tabular}
\par\end{centering}
\caption{$R_{\mathrm{deform}}$ parameters for PbPb at 5.02 TeV. See text.\label{rdeform-parameters} }

\end{table}
In Table \ref{rdeform-parameters}, I report the values of $\left\langle N_{\mathrm{conn}}\right\rangle $
for the different centrality classes, as well as the values for the
parameters. Somewhat unexpectedly, the deformations do not change
much with centrality in the range 0-20\% (M1-M3), explained by a correspondingly
little change in $\left\langle N_{\mathrm{conn}}\right\rangle $.
As a side remark: the average number of Pomerons or the average number
of collisions do change considerably from M3 to M1, but not the connections
number, and it is the latter that counts concerning energy sharing.
\\

Tables like Table \ref{rdeform-parameters} are considered to define the $N_{\mathrm{conn}}$ dependence 
of  $R_{\mathrm{deform}}(x_{\mathrm{PE}})$ for a given system at a given energy, 
first for the $N_{\mathrm{conn}}$ values in the table, but also for arbitrary values via interpolation or extrapolation.
\\

At this point one should discuss some important detail: in order to
compute $R_{\mathrm{deform}}(x_{\mathrm{PE}})$, one needs to do simulations
based on the probability law in Eq. (\ref{proba-interpretation-2}), and
to do so one needs to know the single Pomeron function $G(x^{+},x^{-})$,
which Eq. \ref{proba-interpretation-2}) is based upon. As discussed
in Sec. \ref{======= S-matrix-with-energy-sharing =======}, and actually needed to derive the final formulas
for the laws, one parametrizes the $x^{\pm}$ dependence of $G$
as [see Eq. (\ref{QCDpar})]
\begin{equation}
G(\!x^{+},x^{-},s,b\!)=\sum_{N=1}^{N_{\mathrm{par}}}\alpha_{N}(x^{+}x^{-})^{\beta_{N}},\label{QCDpar-1}
\end{equation}
inspired by asymptotic form of T-matrices (see Appendix \ref{======= asymptotic-behavior =======}).
I argued that this  form provides an excellent fit to numerically
computed expressions of $G_{\mathrm{QCD}}$, based on pQCD, discussed
in detail in Ref. \cite{werner:2023-epos4-heavy}. However, in the present
work, one does not require $G=G_{\mathrm{QCD}}$ anymore -- it is just
a starting point -- and one is free to change parameters to get an appropriate
behavior of elementary quantities as total and elastic cross sections.
At this point, one does not need (yet) to specify the internal structure
of $G$ (the relation between $G$ and $G_{\mathrm{QCD}}$). One only
needs to know the parametric form Eq. (\ref{QCDpar-1}) of $G$, this
is enough to compute the deformations.\\

I now continue to show the results concerning the simulation and parametrization
of $R_{\mathrm{deform}}(x_{\mathrm{PE}})$ for other systems. \\

\begin{figure}[h]
\begin{centering}
\includegraphics[bb=28bp 45bp 595bp 700bp,clip,scale=0.45]
{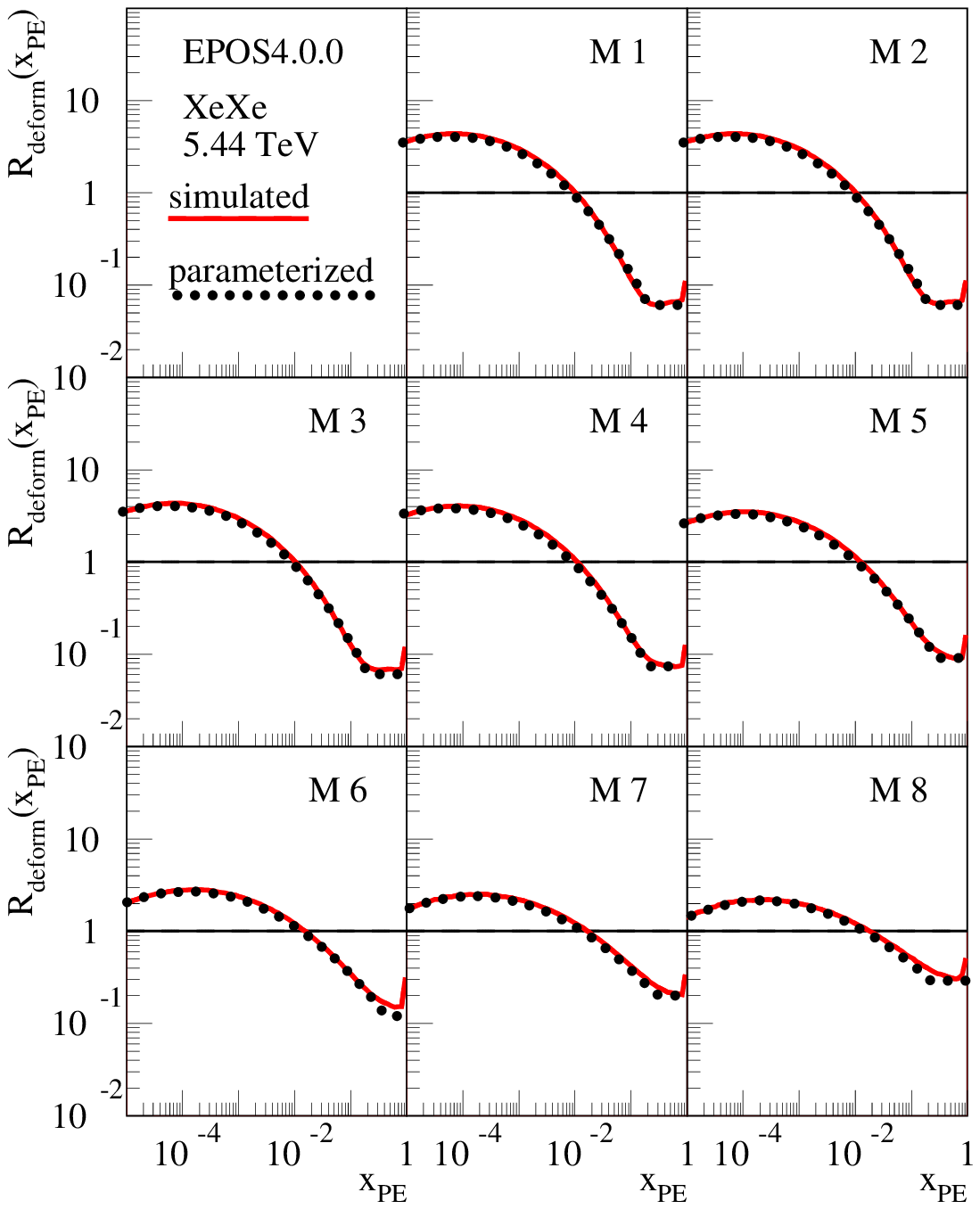}
\par\end{centering}
\centering{}\caption{$R_{\mathrm{deform}}(x_{\mathrm{PE}})$ as a function of $x_{\mathrm{PE}}$
for XeXe at 5.44 TeV, for the centralities M1 to M8.
See text. \label{R-deform-2}}
\end{figure}

In Fig.
\ref{R-deform-2}, I present results for XeXe at 5.44 TeV, for the
centralities M1 to M8, i.e., 0-5\%, 5-10\%, 10-20\%, 30-40\%, 50-60\%, 70-80\%,
80-90\%, 90-100\%. The red solid curves refer to the simulated results,
the black dotted ones to the parametrized curves. These results are
very similar to the 5.02 TeV PbPb ones, and also the table corresponding
to Table \ref{rdeform-parameters} is almost identical. Also here,
the deformations (and the $N_{\mathrm{conn}}$ values vary little
with centrality.
\begin{figure}[h]
\begin{centering}
\includegraphics[bb=28bp 45bp 595bp 700bp,clip,scale=0.45]
{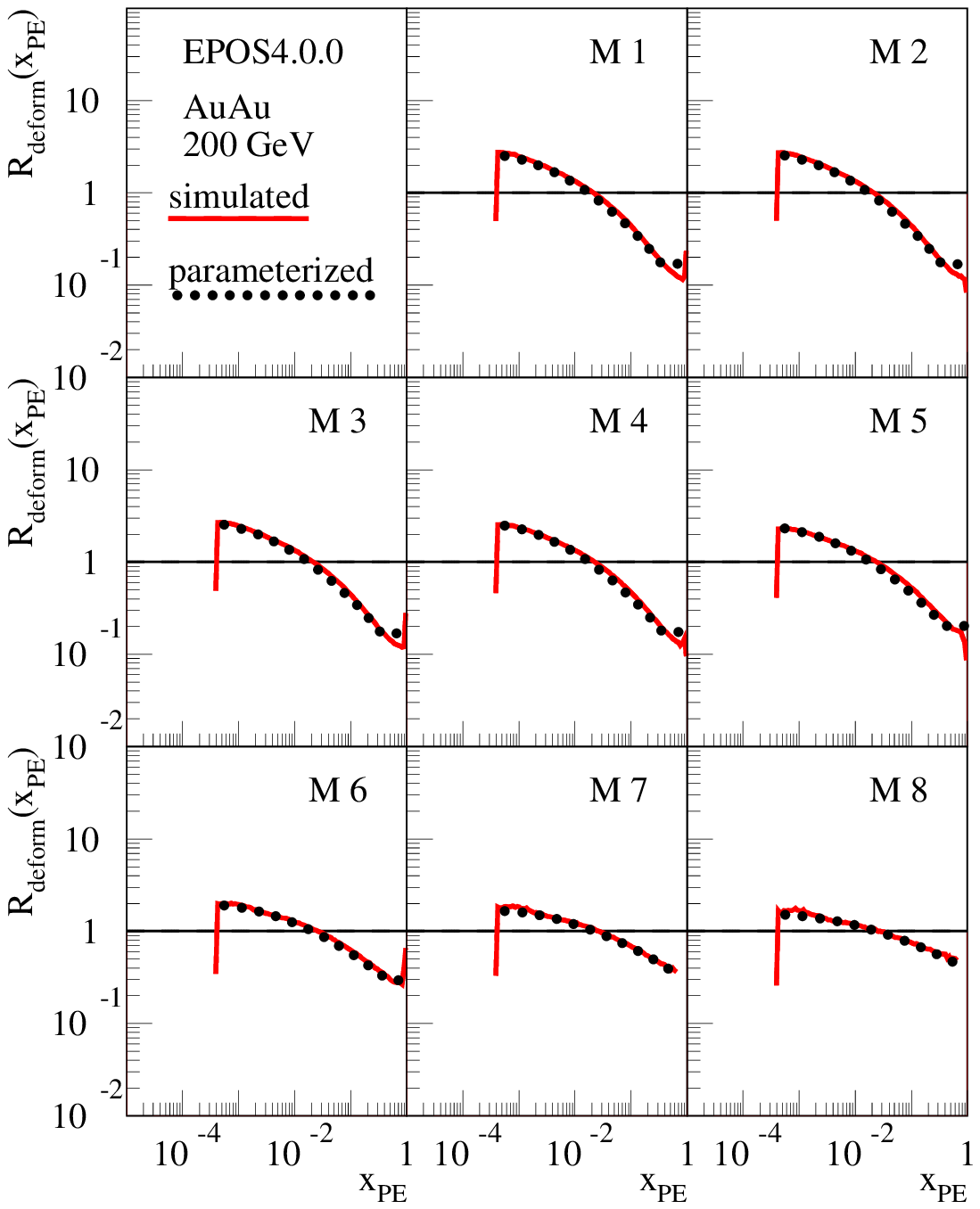}
\par\end{centering}
\centering{}\caption{$R_{\mathrm{deform}}(x_{\mathrm{PE}})$ as a function of $x_{\mathrm{PE}}$
for AuAu at 200 GeV, for the centralities M1 to M8.
See text. \label{R-deform-3}}
\end{figure}
 
\begin{figure}[h]
\begin{centering}
\includegraphics[bb=28bp 45bp 595bp 700bp,clip,scale=0.45]
{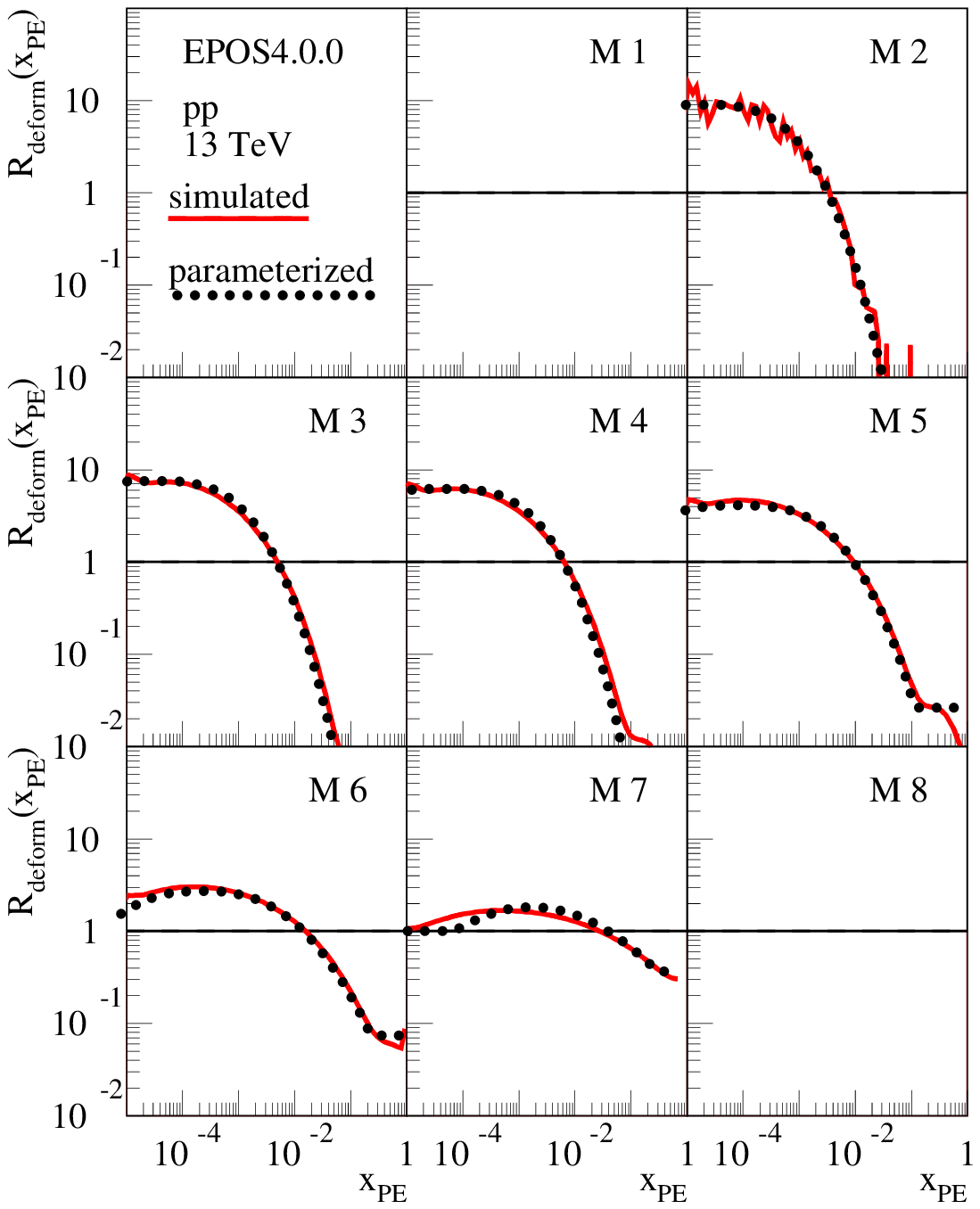}
\par\end{centering}
\centering{}\caption{$R_{\mathrm{deform}}(x_{\mathrm{PE}})$ as a function of $x_{\mathrm{PE}}$
for $pp$ at 13 TeV, for the event classes M1 to M8.
See text. \label{R-deform-4}}
\end{figure}

In Fig. \ref{R-deform-3}, I show results for AuAu at 200 GeV, for the
centralities M1 to M8, i.e., 0-5\%, 5-10\%, 10-20\%, 30-40\%, 50-60\%, 70-80\%,
80-90\%, 90-100\%. As
already observed at higher energies, the deformations are as well
varying only slowly with centrality. Here, the values for $\left\langle N_{\mathrm{conn}}\right\rangle $
evolve between 1 (M8) and 5 (M1). Obviously the covered $x_{\mathrm{PE}}$
range is smaller because the collision energy is smaller. \\

In Fig.
\ref{R-deform-4}, I finally show results for $pp$ at 13 TeV, where
the different event classes M1 to M8 refer to ranges in the number
of Pomerons: 51-60, 31-40, 21-25, 16-20, 10-15, 5-9, 2-4, 1. Contrary
to the heavy-ion results, the values for $\left\langle N_{\mathrm{conn}}\right\rangle $
vary strongly, they go from unity up to 28.5. And correspondingly the deformations are
huge. This will have important consequences, concerning measurable
observables.

\vspace*{1cm}

\subsection{Conclusions}

Let me summarize this section:
\begin{itemize}
\item Reminder: Validity of the AGK theorem means: \textbf{inclusive cross
sections} of collisions of nuclei of mass numbers $A$ and $B$ are
equal to $AB$ times the cross section of a \textbf{single Pomeron}
exchange. The most ``basic'' inclusive cross section is the one
with respect to the variables $x^{\pm}$ (or $x_{\mathrm{PE}}$ and $y_{\mathrm{PE}}$); 
everything else can be derived from this.
\item As a first test of the ``regularized theory'', I check the validity
of AGK. Doing a PbPb simulation at 5.02 TeV, one finds that AGK fails
badly, and indeed also binary scaling does not work at all.
\item To understand the problem,  I study the ratio of inclusive cross sections with respect
to $x_{\mathrm{PE}}$ for $A\!+\!B$ scattering divided by the single Pomerons
result, the latter being the reference curve. Both cross sections are normalized.
The ratio is called the ``deformation
function'' $R_{\mathrm{deform}}(x_{\mathrm{PE}})$.
\item Many simulations are done, for different systems and energies. In
all cases, $R_{\mathrm{deform}}(x_{\mathrm{PE}})$ deviates from unity;
one sees always a suppression at large $x_{\mathrm{PE}}$, and the
effect gets bigger with increasing event activity. 
\item I define for a given Pomeron a quantity called connection number $N_{\mathrm{conn}}$,
counting the number of Pomerons connected to the same projectile or
target, strongly correlated with the deformations: large $N_{\mathrm{conn}}$
means large deformation (due to energy sharing). The variable $N_{\mathrm{conn}}$
represents the ``environment'' of a given Pomeron; it states if
the Pomeron is isolated or competes with others with respect to energy-momentum
sharing. 
\item I study the dependence of the deformations on $\left\langle N_{\mathrm{conn}}\right\rangle $
for different event classes, and one finds simple parametrizations
for $R_{\mathrm{deform}}(x_{\mathrm{PE}})$ in terms of five $N_{\mathrm{conn}}$ dependent
parameters. This allows one to define $R_{\mathrm{deform}}(x_{\mathrm{PE}})$  for arbitrary  $N_{\mathrm{conn}}$ 
values via interpolation and extrapolation.
\end{itemize}

So deformations $R_{\mathrm{deform}}(x_{\mathrm{PE}})\ne1$ represent
the problem (AGK violation), but at least one understands the origin
of the problem (energy-momentum sharing), and one is able to parametrize it.
This will be useful for the next step: the solution of the problem.

\section{The solution via saturation and generalized AGK theorem \label{======= the solution =======} }s

One has  seen in Sec. \ref{======= AGK-violation-deformation =======} that the validity of the AGK
theorem is badly violated, with respect to the variable $x_{\mathrm{PE}}$.
There is nothing one can do about it, it is an unavoidable consequence
of energy-momentum sharing, and the effect can be quantified in terms
deformation functions $R_{\mathrm{deform}}(x^{+},x^{-})$ [in practice
$R_{\mathrm{deform}}(x_{\mathrm{PE}})$], it being the ratio of $A\!+\!B$
inclusive cross sections over the single Pomeron one, using normalized
distributions. Here validity of AGK would mean $R_{\mathrm{deform}}(x^{+},x^{-})=1$,
which is not at all observed; one has always a suppression at large
$x_{\mathrm{PE}}=x^{+}x^{-}$. This invalidity of AGK has important
consequences, like the violation of binary scaling in nuclear collisions.

Let me recall that by AGK theorem I mean some extension of the original one, 
in the sense that an inclusive cross section with respect to some variable $q$ 
in $A\!+\!B$ scattering is equal to $AB$ times the corresponding single Pomeron cross section.

\subsection{The role of saturation}

\begin{figure}[h]
\begin{centering}
\begin{minipage}[c]{0.3\columnwidth}%
\begin{center}
\includegraphics[scale=0.32]
{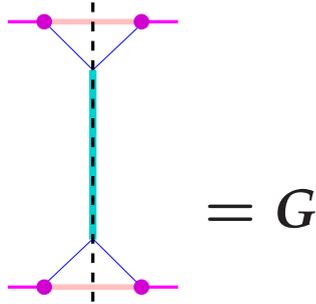}
\par\end{center}%
\end{minipage}%
\begin{minipage}[c]{0.2\columnwidth}%
\noindent \begin{center}
~\vspace*{.1cm}
\par\end{center}
\noindent \begin{center}
{\huge{}
\[
\boldsymbol{\boldsymbol{=G}}
\]
}{\huge\par}
\par\end{center}
\noindent \begin{center}
~\vspace*{.1cm}
\par\end{center}%
\end{minipage}
\par\end{centering}
\centering{}\caption{A single cut Pomeron. \label{single-cut-Pomeron}}
\end{figure}

But does one really need AGK for inclusive cross sections with respect
to the variables $x^{\pm}$ (or $x_{\mathrm{PE}}$)? Not necessarily: one expects binary scaling
to be valid not always, but only for high $p_{t}$ processes, and
the variable $p_{t}$ is related to $x^{\pm}$, but how precisely
depends on the internal structure of $G$. One remembers that the single
cut Pomeron $G$ is the fundamental building block of the multiple
scattering formalism; in the graphs I use simply a vertical line (in
cyan) as shown in Fig. \ref{single-cut-Pomeron},
where the vertical black dashed line represents the cut. All the multiple
scattering formulas depend on $G$, but not on the internal structure.
Only when it comes to statements concerning, for example, $p_{t}$ of
produced partons does one need to specify how $G$ is expressed in terms
of QCD diagrams. So far I am assuming 
\begin{equation}
G=G_{\mathrm{QCD}},\label{G-equal-G-QCD-1}
\end{equation}
where $G_{\mathrm{QCD}}$ is essentially a DGLAP parton ladder, discussed
in detail in Ref. \cite{werner:2023-epos4-heavy}  [see also Fig. \ref{single-pomeron-graph-1}
and the discussion before Eq. (\ref{G-equal-G-QCD})]. But the assumption
in Eq. (\ref{G-equal-G-QCD-1}) is obviously wrong, because it leads
to a strong violation of binary scaling at large $p_{t}$, as shown
in Sec. \ref{======= AGK-violation-deformation =======}.\\

There is another serious problem with Eq. (\ref{G-equal-G-QCD-1}):
As discussed in detail in \cite{werner:2023-epos4-heavy}, the essential
part of $G_{\mathrm{QCD}}$ is a cut parton ladder, based on DGLAP
parton evolutions. But this is certainly not the full story: With
increasing energy, partons with very small momentum fractions $x\ll1$
become increasingly important, since the parton density becomes large,
and therefore the linear DGLAP evolution scheme is not valid anymore,
nonlinear evolution takes over, considering explicitly gluon-gluon
fusion. These phenomena are known as ``saturation'' \cite{Gribov:1983ivg,McLerran:1993ni,McLerran:1993ka,kov95,kov96,kov97,kov97a,jal97,jal97a,kov98,kra98,jal99a,jal99b,jal99}.\\

\begin{figure}[h]
\begin{centering}
(a)$\quad$\hspace*{2.5cm}$\quad$(b)\hspace*{3cm}$\quad$\\
\includegraphics[scale=0.32]
{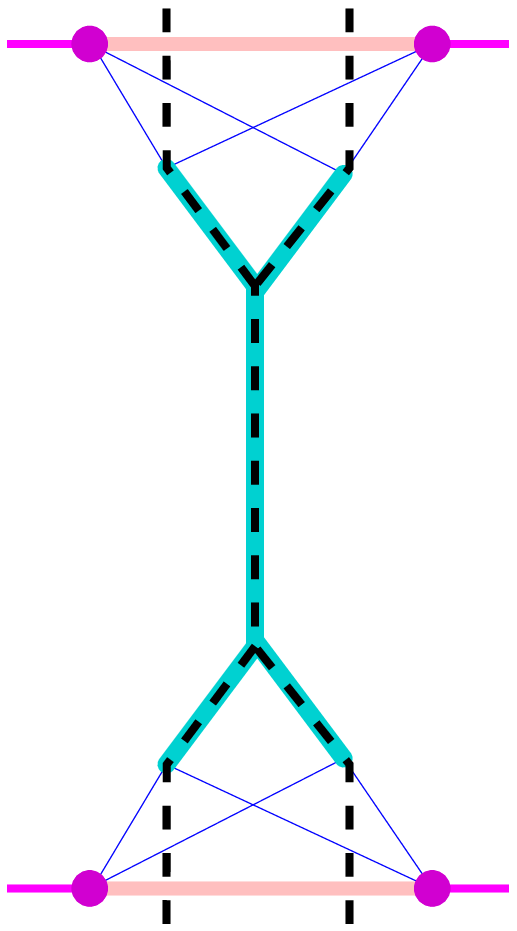}
~~~~~~~~~~\includegraphics[scale=0.32]
{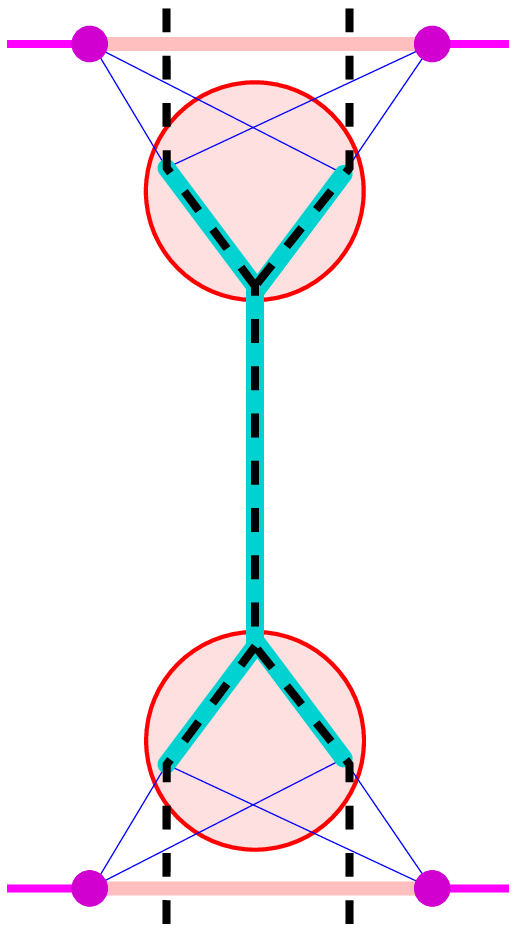}
\par\end{centering}
\centering{}\caption{Cut diagram (a) with nonlinear effects via triple Pomeron vertices
and (b) with the nonlinear effects (inside the red circles) ``summarized''
in the form of saturation scales, which replace these nonlinear parts. \label{cut-triple-Pomeron}}
\end{figure}

At least for scatterings carrying a large value of $x^{+}x^{-}$,
one expects ``nonlinear effects'', which means that two ladders
which evolve first independently and in parallel, finally fuse. And
only after that, the (linear) DGLAP evolution is realized. In the
``Pomeron language'', this corresponds to diagrams with triple (and
more) Pomeron vertices, as sketched in Fig. \ref{cut-triple-Pomeron}(a).
Such nonlinear effects lead to strong destructive interference at
low transverse momentum ($p_{t}$), which may be summarized in terms
of a saturation scale \cite{McLerran:1993ni,McLerran:1993ka}. This
suggests treating these ``saturation phenomena'' not explicitly,
but by introducing saturation scales as the lower limits of the virtualities
for the DGLAP evolutions, as sketched in Fig. \ref{cut-triple-Pomeron}(b).
Also in Refs. \cite{Eskola} and \cite{Sibyll} saturation scales
``summarize'' nonlinear effects, but based on the assumption that the partons 
exactly cover the transverse size of the nucleon (or nucleus). In the present paper, 
the criterion is quite different, as discussed below. \\

So one has two problems -- a wrong association $G=G_{\mathrm{QCD}}$
and a missing treatment of saturation -- but fortunately, the two problems
are connected, and there is an amazingly simple solution based on
saturation scales that solves both problems. \\

\subsection{The solution}

Instead of the ``naive''
assumption $G=G_{\mathrm{QCD}}$, one postulates 

\begin{equation}
G(x^{+},x^{-})=\frac{n}{R_{\mathrm{deform}}(x^{+},x^{-})}G_{\mathrm{QCD}}(Q_{\mathrm{sat}}^{2},x^{+},x^{-}),\label{fundamental-epos4-equation}
\end{equation}\medskip{}

\noindent
such that \textbf{$G$ }itself does not depend on the environment.
By environment I mean, for a given Pomeron in a given configuration
of multiple Pomeron exchanges, the connections of other Pomerons to
the same projectile and target nucleons as the given Pomeron. The
environment is here quantified in terms of $N_{\mathrm{conn}}$. The
quantity $R_{\mathrm{deform}}$ is the $N_{\mathrm{conn}}$ dependent
deformation function discussed in Sec. \ref{======= AGK-violation-deformation =======}, and $n$ is
a constant, not depending on $x_{\mathrm{PE}}$. As discussed earlier,
one first parametrizes $G$ as in Eq. (\ref{QCDpar}), with the parameters
being fixed by comparing simulation results to elementary experimental
data, and then uses Eq. (\ref{fundamental-epos4-equation}) to determine
$Q_{\mathrm{sat}}^{2}$. In this way, $Q_{\mathrm{sat}}^{2}$ depends
on $N_{\mathrm{conn}}$ and on $x^{\pm}$, as does $R_{\mathrm{deform}}$:
\begin{align}
 & R_{\mathrm{deform}}=R_{\mathrm{deform}}(N_{\mathrm{conn}},x^{+},x^{-}),\label{fundamental-2}\\
 & Q_{\mathrm{sat}}^{2}\;=\;Q_{\mathrm{sat}}^{2}(N_{\mathrm{conn}},x^{+},x^{-}),\label{fundamental-3}\\
 & \mathrm{but}\,G\,\mathrm{independent\,of}\,N_{\mathrm{conn}}.\label{fundamental-4}
\end{align}
The $N_{\mathrm{conn}}$ dependence of $Q_{\mathrm{sat}}^{2}$ means
that the low virtuality cutoff for the DGLAP evolutions in $G_{\mathrm{QCD}}$,
is not a constant, but its value depends on the environment in terms
of $N_{\mathrm{conn}}$ and on the energy of the Pomeron. I will refer
to this as ``dynamical saturation scales''.\\

The fundamental relation in Eq. (\ref{fundamental-epos4-equation}), 
together with Eqs. (\ref{fundamental-2}-\ref{fundamental-4})
and in particular the dynamical saturation scales,
will solve  the problems related to the AGK theorem, as will be discussed in the following sections. \\

\begin{figure}[h]
\centering{}

\hspace*{-0.3cm}\includegraphics[bb=20bp 50bp 642bp 570bp,clip,scale=0.32]
{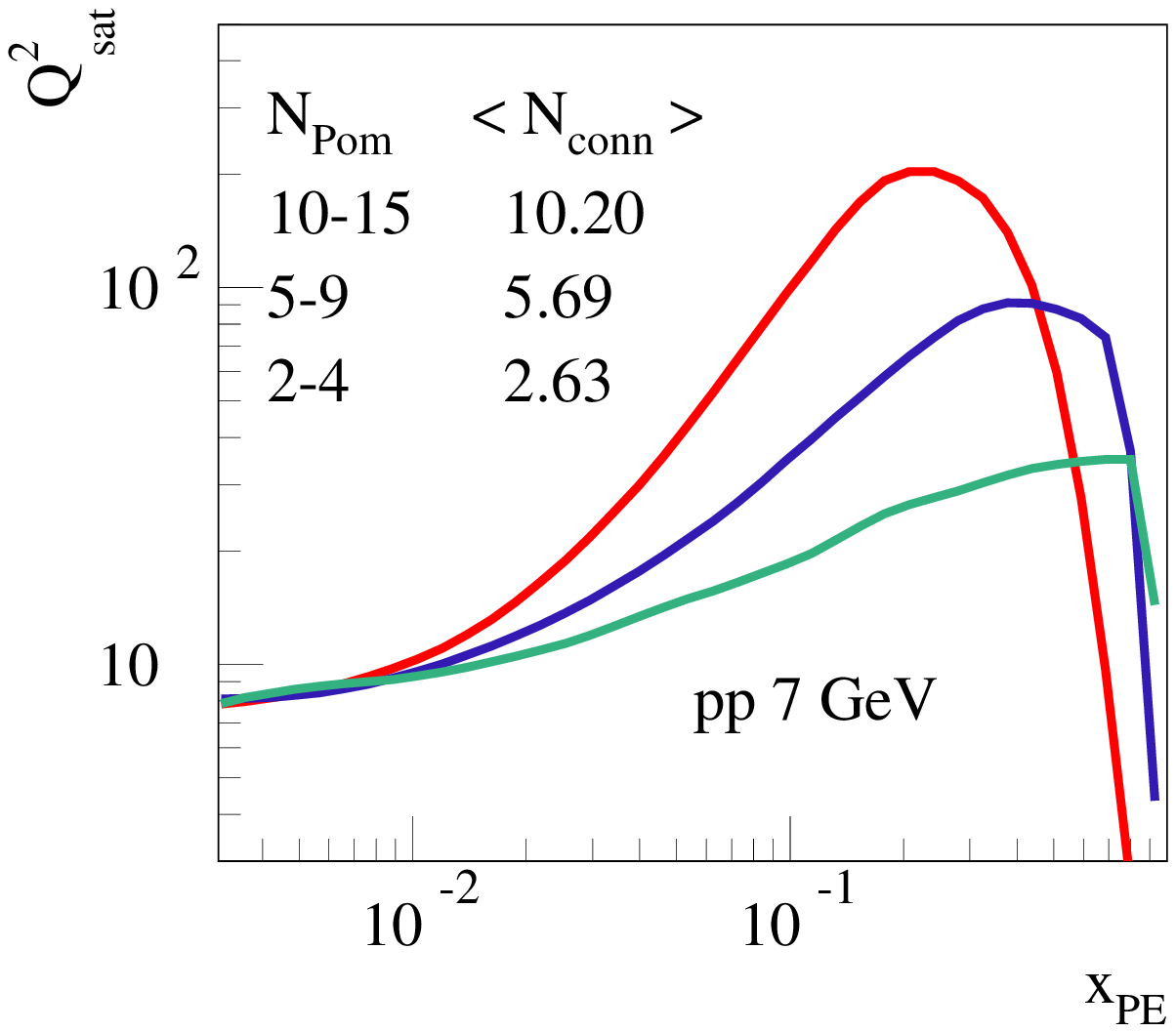} 

\hspace*{-0.3cm}\caption{The saturation scale $Q_{\mathrm{sat}}^{2}$ as a function of $x_{\mathrm{PE}}=x^{+}x^{-}$,
for several $N_{\mathrm{Pom}}$ event classes (from top to bottom,
10-15, 5-9, and 2-4 Pomerons). \label{q2sat-vs-xpe}}
\end{figure}

To get some idea about the $x^{\pm}$ dependence of $Q_{\mathrm{sat}}^{2}$,
obtained from Eq. (\ref{fundamental-epos4-equation}), I show in Fig.
\ref{q2sat-vs-xpe} results for $pp$
scattering at 7 TeV, for several event classes defined via the number
of Pomerons $N_{\mathrm{Pom}}$.
I also indicate the average values of $N_{\mathrm{conn}}$ for the
three event classes. One sees that $\left\langle N_{\mathrm{conn}}\right\rangle $
varies considerably between the different classes, and so does $Q_{\mathrm{sat}}^{2}$.\\

\begin{figure}[h]
\centering{}

\hspace*{-0.3cm}

\hspace*{-0.3cm}\includegraphics[bb=20bp 50bp 642bp 570bp,clip,scale=0.32]
{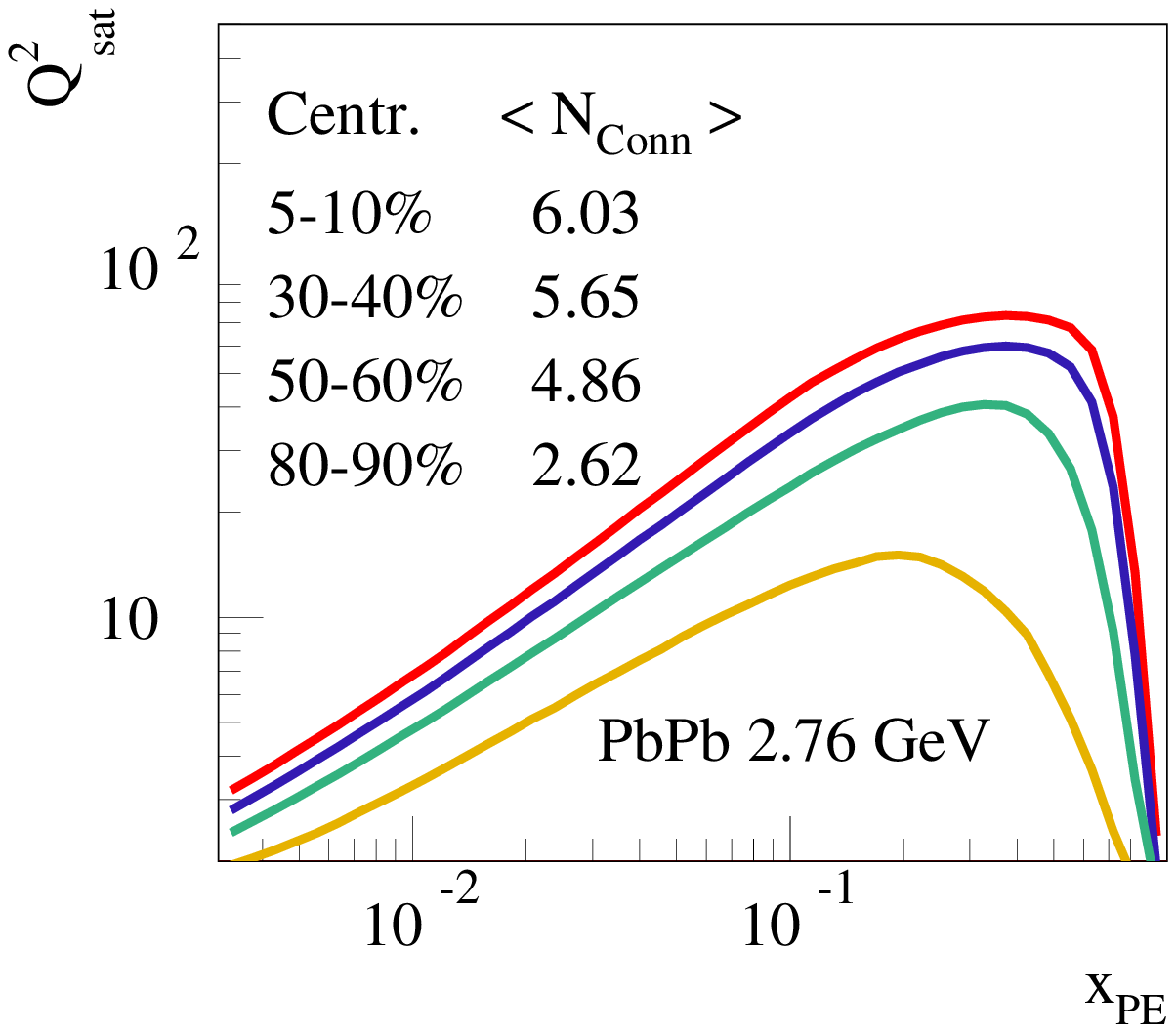}
\caption{The saturation scale $Q_{\mathrm{sat}}^{2}$ as a function of $x_{\mathrm{PE}}$,
for several event classes (from top to bottom: 5-10\%, 10-20\%, 30-40\%,
50-60\% centrality). \label{q2sat-vs-xpe-1}}
\end{figure}

In Fig. \ref{q2sat-vs-xpe-1},
 I show results for PbPb at 2.76 TeV. Here, the variation of the average
$N_{\mathrm{conn}}$ values is quite moderate, and towards central
collisions $Q_{\mathrm{sat}}^{2}$ even ``saturates'' (no variation
anymore). Correspondingly, $Q_{\mathrm{sat}}^{2}$ varies little from
semiperipheral towards central events, 
in particular compared to the strong variation of $Q_{\mathrm{sat}}^{2}$ with the event activity in $pp$ scattering. \\

\subsection{Dynamical saturation scales}

\noindent In order to understand, how these dynamical saturation
scales work, let me investigate the inclusive cross section%
\begin{comment}
\[
P(K)=\prod_{k=1}^{AB}\left[\frac{1}{m_{k}!}\prod_{\mu=1}^{m_{k}}G_{k\mu}\right]\times W_{AB}(\{x_{i}^{+}\},\{x_{j}^{-}\})
\]
see Eq. (\ref{proba-interpretation})

\begin{align*}
 & W_{AB}=\prod_{i=1}^{A}c_{1}(x_{i}^{+})^{\alpha_{\mathrm{remn}}}\prod_{j=1}^{B}c_{1}(x_{j}^{-})^{\alpha_{\mathrm{remn}}}\\
 & \qquad\qquad\prod_{k=1}^{AB}\exp\left(-\tilde{G}(x_{\pi(k)}^{+}x_{\tau(k)}^{-})\right),
\end{align*}
see Eq. (\ref{W-AB-7-1})
\[
x_{i}^{+}=1-\!\!\sum_{\underset{\pi(k)=i}{k=1}}^{AB}\sum_{\mu=1}^{m_{k}}\!x_{k\mu}^{+},\quad x_{j}^{-}=1-\!\!\sum_{\underset{\tau(k)=j}{k=1}}^{AB}\sum_{\mu=1}^{m_{k}}\!x_{k\mu}^{-}.
\]
\end{comment}
{} 
\begin{align}
 & \frac{d^{2}\sigma_{\mathrm{incl}}^{AB\,(N_{\mathrm{conn}})}}{dx^{+}dx^{-}}=\!\!\!\!\sum_{k'=1}^{AB}\;\sum_{\{m_{k}\}\ne0}\;\sum_{\mu'=1}^{m_{k'}}\delta_{N_{\mathrm{conn}}(k',\mu')}^{N_{\mathrm{conn}}}\int\!\!db\!_{A\!B}\!\!\int\!\!dX\!_{A\!B}\nonumber \\
 & \qquad\qquad\quad\Big\{ P(K)\delta(x^{+}-x_{k'\mu'}^{+})\delta(x^{-}-x_{k'\mu'}^{-})\Big\}\label{sigma-incl-AB-2-1-1}
\end{align}
[see Eqs. (\ref{sigma-incl-AB-2-1}) 
and (\ref{b-AB-integration}), (\ref{thickness-function}), (\ref{x-AB-integration-1}), (\ref{proba-interpretation}), and (\ref{W-AB-7-1})].
Here, $K$ refers to a multi-Pomeron configuration $K=\big\{\{m_{k}\},\{x_{k\mu}^{\pm}\}\big\}$,
with $m_{k}$ cut Pomerons per nucleon-nucleon pair $k$ and with Pomeron
light-cone momentum fractions $x_{k\mu}^{\pm}$. In the case of a
single Pomeron in $pp$ scattering, the expression simplifies enormously,
and one gets
\begin{equation}
\frac{d\sigma_{\mathrm{incl}}^{\mathrm{single\,Pom}}}{dx^{+}dx^{-}}=\int d^{2}b\,G(x^{+},x^{-},b)\:W_{11}\left((1-x^{+})(1-x^{-})\right).\label{single-Pom-ref-0}
\end{equation}
Employing Eqs. (\ref{fundamental-epos4-equation}-\ref{fundamental-4}),
using $N_{\mathrm{conn}}=1$ and $R_{\mathrm{deform}}=1$, this expression
may be written as
\begin{align}
 & \int d^{2}b\,G_{\mathrm{QCD}}\left(Q_{\mathrm{sat}}^{2}(1,x^{+},x^{-})\,,x^{+},x^{-}\right)\nonumber \\
 & \qquad\times W_{11}\left((1-x^{+})(1-x^{-})\right).
\end{align}
Since the saturation scale is playing a crucial role in the formalism,
I write the above result for a single Pomeron $pp$ scattering as
\begin{equation}
\frac{d\sigma_{\mathrm{incl}}^{\mathrm{single\,Pom}}}{dx^{+}dx^{-}}=\frac{d\sigma_{\mathrm{incl}}^{\mathrm{single\,Pom}}}{dx^{+}dx^{-}}\left[Q_{\mathrm{sat}}^{2}(1,x^{+},x^{-})\right],\label{single-Pom-ref-1}
\end{equation}
with the general definition (for any $f$)
\begin{align}
 & \frac{d\sigma_{\mathrm{incl}}^{\mathrm{single\,Pom}}}{dx^{+}dx^{-}}\Big[f(x^{+},x^{-})\Big]\label{single-Pom-ref-2}\\
 & =\int d^{2}b\,G_{\mathrm{QCD}}\Bigg(\underbrace{f(x^{+},x^{-})}_{\mathrm{sat.\,scale}}\,,x^{+},x^{-}\Bigg)\nonumber \\
 & \qquad\qquad\qquad\qquad\times W_{11}\left((1-x^{+})(1-x^{-})\right).\nonumber 
\end{align}

\noindent Let me consider the general case $d^{2}\sigma_{\mathrm{incl}}^{AB\,(1)}/dx^{+}dx^{-}$,
i.e., an $A\!+\!B$ scattering with $N_{\mathrm{conn}}=1$,where one has
only isolated Pomerons, and there is never more than one Pomeron connected
to a projectile and a target nucleon, for all nonzero contributions.
Concerning $\int\!\!dX\!_{A\!B}\{...\}$, for given $k'$ and $\mu'$,
the $x_{k'\mu'}^{\pm}$ integration is trivial thanks to the $\delta$
functions, and one gets a factor 
\begin{equation}
G(x^{+},x^{-},b_{k'})\:W_{11}\left((1-x^{+})(1-x^{-})\right),
\end{equation}
multiplied by $P(K')$, where $K'$ is simply the configuration $K$
minus the $\mu'$-th Pomeron of pair $k'$ (with actually $\mu'=1$,
always). The quantity $W$ is defined in Eq. (\ref{W-AB-7-1}). Although
usually not written explicitly, $G$ depends also on the impact parameter,
in this case $b_{k'}=|b+b_{i'}^{A}-b_{j'}^{B}|$, where $i'=\pi(k')$
and $j'=\tau(k')$ are the projectile and target nucleons, respectively,
associated to pair $k'$. The term $P(K')$ does not depend on $x^{+}$
and $x^{-}$, because there are no other Pomerons connected to $i'$
and $j'$. The term $P(K')$ does also not depend on $b_{i'}^{A}$
and $b_{j'}^{B}$. Let me define $\int\!\!dX'\!_{A\!B}$ the integration
except for the $x_{k'\mu'}^{\pm}$ integrations (already done), and $\int\!\!db'\!_{A\!B}$
the impact parameter integrations except for $d^{2}b_{i'}^{A}\,T_A(b_{i'}^{A})\int d^{2}b_{j'}^{B}\,T_B(b_{j'}^{B})$.
Then the integrations $\int\!\!db'\!_{A\!B}\,\int\!\!dX'\!_{A\!B}\,P(K')$
provide a factor independent of $x^{+}$, $x^{-}$, $b_{i'}^{A}$
and $b_{j'}^{B}$. So finally, for the case $N_{\mathrm{conn}}=1$,
one finds
\begin{align}
\frac{d^{2}\sigma_{\mathrm{incl}}^{AB\,(1)}}{dx^{+}dx^{-}} & \propto\sum_{k'=1}^{AB}\;\sum_{\{m_{k}\}\ne0}\delta_{N_{\mathrm{conn}}(k',1)}^{N_{\mathrm{conn}}}\nonumber \\
 & \quad\int d^{2}b_{i'}^{A}\,T_A(b_{i'}^{A})\int d^{2}b_{j'}^{B}\,T_B(b_{j'}^{B})\nonumber \\
 & \quad\int d^{2}b\,G(x^{+},x^{-},|b+b_{i'}^{A}-b_{j'}^{B}|)\nonumber \\
 & \qquad W_{11}\left((1-x^{+})(1-x^{-})\right),
\end{align}

\noindent where the symbol ``$\propto$'' means proportional. A
variable change as $\vec{b}'=\vec{b}+\vec{b}_{i'}^{A}-\vec{b}_{j'}^{B}$,
and then replacing $b'$ by $b$, and using $\int d^{2}b_{i'}^{A}\,T_A(b_{i'}^{A})=1$
and $\int d^{2}b_{j'}^{B}\,T_B(b_{j'}^{B})=1$, allows one to write 
\begin{align}
\frac{d^{2}\sigma_{\mathrm{incl}}^{AB\,(1)}}{dx^{+}dx^{-}} & \propto\sum_{k'=1}^{AB}\;\sum_{\{m_{k}\}\ne0}\delta_{N_{\mathrm{conn}}(k',1)}^{N_{\mathrm{conn}}}\\
 & \int d^{2}b\,G(x^{+},x^{-},b)\:W_{11}\left((1-x^{+})(1-x^{-})\right),\nonumber 
\end{align}
which is up to a factor equal to 
\begin{equation}
\int d^{2}b\,G(x^{+},x^{-},b)\:W_{11}\left((1-x^{+})(1-x^{-})\right).
\end{equation}
Employing Eqs. (\ref{fundamental-epos4-equation}-\ref{fundamental-4}),
using $N_{\mathrm{conn}}=1$ and $R_{\mathrm{deform}}=1$, this expression
becomes identical to what one found for $d\sigma_{\mathrm{incl}}^{\mathrm{single\,Pom}}/dx^{+}dx^{-}$
[see Eqs. (\ref{single-Pom-ref-1}) and (\ref{single-Pom-ref-2})], and one
may write
\begin{equation}
\frac{d^{2}\sigma_{\mathrm{incl}}^{AB\,(1)}}{dx^{+}dx^{-}}\propto\frac{d\sigma_{\mathrm{incl}}^{\mathrm{single\,Pom}}}{dx^{+}dx^{-}}\left[Q_{\mathrm{sat}}^{2}(1,x^{+},x^{-})\right].
\end{equation}
So even in nuclear collisions, when one restricts oneself to $N_{\mathrm{conn}}=1$,
one recovers the cross section of a single isolated Pomeron. \\

In the case of $N_{\mathrm{conn}}>1$, one has [see Eqs. (\ref{R-deform-definition}) and (\ref{single-Pom-ref-0})]
\begin{align}
 & \frac{d^{2}\sigma_{\mathrm{incl}}^{AB\,(N_{\mathrm{conn}})}}{dx^{+}dx^{-}}\propto R_{\mathrm{deform}}(N_{\mathrm{conn}},x^{+},x^{-})\frac{d\sigma_{\mathrm{incl}}^{\mathrm{single\,Pom}}}{dx^{+}dx^{-}}\\
 & \qquad\qquad\qquad=R_{\mathrm{deform}}(N_{\mathrm{conn}},x^{+},x^{-})\nonumber \\
 & \qquad\qquad\qquad\quad\times\int d^{2}b\,G(x^{+},x^{-},b)\:\nonumber \\
 & \qquad\qquad\qquad\quad\times W_{11}\left((1-x^{+})(1-x^{-})\right).\label{sigma-incl-ABN}
\end{align}

\noindent One may now use Eqs. (\ref{fundamental-epos4-equation}) - (\ref{fundamental-4}),
so the rhs of Eq. (\ref{sigma-incl-ABN}) is given as
\begin{align}
 & R_{\mathrm{deform}}(N_{\mathrm{conn}},x^{+},x^{-})\nonumber\\
 & \qquad\times\int d^{2}b\,\frac{n}{R_{\mathrm{deform}}(N_{\mathrm{conn}},x^{+},x^{-})}\nonumber \\
 & \qquad\qquad\times G_{\mathrm{QCD}}\left(Q_{\mathrm{sat}}^{2}(N_{\mathrm{conn}},x^{+},x^{-}),x^{+},x^{-}\right)\nonumber \\
 & \qquad\qquad\qquad\times W_{11}\left((1-x^{+})(1-x^{-})\right).
\end{align}
Here, $R_{\mathrm{deform}}(N_{\mathrm{conn}},x^{+},x^{-})$ cancels
out. Note that this quantity does not depend on $b$. So one finds
\begin{align}
\frac{d^{2}\sigma_{\mathrm{incl}}^{AB\,(N_{\mathrm{conn}})}}{dx^{+}dx^{-}} & \propto\int d^{2}b\,G_{\mathrm{QCD}}\left(Q_{\mathrm{sat}}^{2}(N_{\mathrm{conn}},x^{+},x^{-})\,,x^{+},x^{-}\right)\nonumber \\
 & \qquad\times W_{11}\left((1-x^{+})(1-x^{-})\right),
\end{align}
and using Eq. (\ref{single-Pom-ref-2}), one gets 
\begin{equation}
\frac{d^{2}\sigma_{\mathrm{incl}}^{AB\,(N_{\mathrm{conn}})}}{dx^{+}dx^{-}}\propto\frac{d\sigma_{\mathrm{incl}}^{\mathrm{single\,Pom}}}{dx^{+}dx^{-}}\left[Q_{\mathrm{sat}}^{2}(N_{\mathrm{conn}},x^{+},x^{-})\right],\label{AB-equal-single-Pom}
\end{equation}
where the rhs is again the single Pomeron ($pp$ scattering) expression,
just with $Q_{\mathrm{sat}}^{2}(N_{\mathrm{coll}},x^{+},x^{-})$ as
saturation scale. The crucial point of Eq. (\ref{AB-equal-single-Pom})
is the fact that thanks to Eq. (\ref{fundamental-epos4-equation})
and since $G$ does not depend on $N_{\mathrm{conn}}$, the $R_{\mathrm{deform}}$
expressions disappear. So I have shown the following: 
\begin{itemize}
\item For $A\!+\!B$ scattering as well as for $pp$ scattering, even with large
$N_{\mathrm{conn}}$, the inclusive cross sections with respect to $x^{\pm}$ are always expressed
in terms of  {\bf single Pomeron cross sections}, depending on an $N_{\mathrm{conn}}$
dependent saturation scale (``dynamical saturation scale'').
\end{itemize}
\noindent The $N_{\mathrm{conn}}$ dependence of $x^{\pm}$ distributions
is guided by the saturation scale, and nothing else. \\

\subsection{Generalized AGK theorem}

\begin{figure}[h]
\centering{}\includegraphics[scale=0.3]
{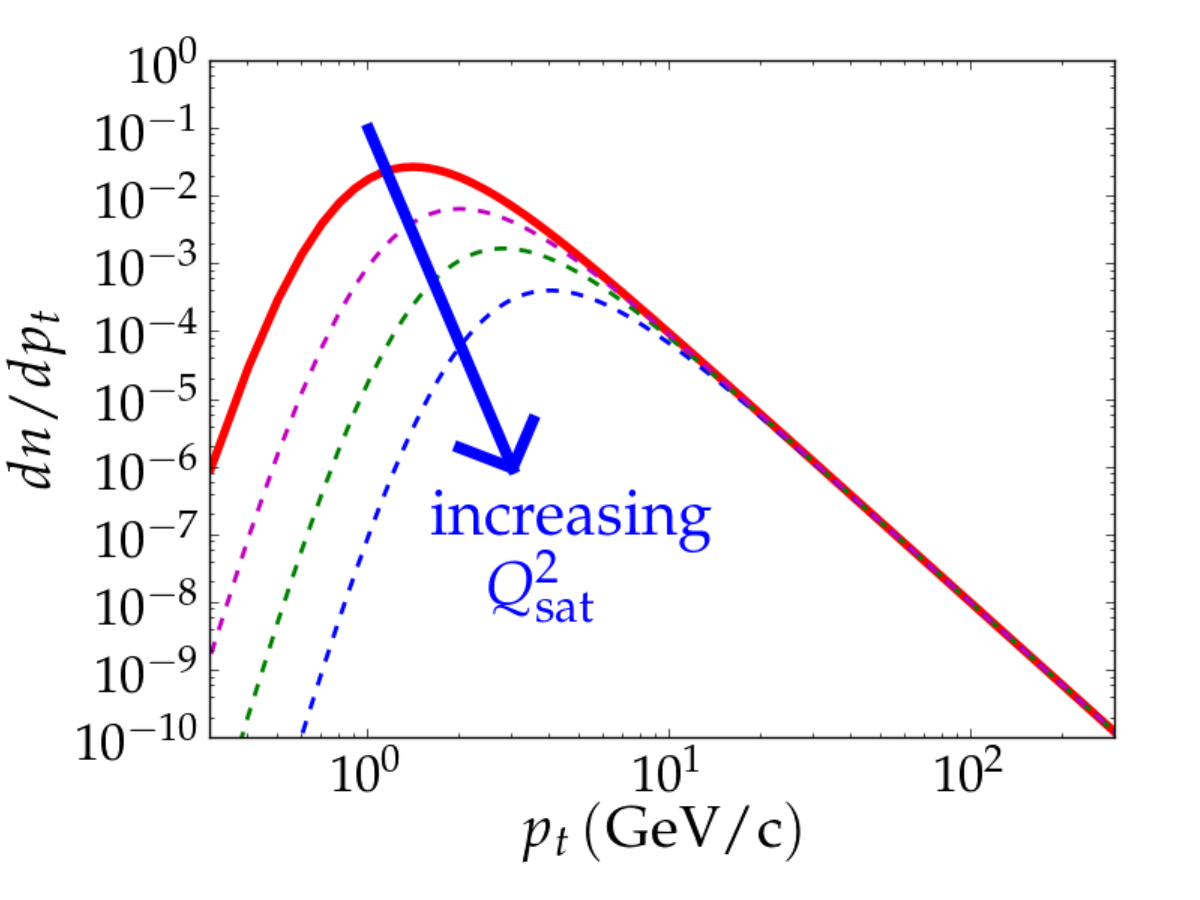}
\caption{Sketch of the suppression of low $p_{t}$ partons with increasing
$Q_{\mathrm{sat}}^{2}$ for a single Pomeron, where the red line corresponds
to some minimum value $Q_{\mathrm{sat\,min}}^{2}$. \label{facto-1}}
\end{figure}

Equations (\ref{AB-equal-single-Pom}) and (\ref{single-Pom-ref-2}) tell
us that also in $A\!+\!B$ scatterings, the partonic structure is given
by $G_{\mathrm{QCD}}$, and therefore also the $p_{t}$ distribution
of the outgoing partons is encoded in the single Pomeron expression
$G_{\mathrm{QCD}}$, for any\textbf{ $N_{\mathrm{conn}}$}. Only the
saturation scales $Q_{\mathrm{sat}}^{2}$ depend on $N_{\mathrm{conn}}$,
and these saturation scales suppress small $p_{t}$ particle production,
but should not affect high $p_{t}$ results, as sketched in Fig. \ref{facto-1}.
To be a bit more quantitative: As discussed in detail in Sec. 3
of Ref. \cite{werner:2023-epos4-heavy}, in the case of $pp$ scattering
with a single Pomeron (of the form $G_{\mathrm{QCD}}$) involved,
one can deduce an inclusive dijet cross section of the form
\begin{align}
 & E_{3}E_{4}\frac{d^{6}\sigma_{\mathrm{incl}}^{\mathrm{single\,Pom}}}{d^{3}p_{3}d^{3}p_{4}}[Q_{0}^{2}]\label{factorization-formula}\\
 & =\sum_{klmn}\int\!\!\int\!\!dx^{+}dx^{-}\,f_{\mathrm{PDF}}^{k}(x^{+},Q_{0}^{2},\mu_{\mathrm{F}}^{2})f_{\mathrm{PDF}}^{l}(x^{-},Q_{0}^{2},\mu_{\mathrm{F}}^{2})\nonumber \\
 & \quad\frac{1}{32s\pi^{2}}\bar{\sum}|\mathcal{M}^{kl\to mn}|^{2}\delta^{4}(p_{1}+p_{2}-p_{3}-p_{4})\frac{1}{1+\delta_{mn}},\nonumber 
\end{align}
with explicit expressions for the parton distribution functions (PDFs)
given in terms of elements (or modules ) of $G_{\mathrm{QCD}}$, like
vertices and evolution functions. I indicate explicitly
the dependence on the low virtuality cutoff $Q_{0}^{2}$. The differential cross
section in Eq. (\ref{factorization-formula})
and the one
in Eq. (\ref{AB-equal-single-Pom}) refer to the same cross section $\sigma_{\mathrm{incl}}^{\mathrm{single\,Pom}}$, 
just in Eq. (\ref{factorization-formula}) one integrates over $x^{\pm}$,
whereas in Eq. (\ref{AB-equal-single-Pom}) the momenta $p_{3}$ and $p_{4}$ are integrated out. 
These momenta
refer to the outgoing partons, from the Born process, and $\mu_{\mathrm{F}}^{2}$
is the factorization scale, being of the same order as the
outgoing momenta. Assuming $f_{\mathrm{PDF}}^{k}(x^{+},Q_{0}^{2},\mu_{\mathrm{F}}^{2})$
to be independent of $Q_{0}^{2}$ for $\mu_{\mathrm{F}}^{2}>Q_{0}^{2}$,
then the only effect of $Q_{0}^{2}$ is a reduction of particles with
$p_{t}^{2}$ below $Q_{0}^{2}.$ In the case of $A\!+\!B$ scattering for given
$N_{\mathrm{conn}}$  [see Eq. (\ref{AB-equal-single-Pom})], one has (up
to a factor) the same formula, just $Q_{0}^{2}$ has to be replaced
by $Q_{\mathrm{sat}}^{2}(N_{\mathrm{coll}},x^{+},x^{-})$, with correspondingly
a reduction of particle production below this scale. The minimum bias
(MB) inclusive cross section may be written as a superposition of
the different contribution for given values of $N_{\mathrm{conn}}$
with weights $w^{(N_{\mathrm{conn}})}$,
\begin{align}
 & E_{3}E_{4}\frac{d^{6}\sigma_{\mathrm{incl}}^{AB\,(\mathrm{MB})}}{d^{3}p_{3}d^{3}p_{4}}\propto\sum_{N_{\mathrm{conn}}=1}^{\infty}w^{(N_{\mathrm{conn}})}\,\\
 & \qquad\times\,E_{3}E_{4}\frac{d^{6}\sigma_{\mathrm{incl}}^{\mathrm{single\,Pom}}}{d^{3}p_{3}d^{3}p_{4}}\left[Q_{\mathrm{sat}}^{2}(N_{\mathrm{conn}},x^{+},x^{-})\right].\nonumber 
\end{align}
If one is only interested in $p_{t}^{2}$ values bigger than the saturation
scales, then one may replace the saturation scales with some constant value, say $Q_{0}^{2}$, and using $\sum_{N_{\mathrm{conn}}=1}^{\infty}w^{(N_{\mathrm{conn}})}=1$,
one finds the ``generalized AGK theorem'':
\begin{align}
 & E_{3}E_{4}\frac{d^{6}\sigma_{\mathrm{incl}}^{AB\,(\mathrm{MB})}}{d^{3}p_{3}d^{3}p_{4}}\label{binary-scaling-2}\\
 & \qquad=AB\times E_{3}E_{4}\frac{d^{6}\sigma_{\mathrm{incl}}^{\mathrm{single\,Pom}}}{d^{3}p_{3}d^{3}p_{4}}\left[Q_{0}^{2}\right],\:\mathrm{(at\,large}\,p_{t}),\nonumber 
\end{align}
where the ``normalization constant'' $n$ in Eq. (\ref{fundamental-epos4-equation})
has been used to ensure the factor $AB$. The term ``large $p_{t}$''
means $p_{t}^{2}$ values bigger than all the ``relevant'' $Q_{\mathrm{sat}}^{2}$
values (with non-negligible weights $w^{(N_{\mathrm{conn}})}$). 

\begin{figure}[h]
\centering{}\includegraphics[bb=0bp 30bp 595bp 550bp,clip,scale=0.25]
{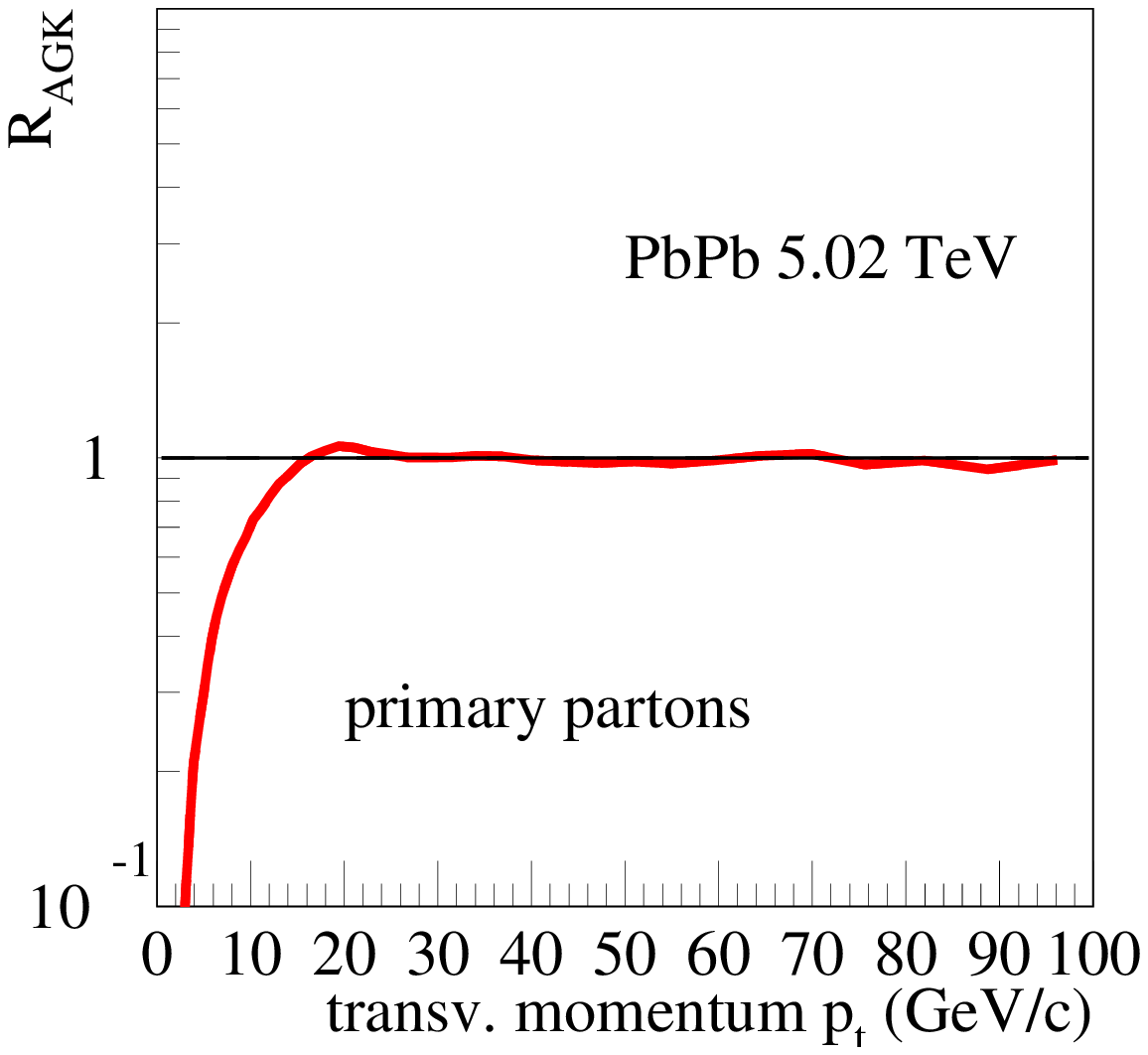}
\caption{The ratio $R_{\mathrm{AGK}}$ (see text) for minimum bias PbPb collisions
at 5.02 TeV. \label{r-deform-1}}
\end{figure}

In the following, it will be  shown that AGK really works
in practice. One defines $p_{t}$ to be the transverse momentum of
particle 3 in Eq. (\ref{binary-scaling-2}), and one integrates over
the longitudinal momentum of particle 3 and the three momentum components
of particle 4. Then one computes the ratio of the full Monte Carlo simulation
over $AB$ times the single Pomeron distribution,
\begin{equation}
R_{\mathrm{AGK}}=\left.\frac{d\sigma_{\mathrm{incl}}^{AB\,(\mathrm{MB})}}{dp_{t}}\right|_{\mathrm{Full\,MC}}\,\Bigg/\,\left\{ AB\frac{d\sigma_{\mathrm{incl}}^{single\,Pom}}{dp_{t}}\left[Q_{0}^{2}\right]\right\} ,
\end{equation}
showing the result in Fig. \ref{r-deform-1}.
The single Pomeron result is calculated numerically (without Monte
Carlo). The simulations are done  for minimum bias PbPb collisions at 5.02 TeV.
One can see that the ratio is close to unity for large values of $p_{t}$,
whereas low $p_{t}$ values are suppressed. In other words, AGK holds
at high $p_{t}$! Not shown here, but also in $pp$, the full simulation
over the single Pomeron reference curve is close to unity at large
$p_{t}$ and  one gets therefore $R_{\mathrm{AA}}\approx1$.\\

\subsection{Conclusion}

\noindent Let me summarize this (crucial) section: 
\begin{itemize}
\item It is first recalled that the validity of the AGK theorem (needed to get binary
scaling) is badly violated, which manifests itself by $R_{\mathrm{deform}}(x^{+},x^{-})\ne1$,
and one understands that this is unavoidable.
\item It is also recalled that binary scaling is not always expected, but only
at high $p_{t}$ (concerning the outgoing partons), and that the behavior
of $p_{t}$ distributions depends on the internal structure of the single
Pomeron $G$. So far, I was using $G=G_{\mathrm{QCD}}$, with $G_{\mathrm{QCD}}$
(explained in Ref. \cite{werner:2023-epos4-heavy}) essentially based on
DGLAP evolutions preceding a hard QCD scattering. This association
seems to be wrong.
\item It is pointed out that a well-known feature is completely missing --
saturation -- and this problem seems to be related to the before-mentioned
problem of a wrong association $G=G_{\mathrm{QCD}}$.
\item It is  proposed to solve both problems, by postulating 
the association $G=k\times G_{\mathrm{QCD}}(Q_{\mathrm{sat}}^{2},x^{+},x^{-})$,
with $k$ being inversely proportional to the deformation function,
which defines a saturation scale $Q_{\mathrm{sat}}^{2}$, depending
on the ``environment'' in terms of the Pomeron connection number
$N_{\mathrm{conn}}$, which replaces the virtuality cutoff $Q_{0}^{2}$
usually used in DGLAP evolutions. In this way, one incorporates saturation.
\item One can prove that
for $A\!+\!B$ scattering, inclusive cross sections -- with
respect to $x^{\pm}$ -- are always expressed in terms
of the \textbf{single Pomeron} cross section, but depending on an
$N_{\mathrm{conn}}$-dependent saturation scale.%
\item One can prove that
for minimum bias $A\!+\!B$ scattering, the inclusive cross section --
with respect to the transverse momenta of the outgoing partons, for large transverse
momenta -- is equal to $AB$ times the one of the corresponding
\textbf{single Pomeron }cross section. One refers to this as the ``\textbf{generalized
AGK theorem}'', valid at high $p_{t}$, in a scenario with energy
sharing. %
\end{itemize}
As a final remark: Within a rigorous parallel scattering
scenario (which seems mandatory), and respecting energy conservation
(which seems mandatory as well), the only way to not get in contradiction
with factorization and binary scaling seems to be the consideration
of saturation via $G=k\times G_{\mathrm{QCD}}(Q_{\mathrm{sat}}^{2})$
with $k$ being inversely proportional to the deformation function.
In this sense, parallel scattering, energy conservation, saturation,
and factorization (and binary scaling in $A\!+\!B$) are deeply connected.%
\begin{comment}
\noindent Having very narrow $y_{\mathrm{PE}}$ distributions, one
may use the approximation $y_{\mathrm{PE}}=0$, $x^{\pm}=\sqrt{x_{\mathrm{PE}}}e^{\pm y_{\mathrm{PE}}}$
$=\sqrt{x_{\mathrm{PE}}}$), and one gets
\begin{equation}
\frac{d^{2}\sigma_{\mathrm{incl}}^{AB\,(1)}}{dx_{\mathrm{PE}}}\!\propto\!\int d^{2}b\,G(x^{+},x^{-},b)\:W_{11}\left((1-x^{+})(1-x^{-})\right).
\end{equation}
\end{comment}

\section{High- and low-$\boldsymbol{\boldsymbol{p_{t}}}$ domain \label{======= high-and-low pt  =======}}

\begin{figure}[h]
\begin{centering}
\noindent\begin{minipage}[c]{0.1\columnwidth}%
\begin{center}
{\huge{}$\sum$}
\par\end{center}%
\end{minipage}\hspace*{-0.5cm}%
\begin{minipage}[c]{0.37\columnwidth}%
\begin{center}
\includegraphics[scale=0.3]
{Home/2/komm/conf/99aktuell/epos/poms1d}
\par\end{center}%
\end{minipage}$\quad\quad$%
\begin{minipage}[c]{0.22\columnwidth}%
\begin{center}
{\huge{}$\boldsymbol{\boldsymbol{=AB\,\times}}$}
\par\end{center}%
\end{minipage}%
\begin{minipage}[c]{0.22\columnwidth}%
\begin{center}
\includegraphics[scale=0.3]
{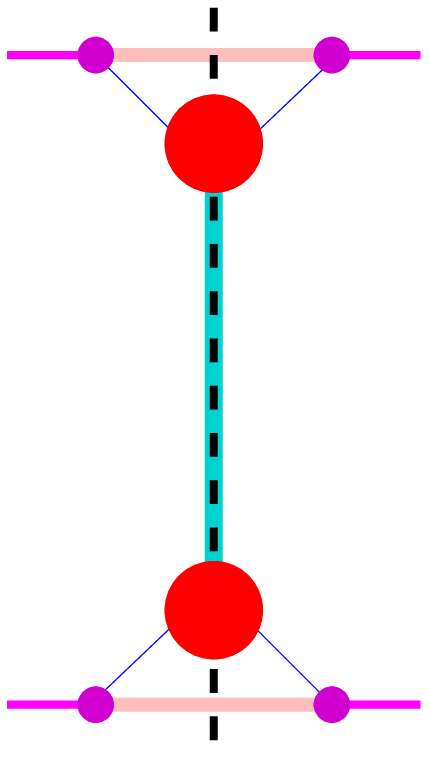}
\par\end{center}%
\end{minipage}
\par\end{centering}
\centering{}\caption{The only diagram contributing to inclusive cross sections in $A\!+\!B$ scattering:
a single cut Pomeron (multiplied by $AB$). \label{cut-diagram-2-1}}
\end{figure}

In Sec. \ref{======= the solution =======}, I have discussed in great detail the ``asymptotic
behavior'', i.e., the fact that one recovers in the new formalism
the validity of the AGK theorem \textbf{at high transverse momenta},
saying that for \textbf{inclusive cross sections} 
everything breaks down to the single Pomeron case (see Fig. \ref{cut-diagram-2-1}),
although one has
in reality many scatterings happening in parallel.

\begin{figure}[h]
\centering{}(a)$\;$ \hspace*{6cm}$\;$\\
 \includegraphics[bb=0bp 0bp 460bp 405bp,scale=0.3]
{Home/2/komm/conf/99aktuell/epos/ldst2h7c}\\
(b)$\;$ \hspace*{6cm}$\;$\\
$\qquad\qquad$\includegraphics[scale=0.3]
{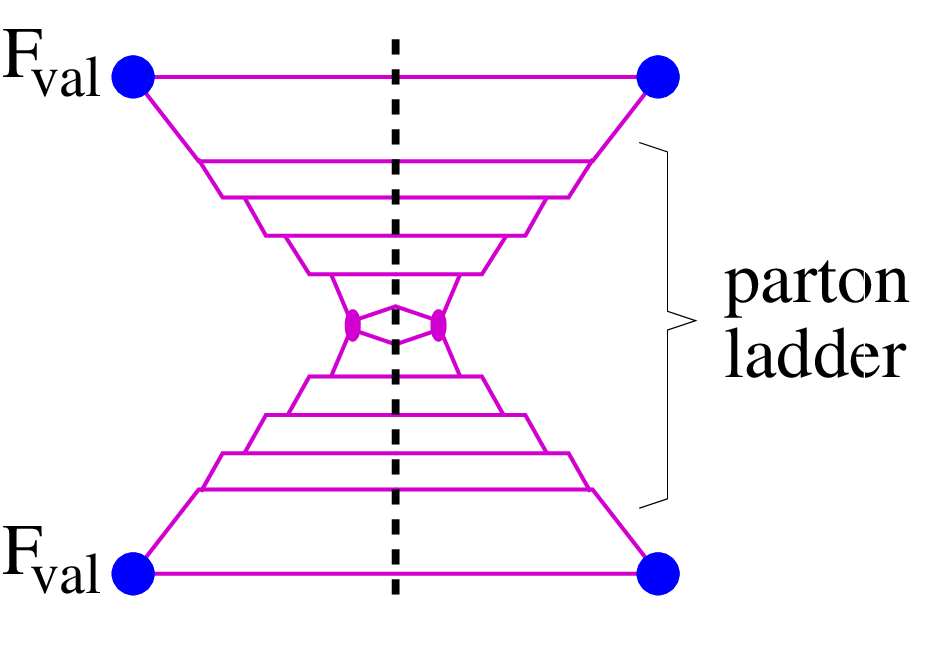}
\caption{The contributions (a) $G_{\mathrm{QCD}}^{\mathrm{sea-sea}}$ and (b) $G_{\mathrm{QCD}}^{\mathrm{val-val}}$. \label{G-sea-sea}}
\end{figure}

\begin{figure}[h]
\centering{}%
\begin{minipage}[c]{0.25\columnwidth}%
{\LARGE{}$f_{\mathrm{PDF}}=$}{\LARGE\par}

~%
\end{minipage}%
\begin{minipage}[c]{0.28\columnwidth}%
\includegraphics[bb=0bp 0bp 220bp 205bp,clip,scale=0.3]
{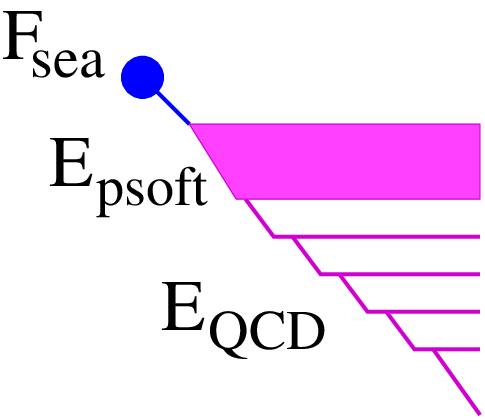}%
\end{minipage}%
\noindent\begin{minipage}[c]{0.1\columnwidth}%
~
\begin{center}
{\LARGE{}+}{\LARGE\par}
\par\end{center}
~%
\end{minipage}%
\begin{minipage}[c]{0.2\columnwidth}%
\includegraphics[bb=0bp 0bp 220bp 195bp,clip,scale=0.3]
{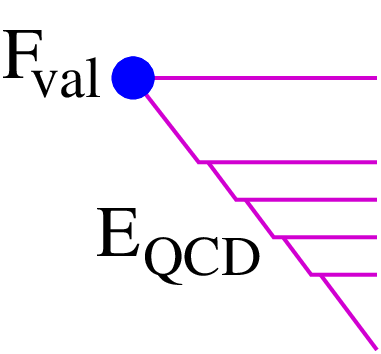}

~%
\end{minipage}\caption{The two contributions of the parton distribution function (a common
remnant vertex factor $V$ is not shown).\label{G-sea-sea-2}}
\end{figure}

To compute the inclusive cross section versus $p_{t}$ (of the outgoing
partons), one needs to consider the internal structure of $G_{\mathrm{QCD}}$,
which is discussed in detail in Secs. 2 and 3 of Ref. \cite{werner:2023-epos4-heavy},
where I show that $G_{\mathrm{QCD}}$ is given as a sum of several
contributions: There is the ``sea-sea'' contribution, where one
has a ``pseudosoft block'' preceding the first perturbative parton,
as indicated in Fig. \ref{G-sea-sea}(a). Here, a sea quark or a gluon
is the first parton entering the parton ladder. The vertices $F_{\text{sea}}$
represent couplings to the projectile and target nucleons.
\begin{figure}[h]
\centering{}(a)$\;$ \hspace*{3.7cm}$\;$(b)$\;$\hspace*{2.7cm}$\quad$\\
\begin{minipage}[c]{0.5\columnwidth}%
\includegraphics[bb=0bp 0bp 220bp 405bp,scale=0.3]
{Home/2/komm/conf/99aktuell/epos/ldst2h7d}%
\end{minipage}%
\begin{minipage}[c]{0.4\columnwidth}%
$\qquad$\\
 \includegraphics[scale=0.3]
{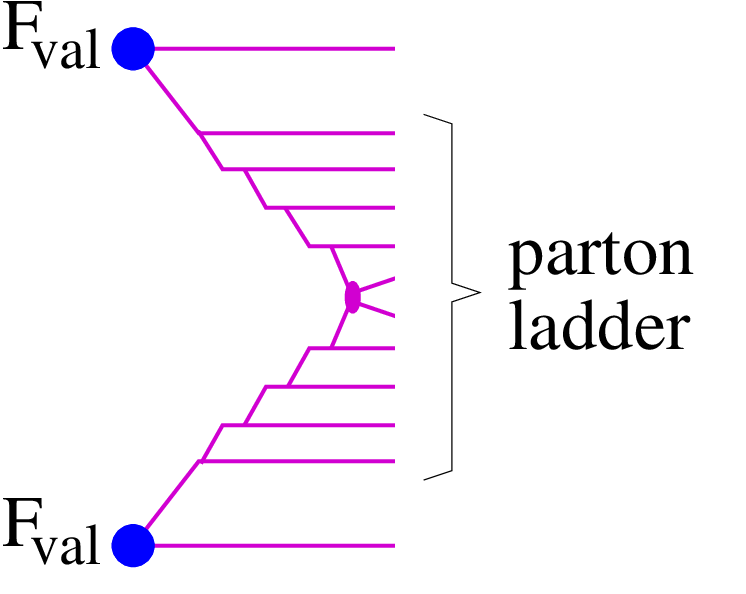}%
\end{minipage}\caption{The two inelastic processes corresponding to the two cut diagrams
of Fig. \ref{G-sea-sea}.\label{G-sea-sea-1}}
\end{figure}
Then there is the ``val-val'' contribution where on both sides a
valence quark is the first parton entering the parton ladder, as shown
in Fig. \ref{G-sea-sea}(b). The vertices $F_{\text{val}}$ represent
couplings to the projectile and target nucleons. In both cases, the
central part is a parton ladder based on DGLAP parton evolutions \cite{GribovLipatov:1972,AltarelliParisi:1977,Dokshitzer:1977}.
The inelastic processes corresponding to the two cut diagrams of Fig.
\ref{G-sea-sea} are shown in Fig. \ref{G-sea-sea-1}. In addition
to ``sea-sea'' and ``val-val'', one has also the mixed contributions
``sea-val'' and ``val-sea''. In Ref. \cite{werner:2023-epos4-heavy},
one finds explicit expressions for the four cases, expressed in terms
of the modules $F_{\text{sea}}$, $F_{\text{val}}$, $E_{\mathrm{psoft}}$,
$E_{\mathrm{QCD}}$, and ``Born'' (the QCD matrix elements of the Born
process). As shown in Ref. \cite{werner:2023-epos4-heavy}, on may
rearrange the integrations such that one may define parton distribution
functions $f_{\mathrm{PDF}}$, corresponding to the sum of
two diagrams as shown in Fig. \ref{G-sea-sea-2}, which allows one to
write jet cross sections as an integral over PDFs and a QCD matrix
element [see Eq. (\ref{factorization-formula})], which amounts to
factorization.
\begin{figure}[h]
\centering{}\includegraphics[bb=20bp 20bp 590bp 810bp,clip,scale=0.42]
{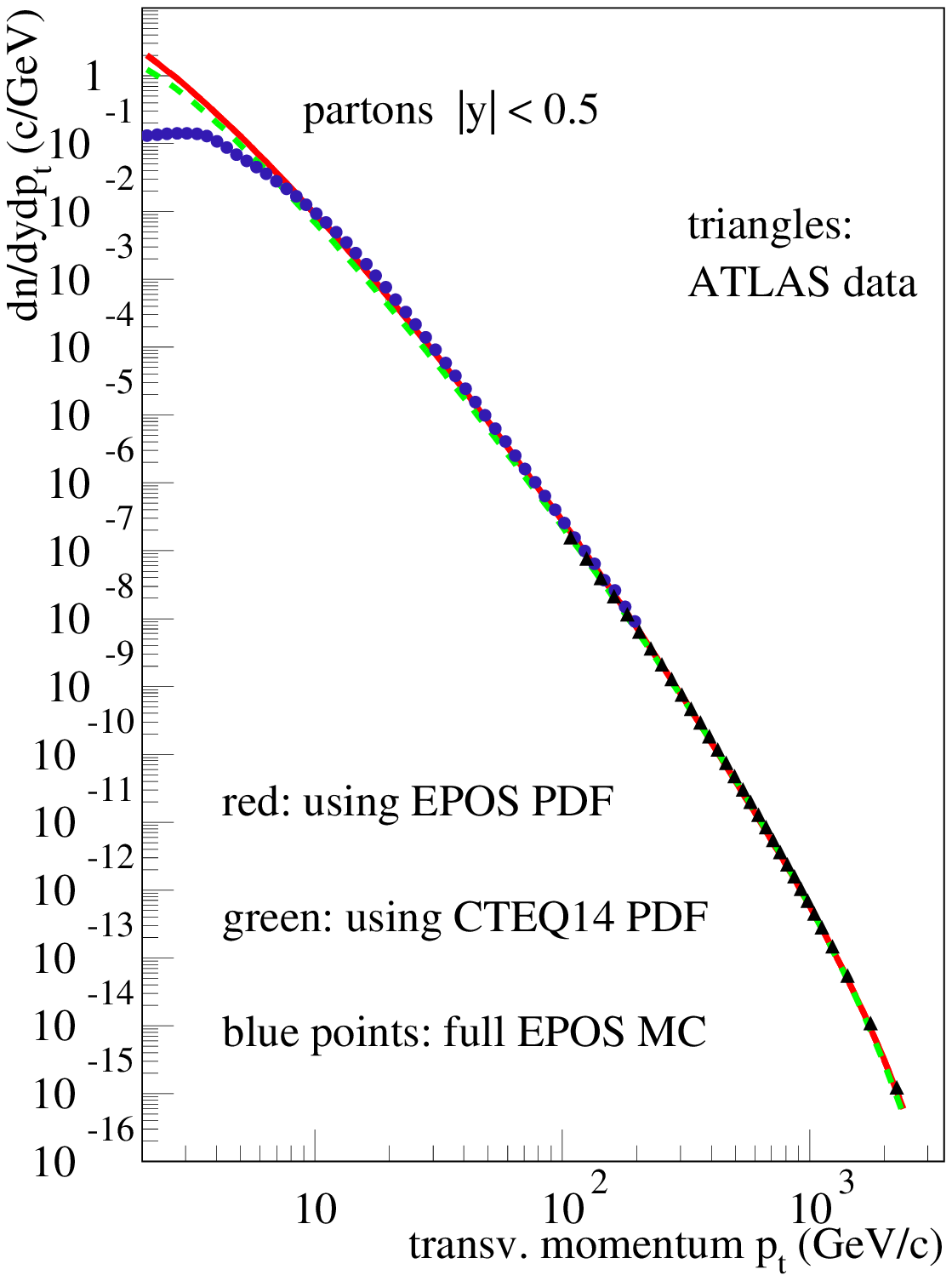}\\
 \caption{Parton yield $dn/dp_{t}dy$ for $pp$ at 13 TeV. I show results based
on EPOS PDFs (red solid line), CTEQ PDFs (green dashed line), the full
EPOS simulation (blue circles), and experimental data from ATLAS (black
triangles). \label{parton-pt-spectra}}
\end{figure}
I use these PDFs to compute the jet (parton) cross section for pp
at 13 TeV, using Eq. (\ref{factorization-formula}), integrating out
the momentum of the second parton and the azimuthal angle of the first
parton, which finally gives (see Ref. \cite{werner:2023-epos4-heavy})
\begin{align}
 & \frac{d^{2}\sigma}{dy\,dp_{t}^{2}}=\sum_{klmn}\int dx\,f_{\mathrm{PDF}}^{k}(x_{1},\mu_{\mathrm{F}}^{2})f_{\mathrm{PDF}}^{l}(x_{2},\mu_{\mathrm{F}}^{2})\nonumber \\
 & \qquad\times\frac{\pi\alpha_{s}^{2}}{s^{2}}\left\{ \begin{array}{c}
\frac{1}{g^{4}}\end{array}\bar{\sum}|\mathcal{M}^{kl\to mn}(s,t)|^{2}\right\} \frac{1}{1\!+\!\delta_{mn}}\,x_{1}x_{2}\frac{1}{x}\,,\label{jet-differential-xsection-1}\\
 & \mathrm{\quad with}\;x_{1}=x+\begin{array}{c}
\frac{p_{t}}{\sqrt{s_{\mathrm{pp}}}}\end{array}e^{y},\,x_{2}=\begin{array}{c}
\frac{x_{1}}{x}\,\frac{p_{t}}{\sqrt{s_{\mathrm{pp}}}}\end{array}e^{-y},\nonumber \\
 & \qquad\qquad s=x_{1}x_{2}\sqrt{s_{\mathrm{pp}}},\,t=-p_{t}x_{1}\sqrt{s_{\mathrm{pp}}}e^{-y},
\end{align}
with $\left\{ ...\right\}$ being the form in which the squared matrix elements
are usually tabulated, with $\alpha_{s}=g^{2}/4\pi$. I define the
parton yield $dn/dp_{t}dy$ as the cross section $d\sigma/dy\,dp_{t}^{2}$,
divided by the inelastic $pp$ cross section, times $2\,p_{t}$, showing
the result in Fig. \ref{parton-pt-spectra}. In addition to the results based
on EPOS PDFs (red solid line), I show the corresponding curves based on 
"The Coordinated Theoretical-Experimental Project on QCD" (CTEQ) 
PDFs \cite{Dulat_2016-CTEQ-PDF}
(green dashed line), the full EPOS simulation (blue circles), and
experimental data from ATLAS \cite{ATLAS:2017ble} (black triangles).
At large values of $p_{t}$, all the different distributions agree,
whereas at low $p_{t}$ the EPOS Monte Carlo simulation results (using
the full multiple scattering scenario) are significantly below the
PDF results, as expected due to screening effects.\\

\begin{figure}[h]
\noindent \centering{}\includegraphics[scale=0.25]
{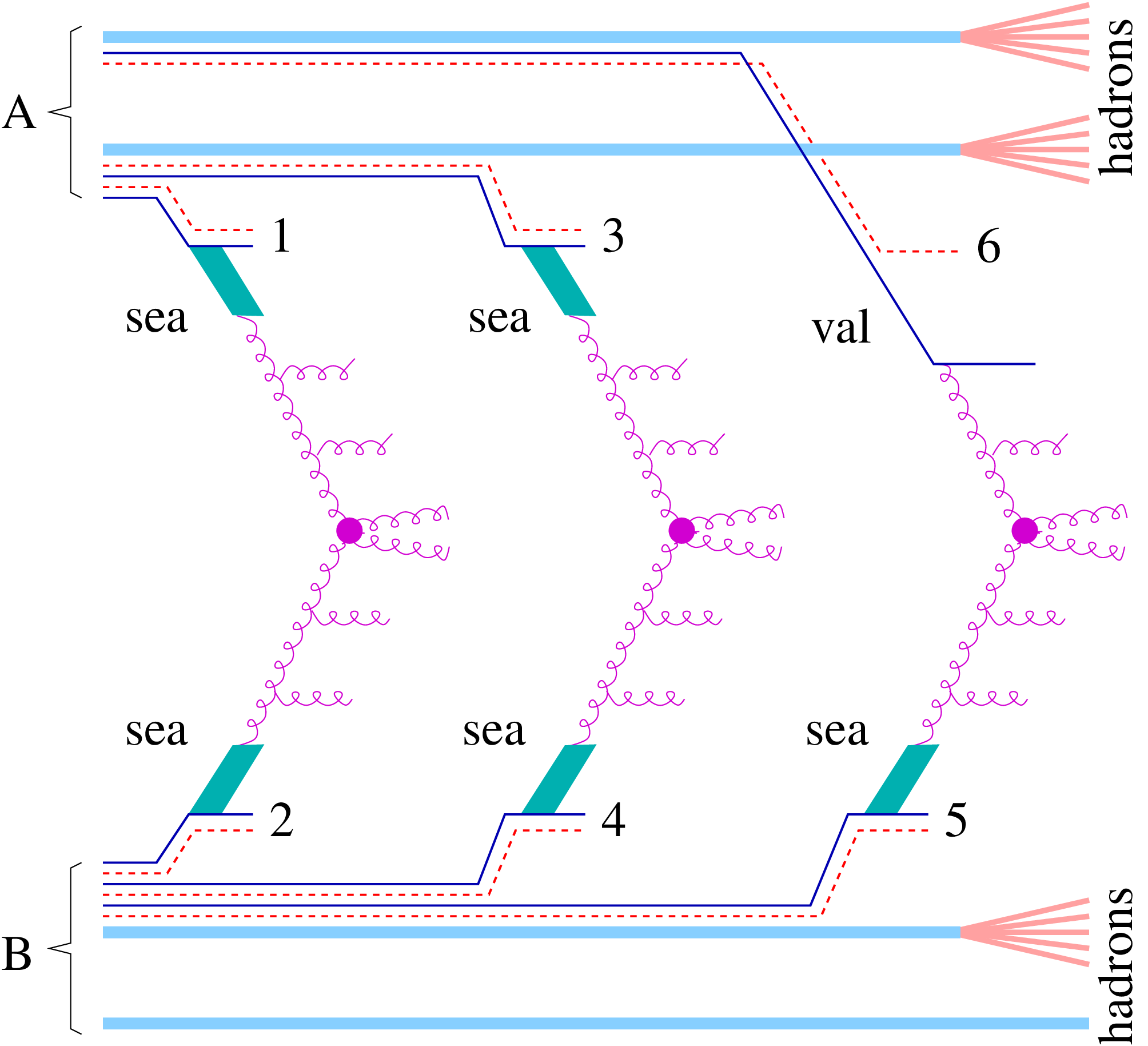}
\caption{Partonic configuration of two colliding nuclei, each one composed
of two nucleons, with three scatterings (from three cut Pomerons).
Dark blue lines mark active quarks, red dashed lines active antiquarks,
and light blue thick lines projectile and target remnants. One of
the target nucleons is just a spectator.\label{partonic-configuration}}
\end{figure}

The possibility of using factorization is extremely useful, when one
is interested in rare processes such as particle production at large transverse
momentum, and it is very important that in EPOS4 one recovers factorization
at large $p_{t}$, since it is a crucial element and its violation
would simply disqualify the model. However, the majority of all particles
are produced at low $p_{t}$, and even when one is interested in high
$p_{t}$ particles, one needs to worry about possible interactions
with the ``bulk'' of low $p_{t}$ particles. And here, one needs
to employ the full multiple scattering machinery, and this concerns
$pp$ and nuclear scatterings. Multiple Pomeron configurations are
generated using the probability law [see Eqs. (\ref{proba-interpretation}) and (\ref{W-AB-7-1})]
\begin{equation}
P(K)=\prod_{k=1}^{AB}\left[\frac{1}{m_{k}!}\prod_{\mu=1}^{m_{k}}G_{k\mu}\right]\times W_{AB}(\{x_{i}^{+}\},\{x_{j}^{-}\}),\label{proba-interpretation-2-1}
\end{equation}
with $G_{k\mu}\!=G\!\left(\!x_{k\mu}^{+},x_{k\mu}^{-},s,b_{k}\!\right)$ [see Eq. (\ref{G-k-mu})]
being expressed in terms of $G_{\mathrm{QCD}}$ via the
fundamental equation (\ref{fundamental-epos4-equation}), by introducing
dynamical saturation scales, as discussed in Sec. \ref{======= the solution =======}.
In Sec. 3 of Ref. \cite{werner:2023-epos4-heavy}, it is explained
in detail how to generate partons, based on the explicit expressions
for $G_{\mathrm{QCD}}$, which will allow to obtain ``partonic configurations'',
as shown in Fig. \ref{partonic-configuration}
for the example of a collision of two nuclei $A$ and $B$,
where (for simplicity) each nucleus is composed of two nucleons. 
Dark blue lines mark active
quarks, red dashed lines active antiquarks, and light blue thick lines
projectile and target remnants [nucleons minus the active (anti)quarks].
There are two scatterings of sea-sea type, and one of val-sea
type. One considers each incident nucleon as a reservoir of three
valence quarks plus quark-antiquark pairs. The ``objects'' which
represent the ``external legs'' of the individual scatterings are
colorwise ``white'': quark-antiquark pairs in most cases as shown
in the figure, but one may as well have quark-diquark pairs, or even
antiquark-antidiquark pairs -- in any case, a $3$ and
a $\bar{3}$ color representation. 

The transition from partonic configurations (as in Fig. \ref{partonic-configuration})
to strings, based on color flow diagrams, is discussed in detail in
Sec. 4 of Ref. \cite{werner:2023-epos4-heavy}. A short summary
will be given in the following.

Let me for simplicity consider the quark-antiquark option for the
$3$ and the $\bar{3}$ color representations, and first of all look
at the ``sea'' cases (on the projectile or the target side) of Fig.
\ref{partonic-configuration}. In each case, a quark-antiquark pair
is emitted as final state time-like (TL) parton pair (marked 1, 2, 3, 4, and 5)
and a spacelike (SL) ``soft Pomeron'' (indicated by a thick cyan
line), which is meant to be similar to the QCD evolution, but emitting
only soft gluons, which one does not treat explicitly. Then emerging
from this soft Pomeron, one sees a first perturbative SL gluon (another
possibility is the emission of a quark), which initiates the partonic
cascade. In the case of ``val'', one also has a quark-antiquark
pair as an external leg, but here first an antiquark is emitted as TL
final particle (marked 6), plus an SL quark starting the partonic
cascade.

In the case of multiple scattering as in Fig. \ref{partonic-configuration},
the projectile and target remnants remain colorwise white, but they
may change their flavor content during the multiple collision process.
The quark-antiquark pair ``taken out'' for a collision (the ``external
legs'' for the individual collisions), maybe $u-\bar{s}$, then
the remnant for an incident proton has flavor $uds$. In addition,
the remnants get massive, much more than simply resonance excitation.
One may have remnants with masses of $10\,\mathrm{GeV/c^{2}}$ or
more, which contribute significantly to particle production (at large
rapidities).

\begin{figure}[h]
\noindent \centering{}\includegraphics[scale=0.25]
{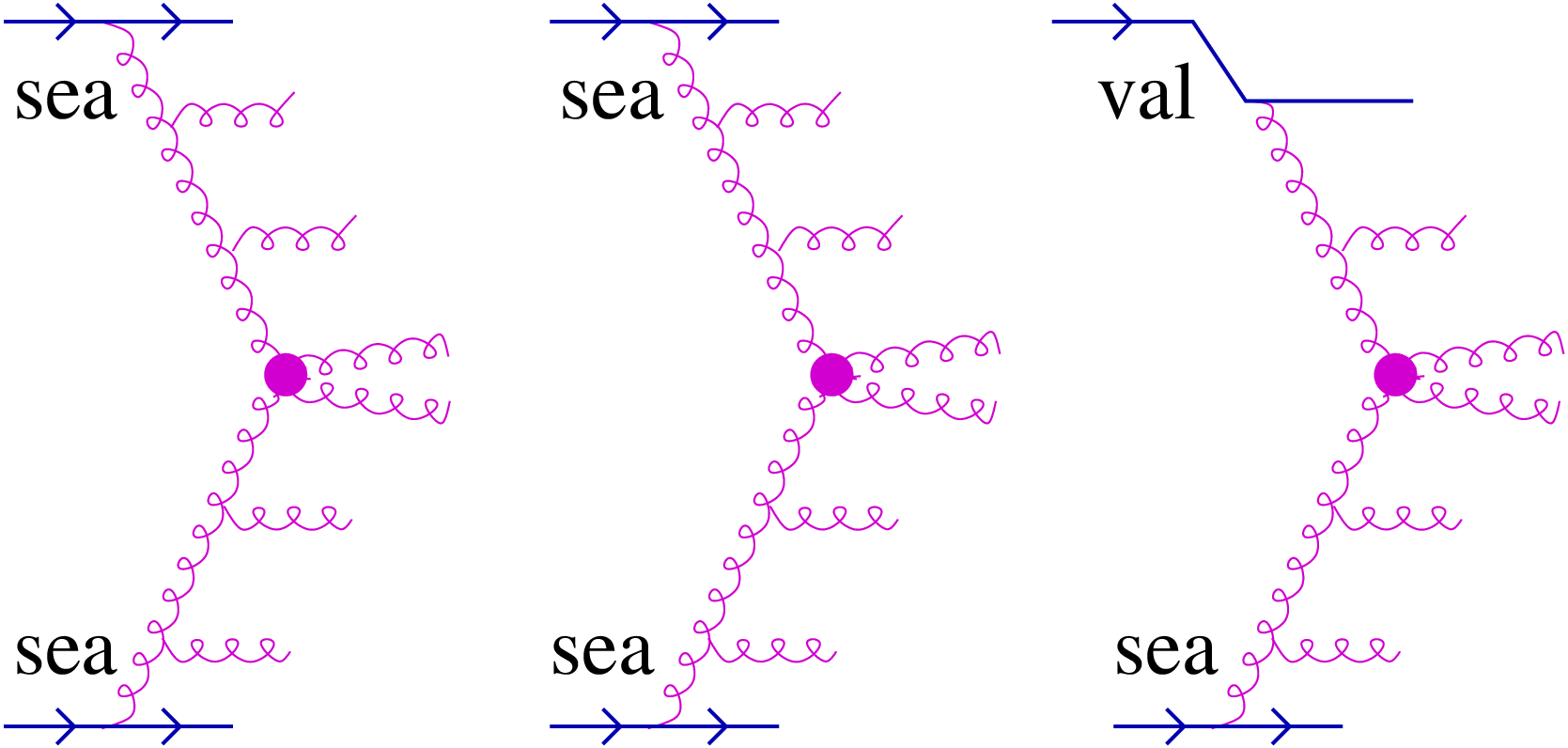}
\caption{Configuration colorwise equivalent to the one of Fig. \ref{partonic-configuration}.
The outgoing antiquarks are drawn as incoming quarks (arrows towards
vertices). \label{partonic-configuration-1}}
\end{figure}

\begin{figure}[h]
\noindent \centering{}\includegraphics[scale=0.25]
{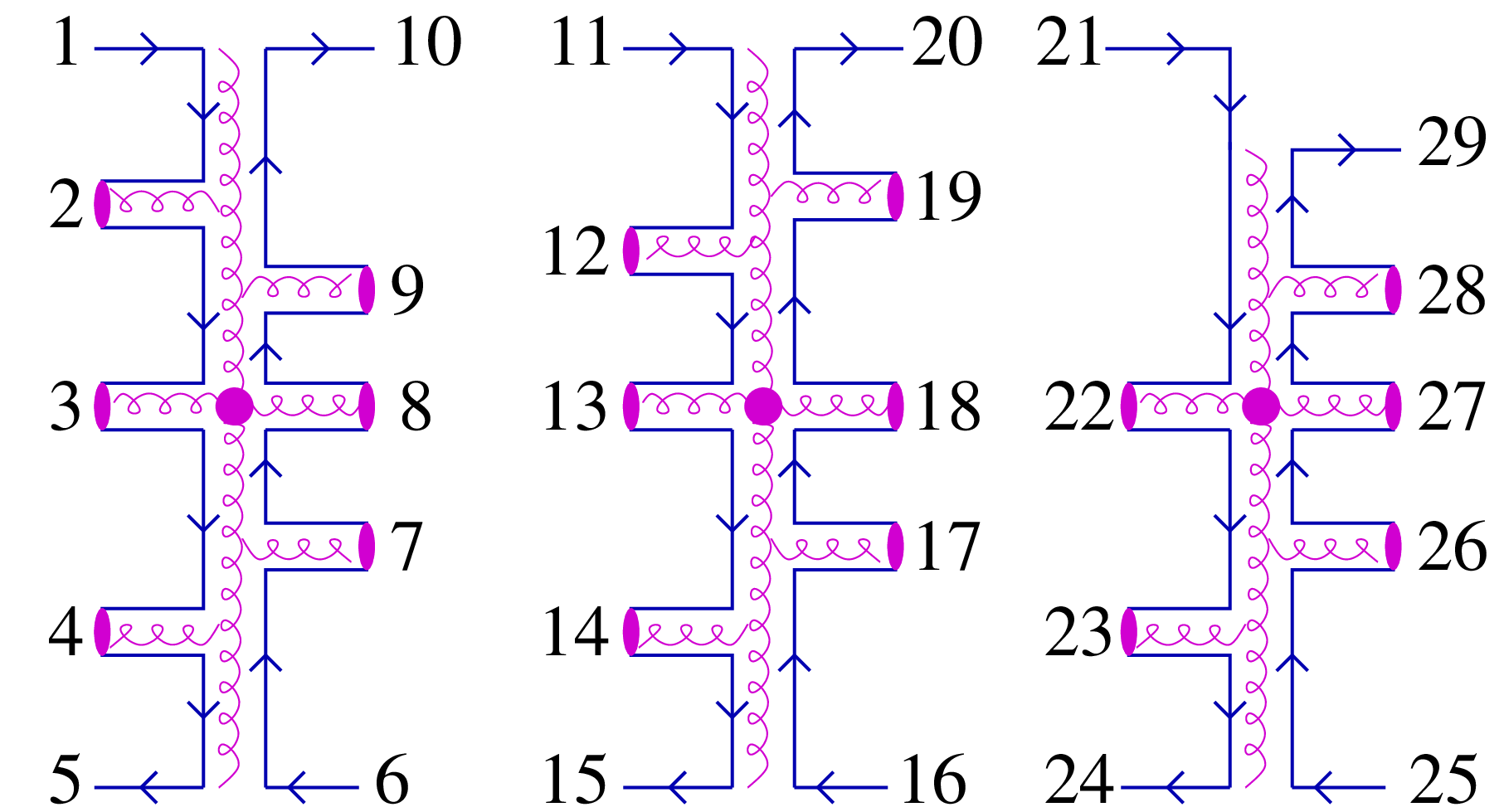}
\caption{A possible color flow diagram for the three scatterings of Fig. \ref{partonic-configuration-1}.
\label{partonic-configuration-1-1}}
\end{figure}

In the following, I discuss the color flow for a given configuration,
for example, the one in Fig. \ref{partonic-configuration}. Since
the remnants are by construction white, one does not need to worry
about them, one just considers the rest of the diagram. In addition,
colorwise, the ``soft Pomeron'' part behaves as a gluon. Finally,
I use the following convention for the SL partons which are immediately
emitted: one uses for the quarks an array away from the vertex, and
for the antiquarks an array towards the vertex. The diagram equivalent
to Fig. \ref{partonic-configuration} is then the one shown in Fig.
\ref{partonic-configuration-1}. 
Based on Fig. \ref{partonic-configuration-1}, considering the fact
that in the parton evolution and the Born process, the gluons are
emitted randomly to the right or to the left, I show in Fig. \ref{partonic-configuration-1-1}
a possible color flow diagram for the three scatterings. Horizontal
lines refer to TL partons, which later undergo a time-like cascade,
while the vertical lines refer to spacelike intermediate partons.
I added integers just to mark the different TL partons. For the leftmost
scattering, starting from one ``end'', say ``1'', one follows
the color flow to ``5'', and then starting from ``6'' to ``10'',
so one gets two chains: 1-2-3-4-5 and 6-7-8-9-10. The end partons
of each chain are always quarks or antiquarks, the inner partons are
gluons. Similar chains are obtained for the second scattering, 11-12-13-14-15
and 16-17-18-19-20, and for the third scattering, 21-22-23-24 and
25-26-27-28-29. 

\begin{figure}[h]
\noindent \centering{}\includegraphics[scale=0.25]
{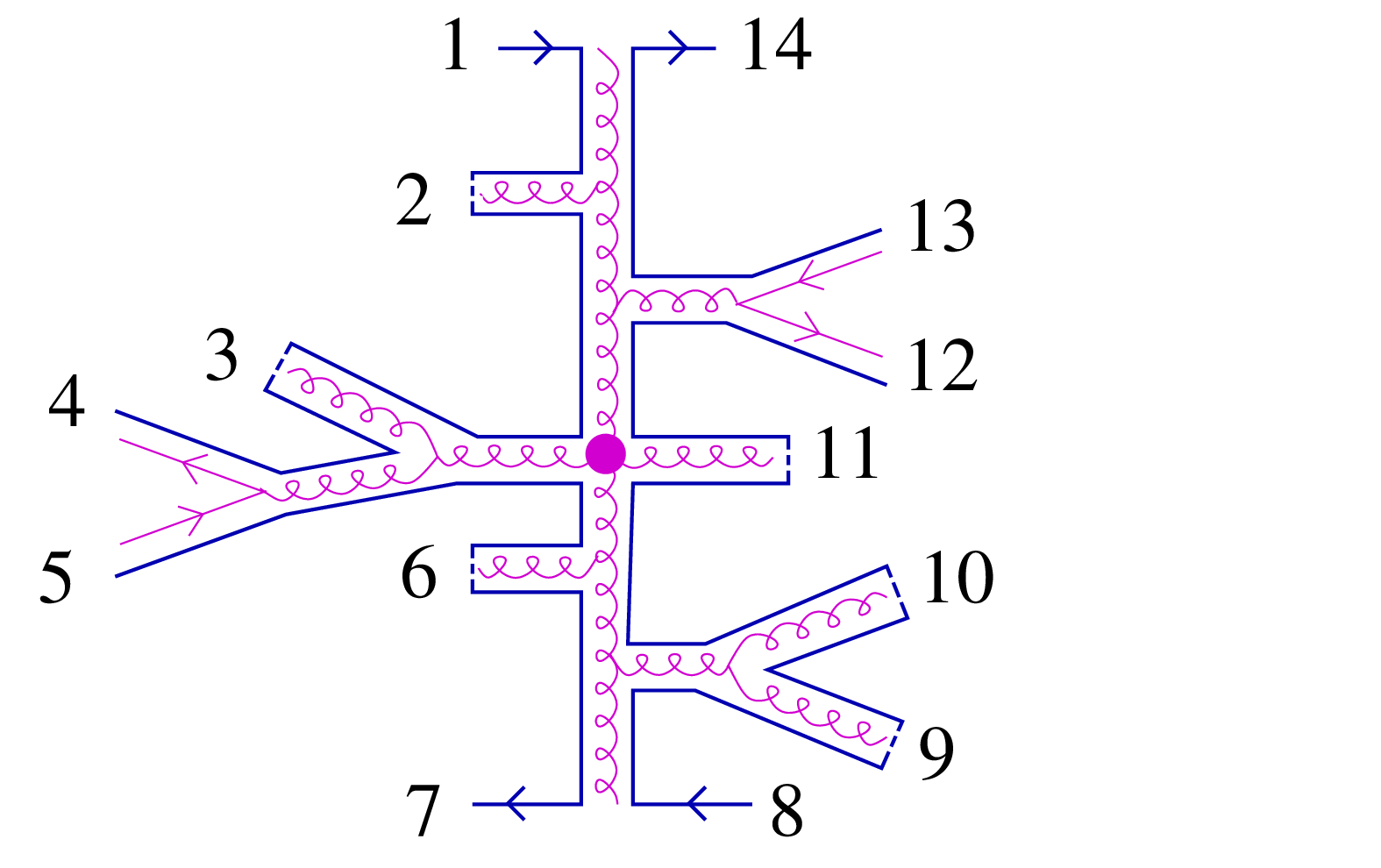}
\caption{A color flow diagram with SL and TL cascades.\label{complete-example-1}}
\end{figure}

In the above example, I considered (for simplicity) only gluons,
and no timelike cascade. In general, the situation is a bit more complicated,
as shown in Fig. \ref{complete-example-1}.
Here, one of the gluons emitted in the Born process splits into a
gluon (3), a quark (4), and an antiquark (5). Two gluons emitted
in the SL cascade, split into the partons 9,10 and 12,13. One again
follows the color flow, always starting and ending with a single arrow,
and one identifies the following chains: 1-2-3-4, then 5-6-7, then
8-9-10-11-12, and finally 13-14. 

All these chains of partons will be mapped (in a unique fashion) to
kinky strings, where each parton corresponds to a kink, and the parton
four-momentum defines the kink properties, as already done in earlier
EPOS versions as described in Ref. \cite{Drescher:2000ha}, where one also
discusses the string decay.

\section{Secondary interactions: The role of core, corona, and remnants, at
RHIC and LHC energies \label{======= Secondary-interactions =======}}

The S-matrix part discussed sofar concerns ``primary scatterings'',
happening instantaneously at $t=0$. As a result, one obtains multiple
Pomeron configurations, which translates into complex partonic configurations,
and eventually into kinky strings and remnants, as discussed in 
Sec. \ref{======= high-and-low pt  =======} (for details see Ref. \cite{werner:2023-epos4-heavy}).
String decay traditionally produces string segments which correspond
to hadrons. But one considers the possibility of having a dense environment,
and here the string segments cannot \textquotedblleft evolve\textquotedblright{}
into hadrons. So one uses the term \textquotedblleft prehadrons\textquotedblright{}
for these segments, and they either \textquotedblleft fuse\textquotedblright{}
to produce the core, or become hadrons if they escape the core. A
similar argument is used for excited remnants, which may decay into
hadrons (see Fig. \ref{partonic-configuration}), but again this may
happen in a dense area, so one names them ``prehadrons'' as well. Based
on these prehadrons from string or remnant decay, a core-corona procedure will be employed, which
allows identifying the ``core'', which will then be treated as a
fluid that evolves and eventually decays into hadrons, which still
may collide with each other. In the present paper, I focus on PbPb
collisions at 5.02 TeV and AuAu scattering at 200 GeV. In
Sec. 3 of Ref. \cite{werner:2023-epos4-micro}, one shows
explicitly partonic configurations (similar as Fig. \ref{partonic-configuration})
but for both LHC and at RHIC energies. They are similar, but with decreasing
energy, it becomes simply more and more likely that the Pomerons are
replaced by purely soft ones. Also, the Pomerons get less energetic,
producing fewer particles. A dedicated paper on lower RHIC energies
is in preparation.

Let me discuss some more details about core and corona.
Based on the prehadrons from strings and remnants, one employs a so-called
core-corona procedure (introduced in Ref. \cite{Werner:2007bf}, updated
in Ref. \cite{Werner:2013tya}), at some given (early) proper time $\tau_{0}$,
to separate ``core prehadrons'' from ``corona prehadrons''. The
former constitute bulk matter and will be treated via hydrodynamics;
the latter become simply hadrons and propagate with reduced energy
(due to the energy loss). For details, see Sec. 3 of Ref. \cite{werner:2023-epos4-micro}.
\begin{figure}[h]
\centering{}\includegraphics[bb=45bp 30bp 570bp 790bp,clip,scale=0.43]
{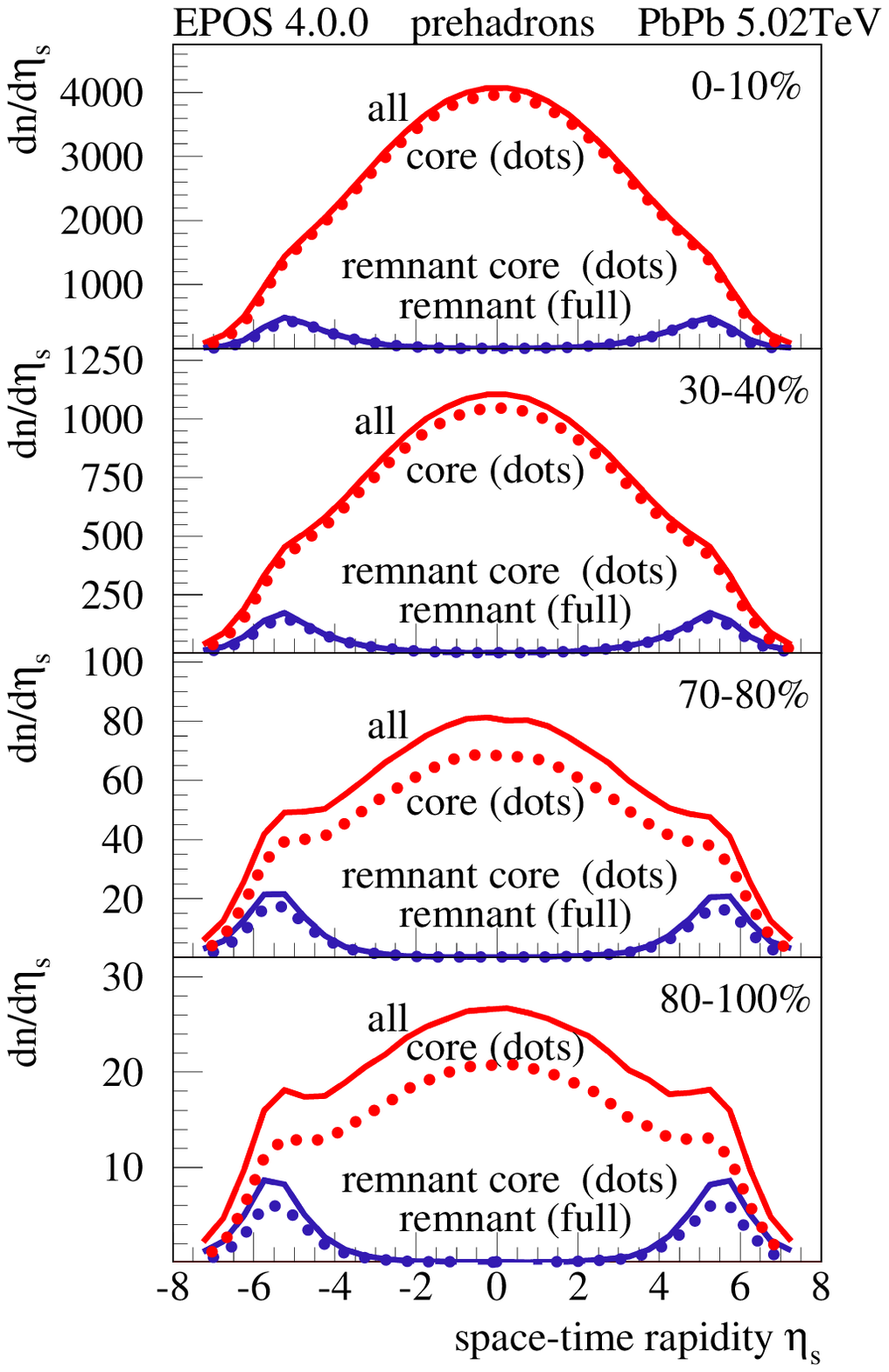}
\caption{The prehadron yield as a function of space-time rapidity, for different
centralities in PbPb collisions at 5.02 TeV. The curves refer to
all prehadrons (red solid curve), all core prehadrons (red dotted curve), prehadrons
from remnant decay (blue solid curve), and core prehadrons from remnant decay
(blue dotted curve). \label{prehadrons-aa-1}}
\end{figure}
\begin{figure}[h]
\centering{}\includegraphics[bb=45bp 30bp 570bp 790bp,clip,scale=0.43]
{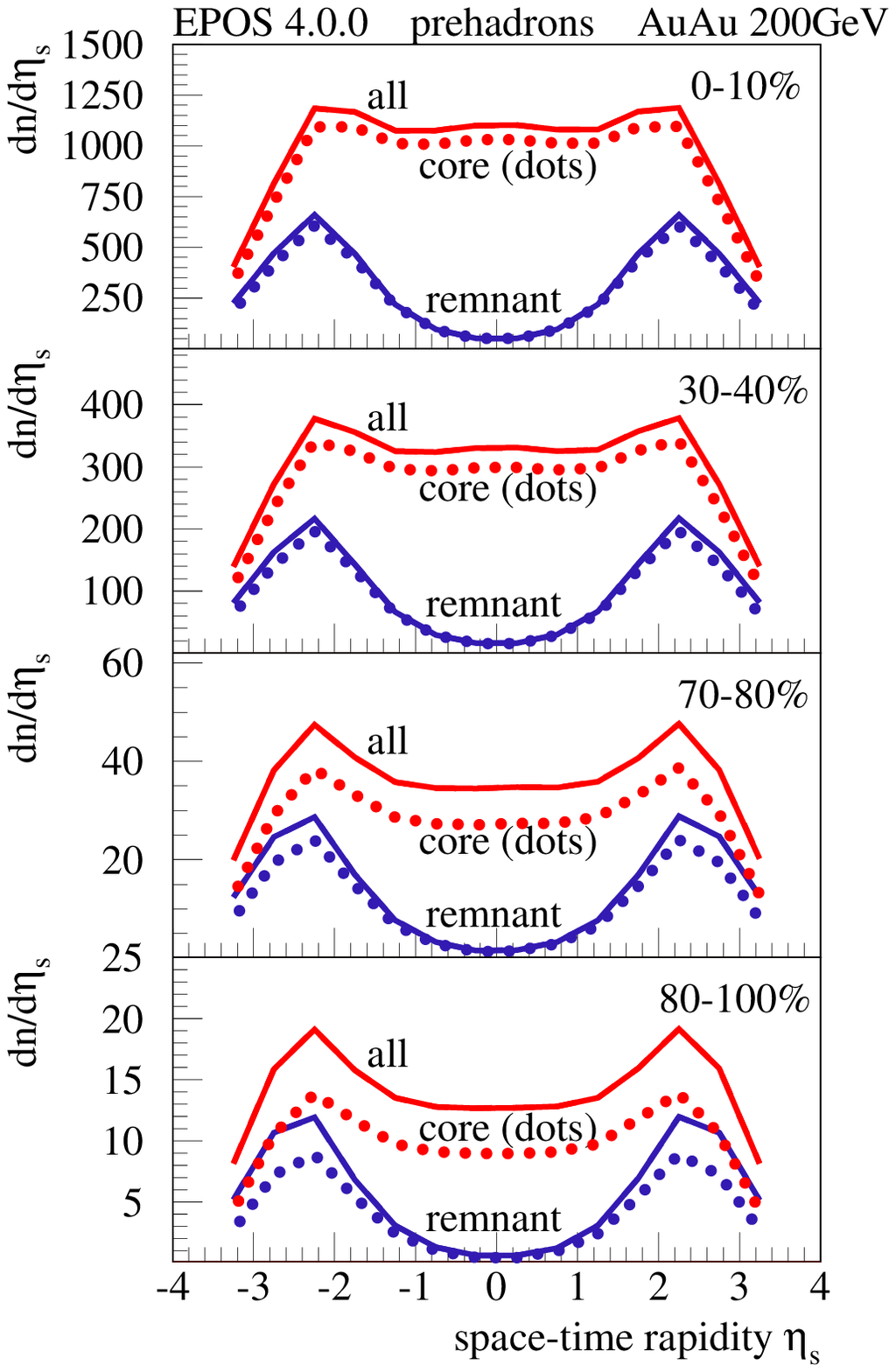}
\caption{Same as Fig. \ref{prehadrons-aa-1}, but for AuAu at 200 GeV. \label{prehadrons-aa}}
\end{figure}

In the following, it will be tried to understand the relative importance
of the core part, and of the fraction coming from remnant decay. In
Fig. \ref{prehadrons-aa-1}, I show results for different centralities
in PbPb collisions at 5.02 TeV, namely (from top to bottom), 0-10\%,
30-40\%, 70-80\%, and 80-100\% (based on the distribution of the impact
parameter). I plot four different curves: all prehadrons (red solid curve),
all core prehadrons (red dotted curve), prehadrons from remnant decay (blue
solid curve), and core prehadrons from remnant decay (blue dotted curve). The remnant
contributions show up preferentially at large rapidities, and in all
cases, they do contribute to the core. Comparing the red solid and
dotted curves, one sees that the core fraction (ratio of $dn/d\eta_{s}$
of the core contribution over all) is in all cases substantial: 0.97
for 0-10\%, 0.95 for 30-40\%, 0.85 for 70-80\%, and 0.78 for 80-100\%.
One also sees that this ``core dominance'' extends over a wide rapidity
range.

In Fig. \ref{prehadrons-aa}, I show results for different centralities
in AuAu collisions at 200 GeV. One sees that the core still dominates,
but the core fractions are significantly smaller compared to PbPb
at 5.02 TeV. Since the overall yields decrease with decreasing energy,
the relative importance of the remnant contributions (remnant over
all) increases. But the remnant contribution at central rapidities
remains small. 

Having identified core prehadrons, one computes the corresponding
energy-momentum tensor $T^{\mu\nu}$ and the flavor flow vector at
some space-time position $x$ at initial proper time $\tau=\tau_{0}$
as 
\begin{equation}
T^{\mu\nu}(x)=\sum_{i}\frac{p_{i}^{\mu}p_{i}^{\nu}}{p_{i}^{0}}g(x-x_{i})
\end{equation}
and 
\begin{equation}
N_{q}^{\mu}(x)=\sum_{i}\frac{p_{i}^{\mu}}{p_{i}^{0}}\,q_{i}\,g(x-x_{i}),
\end{equation}
with $q_{i}\in{u,d,s}$ being the net flavor content and $p_{i}$
the four-momentum of prehadron $i$. The function $g$ is some Gaussian
smoothing kernel (see Ref. \cite{werner:2023-epos4-micro} for more details).
The Lorentz transformation into the comoving frame provides the energy
density $\varepsilon$ and the flow velocity components $v^{i}$,
which will be used as the initial condition for a hydrodynamical evolution
\cite{Werner:2013tya,Karpenko_2014}. This is based on the hypothesis
that equilibration happens rapidly and affects essentially the space
components of the energy-momentum tensor.

\begin{figure}
\centering{}\includegraphics[bb=15bp 40bp 540bp 790bp,clip,scale=0.41]
{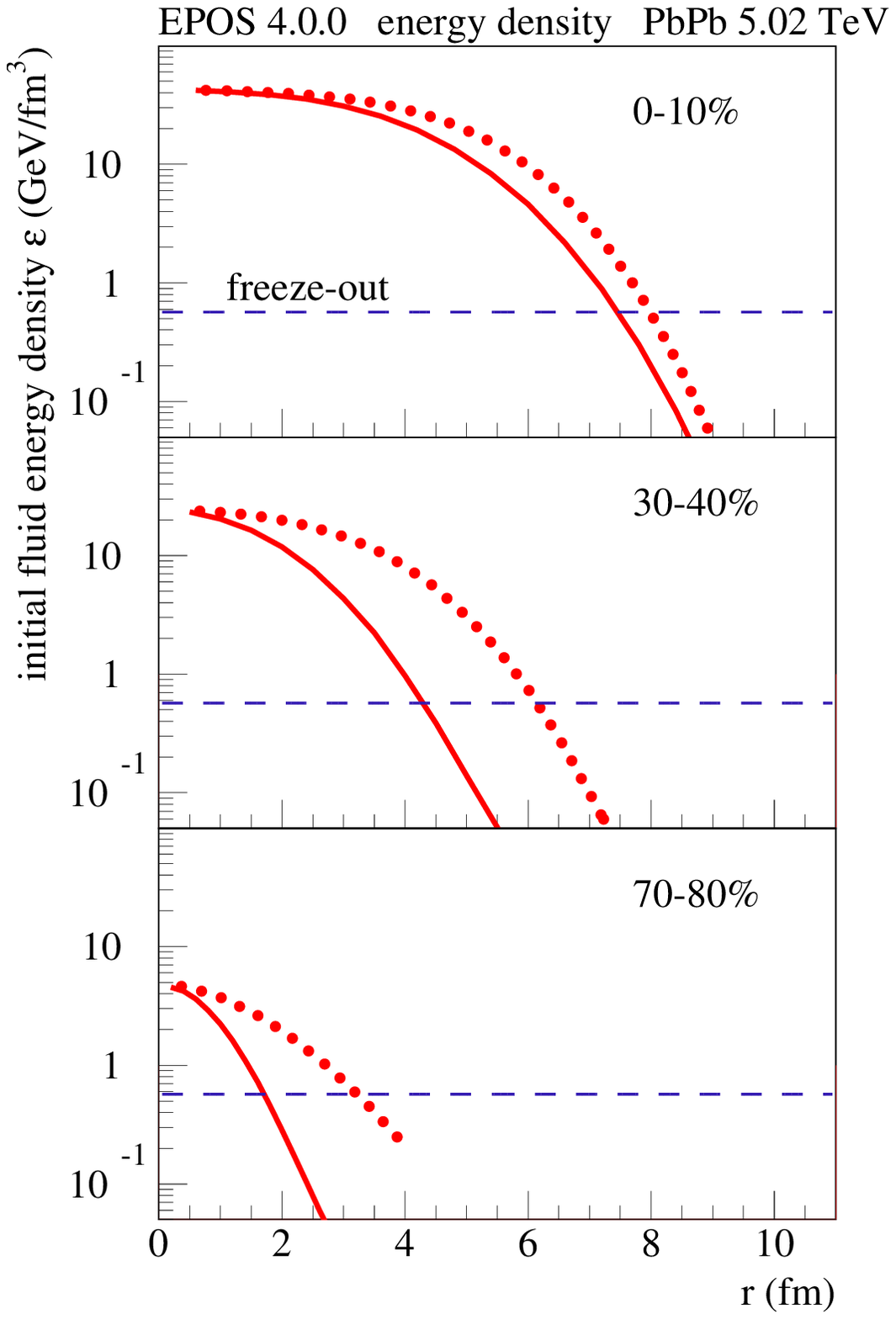}
\medskip{}

\centering{}\includegraphics[bb=15bp 40bp 540bp 790bp,clip,scale=0.41]
{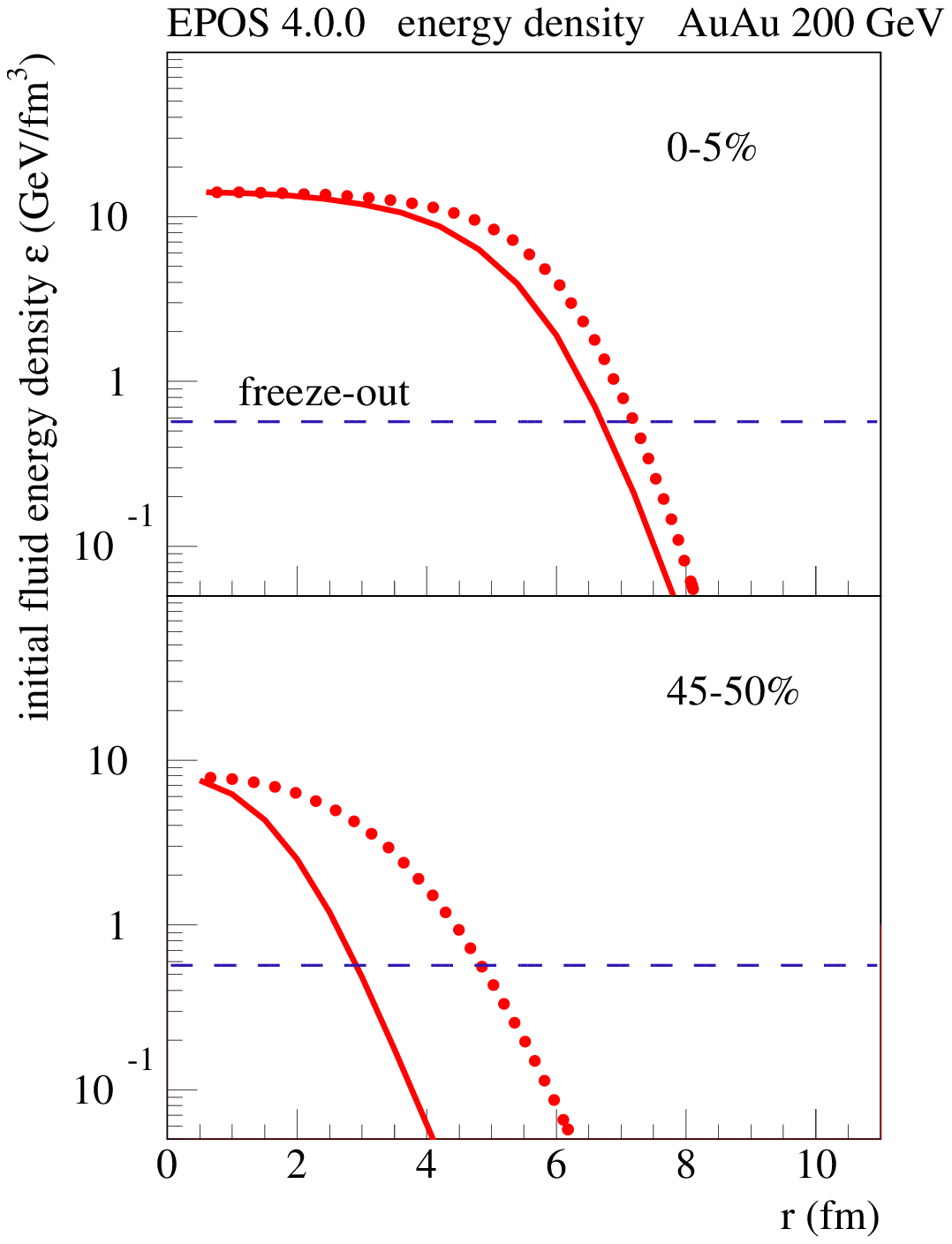}
\caption{Energy density at the initial proper time $\tau_{0}$ as a function
of the transverse coordinate $r$, for an azimuthal angle $\phi=0$
(red solid curve) and $\phi=\pi/2$ (red dotted curve). The blue dashed lines represent
the freeze-out energy density. I show results for different centralities
(defined via impact parameter) in PbPb collisions at 5.02 TeV (upper
plot) and in AuAu collisions at 200 GeV (lower plot). \label{energy-density}}
\end{figure}

In Fig. \ref{energy-density}, I show the energy density at the initial
proper time $\tau_{0}$ as a function of the transverse coordinate
$r$ for different centralities (defined via impact parameter) in
PbPb collisions at 5.02 TeV (upper plot) and in AuAu collisions at
200 GeV (lower plot). The blue dashed lines represent the freeze-out
energy density. For each event, one determines (based on the energy
density distribution) the event plane angle $\psi$ and rotates the
system accordingly (to have after rotation event plane angles zero).
The plots in Fig. \ref{energy-density} represent averages over such
rotated events, the solid lines correspond to azimuthal angles $\phi=0$,
tand the dotted lines to $\phi=\pi/2$. The difference between the two
lines reflects the azimuthal asymmetry. Even in peripheral PbPb collisions,
there is some core production, and one gets actually an energy density
of about $\mathrm{4-5\,GeV/fm^{3}}$ for 70-80\% centrality, but the
radial extension is small, and the lifetime as well. 
The numerical values in $\mathrm{fm/c}$ used for $\tau_{0}$ are between 1 (peripheral) and 1.5 (central) for PbPb 
and always 1 for AuAu.

It follows a viscous~hydrodynamic~expansion. Starting from the initial
proper time $\tau_{0}$, the core part of the system evolves according
to the equations of relativistic viscous hydrodynamics \cite{Werner:2013tya,Karpenko_2014},
where one uses presently $\eta/s=0.08$. The ``core-matter'' hadronizes
on some hypersurface defined by a constant energy density $\epsilon_{H}$
(presently $0.57\mathrm{\,GeV/fm^{3}}$). In earlier versions \cite{Werner:2011-hydro-pp-900GeV},
one used a so-called Cooper-Frye procedure. This is problematic in
particular for small systems: not only do energy and flavor conservation
become important, but one also encounters problems due to the fact
that one gets small ``droplets'' with huge baryon chemical potential,
with strange results for heavy baryons. In EPOS4, one uses systematically
microcanonical hadronization, as discussed in Ref. \cite{werner:2023-epos4-micro}.
After the hadronization of the fluid, the created hadrons as well
as the corona prehadrons (having been promoted to hadrons) may still
interact via hadronic scatterings, and here one uses UrQMD \cite{Bas98,Ble99}.

In the following, I will study core and corona contributions to hadron
production. I will distinguish the following: 
\begin{description}
\item [{(A)}] The \textbf{core+corona} contribution: primary interactions
(S-matrix approach for parallel scatterings), plus core-corona separation,
hydrodynamic evolution, and microcanonical hadronization, but without
hadronic rescattering. 
\item [{(B)}] The \textbf{core} contribution: as case A, but considering
only core particles. 
\item [{(C)}] The \textbf{corona} contribution: as case A, but considering
only corona particles. 
\item [{(D)}] The \textbf{full} EPOS4 scheme: as case A, but in addition
hadronic rescattering. 
\end{description}
\noindent In cases A, B, and C, one needs to exclude the hadronic
afterburner, because the latter affects both core and corona particles,
so in the full approach, the core and corona contributions are not
visible anymore.

\begin{figure}[h]
\centering{}\includegraphics[bb=30bp 30bp 570bp 620bp,clip,scale=0.47]
{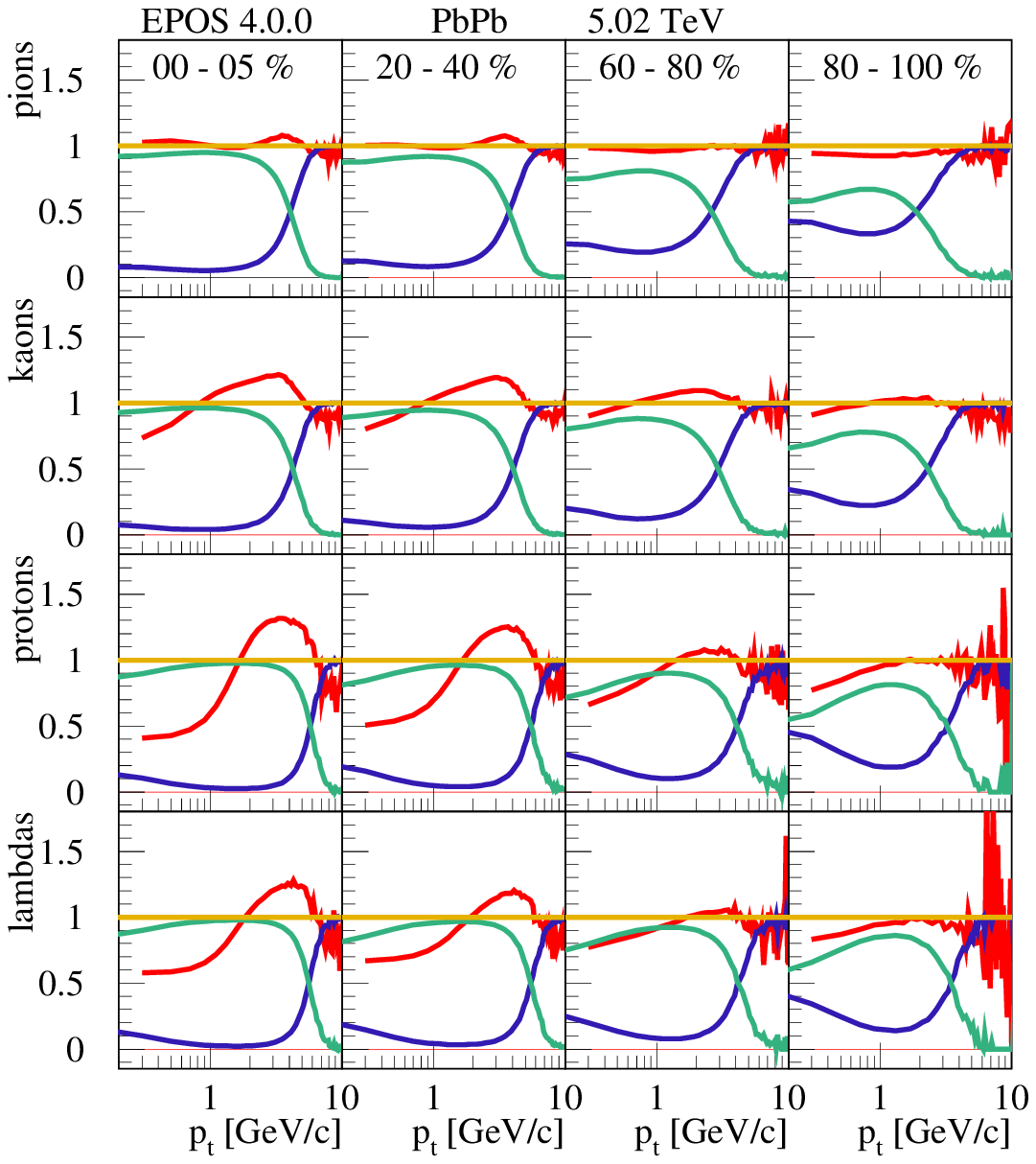}

\centering{}\includegraphics[bb=30bp 30bp 570bp 620bp,clip,scale=0.47]
{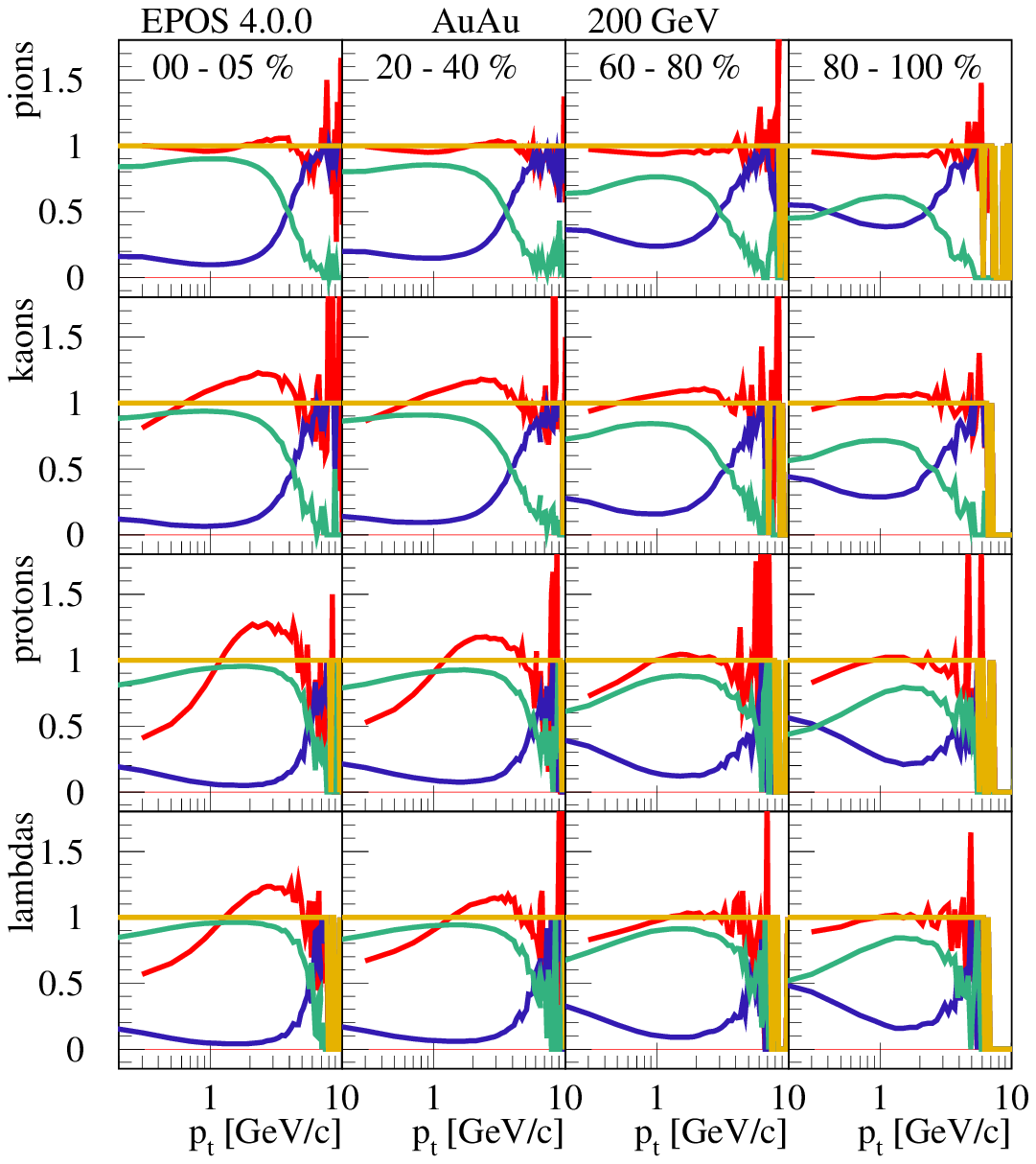}

\caption{The X/``core+corona'' ratio as a function of $p_t$ (for $|\eta|<1$), 
with X being the ``corona'' contribution
(blue), the ``core'' (green), and the ``full'' contribution (red),
for four centrality classes and four different particle species, for
PbPb at 5.02 TeV (upper plot) and AuAu at 200 GeV (lower plot).\label{core-corona-1}}
\end{figure}

In Fig. \ref{core-corona-1} (upper plot),
 I show ratios X/``core+corona'' versus $p_{t}$, with X being the ``corona''
contribution (blue), the ``core'' (green), and the ``full'' contribution
(red), for PbPb collisions at 5.02 TeV, for (from top to bottom) pions
($\pi^{\pm}$), kaons ($K^{\pm}$), protons ($p$ and $\bar{p}$),
and lambdas ($\Lambda$ and $\bar{\Lambda}$). The four columns represent
four different centrality classes, namely 0-5\%, 20-40\%, 60-80\%, and
80-100\%. Looking at the green (core) and blue (corona)
curves, one observes that the core contribution increases with centrality,
but it also increases with the hadron mass (from top to bottom). Concerning
the $p_{t}$ dependence, one observes a maximum of the green core
curves around 1-2$\,\mathrm{GeV/c}$; at very low $p_{t}$ the core
contribution goes down, so even at very small $p_{t}$ values the
corona contributes. The crossing of the green core and the blue corona
curves (core = corona) occurs between around 2 GeV/c (mesons, peripheral)
and 5 GeV/c (baryons, central). The red curve, ``full'' over ``core+corona'', 
represents the effect of the hadronic cascade in the
``full'' case. The pions are not much affected, but for kaons and
even more for protons and lambdas, rescattering makes the spectra
harder. One should keep in mind that rescattering involves particles
from fluid hadronization, but also corona particles from hard processes.
Concerning the baryons, rescattering reduces (considerably) low $p_{t}$
yields, due to baryon-antibaryon annihilation. 

In Fig. \ref{core-corona-1}
(lower plot), I show the corresponding results for AuAu collisions
at 200 GeV, being similar compared to PbPb at 5.02 TeV/, but the ``core''
contributions are weaker.

\section{Results \label{=======results=======}}

In this section, I show simulation results compared to data. I will
not add too many comments to each curve, the main purpose is to check
if the concepts discussed in the previous sections give a coherent
picture (and reproduce the data) or not.

Although the $p_{t}$ spectra cover usually many orders of magnitude,
I have chosen the dimensions such that differences of 10\% between
data and simulations are always visible.

I will show many results for AuAu collisions at 200 GeV and some results
for PbPb at 5.02 TeV. From the theory point of view, the high energy
case contains in principle everything (as discussed in the preceding
sections), one does not need to add ``features'' at lower energies,
simply certain phenomena ``die out'' when reducing the energy: The
number of Pomerons per nucleon-nucleon collision gets smaller, the
Pomerons get less energetic, and the remnant contributions get relatively
more important. 

\subsection{Spectra for AuAu at 200 GeV \label{-------spectra-200-gev-------}}

In this section, I will show particle spectra for AuAu collisions
at 200 GeV. From Fig. \ref{prehadrons-aa}, one knows that at this
energy the remnant contribution is important, in particular beyond space-time rapidities $|\eta_{s}|$ of 2 ($\eta_{s}$
is numerically similar to the pseudorapidity $\eta$). One also
knows from Fig. \ref{core-corona-1}, that at central rapidity the
core largely dominates, up to $p_{t}$ values of 3-5 GeV/c. 

\begin{figure}[h]
\centering{}\includegraphics[bb=20bp 20bp 700bp 600bp,clip,scale=0.33]
{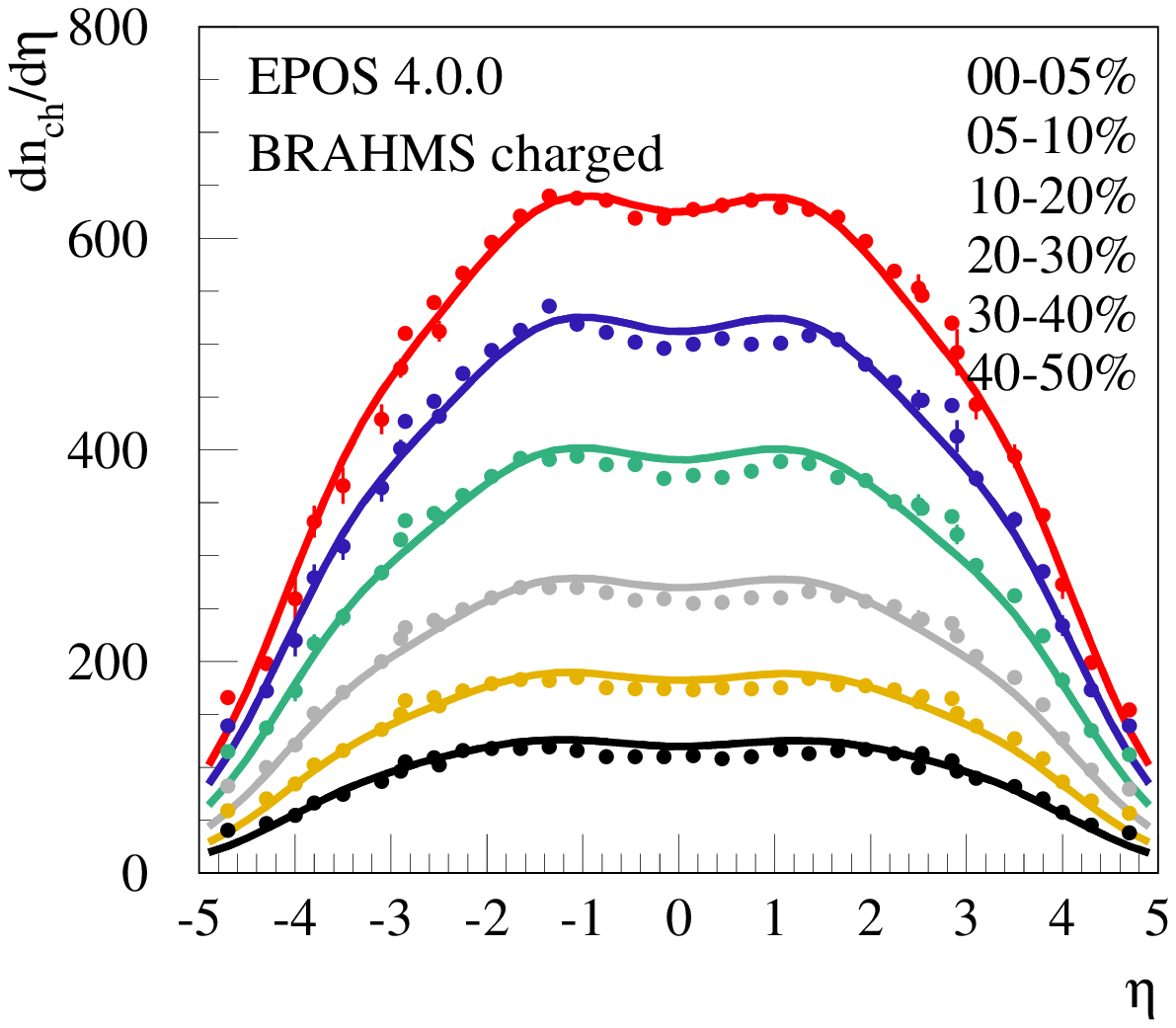}
\caption{Pseudorapidity distributions of charged particles in AuAu collisions
at 200 GeV, comparing EPOS4 simulations (lines) to BRAHMS data (points).
\label{rapidity-charged}}
\end{figure}
\begin{figure}[h]
\centering{}\includegraphics[bb=30bp 35bp 450bp 580bp,clip,scale=0.6]
{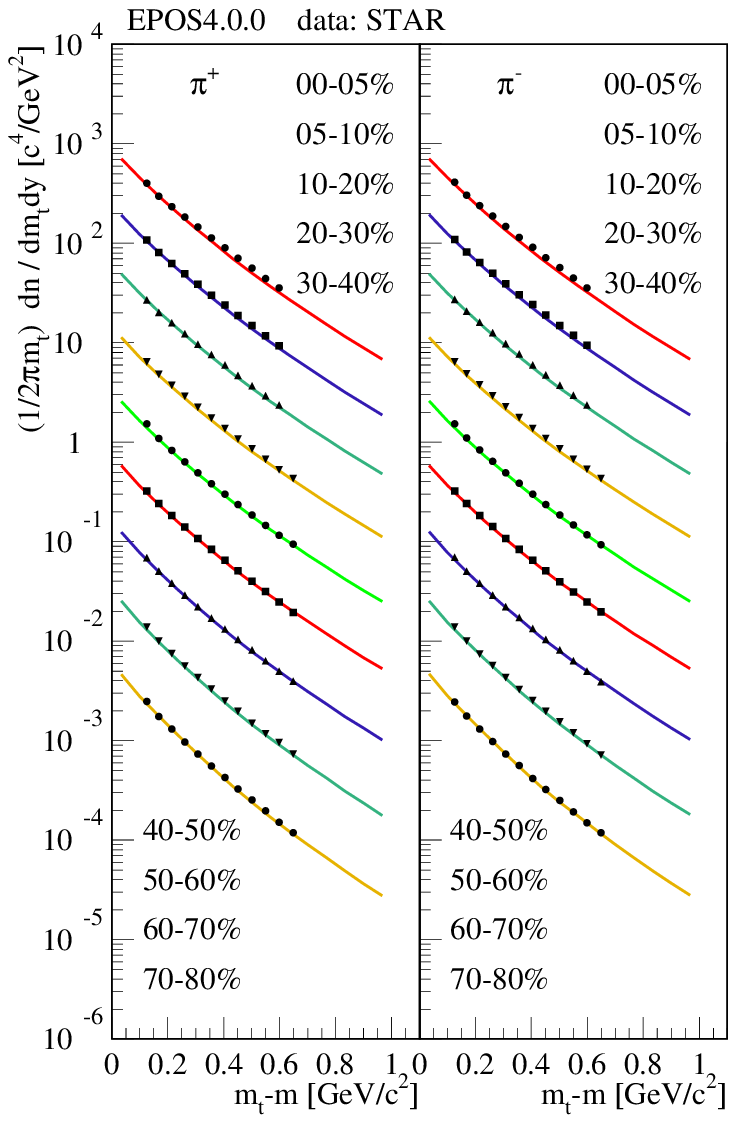}
\caption{Transverse mass distributions of $\pi^{+}$, $\pi^{-}$ in AuAu collisions
at 200 GeV for different centrality classes. EPOS4 simulations (lines)
are compared to data from STAR. From top to bottom, one multiplies
the curves by $3^{-i}$ ($i=0,1,2,3,...$). \label{200-transverse-mass-1}}
\end{figure}
In Fig.
\ref{rapidity-charged}, 
I show pseudorapidity distributions of charged particles in AuAu collisions
at 200 GeV, comparing EPOS4 simulations (lines) to BRAHMS data \cite{BRAHMS:2001}
(points). I show results for different centralities, from top to bottom:
0-5\%, 5-10\%, ..., 40-50\%. 

\begin{figure}[H]
\centering{}\includegraphics[bb=30bp 35bp 450bp 580bp,clip,scale=0.6]
{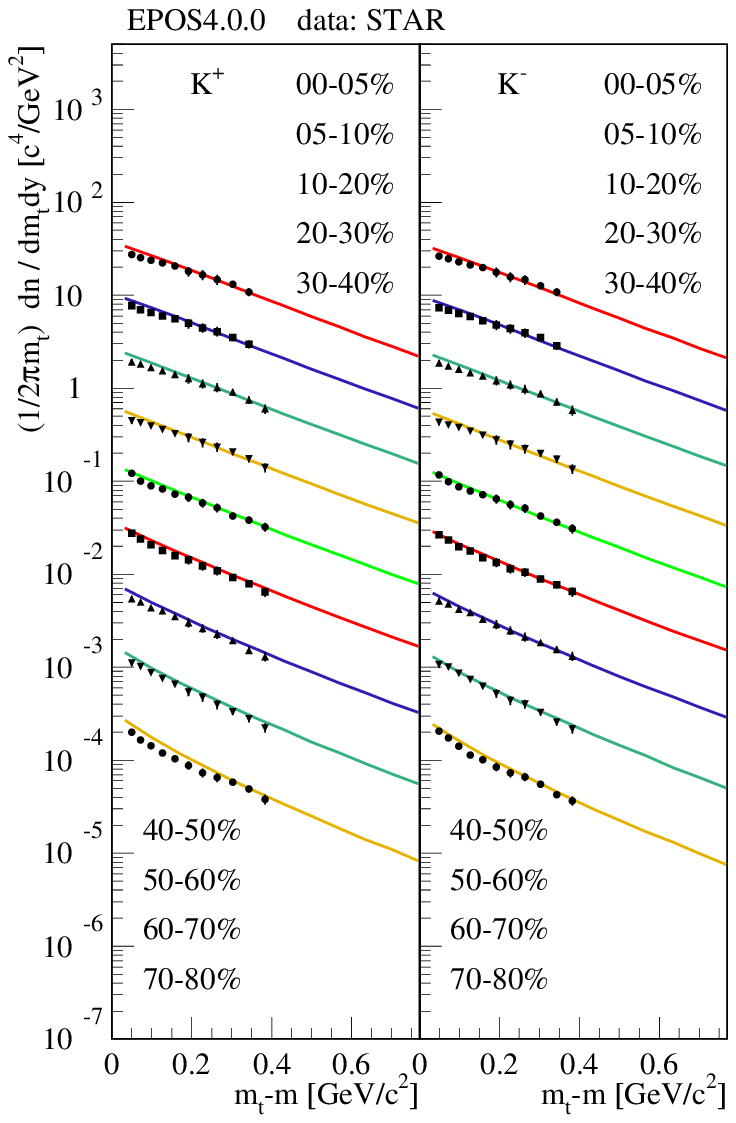}\\
 \centering{}\includegraphics[bb=30bp 30bp 450bp 580bp,clip,scale=0.6]
{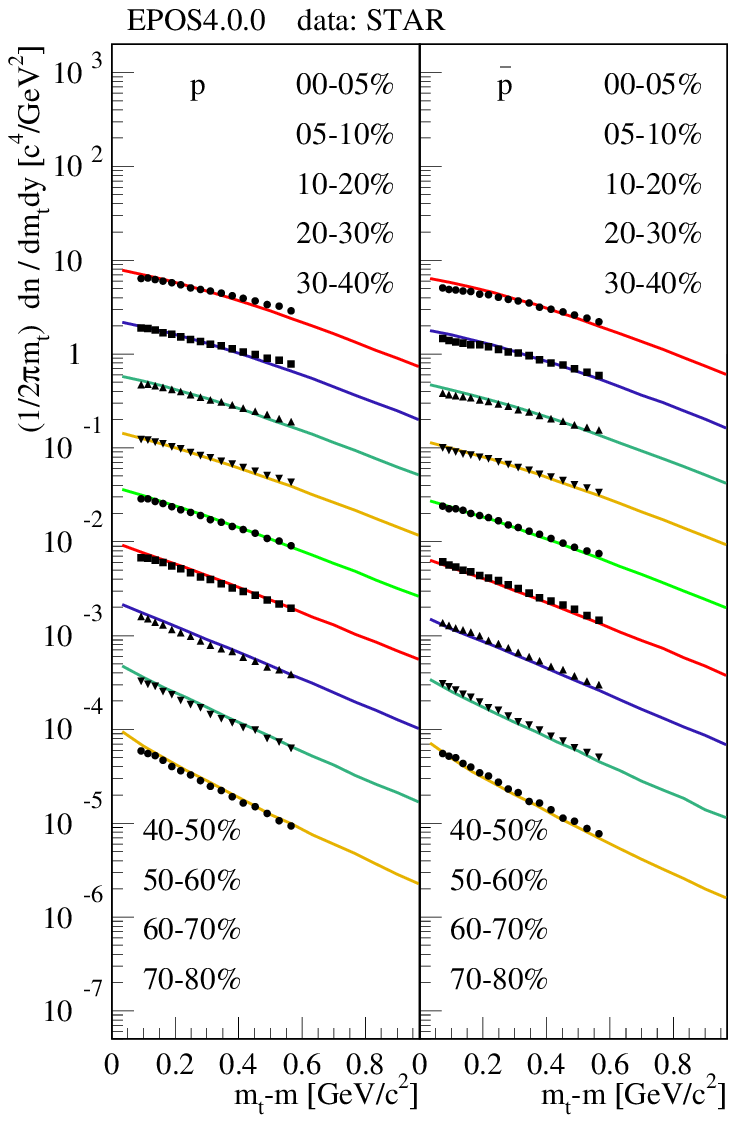}
\caption{Same as Fig. \ref{200-transverse-mass-1}, but for $K^{+}$, $K^{-}$,
$p$, and $\bar{p}$. \label{200-transverse-mass-2}}
\end{figure}

In Figs. \ref{200-transverse-mass-1}
and \ref{200-transverse-mass-2}, I show transverse mass distributions
of $\pi^{+}$, $\pi^{-}$, $K^{+}$, $K^{-}$, $p$, and $\bar{p}$ for
different centrality classes (as indicated in the figures), comparing
EPOS4 simulations (lines) to data from STAR \cite{STAR:2004}. From
top to bottom (for each subplot), the curves are multiplied by $3^{-i}$
($i=0,1,2,3,...$). \\

\begin{figure}[H]
\centering{}\includegraphics[bb=30bp 35bp 450bp 580bp,clip,scale=0.6]
{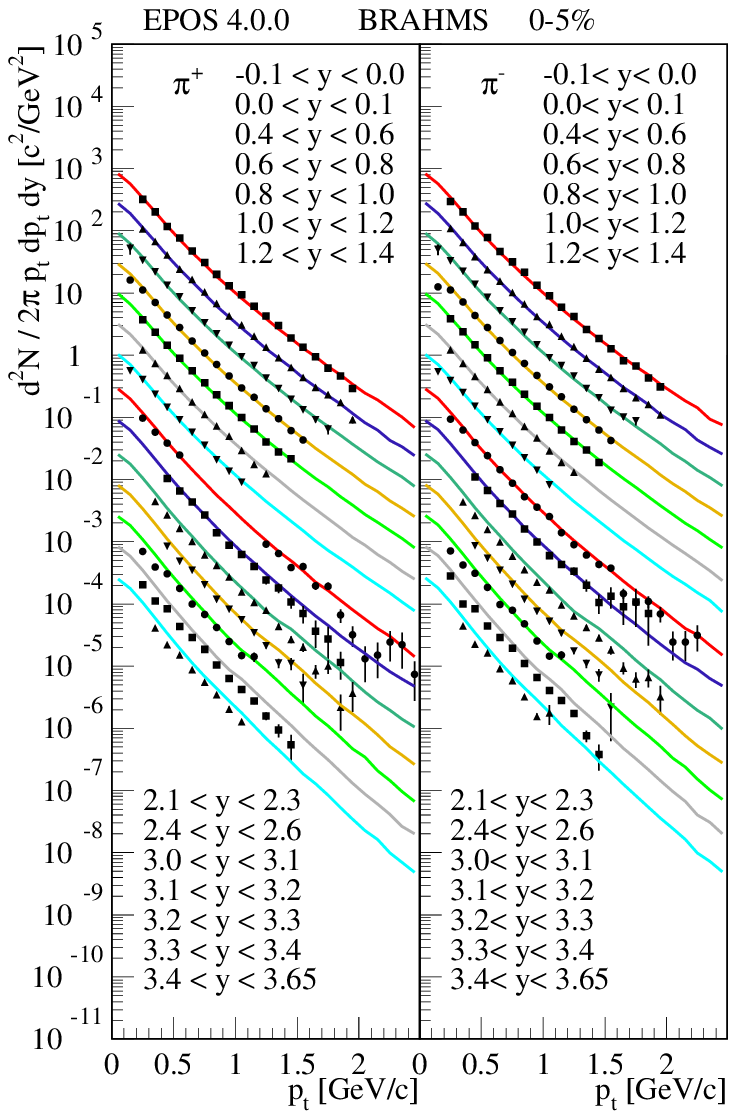}\\
 \centering{}\includegraphics[bb=30bp 35bp 450bp 580bp,clip,scale=0.6]
{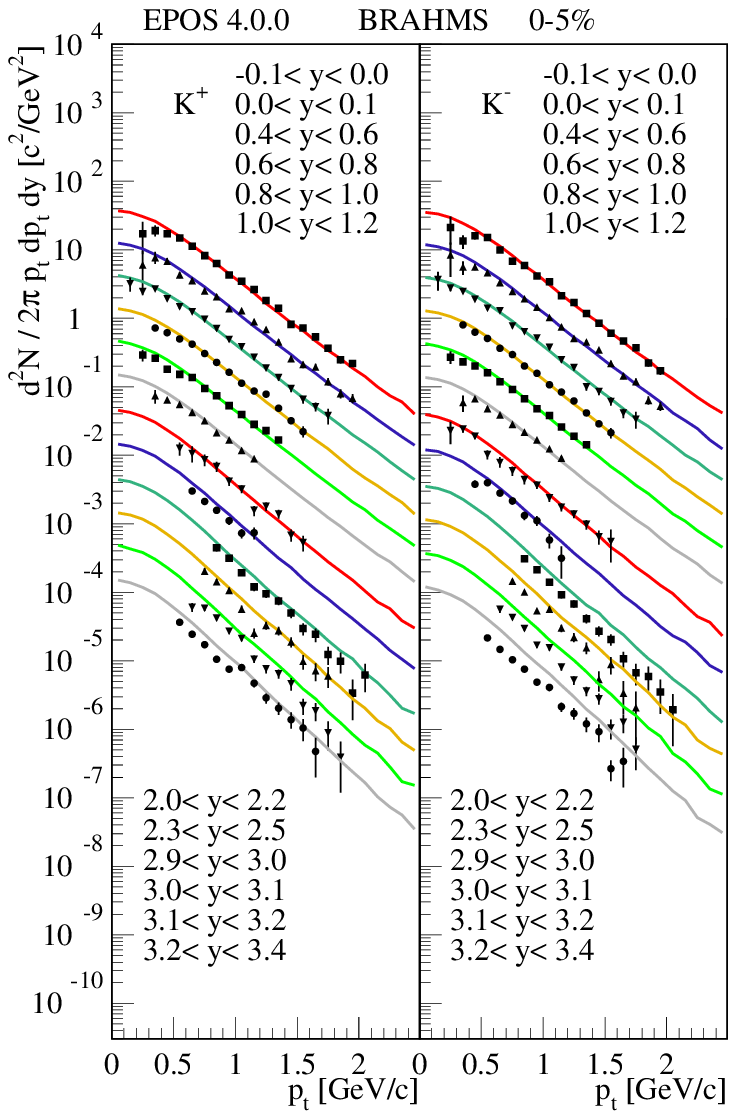}
\caption{Transverse momentum distributions of $\pi^{+}$, $\pi^{-}$, $K^{+}$, and
$K^{-}$ in central (0-5\%) AuAu collisions at 200 GeV for different
rapidities. EPOS4 simulations (lines) are compared to data from BRAHMS.
From top to bottom, the curves are multiplied by $3^{-i}$ ($i=0,1,2,3,...$).
\label{200-transverse-momentum-rap-1}}
\end{figure}

In Fig. \ref{200-transverse-momentum-rap-1}, I
show transverse momentum distributions of $\pi^{+}$, $\pi^{-}$,
$K^{+}$, and $K^{-}$ in central (0-5\%) AuAu collisions at 200 GeV for
different rapidities (as indicated in the plots). EPOS4 simulations
(lines) are compared to data from BRAHMS \cite{BRAHMS:2005}. From
top to bottom, the curves are multiplied by $3^{-i}$ ($i=0,1,2,3,...$).

\begin{figure}[h]
\centering{}\includegraphics[bb=30bp 100bp 450bp 580bp,clip,scale=0.6]
{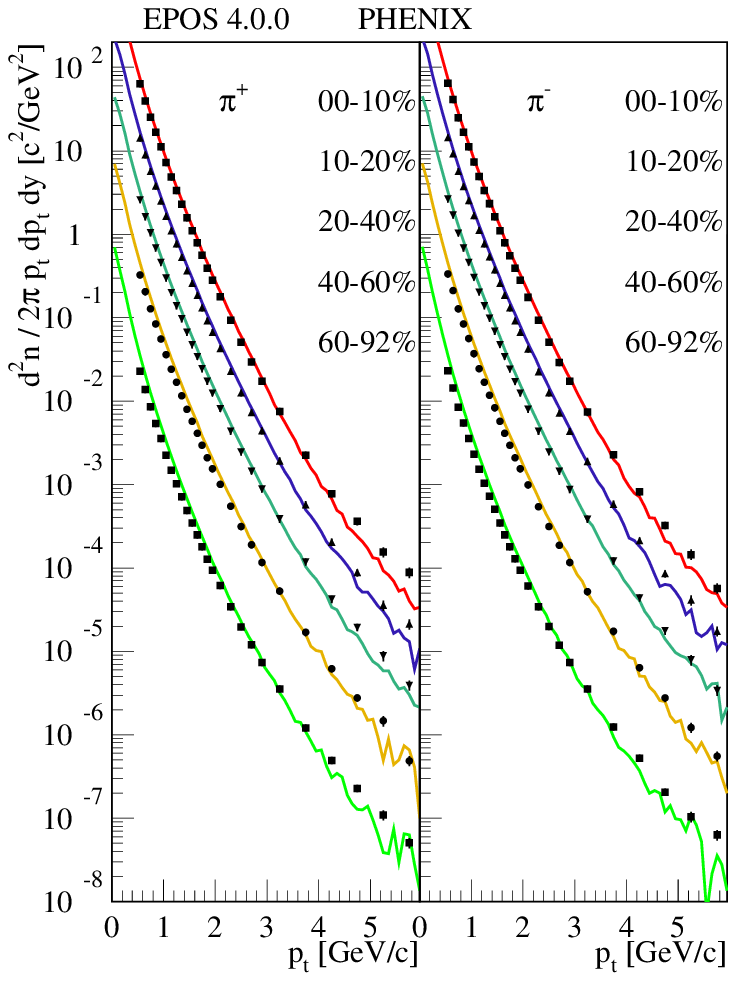}
\caption{Transverse momentum distributions of $\pi^{+}$, $\pi^{-}$ in AuAu
collisions at 200 GeV at central rapidity for different centralities.
EPOS4 simulations (lines) are compared to data from PHENIX \cite{PHENIX:2013}.
From top to bottom, the curves are multiplied by $3^{-i}$ ($i=0,1,2,3,4$).
\label{200-transverse-momentum-phe-1}}
\end{figure}

\begin{figure}[h]
\centering{}\includegraphics[bb=30bp 100bp 450bp 580bp,clip,scale=0.6]
{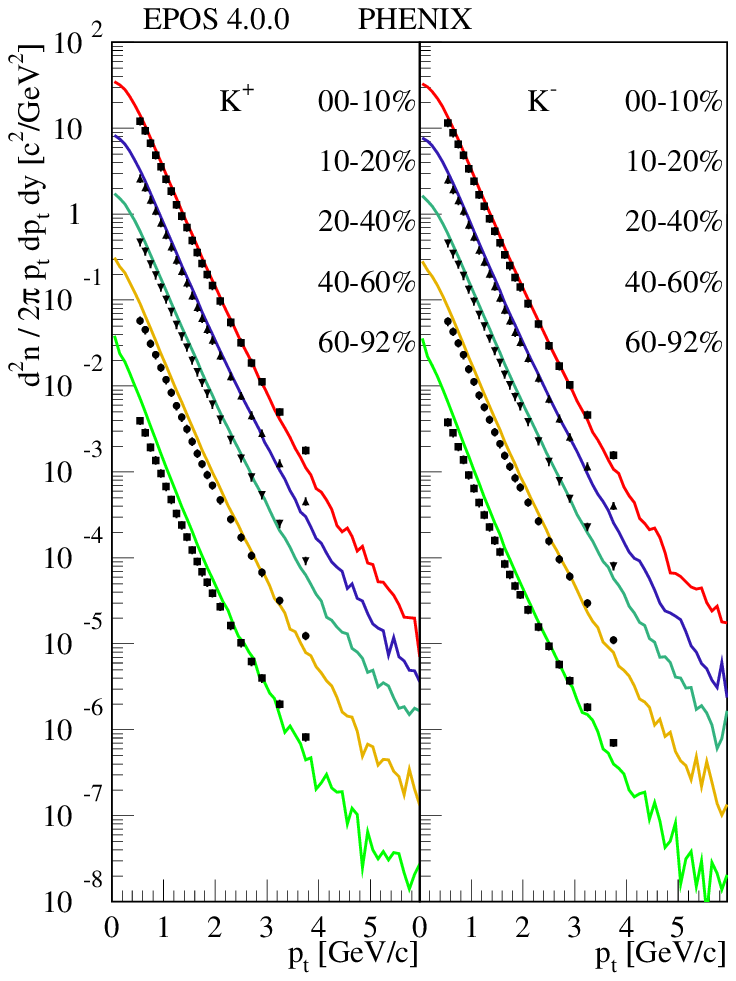}\\
 \centering\includegraphics[bb=30bp 100bp 450bp 580bp,clip,scale=0.6]
{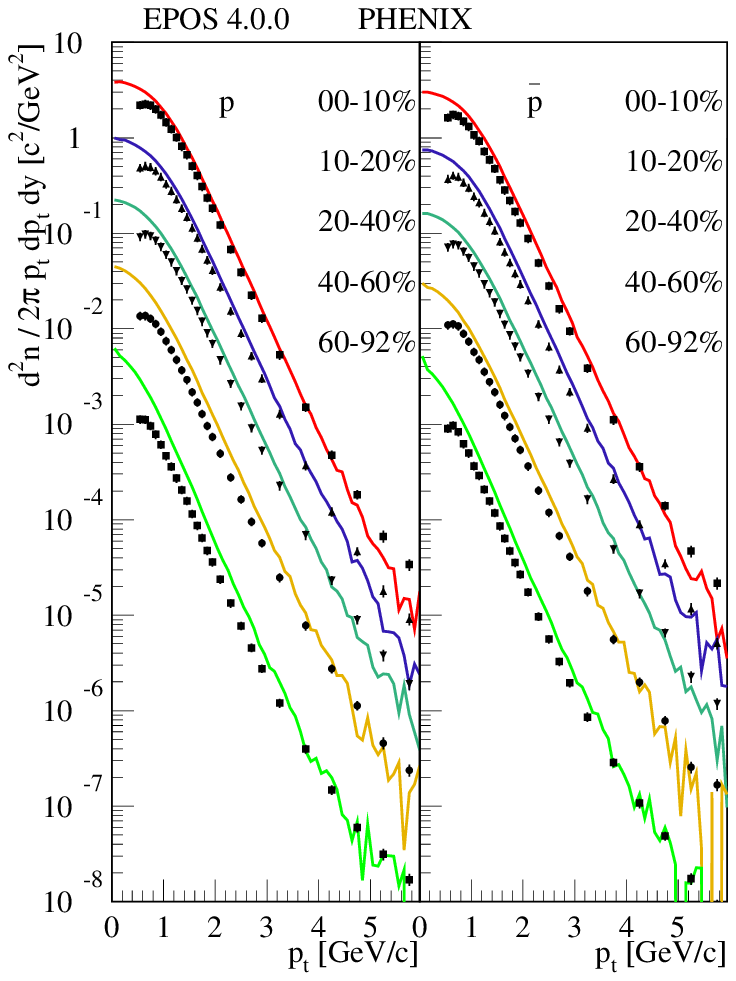}
\caption{Same as Fig. \ref{200-transverse-momentum-phe-1}, but for $K^{+}$,
$K^{-}$, $p$, and $\bar{p}$. \label{200-transverse-momentum-phe-2}}
\end{figure}

In Figs. \ref{200-transverse-momentum-phe-1} and \ref{200-transverse-momentum-phe-2},
I plot transverse momentum distributions of $\pi^{+}$, $\pi^{-}$,
$K^{+}$, $K^{-}$, $p$, and $\bar{p}$ at central rapidity for different
centralities. EPOS4 simulations (lines) are compared to data from PHENIX
\cite{PHENIX:2013}. From top to bottom, the curves are multiplied
by $3^{-i}$ ($i=0,1,2,3,4$).\\

\begin{figure}[h]
\centering\includegraphics[bb=30bp 100bp 450bp 580bp,clip,scale=0.6]
{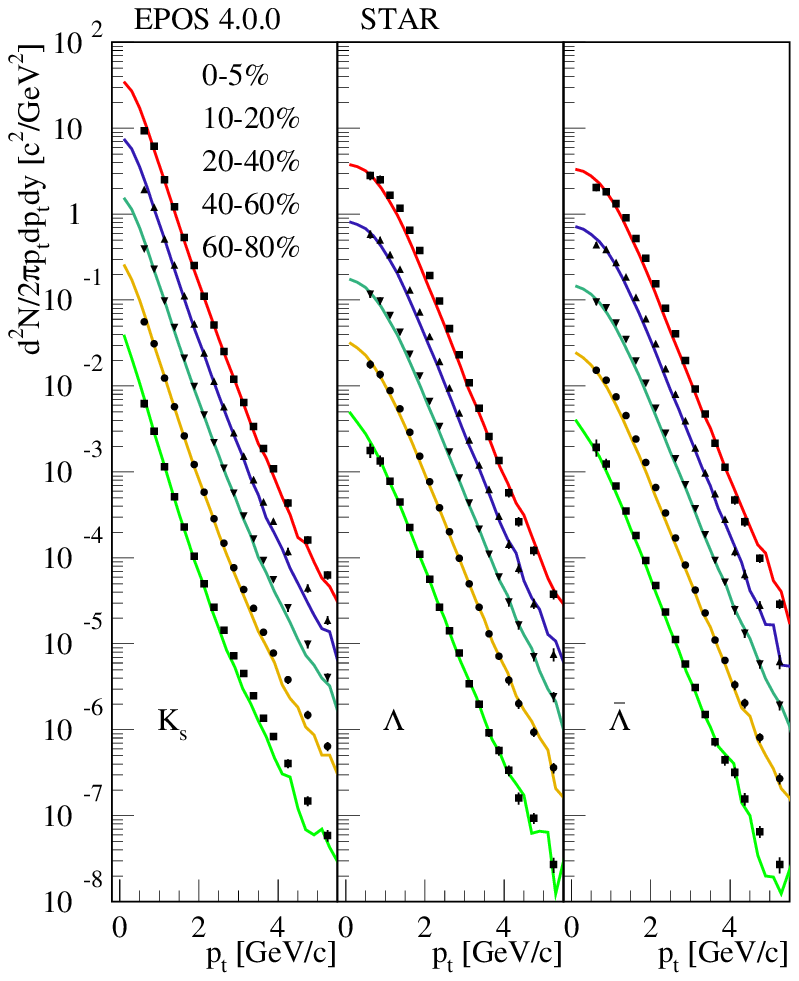} 

\centering{}\includegraphics[bb=30bp 100bp 450bp 580bp,clip,scale=0.6]
{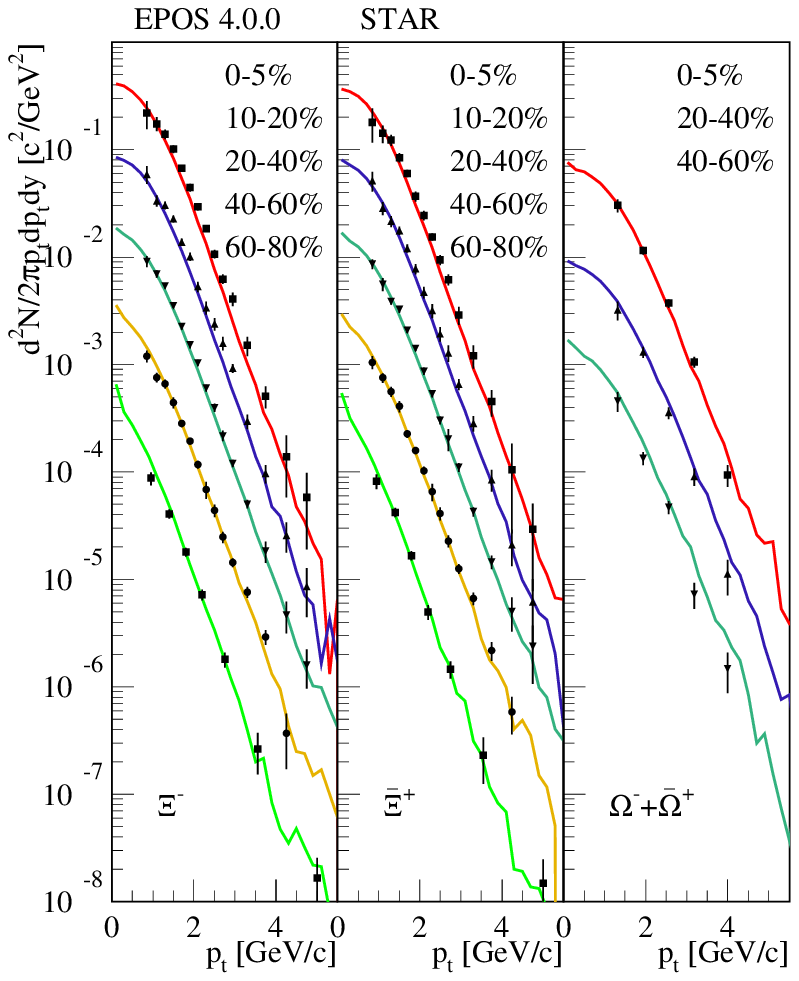}
\caption{Transverse momentum distributions of $K_{0}$, $\Lambda$, $\bar{\Lambda}$
$\Xi^{-}$, $\bar{\Xi}^{+}$, $\Omega$ in AuAu collisions at 200
GeV at central rapidity for different centralities. EPOS4 simulation
(lines) are compared to data from STAR. From top to bottom, one multiplies
the curves by $3^{-i}$, $i=0,1,2,3,4$. \label{200-transverse-momentum-star-1}}
\end{figure}

In Fig. \ref{200-transverse-momentum-star-1}, I show transverse
momentum distributions of $K_{0}$, $\Lambda$, $\bar{\Lambda}$ $\Xi^{-}$,
$\bar{\Xi}^{+}$, and $\Omega$ at central rapidity for different centralities.
EPOS4 simulations (lines) are compared to data from STAR \cite{STAR:2012}
(for $K_{0}$, $\Lambda$, $\bar{\Lambda}$ ) and \cite{STAR:2007}.
From top to bottom, the curves are multiplied by $3^{-i}$ ($i=0,1,2,3,4$).
\\

In general, the simulation results are close to the data, concerning
identified particles as pions, kaons, and protons, as well as
hyperons. \\

Concerning protons and antiprotons in Fig. \ref{200-transverse-momentum-phe-2},
the data drop remarkably at low $p_{t}$ which is not seen in the
simulations, and as well not in other proton and antiproton data
(see, for example, Fig. \ref{200-transverse-mass-2}).\\

In all cases, one suppresses (or not) feed-down from decays, to be
consistent with the corresponding experimental data.

\clearpage

\subsection{Flow harmonics for AuAu at 200 GeV \label{-------flow-200-gev--------}}

The so-called flow harmonics $v_{n}$ are important observables, characterizing
anisotropic azimuthal flow, defined as
\begin{equation}
v_{n}=\left\langle \cos\left(n(\phi-\psi_{n})\right)\right\rangle ,
\end{equation}
with $n$ being the order of the flow harmonic, $\phi$ the azimuthal
angle and $\psi_{n}$ the event plane angle of harmonic $n$. Several
methods have been developed over the years to determine $v_{n}$
from the momentum vectors of the observed particles. In the following,
I will compare EPOS4 $v_{n}$ results with experimental data, employing
exactly the same methods as used in the experiments, with details
being found in the corresponding citations.

\begin{figure}[h]
\centering{}\includegraphics[bb=20bp 30bp 595bp 760bp,clip,scale=0.4]
{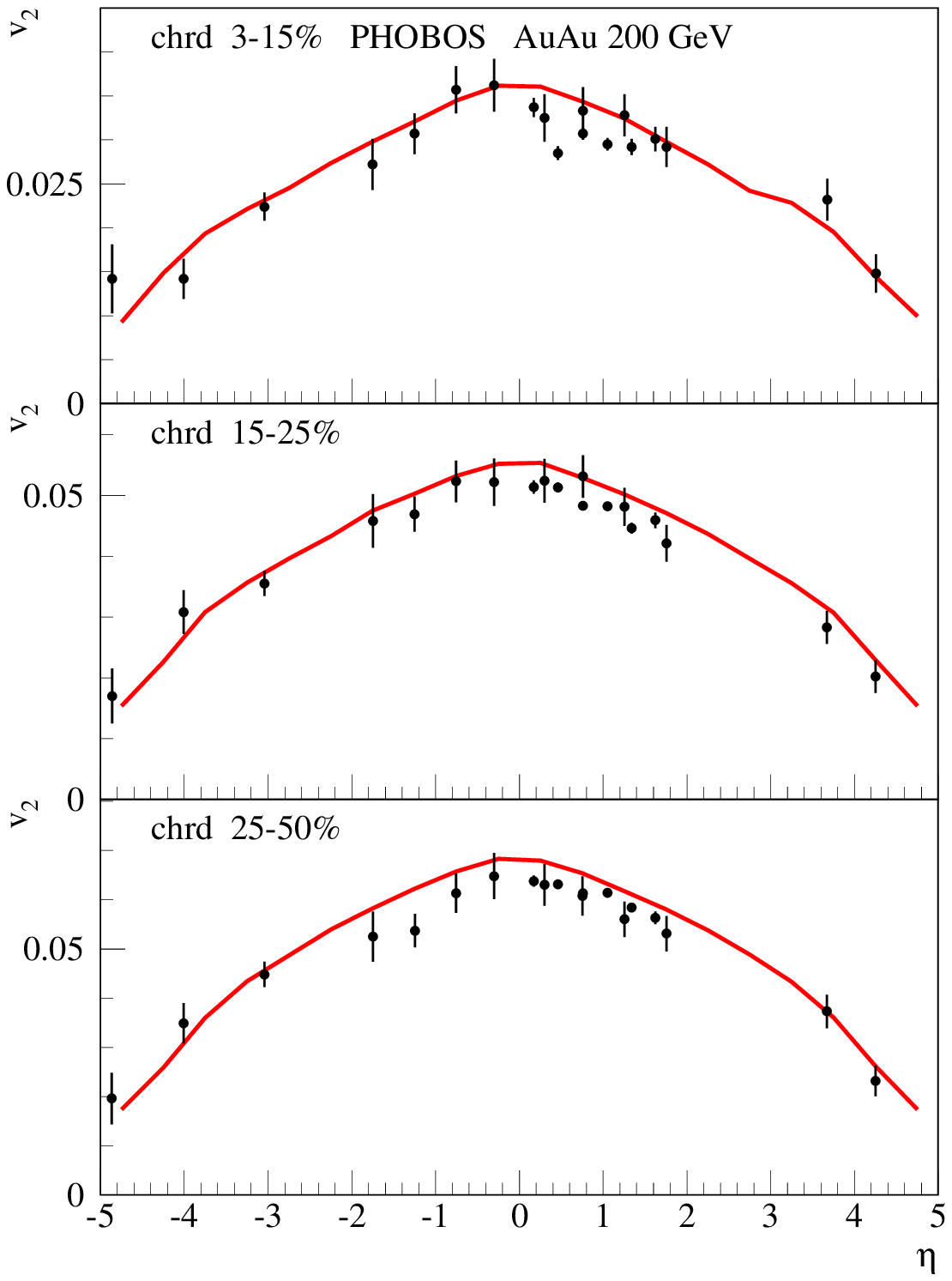}
\caption{Pseudorapidity dependence of $v_{2}$ of charged particles for different multiplicity classes
in AuAu collisions at 200 GeV. I compare the simulations (red lines)
with data from PHOBOS (black points). \label{v2-auau-1}}
\end{figure}

\begin{figure}[h]
\centering{}\includegraphics[bb=20bp 30bp 595bp 500bp,clip,scale=0.4]
{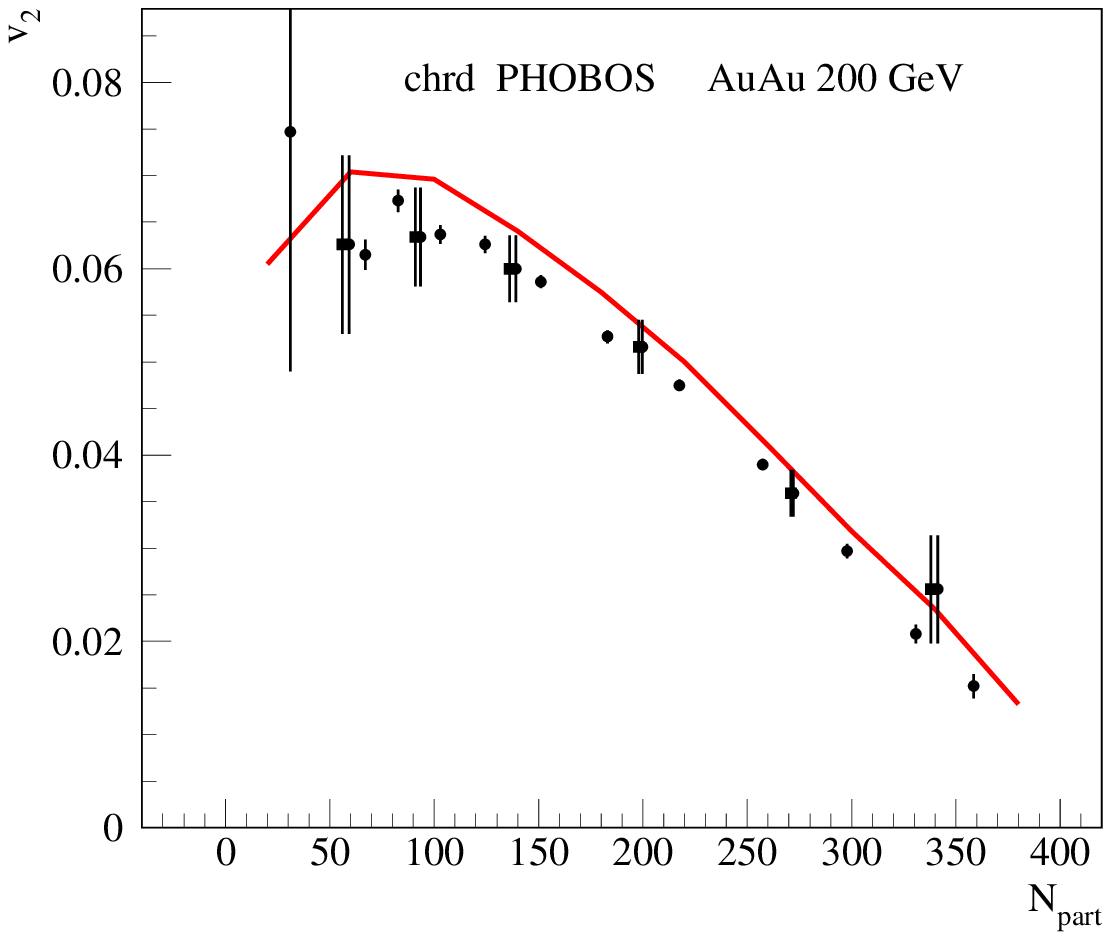}
\caption{Centrality dependence (using $N_{\mathrm{part}}$) of $v_{2}$ of charged particles in
AuAu collisions at 200 GeV. I compare the simulations (red lines)
with data from PHOBOS (black points). \label{v2-auau-2}}
\end{figure}

\begin{figure}[h]
\centering{}\includegraphics[bb=20bp 30bp 595bp 800bp,clip,scale=0.4]
{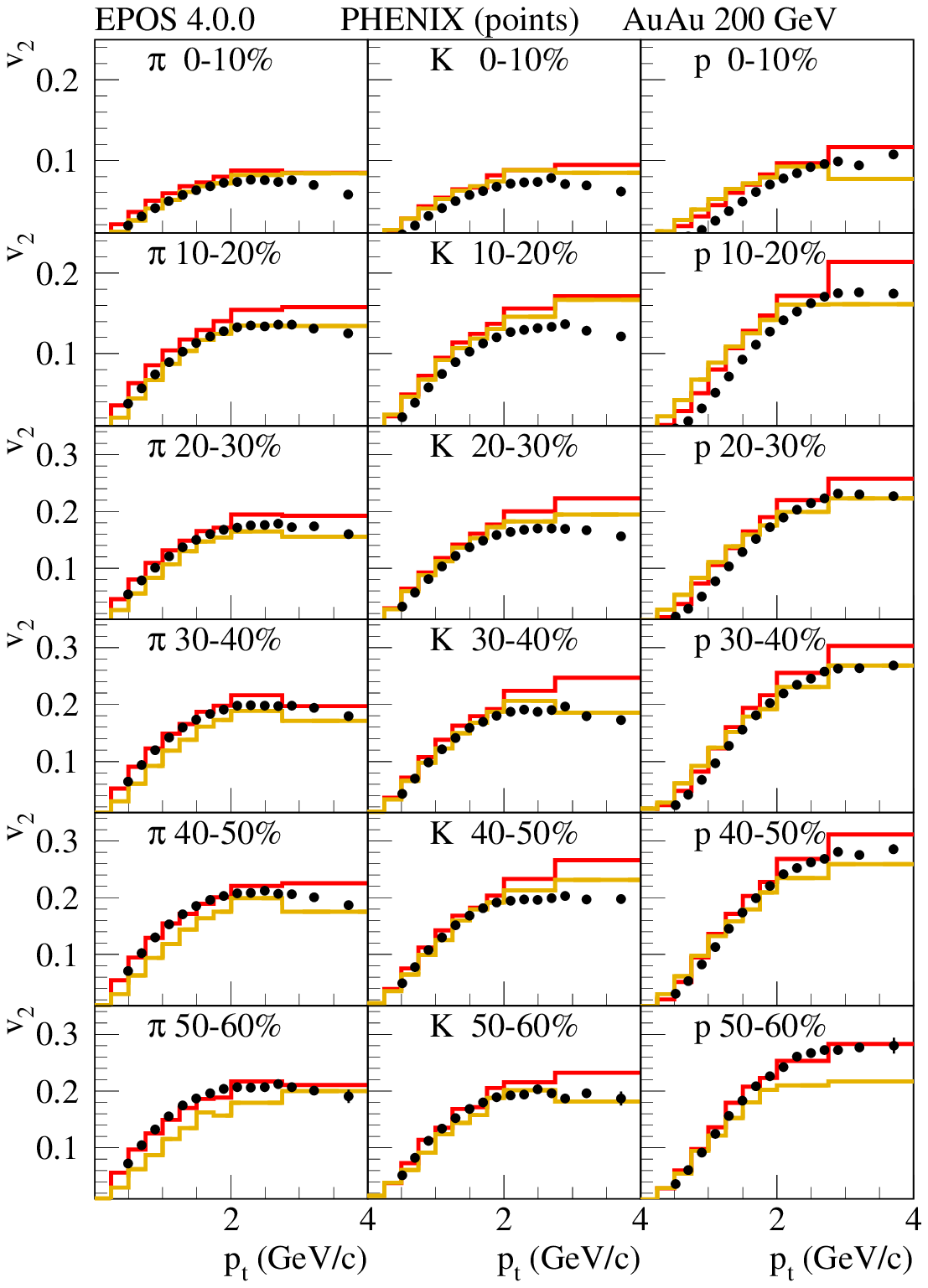}
\caption{Transverse momentum dependence of $v_{2}$ for pions (left column),
kaons (middle column) and protons (right column) in AuAu collisions
at 200 GeV, for different centralities (from top to bottom): 0-10\%,
10-20\%, 20-30\%, 30-40\%, 40-50\%, and 50-60\%. I compare the full
simulations (red lines) and those without hadronic cascade (yellow
lines) with data from PHENIX (black points). \label{v2-auau-3}}
\end{figure}

\begin{figure}[h]
\centering{}\includegraphics[bb=20bp 30bp 595bp 780bp,clip,scale=0.4]
{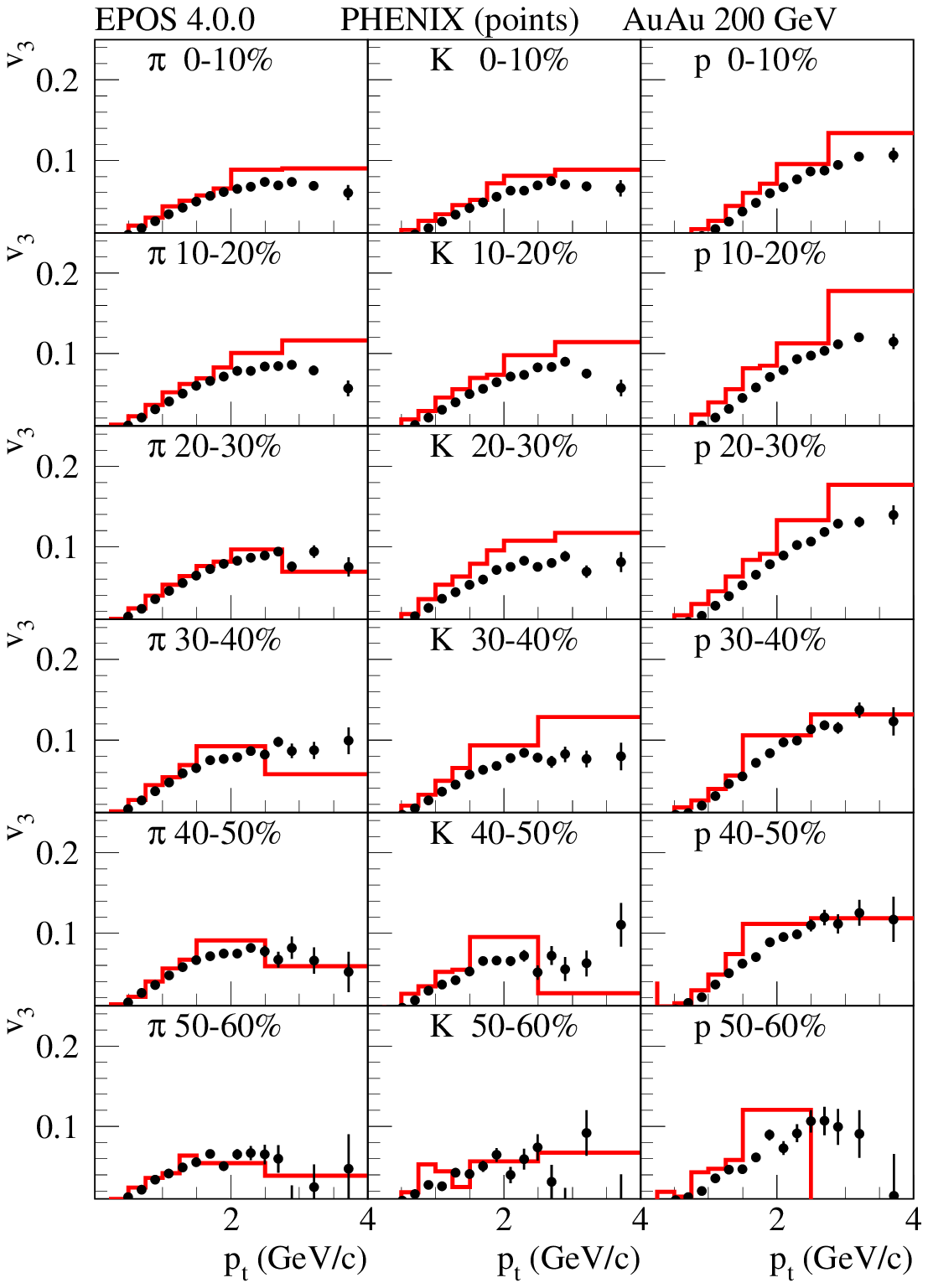}

\centering{}\includegraphics[bb=20bp 30bp 595bp 720bp,clip,scale=0.4]
{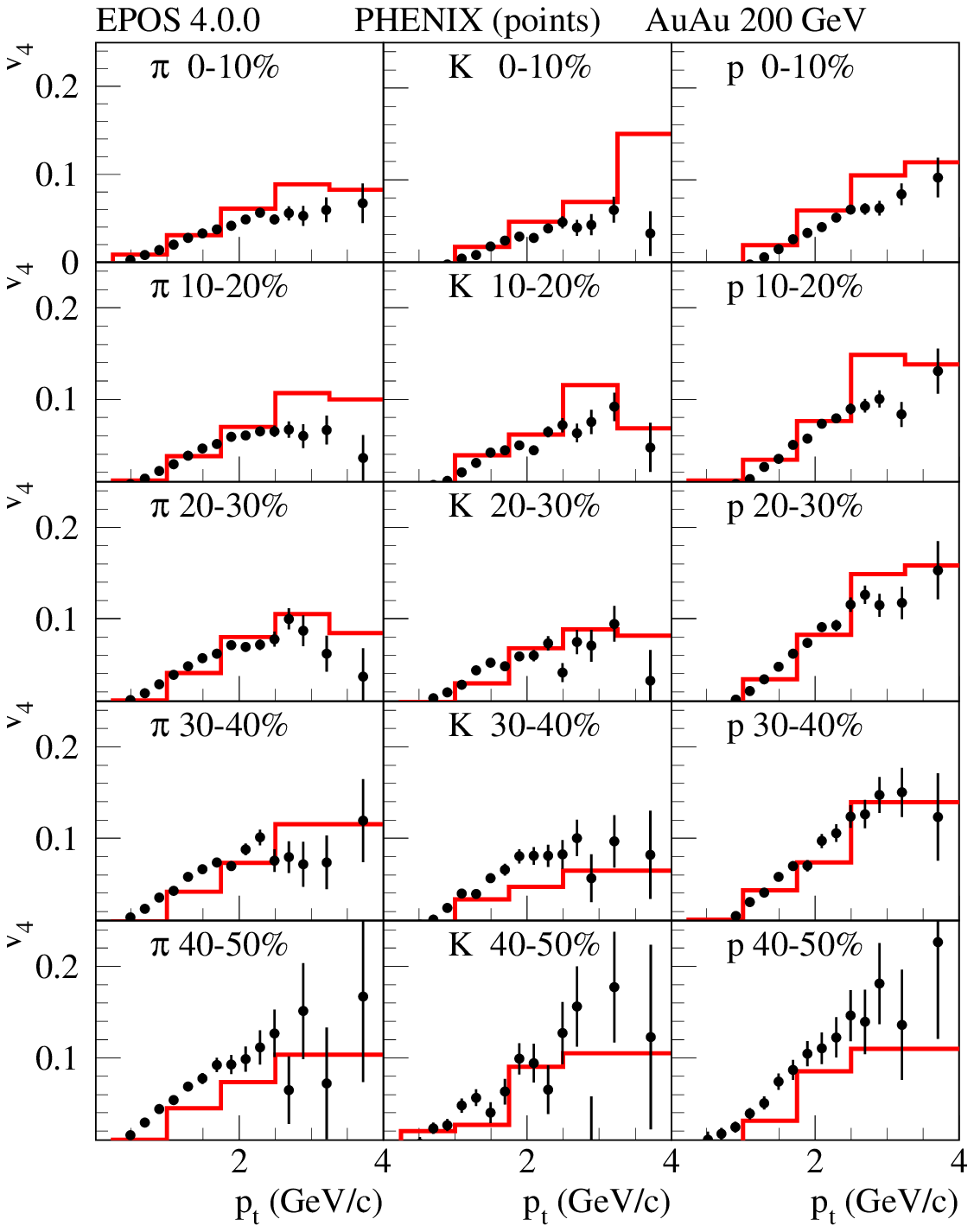}

\caption{Transverse momentum dependence of $v_{3}$ (upper plot) and $v_{4}$
(lower plot) for pions (left column), kaons (middle column) and protons
(right column) in AuAu collisions at 200 GeV, for different centralities.
I compare the full simulations (red lines) with data from PHENIX (black
points). \label{v2-auau-4}}
\end{figure}

In Fig. \ref{v2-auau-1},
I plot the pseudorapidity dependence of $v_{2}$ for different multiplicity
classes in AuAu collisions at 200 GeV. I compare the simulations
(red lines) with data from PHOBOS \cite{PHOBOS:2004vcu} (black points).
In Fig. \ref{v2-auau-2},
I plot the centrality dependence (using $N_{\mathrm{part}}$) of $v_{2}$
in AuAu collisions at 200 GeV. I compare the simulations (red lines)
with data from PHOBOS \cite{PHOBOS:2004vcu} (black points). In Fig.
\ref{v2-auau-3},
I plot the transverse momentum dependence of $v_{2}$ for pions (left
column), kaons (middle column) and protons (right column) in AuAu
collisions at 200 GeV, for different centralities (from top to bottom):
0-10\%, 10-20\%, 20-30\%, 30-40\%, 40-50\%, and 50-60\%. I compare
the full simulations (red lines) and those without hadronic cascade
(yellow lines) with data from PHENIX \cite{PHENIX:2014uik} (black
points). In Fig. \ref{v2-auau-4},
I show results for $v_{3}$ (upper plot) and $v_{4}$ (lower plot),
comparing full simulations (red lines) with data from PHENIX \cite{PHENIX:2014uik}
(black points).

\newpage

\subsection{Spectra for PbPb at 5.02 TeV \label{-------spectra-3-5-tev----------1}}

\begin{figure}[H]
\centering{}\includegraphics[bb=30bp 35bp 450bp 580bp,clip,scale=0.6]
{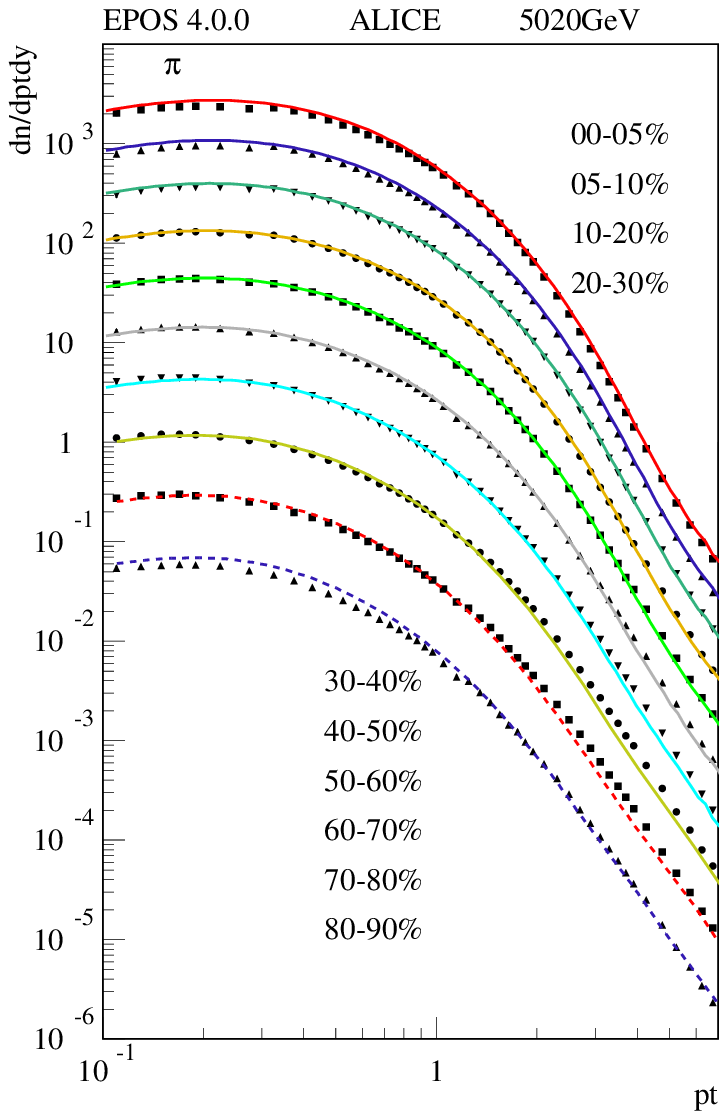}\\
 \caption{Transverse momentum distributions of pions in PbPb collisions at 5.02
TeV for different centrality classes. EPOS4 simulations (lines) are
compared to data from ALICE. From top to bottom, one multiplies the
curves by $2^{-i}$ ($i=0,1,2,3,4, ...$). \label{transverse momentum 16-1}}
\end{figure}

\begin{figure}[H]
\centering{}\includegraphics[bb=30bp 35bp 450bp 580bp,clip,scale=0.6]
{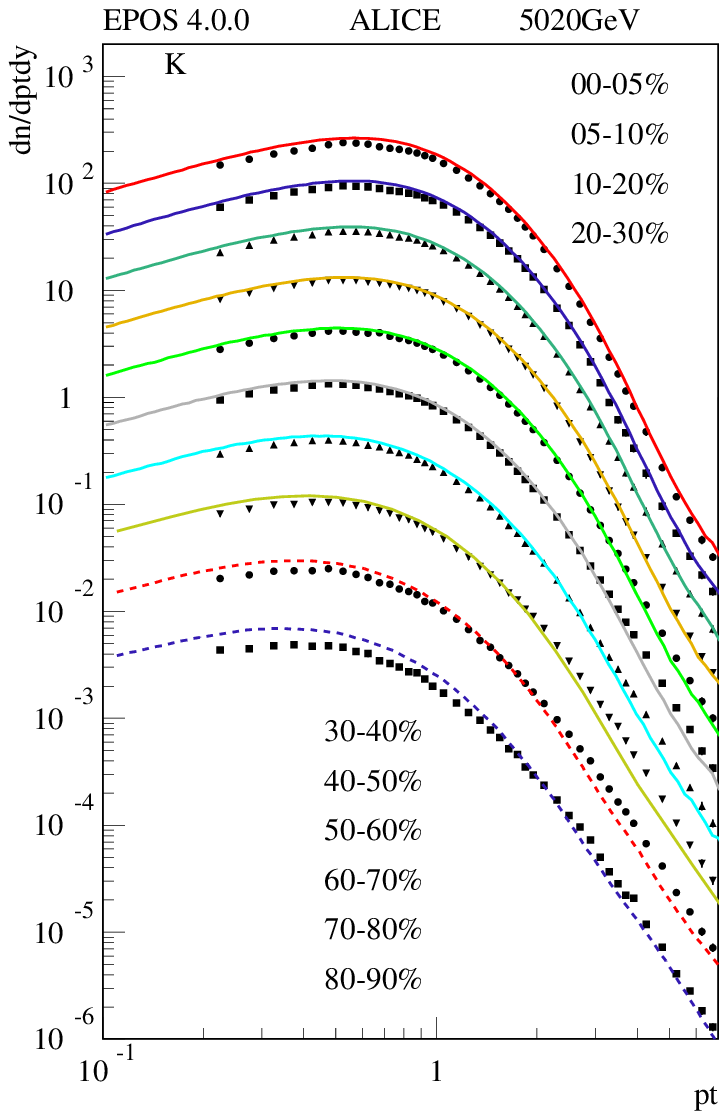}

\centering{}\includegraphics[bb=30bp 30bp 450bp 580bp,clip,scale=0.6]
{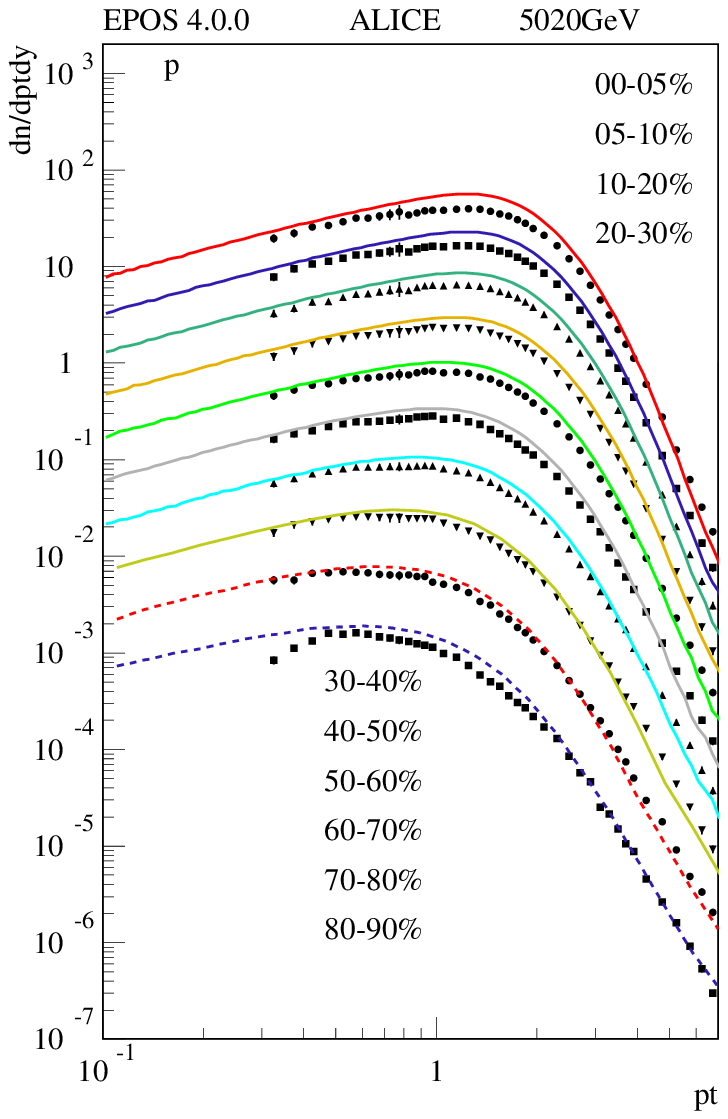}
\caption{Same as Fig. \ref{transverse momentum 16-1}, but for kaons (upper
plot) and protons (lower plot). \label{transverse momentum 16-2}}
\end{figure}

I will show some $p_{t}$ spectra of identified particles, in PbPb
collisions at 5.02 TeV. In Figs. \ref{transverse momentum 16-1} and
\ref{transverse momentum 16-2}, I show transverse momentum distributions
at central rapidity of pions, kaons, and protons in PbPb collisions
at 5.02 TeV for different centrality classes (from top to bottom:
0-5\%, 5-10\%, ..., 80-90\%). EPOS4 simulations (lines) are compared
to data from ALICE \cite{ALICE:2020-5T-pikap}. From Fig. \ref{core-corona-1},
one knows that corona particles dominate beyond 2-3 GeV/c for peripheral
collisions and beyond 4-5 GeV/c for central ones. They are affected
by the energy loss parameter in the core-corona procedure, and allow one
to determine it. Concerning the kaons and protons, one clearly sees
the ``flow effect'', i.e. an increase at intermediate $p_{t}$,
compared to pions.

The picture looks consistent, although there is a systematic excess
of protons at intermediate $p_{t}$. It should be emphasized that
EPOS4.0.0 has been ``tuned'' to thousands of experimental results
(with only a small fraction shown in this paper), and the aim is to
see to what extent one gets a consistent overall picture, and not
to optimize a particular curve.

\subsection{Flow harmonics for PbPb at 2.76 and 5.02 TeV \label{-------flow-3-5-tev---------}}

\begin{figure}[H]
\centering{}\includegraphics[bb=20bp 30bp 595bp 800bp,clip,scale=0.4]
{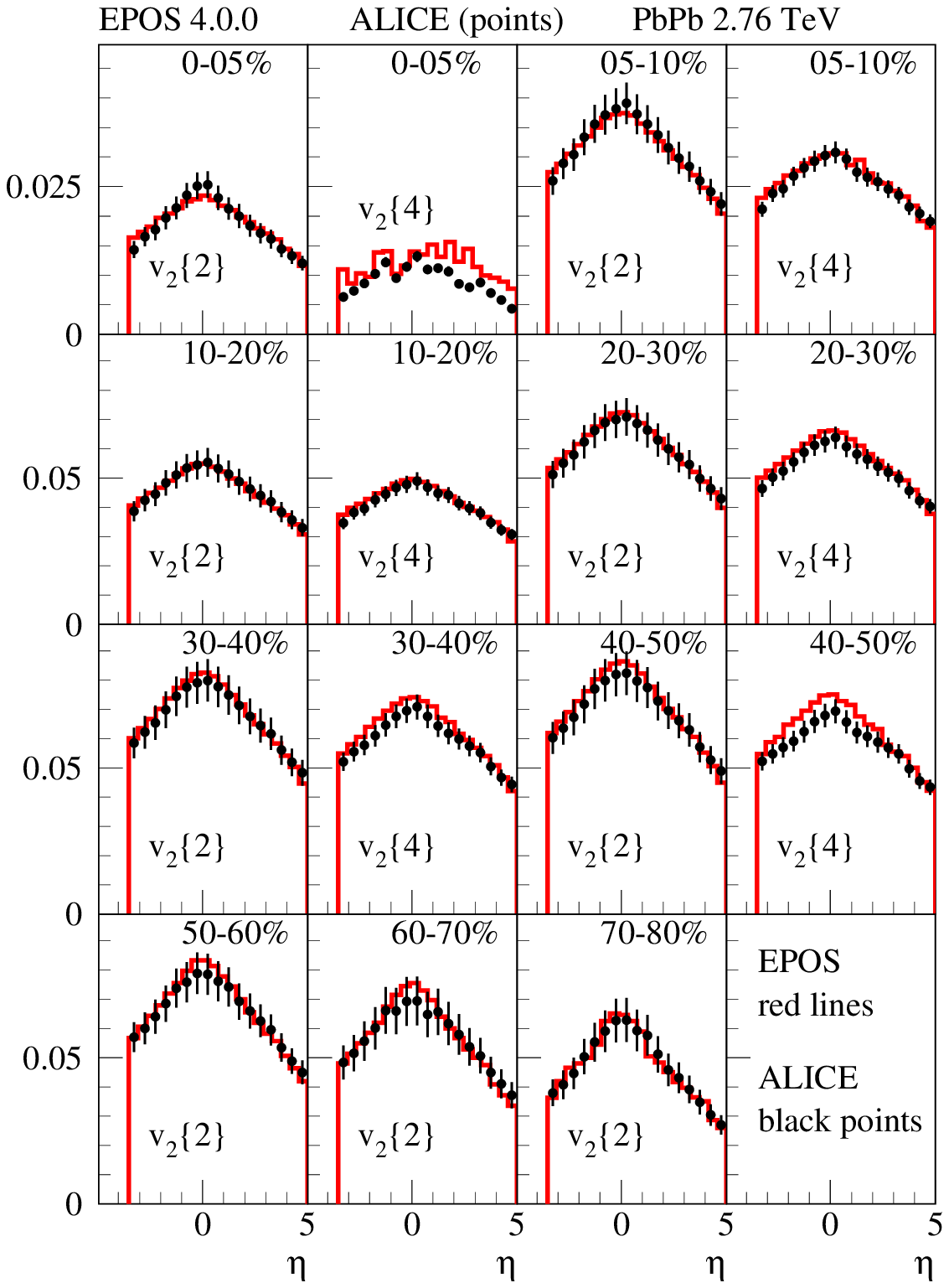}%
\begin{comment}
\begin{center}
\includegraphics[bb=20bp 30bp 595bp 800bp,clip,scale=0.48]
{Home/2/komm/conf/99aktuell/w22flow/PS/z-sel1-02.eps}
\par\end{center}
\end{comment}
\caption{Pseudorapidity dependence of $v_{2}\left\{ 2\right\} $ and $v_{2}\left\{ 4\right\} $
for different multiplicity classes (using percentiles) in PbPb collisions
at 2.76 GeV. I compare the EPOS4 simulations (red lines) with data
from ALICE (black points). \label{v2-pbpb-1}.}
\end{figure}

\begin{figure}[H]
\centering{}\includegraphics[bb=20bp 30bp 595bp 800bp,clip,scale=0.4]
{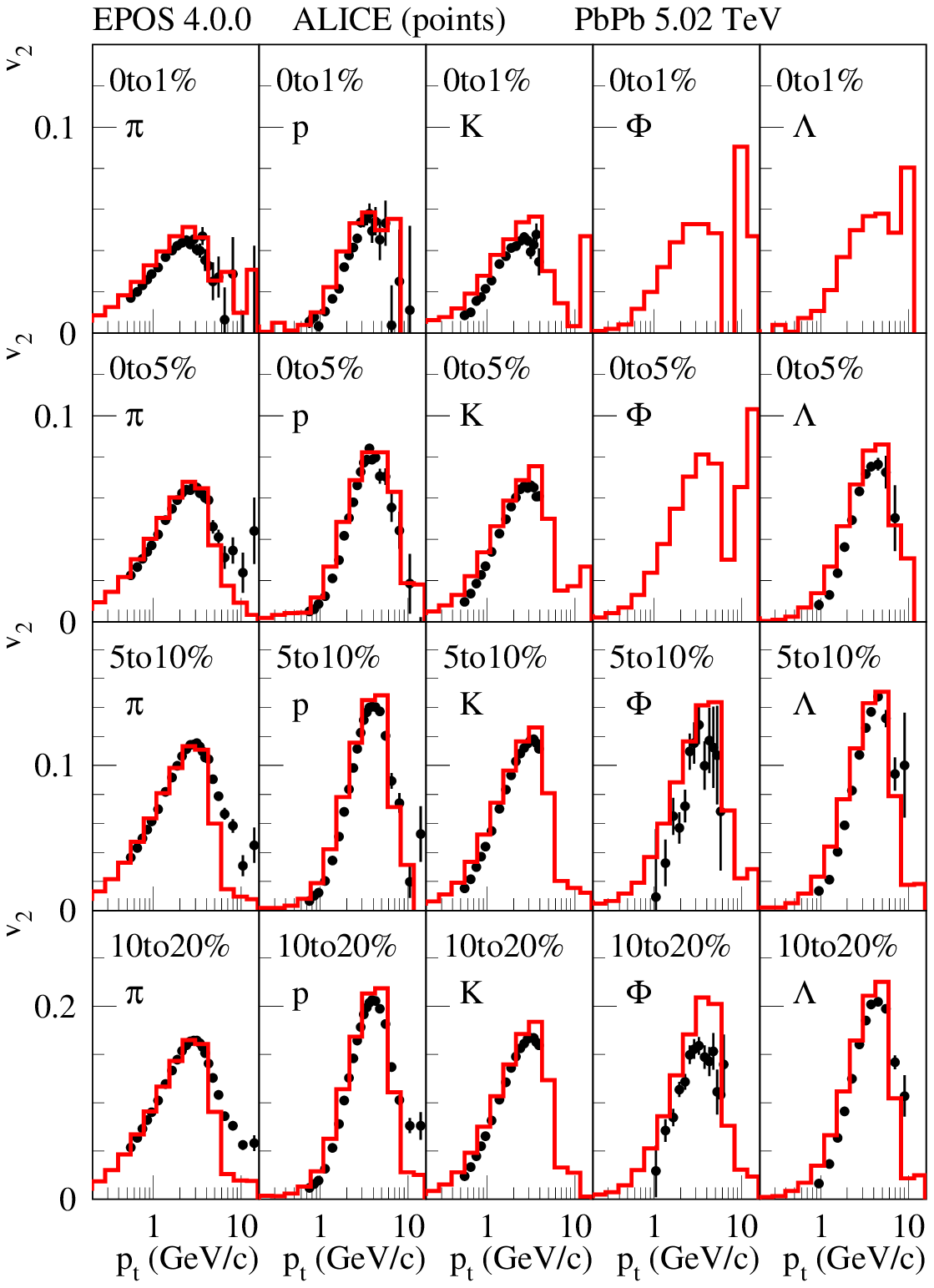}

\centering{}\includegraphics[bb=20bp 30bp 595bp 800bp,clip,scale=0.4]
{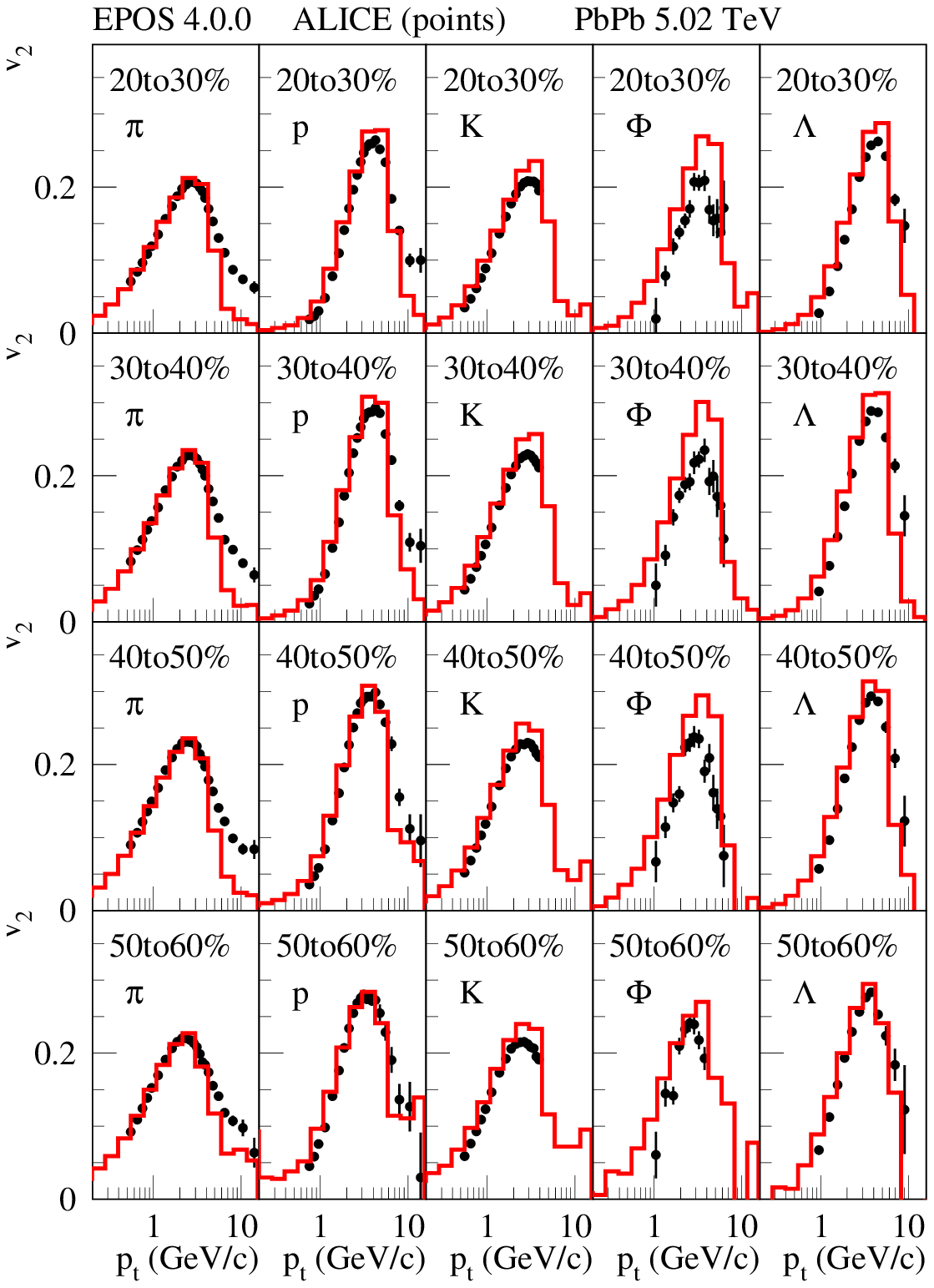}
\caption{$p_{t}$ dependence of $v_{2}$ for different multiplicity classes
(using percentiles) in PbPb collisions at 5.02 TeV, for (from left
to right) pions, protons, kaons ($K_{s})$, $\phi$ mesons, and lambdas.
I compare the simulations (red lines) with data from ALICE (black
points). \label{fig: v2 PbPb5-1}.}
\end{figure}

\begin{figure}[h]
\centering{}\includegraphics[bb=20bp 30bp 595bp 800bp,clip,scale=0.4]
{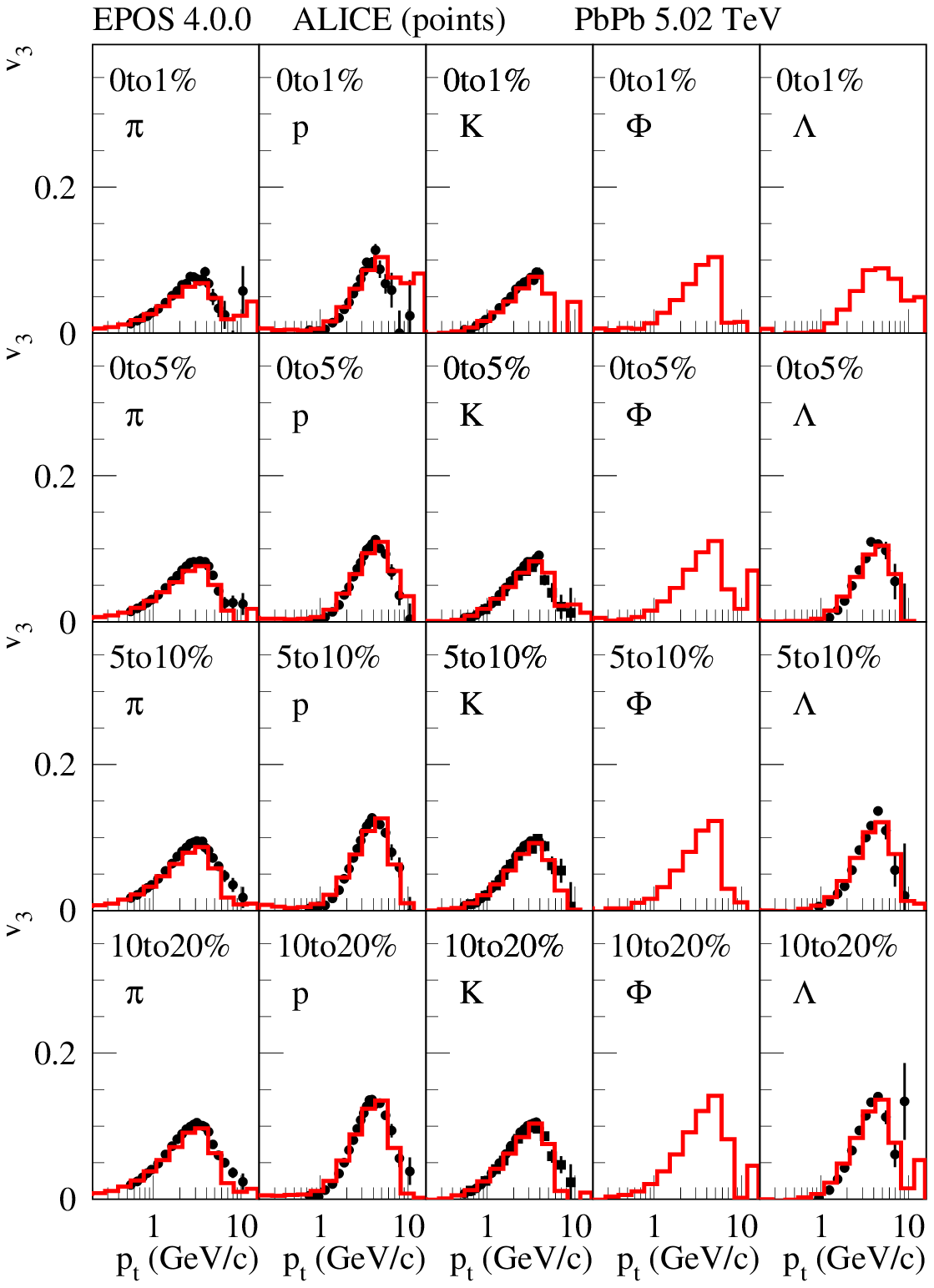}

\centering{}\includegraphics[bb=20bp 30bp 595bp 800bp,clip,scale=0.4]
{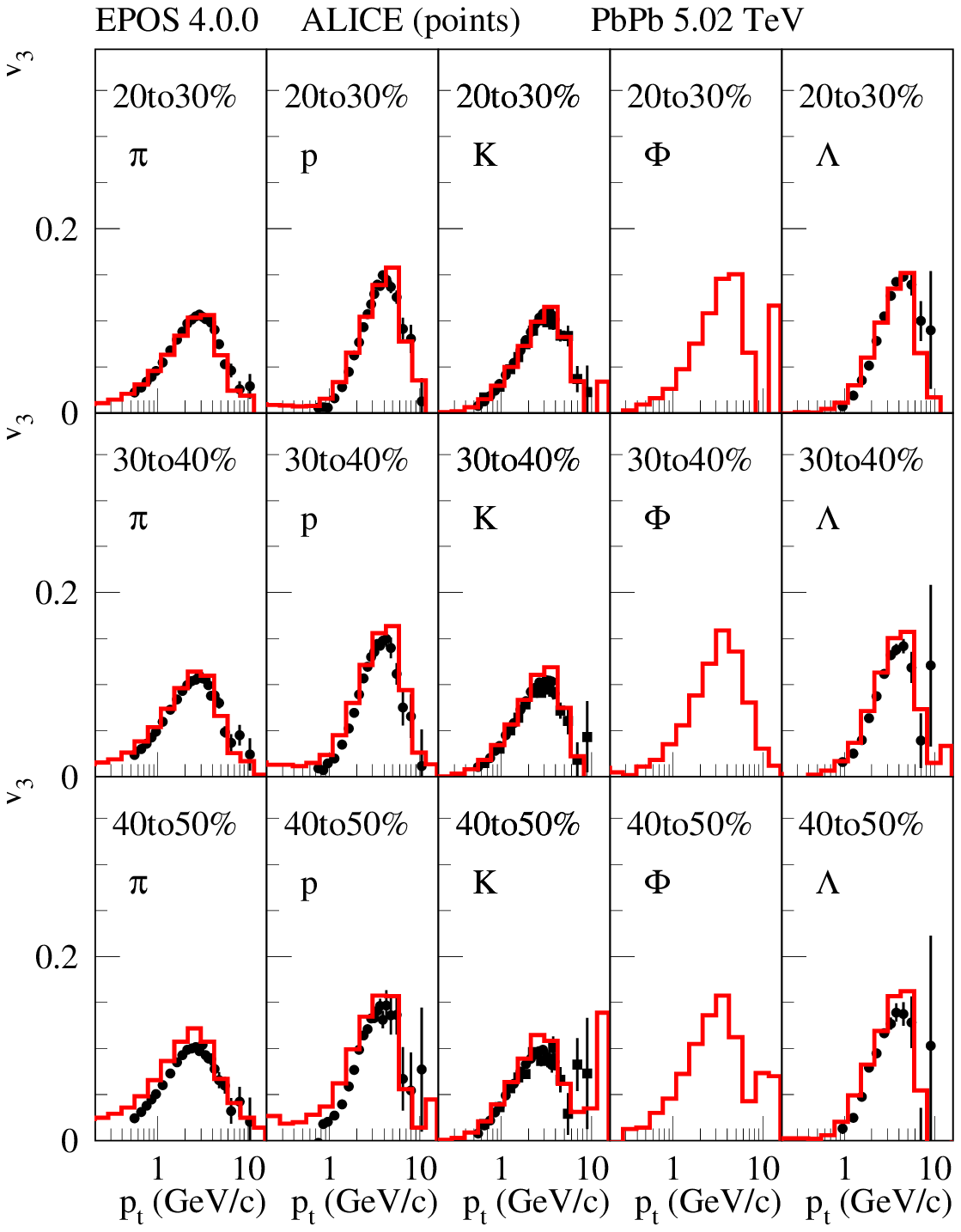}
\caption{$p_{t}$ dependence of $v_{3}$ for different multiplicity classes
in PbPb collisions at 5.02 TeV. I compare the simulations (red lines)
with data from ALICE (black points). \label{fig: v2 PbPb5-2}.}
\end{figure}

In the following, I show results for the flow harmonics $v_{2}$ and
$v_{3}$ in PbPb collisions at LHC energies. In Fig. \ref{v2-pbpb-1},
I plot the pseudorapidity dependence of $v_{2}\left\{ 2\right\} $
and $v_{2}\left\{ 4\right\} $ (see Ref. \cite{ALICE:2016-eta-flow} for
the definitions) for different multiplicity classes (using percentiles)
in PbPb collisions at 2.76 GeV. I compare the EPOS4 simulations (red
lines) with data from ALICE \cite{ALICE:2016-eta-flow} (black points).
In Fig. \ref{fig: v2 PbPb5-1}, I plot the $p_{t}$ dependence of
$v_{2}$ for different multiplicity classes (using percentiles) in
PbPb collisions at 5.02 TeV, for (from left to right) pions, protons,
kaons ($K_{s})$, $\phi$ mesons, and lambdas. I compare the simulations
(red lines) with data from ALICE \cite{ALICE:2018yph} (black points).
In Fig. \ref{fig: v2 PbPb5-2}, I show the corresponding plots for
$v_{3.}$ The simulations give in general a decent description of
the data, but the simulation results for $v_{2}$ at high $p_{t}$
are significantly below the data, pointing to some possible problem
related to parton energy loss, which is presently realized as energy
loss of prehadrons during the core-corona procedure. Some real parton
energy loss procedure in the EPOS4 framework is presently developed.

\section{Summary and conclusion \label{======= conclusion =======}}

It was  first recalled that it  is of fundamental importance to realize
that multiple nucleon-nucleon scatterings in high-energy nucleus-nucleus
($A\!+\!B$) collisions must happen in parallel, and not sequentially.
There is nothing like a first, second, third, etc., collision; they are
all equal and simultaneous. In this context, I presented a discussion
about time scales, and the corresponding applicability of this ``parallel
scattering scenario'' as a function of the collision energy. It was
estimated that beyond a collision energy of 24 GeV the ``parallel
scattering scenario'' is mandatory. The term ``parallel scattering''
actually refers to both, nucleon-nucleon scatterings and parton-parton
scatterings in each nucleon-nucleon collision. I then reviewed early
work on parallel scattering based on S matrix theory, using the concept
of subprocesses referred to as Pomerons, and I showed the importance
of the AGK theorem, which allows one to make the link between the multiple
scattering approach and simple geometric properties as binary scaling.
But this early work considers an ``infinite energy limit'', so energy
sharing among the Pomerons is not an issue. However, in a realistic
scenario -- in particular when it is used as a basis for
event-by-event Monte Carlo procedures -- it is mandatory to include energy-momentum
sharing.

I showed how to implement energy-momentum sharing in the S matrix
approach, and how to prove the validity of the AGK theorem, which
means that the inclusive cross section (with respect to the Pomeron
energy) for $A\!+\!B$ scattering is $AB$ times the inclusive cross
section of a single Pomeron. As a direct consequence, one gets binary
scaling, which means that the inclusive cross section for $A\!+\!B$
scattering is $AB$ times the inclusive cross section for $pp$ scattering.
Starting from the expression for the inelastic $A\!+\!B$ scattering
cross section, I showed how to derive a probability law $\sum_{\{m_{k}\}}\int dX\!_{A\!B}P(K)=1$,
with a ``configuration'' $K=\big\{\{m_{k}\},\{x_{k\mu}^{\pm}\}\big\}$
representing $m_{k}$ cut Pomerons per pair $k$, with light-cone
momentum fractions $x_{k\mu}^{\pm}$, where $P$ is given as a product
of single Pomeron expressions $G$ times some known function $W_{AB}$
(being the result of integrating out all elastic scatterings, i.e.,
uncut Pomerons. The symbol $G$ refers to a cut diagram representing
a single scattering. The calculation $W_{AB}$ (multidimensional integral)
could be done, showing that it is not strictly non-negative, which
ruins the probability interpretation of $P$.

I showed that is possible to regularize $W_{AB}$ to make it strictly
non-negative, which allows recovering the probability interpretation
of $P$, and which allows generating multiple scattering (or multiple
Pomeron) configurations. In order to generate partons, one needs to
specify the internal structure of the Pomeron $G$. I use the simplest
choice, namely $G=G_{\mathrm{QCD}}$, where $G_{\mathrm{QCD}}$ refers
to a pQCD diagram, with the main element being a parton-parton scattering
graph with parton evolutions on both sides and QCD matrix element
in the middle. Computing the inclusive cross section (with respect
to the variable $p_{t}$) in $A\!+\!B$ scattering, one finds that
the AGK theorem is badly violated (which as a consequence ruins binary
scaling). So the model is not usable as it is.

In order to understand the problem, I defined a quantity $R_{\mathrm{deform}}(x^{+},x^{-})$,
referred to as the ``deformation function'', being the ratio of the
normalized inclusive $A\!+\!B$ cross section over the single Pomeron
one, considering the Pomeron light-cone momentum fractions $x^{\pm}$
as variables. One needs $R_{\mathrm{deform}}(x^{+},x^{-})=1$ in order
to validate the AGK theorem, at least in the case $G=G_{\mathrm{QCD}}$.
But $R_{\mathrm{deform}}$ is far from unity, which is the origin
of the violation of AGK. I investigated $R_{\mathrm{deform}}$ more
in detail, and it was understood that a quantity called \textquotedblleft connection
number\textquotedblright{} $N_{\mathrm{conn}}$, which counts the
number of Pomerons connected to a given pair $i,j$ of projectile and target
nucleons, plays an important role. The bigger $N_{\mathrm{conn}}$,
the bigger the number of Pomerons which need to share the given initial
energy of the pair, and therefore, with increasing $N_{\mathrm{conn}}$, 
the deformation function $R_{\mathrm{deform}}$ decreases more and
more below unity, at large values of $x^{\pm}$. This is unavoidable,
and therefore a violation of AGK is unavoidable. At least, one could
``quantify the problem'', being able to find a simple parametrization
of $R_{\mathrm{deform}}$, for given values of $N_{\mathrm{conn}}$,
for all systems and all energies, with tabulated parameters. So in
the following, $R_{\mathrm{deform}}$ could be considered to be known.

At this point, I realized that there were two problems: (a) the assumption
$G=G_{\mathrm{QCD}}$, making the link between the ``Pomeron approach''
and QCD, seemed to be wrong; (b) saturation was not implemented, which
has been known since a long time to play an important role. But fortunately,
the two problems are connected, and there is an amazingly simple solution
based on saturation scales that solves both problems. Instead of the
\textquotedblleft naive\textquotedblright{} assumption $G=G_{\mathrm{QCD}}$,
I postulated 
\begin{equation}
G(x^{+},x^{-})=\frac{n}{R_{\mathrm{deform}}(x^{+},x^{-})}G_{\mathrm{QCD}}(Q_{\mathrm{sat}}^{2},x^{+},x^{-}),\label{fundamental-epos4-equation-1}
\end{equation}
such that \textbf{$G$ }itself does not depend on the environment,
in terms of $N_{\mathrm{conn}}$, which means that the $N_{\mathrm{conn}}$
dependence of $R_{\mathrm{deform}}$ is ``absorbed'' by the saturation
scale $Q_{\mathrm{sat}}^{2}$, which in this way depends on $N_{\mathrm{conn}}$
and also on $x^{\pm}$. Based on Eq. (\ref{fundamental-epos4-equation-1}),
I could prove the validity of the AGK theorem, saying that the inclusive
cross section (with respect to $p_{t}$) for $A\!+\!B$ scattering
is $AB$ times the inclusive cross section of a single Pomeron \textendash{}
but only at large $p_{t}$ (bigger than the relevant saturation scales).
In this way, I was finally able to connect the multiple Pomeron approach
(for parallel scatterings) and pQCD, by introducing saturation scales.
The latter seem indispensable for getting a consistent picture. 

I discussed consequences of having, on one hand, a multiple scattering
approach for instantaneous parallel scatterings, and on the other
hand, the validity of the AGK theorem. The latter allows one to define
parton distribution functions and do simple calculations of inclusive
cross sections at very high $p_{t}$ (the same way as models based
on factorization do). The former does allow one to do much more, treating
problems far beyond the factorization scheme, by using the full multiple
scattering machinery. 

An important application is the study of collective effects in $A\!+\!B$ or $pp$ scattering.
Form the multiple (parallel) scattering approach (referred to as primary interactions) 
one gets  first of all a (more or less)
large number of prehadrons, which are then subject to secondary interactions,
in the form of core-corona separation, hydrodynamic evolution of the
core, and microcanonical decay of the plasma. All this has been discussed
briefly, not being the main subject of this paper. Finally, I showed
some results on heavy-ion collisions at LHC and RHIC energies, covering
$p_{t}$ spectra and flow harmonics of identified hadrons, in order to show to what extent the model works.

\subsection*{Acknowledgements}

K.W. thanks Tanguy Pierog for many contributions during the past decade
and in particular the important proposal (and first implementation)
in 2015 of using a Pomeron definition as $G\propto G_{\mathrm{QCD}}(Q_{\mathrm{sat}}^{2})$
with a parametrized $G$. He also made a first attempt to take into
account the ``deformation'' by using rescaled longitudinal momentum
arguments.

~

\clearpage{}

\appendix
\noindent \textbf{\LARGE{}Appendix}{\LARGE\par}

\section{Asymptotic behavior of T-matrices for $pp$ scattering \label{======= asymptotic-behavior =======}}

\begin{figure}[h]
\begin{centering}
(a)\hspace*{3.7cm}(b)$\qquad\qquad\qquad\qquad$\\
\includegraphics[scale=0.35]
{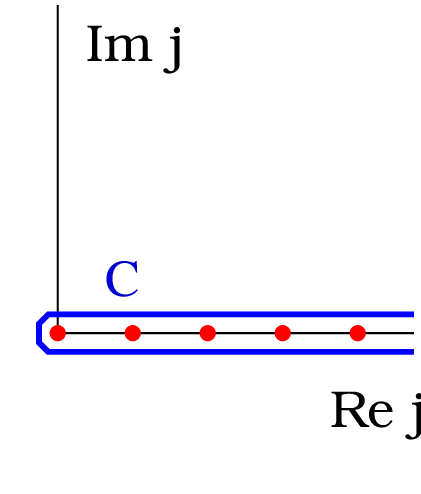}
$\qquad\qquad$\includegraphics[scale=0.35]
{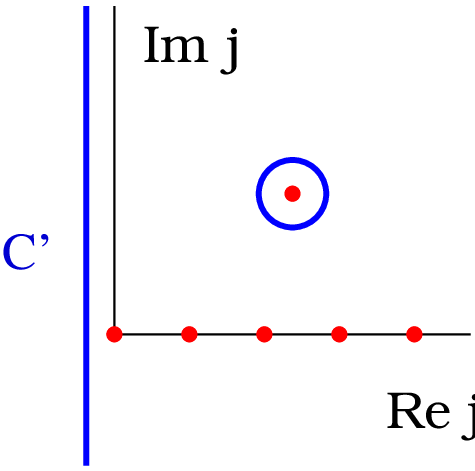}
\caption{Contour integrations in $j$ plane.\label{contour-integrations}}
\par\end{centering}
\end{figure}

\begin{figure}[h]
\centering{}\includegraphics[scale=0.27]
{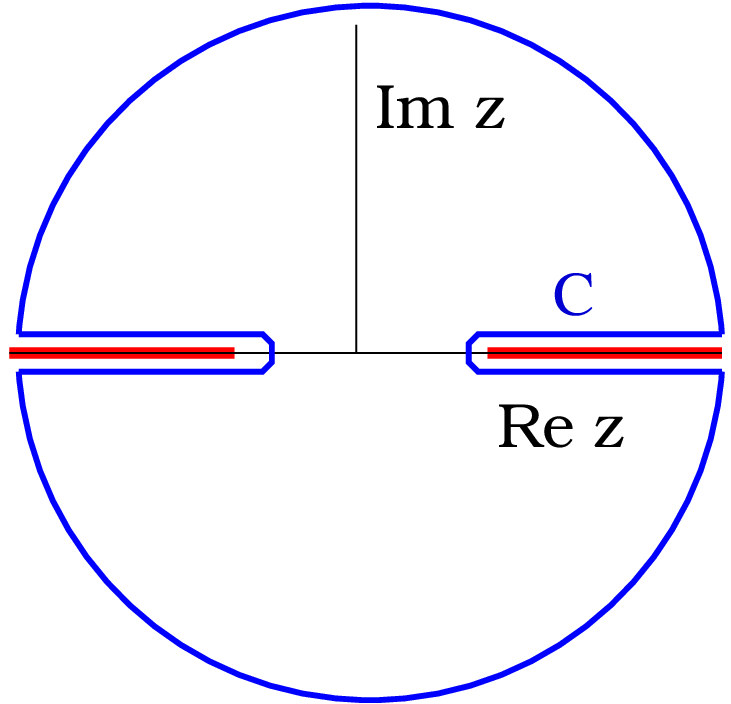}
\caption{Contour integration in $z$ plane.\label{contour-integrations-1}}
\end{figure}

Consider a reaction (all particles having mass $m$) 
\begin{equation}
1+2\to3+4
\end{equation}
in the $x-z$ plane. In the c.m. system, one has ($P=|\vec{p}|$) 
\begin{equation}
P_{1}=P_{2}=P_{3}=P_{4},
\end{equation}
and
\begin{equation}
E_{1}=E_{2}=E_{3}=E_{4}.
\end{equation}
The vectors $(E,p_{x},p_{z})$ are given as (for particles 1,2,3,4):

\begin{equation}
(E,p_{x},p_{z})=\left\{ \begin{array}{c}
(\sqrt{P^{2}+m^{2}},0,P)\qquad\qquad\qquad\\
(\sqrt{P^{2}+m^{2}},0,-P)\qquad\qquad\quad\\
(\sqrt{P^{2}+m^{2}},P\sin\theta,P\cos\theta)\;\quad\\
(\sqrt{P^{2}+m^{2}},-P\sin\theta,-P\cos\theta)\;
\end{array}\right.,
\end{equation}
which gives
\begin{equation}
s=4\left(P^{2}+m^{2}\right)\qquad\qquad\qquad\qquad\qquad\qquad\quad
\end{equation}
\begin{equation}
t=0-P^{2}\sin^{2}\theta-P^{2}\left(1-\cos\theta\right)^{2}=-2P^{2}(1-\cos\theta)
\end{equation}
\begin{equation}
u=0-P^{2}\sin^{2}\theta-P^{2}\left(1+\cos\theta\right)^{2}=-2P^{2}(1+\cos\theta)
\end{equation}
Defining $z=\cos\theta$, one gets 
\begin{equation}
t=2(\frac{s}{4}-m^{2})(z-1),
\end{equation}
\begin{equation}
u=2(\frac{s}{4}-m^{2})(-z-1),
\end{equation}
which gives for zero masses the relation 
\begin{equation}
z=1+\frac{2t}{s}.\label{z-t-s-relation}
\end{equation}
One may always expand the T-matrix (partial wave expansion)
\begin{equation}
\boldsymbol{\mathrm{T}}(s,t)=\sum_{j=0}^{\infty}(2j+1)\mathcal{T}(j,s)P_{j}(z)
\end{equation}
with 
\begin{equation}
\mathcal{T}(j,s)=\frac{1}{2}\int_{-1}^{1}dz\,\boldsymbol{\mathrm{T}}(s,t)P_{j}(z)
\end{equation}
due to the orthogonality property of Legendre polynomials,
\begin{equation}
\frac{2n+1}{2}\int_{-1}^{1}dx\,P_{n}(x)P_{m}(x)=\delta_{mn}.
\end{equation}
One analytically continues to the unphysical region of the $1+2\to3+4$
process, with
\begin{equation}
\boldsymbol{\mathrm{T}}_{1+2\to3+4}(s<0,t>0),
\end{equation}
corresponding to the physical region of the $1+\bar{3}\to\bar{2}+4$
($t$-channel) process, with
\begin{equation}
\boldsymbol{\mathrm{T}}_{1+\bar{3}\to\bar{2}+4}(t>0,s<0),
\end{equation}
and after exchanging $s$ and $t$, one gets 
\begin{equation}
\boldsymbol{\mathrm{T}}_{1+\bar{3}\to\bar{2}+4}(s>0,t<0).
\end{equation}
With this new definition of $\boldsymbol{\mathrm{T}}$, the partial
wave expansion reads
\begin{equation}
\boldsymbol{\mathrm{T}}(s,t)=\sum_{j=0}^{\infty}(2j+1)\mathcal{T}(j,t)P_{j}(z),
\end{equation}
with
\begin{equation}
\mathcal{T}(j,t)=\frac{1}{2}\int_{-1}^{1}dz\,\boldsymbol{\mathrm{T}}(s,t)P_{j}(z),\label{partial-wave}
\end{equation}
and with
\begin{equation}
z=1+\frac{2s}{t}.
\end{equation}
The Watson-Sommerfeld transform \cite{Watson:1918,Sommerfeld:1949}
amounts to writing the partial wave expansion as
\begin{equation}
\boldsymbol{\mathrm{T}}(s,t)=\frac{1}{2i}\int_{C}dj\,\frac{1}{\sin\pi j}(2j+1)\mathcal{T}(j,t)P_{j}(z),
\end{equation}
with a contour integration in the complex j plane, as shown in Fig.
\ref{contour-integrations}(a). Opening the contour [see Fig. \ref{contour-integrations}(b)]
to integrate along the imaginary axis (with $\mathrm{Re}j=-\frac{1}{2}$),
one picks up poles of $\mathcal{T}$ at $j=\alpha_{n}(t)$, with residues
$\beta'_{n}(t)$, so one gets
\begin{equation}
\boldsymbol{\mathrm{T}}(s,t)=\frac{1}{2i}\int_{C'}dj\,\frac{2j+1}{\sin\pi j}\mathcal{T}(j,t)P_{j}(z)
\end{equation}
\[
+\sum\underbrace{\pi\frac{2\alpha_{n}(t)+1}{\sin\pi\alpha_{n}(t)}\beta'_{n}(t)}_{\beta_{n}(t)}P_{\alpha_{n}(t)}(z).
\]
As one will see, the partial wave amplitudes have contributions which
alternate in sign, so there is a factor
\begin{equation}
(-1)^{j}=\exp(i\pi j)
\end{equation}
which diverges on the imaginary axis. The problem will be solved by
separating even and odd terms. To see this problem, let me first investigate
$\mathcal{T}(j,t)$. One considers $\boldsymbol{\mathrm{T}}(s,t)$
as a function of $z$ and $t$, and writes (Cauchy)
\begin{equation}
\boldsymbol{\mathrm{T}}(z,t)=\frac{1}{2\pi i}\int_{c}\frac{\boldsymbol{\mathrm{T}}(z',t)}{z'-z}dz'
\end{equation}
where one chooses $C$ as in Fig. \ref{contour-integrations-1}.
One finds (assuming the semicircles do not contribute)

\begin{equation}
\boldsymbol{\mathrm{T}}(z,t)=\frac{1}{2\pi i}\int_{z_{0}^{+}}^{\infty}\frac{\mathrm{disc}\,T(z',t)}{z'-z}dz'
\end{equation}
\[
+\frac{1}{2\pi i}\int_{-\infty}^{-z_{0}^{-}}\frac{\mathrm{disc}\,T(z',t)}{z'-z}dz'
\]
This so-called dispersion relation %
\begin{comment}
more details see conf/19lectureKarls/lecture/karls.lyx
\end{comment}
may be inserted into Eq. (\ref{partial-wave}):
\begin{equation}
\mathcal{T}(j,t)=\frac{1}{2\pi i}\int_{z_{0}}^{\infty}\mathrm{disc}\,\boldsymbol{\mathrm{T}}(z',t)\frac{1}{2}\int_{-1}^{1}\frac{P_{j}(z)}{z'-z}dzdz'
\end{equation}
\[
+\frac{1}{2\pi i}\int_{-\infty}^{-z_{0}}\mathrm{disc}\,\boldsymbol{\mathrm{T}}(z',t)\frac{1}{2}\int_{-1}^{1}\frac{P_{j}(z)}{z'-z}dzdz'.
\]
One may use ``Neumanns formula''
\begin{equation}
\frac{1}{2}\int_{-1}^{1}\frac{P_{j}(z)}{z'-z}dz=Q_{j}(z'),
\end{equation}
with Legendre functions $Q_{j}$ of the second kind, and get
\begin{equation}
\mathcal{T}(j,t)=\frac{1}{2\pi i}\int_{z_{0}}^{\infty}\mathrm{disc}\,\boldsymbol{\mathrm{T}}(z',t)Q_{j}(z')dz'
\end{equation}
\[
\qquad\qquad+\frac{1}{2\pi i}\int_{z_{0}}^{\infty}\mathrm{disc}\,\boldsymbol{\mathrm{T}}(-z',t)Q_{j}(-z')dz'
\]
Using the symmetry property $Q_{j}(-z)=(-1)^{j+1}Q_{j}(z)$ and the
definition $D_{j}^{\pm}=\frac{1}{2\pi i}\int_{z_{0}}^{\infty}\mathrm{disc}\,\boldsymbol{\mathrm{T}}(\pm z',t)Q_{j}(z')dz'$,
one gets 
\begin{equation}
\mathcal{T}(j,t)=\Big\{ D_{j}^{+}+\mathrm{(-1)^{j+1}}D_{j}^{-}\Big\},
\end{equation}
which shows the above-mentioned $(-1)^{j}$ problem. Writing 

\begin{equation}
\mathcal{T}(j,t)=\sum_{\eta\in\{-1,1\}}\frac{1+\eta(-1)^{j}}{2}\Big\{ D_{j}^{+}+\mathrm{(-1)^{j+1}}D_{j}^{-}\Big\}
\end{equation}
allows one to separate even and odd terms, 
\begin{equation}
\mathcal{T}^{1}(j,t)=D_{j}^{+}-D_{j}^{-},\;\mathcal{T}^{-1}(j,t)=D_{j}^{+}+D_{j}^{-},
\end{equation}
as
\begin{equation}
\mathcal{T}(j,t)=\sum_{\eta\in\{-1,1\}}\frac{1+\eta(-1)^{j}}{2}\,\mathcal{T}^{\eta}(j,t)
\end{equation}
with a so-called signature $\eta$. The Watson-Sommerfeld transform
then reads

\[
\boldsymbol{\mathrm{T}}(s,t)=\sum_{\eta}\frac{1}{2i}\int_{C'}dj\,\begin{array}{c}
\frac{1+\eta(-1)^{j}}{2}\end{array}\,\frac{1}{\sin\pi j}(2j+1)\mathcal{T}^{\eta}(j,t)P_{j}(z)
\]
\begin{equation}
+\sum_{\eta}\sum_{n_{\eta}}\frac{1+\eta(-1)^{\alpha_{n_{\eta}}(t)}}{2}\,\beta_{n_{\eta}}(t)\,P_{\alpha_{n_{\eta}}(t)}(z),
\end{equation}
and there is no $(-1)^{j}$ problem any more. The poles $\alpha_{n}(t)$
are called 
\begin{itemize}
\item even signature Regge poles ($\eta=+1$) or
\item odd signature Regge poles ($\eta=-1$).
\end{itemize}
Asymptotically, for $s\gg|t|$, i.e., $|z-1|=|\frac{2s}{t}|\gg1$,
the Legendre polynomial 
\begin{equation}
P_{j}(z)=\sum_{k=0}^{j}\left(\begin{array}{c}
j\\
k
\end{array}\right)\left(\begin{array}{c}
j+k\\
k
\end{array}\right)\left(\frac{z-1}{2}\right)^{k}
\end{equation}
is dominated by its leading term,
\begin{equation}
P_{j}(z)\to\left(\begin{array}{c}
2j\\
j
\end{array}\right)\left(\frac{z-1}{2}\right)^{j},
\end{equation}
so one has in this limit
\begin{equation}
P_{j}(z)=P_{j}(1+\frac{2s}{t})\to\frac{\Gamma(2j+1)}{\Gamma^{2}(j+1)}\left(\frac{s}{t}\right)^{j}.
\end{equation}
The contour integration (along $j=-\frac{1}{2}+iy)$, for $s\gg t$
vanishes, because of factor $(s/t)^{-1/2}$, and one gets\textbf{
}(absorbing the $\Gamma$ functions and $t^{-j}$ into $\beta$),
for $s\gg t$:

\begin{equation}
\boldsymbol{\mathrm{T}}(s,t)=\sum_{\eta}\sum_{n_{\eta}}\frac{1+\eta(-1)^{\alpha_{n_{\eta}}(t)}}{2}\,\beta_{n_{\eta}}(t)\,s^{\alpha_{n_{\eta}}(t)}
\end{equation}
With $\alpha(t)$ being the rightmost (leading) pole, one gets

\begin{equation}
\boldsymbol{\mathrm{T}}(s,t)=\frac{1+\eta e^{-i\pi\alpha(t)}}{2}\beta(t)\,s^{\alpha(t)}
\end{equation}
viewed at the time to correspond to the exchange of some hypothetical
``particle'' called Reggeon or Pomeron. Absorbing $(1+\eta e^{-i\pi\alpha(t)})/2$
into $\beta$, and (because of $t\ll s$)
\[
\alpha(t)=\alpha(0)+\alpha't\,,
\]
and neglecting the $t$ dependence of $\beta$, one gets finally\medskip{}
\begin{equation}
\boldsymbol{\mathrm{T}}(s,t)=\beta\,s^{\alpha(0)+\alpha't},\label{t-matrix-regge-pole-expression}
\end{equation}
\smallskip{}

\noindent
with $\alpha(0)$ called intercept, and $\alpha'$ called slope.
This is referred to as Regge pole expression for the T-matrix
\cite{Regge:1959}.

\section{Explicit formulas for W$^{\boldsymbol{AB}}$ \label{======= W-AB =======}}

The $W_{AB}$ expression is given as [see Eqs. (\ref{W-1}) and (\ref{W-2})]
\begin{align}
W_{AB} & =\sum_{l_{1}=0}^{\infty}\!\!\ldots\!\!\sum_{l_{AB}=0}^{\infty}\int d\tilde{X}\!_{A\!B}\prod_{k=1}^{AB}\left[\frac{1}{l_{k}!}\prod_{\lambda=1}^{l_{k}}-G'_{k\lambda}\right]\label{W-1-1}\\
 & \prod_{i=1}^{A}V(x_{i}^{+}-\!\!\sum_{\underset{\pi(k)=i}{k=1}}^{AB}\sum_{\lambda=1}^{l_{k}}\!\tilde{x}_{k\lambda}^{+})\prod_{j=1}^{B}V(x_{j}^{-}-\!\!\sum_{\underset{\tau(k)=j}{k=1}}^{AB}\sum_{\lambda=1}^{l_{k}}\!\tilde{x}_{k\lambda}^{-}),\nonumber 
\end{align}
with
\begin{equation}
x_{i}^{+}=1-\!\!\sum_{\underset{\pi(k)=i}{k=1}}^{AB}\sum_{\mu=1}^{m_{k}}\!x_{k\mu}^{+},\quad x_{j}^{-}=1-\!\!\sum_{\underset{\tau(k)=j}{k=1}}^{AB}\sum_{\mu=1}^{m_{k}}\!x_{k\mu}^{-}.\label{W-2-1}
\end{equation}
Using $G=G_{\mathrm{QCD}}$ [see Eq. (\ref{G-equal-G-QCD}) and the
discussion before], one can show \cite{Drescher:2000ha} that one
may obtain an almost perfect fit of the numerically computed functions
$G_{\mathrm{QCD}}$, of the form
\begin{align}
 & G_{\mathrm{QCDpar}}(\!x^{+},x^{-},s,b\!)=\sum_{N=1}^{N_{\mathrm{par}}}\alpha_{N}(x^{+}x^{-})^{\beta_{N}},\label{fit-1}\\
 & \qquad\alpha_{N}=\left(\alpha_{D_{N}}+\alpha_{D_{N}}^{*}\right)\,s^{\left(\beta_{D_{N}}+\gamma_{D_{N}}b^{2}\right)}\,e^{-\frac{b^{2}}{\delta_{D_{N}}}}\,,\label{fit-2}\\
 & \qquad\beta_{N}=\beta_{D_{N}}+\beta_{D_{N}}^{*}+\gamma_{D_{N}}b^{2}-\alpha_{\mathrm{part}}\,,\label{fit-3}
\end{align}
with $\alpha_{D_{N}}^{*}\neq0$ and $\beta_{D_{N}}^{*}\neq0$ only
if $\alpha_{D_{N}}=0$, and with $N_{\mathrm{par}}$ being a small
number (4). This parametric form has been inspired by the asymptotic
expressions for T-matrices (see Appendix \ref{======= asymptotic-behavior =======}).
So I will use $G=G_{\mathrm{QCDpar}}$ in the following.
Furthermore, the vertices are parametrized as
\begin{equation}
V(x)=x^{\alpha_{\mathrm{remn}}}.\label{fit-4}
\end{equation}

\subsection{Doing the integrals}

In the following discussion, I use explicitly the (current) value
$N_{\mathrm{par}}=4$, for simplicity, but the results can be easily
adapted to other values. Using $G=G_{\mathrm{QCDpar}}$ and $V(x)=x^{\alpha_{\mathrm{remn}}}$,
and defining $G_{Nk\lambda}=\alpha_{N}(\tilde{x}_{k\lambda}^{+}\tilde{x}_{k\lambda}^{-})^{\beta_{N}}$,
one gets 
\begin{align}
W_{AB} & =\sum_{l_{1}=0}^{\infty}\!\!\ldots\!\!\sum_{l_{AB}=0}^{\infty}\int d\tilde{X}\!_{A\!B}\prod_{k=1}^{AB}\left[\frac{1}{l_{k}!}\prod_{\lambda=1}^{l_{k}}\sum_{N=1}^{4}-G_{Nk\lambda}\right]\label{W-1-2}\\
 & \prod_{i=1}^{A}(x_{i}^{+}-\!\!\sum_{\underset{\pi(k)=i}{k=1}}^{AB}\sum_{\lambda=1}^{l_{k}}\!\tilde{x}_{k\lambda}^{+})^{\alpha_{\mathrm{remn}}}\prod_{j=1}^{B}(x_{j}^{-}-\!\!\sum_{\underset{\tau(k)=j}{k=1}}^{AB}\sum_{\lambda=1}^{l_{k}}\!\tilde{x}_{k\lambda}^{-})^{\alpha_{\mathrm{remn}}}.\nonumber 
\end{align}
One has  for any $k$, using simply $l=l_{k}$,
\begin{equation}
\sum_{l=0}^{\infty}\!\ldots\!\frac{1}{l!}\prod_{\lambda=1}^{l}\sum_{N=1}^{4}-G_{Nk\lambda}=\sum_{l=0}^{\infty}\!\ldots\!\frac{1}{l!}\sum_{N_{1}=1}^{4}\!\!\ldots\!\!\sum_{N_{l}=1}^{4}\prod_{\lambda=1}^{l}-G_{N_{\lambda}k\lambda}.
\end{equation}
Keeping in mind that this is an integrand, and one may exchange integration
variables $\tilde{x}_{k\lambda}^{\pm}$ and $\tilde{x}_{k\lambda'}^{\pm}$~,
one may exchange for a given value of $N$ the terms $G_{Nk\lambda}$
and $G_{Nk\lambda'}$ without changing the result, which means that
in the sum over $\prod_{\lambda=1}^{l}-G_{N_{\lambda}k\lambda}$ there
are {\footnotesize{}$\left(\!\!\begin{array}{c}
l!\\
r_{1}!r_{2}!r_{3}!r_{4}!
\end{array}\!\!\right)$} identical expressions with $r_{N}$ times $N_{\lambda}=N$, for
$N=1,...,4$, which one may write as 
\begin{equation}
\prod_{\lambda=1}^{r_{1}}-G_{1k\lambda}\!\!\!\prod_{\lambda=r_{1}+1}^{r_{1}+r_{2}}-G_{2k\lambda}\!\!\!\!\!\!\prod_{\lambda=r_{1}+r_{2}+1}^{r_{1}+r_{2}+r_{3}}-G_{3k\lambda}\!\!\!\!\!\!\prod_{\lambda=r_{1}+r_{2}+r_{3}+1}^{r_{1}+r_{2}+r_{3}+r_{4}}\!\!\!-G_{4k\lambda},
\end{equation}
with the constraint $r_{1}+r_{2}+r_{3}+r_{4}=l$, which will be released
when one does the sum $\sum_{l=0}^{\infty}$. Putting all together,
one gets (using $l_{k}=l$ and $r_{Nk}$ instead of $r_{N}$
\begin{align}
 & \sum_{l_{k}=0}^{\infty}\!\ldots\!\frac{1}{l_{k}!}\prod_{\lambda=1}^{l_{k}}\sum_{N=1}^{4}-G_{Nk\lambda}=\sum_{r_{1k}=0}^{\infty}\cdots\sum_{r_{4k}=0}^{\infty}\,\ldots\,\\
 & \qquad\frac{1}{r_{1k}!\ldots r_{4k}!}\prod_{\lambda=1}^{r_{1k}}-G_{1k\lambda}\cdots\prod_{\lambda=r_{1k}+r_{2k}+r_{3k}+1}^{r_{1k}+r_{2k}+r_{3k}+r_{4k}}\!\!\!-G_{Nk\lambda}.\nonumber 
\end{align}
Inserted into Eq. (\ref{W-1-2}), using $G_{Nk\lambda}=\alpha_{N}(\tilde{x}_{k\lambda}^{+}\tilde{x}_{k\lambda}^{-})^{\beta_{N}}$,
one gets
\begin{align}
 & W_{AB}=\sum_{r_{11}=0}^{\infty}\!\!\!\cdots\!\!\!\sum_{r_{41}=0}^{\infty}\cdots\sum_{r_{1\,AB}=0}^{\infty}\!\!\!\cdots\!\!\!\sum_{r_{4\,AB}=0}^{\infty}\label{W-1-3}\\
 & \;\int d\tilde{X}\!_{A\!B}\prod_{k=1}^{AB}\Bigg[\frac{1}{r_{1k}!\ldots r_{4k}!}\prod_{\lambda=1}^{r_{1k}}-\alpha_{1}(\tilde{x}_{k\lambda}^{+}\tilde{x}_{k\lambda}^{-})^{\beta_{1}}\cdots\nonumber \\
 & \qquad\qquad\qquad\cdots\prod_{\lambda=r_{1k}+r_{2k}+r_{3k}+1}^{r_{1k}+r_{2k}+r_{3k}+r_{4k}}\!\!\!-\alpha_{4}(\tilde{x}_{k\lambda}^{+}\tilde{x}_{k\lambda}^{-})^{\beta_{4}}\Bigg]\nonumber \\
 & \prod_{i=1}^{A}(x_{i}^{+}-\!\!\sum_{\underset{\pi(k)=i}{k=1}}^{AB}\sum_{\lambda=1}^{l_{k}}\!\tilde{x}_{k\lambda}^{+})^{\alpha_{\mathrm{remn}}}\prod_{j=1}^{B}(x_{j}^{-}-\!\!\sum_{\underset{\tau(k)=j}{k=1}}^{AB}\sum_{\lambda=1}^{l_{k}}\!\tilde{x}_{k\lambda}^{-})^{\alpha_{\mathrm{remn}}}.\nonumber 
\end{align}
This will allow one to separate the $x^{+}$ and $x^{-}$ integrations.
I will use
\begin{align}
\int d\tilde{X}\!_{A\!B} & =\int\,\prod_{k=1}^{AB}\bigg(\prod_{\lambda=1}^{l_{k}}d\tilde{x}_{k\lambda}^{+}d\tilde{x}_{k\lambda}^{-}\bigg)\nonumber \\
 & =\underbrace{\int\,\prod_{k=1}^{AB}\bigg(\prod_{\lambda=1}^{l_{k}}d\tilde{x}_{k\lambda}^{+}\bigg)}_{d\tilde{X}_{AB}^{+}}\;\underbrace{\int\,\prod_{k=1}^{AB}\bigg(\prod_{\lambda=1}^{l_{k}}d\tilde{x}_{k\lambda}^{-}\bigg)}_{d\tilde{X}_{AB}^{-}},
\end{align}
where the upper limit $l_{k}$ is defined as 
\begin{equation}
l_{k}=r_{1k}+r_{2k}+r_{3k}+r_{4k}.
\end{equation}
In addition, I define four intervals of integers, as 
\begin{align}
 & I_{1k}=[1\,..\,r_{1k}]\,,\\
 & I_{2k}=[r_{1k}+1\,..\,r_{1k}+r_{2k}]\,,\\
 & I_{3k}=[r_{1k}+r_{2k}+1\,..\,r_{1k}+r_{2k}+r_{3k}]\,,\\
 & I_{4k}=[r_{1k}+r_{2k}+r_{3k}+1\,..\,r_{1k}+r_{2k}+r_{3k}+r_{4k}]\,,
\end{align}
such that
\begin{equation}
[1\,..\,l_{k}]=I_{1k}\cup I_{2k}\cup I_{3k}\cup I_{4k}.
\end{equation}
Although writing the Pomeron expression as a sum of four terms is
a purely mathematical operation, it is nevertheless useful for the
discussion to associate the indices in the interval $I_{Nk}$ to ``Pomerons
of type $N$''. This allows one to consider $r_{Nk}$ as the number of
Pomerons of type $N$ associated to nucleon-nucleon pair $k$. With
the above definitions and interpretations, one gets
\begin{align}
W_{AB} & =\sum_{r_{11}=0}^{\infty}\!\!\!\cdots\!\!\!\sum_{r_{41}=0}^{\infty}\cdots\sum_{r_{1\,AB}=0}^{\infty}\!\!\!\cdots\!\!\!\sum_{r_{4\,AB}=0}^{\infty}\nonumber \\
 & \prod_{k=1}^{AB}\frac{(-\alpha_{1})^{r_{1k}}}{r_{1k}!}\frac{(-\alpha_{2})^{r_{2k}}}{r_{2k}!}\frac{(-\alpha_{3})^{r_{3k}}}{r_{3k}!}\frac{(-\alpha_{4})^{r_{4k}}}{r_{4k}!}\nonumber \\
 & \quad U^{+}\Big(\{r_{Nk}\},\{x_{i}^{+}\}\Big)\:U^{-}\Big(\{r_{Nk}\},\{x_{i}^{-}\}\Big)\,,\label{W-AB-2}
\end{align}
seen as the sum over all possible numbers of Pomerons of all possible
types (1,2,3,4), with
\begin{align}
 & U^{+}\Big(\{r_{Nk}\},\{x_{i}^{+}\}\Big)\nonumber \\
 & =\int d\tilde{X}_{AB}^{+}\prod_{k=1}^{AB}\Bigg[\prod_{\lambda\in I_{1k}}(\tilde{x}_{k\lambda}^{+})^{\beta_{1}}\cdots\!\!\!\!\prod_{\lambda\in I_{4k}}\!\!(\tilde{x}_{k\lambda}^{+})^{\beta_{4}}\Bigg]\nonumber \\
 & \qquad\qquad\prod_{i=1}^{A}(x_{i}^{+}-\!\!\sum_{\underset{\pi(k)=i}{k=1}}^{AB}\sum_{\lambda=1}^{l_{k}}\!\tilde{x}_{k\lambda}^{+})^{\alpha_{\mathrm{remn}}}.
\end{align}
The expression for $U^{-}\Big(\{r_{Nk}\},\{x_{i}^{-}\}\Big)$ is identical,
just with $x_{j}^{-}$ instead of $x_{i}^{+}$, with $\prod_{j=1}^{B}$
instead of $\prod_{i=1}^{A}$, and with $\tau(k)=j$ instead of $\pi(k)=i$.
Using 
\begin{equation}
\prod_{k=1}^{AB}=\prod_{i=1}^{A}\prod_{\underset{\pi(k)=i}{k=1}}^{AB},
\end{equation}
and defining 
\begin{equation}
\epsilon_{\lambda k}=\left\{ \begin{array}{c}
\beta_{1}\:\mathrm{for}\:\lambda\in I_{1k}\\
\beta_{2}\:\mathrm{for}\:\lambda\in I_{2k}\\
\beta_{3}\:\mathrm{for}\:\lambda\in I_{3k}\\
\beta_{4}\:\mathrm{for}\:\lambda\in I_{4k}
\end{array}\right.,\label{epsilon-lambda-k}
\end{equation}
one may separate the contributions for the different nucleons $i$
as 
\[
U^{+}\Big(\{r_{Nk}\},\{x_{i}^{+}\}\Big)=\prod_{i=1}^{A}U_{i}^{+}\Big(\{r_{Nk}\},x_{i}^{+}\Big)
\]
with an expression for given $i$,
\begin{align}
 & U_{i}^{+}\Big(\{r_{Nk}\},x_{i}^{+}\Big)=\int\!\prod_{\underset{\pi(k)=i}{k=1}}^{AB}\!\bigg(\!\prod_{\lambda=1}^{l_{k}}d\tilde{x}_{k\lambda}^{+}\bigg)\\
 & \qquad\prod_{\underset{\pi(k)=i}{k=1}}^{AB}\Bigg[\prod_{\lambda=1}^{l_{k}}(\tilde{x}_{k\lambda}^{+})^{\epsilon_{\lambda k}}\Bigg](x_{i}^{+}-\!\!\sum_{\underset{\pi(k)=i}{k=1}}^{AB}\sum_{\lambda=1}^{l_{k}}\!\tilde{x}_{k\lambda}^{+})^{\alpha_{\mathrm{remn}}}.\nonumber 
\end{align}
Let me rename, for given $i$, the $\tilde{x}_{k\lambda}^{+}$ linked
to nucleon $i$ as $x_{1},\,x_{2},\,\ldots,\,x_{L}$, the $\epsilon_{\lambda k}$
as $\epsilon_{1},\,\epsilon_{2},\,\ldots,\epsilon_{L}$, where $L$
is per definition the number of Pomerons linked to $i$, and let us
define $x=x_{i}^{+}$. Then one gets 
\begin{equation}
U_{i}^{+}\Big(\{r_{Nk}\},x\Big)=\int\!\!\prod_{\Lambda=1}^{L}dx_{\Lambda}\,\prod_{\Lambda=1}^{L}\,(x_{\Lambda})^{\epsilon_{\Lambda}}(x-\!\!\sum_{\Lambda=1}^{L}\!x_{\Lambda})^{\alpha_{\mathrm{remn}}}.
\end{equation}
I define new variables,
\begin{align}
 & u_{\Lambda}=\frac{x_{\Lambda}}{x-x_{1}-\ldots-x_{\Lambda-1}},\\
 & du_{\Lambda}=\frac{dx_{\Lambda}}{x-x_{1}-\ldots-x_{\Lambda-1}},
\end{align}
which have the property

\begin{equation}
\prod_{a=1}^{\Lambda-1}(1-u_{a})=\prod_{a=1}^{\Lambda-1}\frac{x-\ldots-x_{a}}{x-\ldots-x_{a-1}}=\frac{x-\ldots-x_{\Lambda-1}}{x},
\end{equation}
and therefore%
\begin{comment}
\[
x_{\Lambda}=u_{\Lambda}(x-x_{1}-\ldots-x_{\Lambda-1})=u_{\Lambda}x\prod_{a=1}^{\Lambda-1}(1-u_{a})\,;\,dx_{\Lambda}=du_{\Lambda}(x-x_{1}-\ldots-x_{\Lambda-1})=du_{\Lambda}x\prod_{a=1}^{\Lambda-1}(1-u_{a})
\]
\end{comment}
{} 
\begin{equation}
x_{\Lambda}=xu_{\Lambda}\prod_{a=1}^{\Lambda-1}(1-u_{a}),\quad dx_{\Lambda}=xdu_{\Lambda}\prod_{a=1}^{\Lambda-1}(1-u_{a}).
\end{equation}
This leads to%
\begin{comment}
\begin{equation}
U_{i}^{+}\Big(\{r_{Nk}\},x\Big)=\int\!\!\prod_{\Lambda=1}^{L}dx_{\Lambda}\,\prod_{\Lambda=1}^{L}\,(x_{\Lambda})^{\epsilon_{\Lambda}}(x-\!\!\sum_{\Lambda=1}^{L}\!x_{\Lambda})^{\alpha_{\mathrm{remn}}}.
\end{equation}
\textemdash \textemdash \textemdash \textemdash \textemdash \textemdash \textemdash \textemdash \textemdash \textemdash -
\[
=\int\prod_{\Lambda=1}^{L}\left[xdu_{\Lambda}\prod_{a=1}^{\Lambda-1}(1-u_{a})\,\right]\quad\prod_{\Lambda=1}^{L}\left[xu_{\Lambda}\prod_{a=1}^{\Lambda-1}(1-u_{a})\right]^{\epsilon_{\Lambda}}\quad\left(x\prod_{\Lambda=1}^{L}(1-u_{\Lambda})\right)^{\alpha_{\mathrm{remn}}}
\]
\[
=\int\prod_{\Lambda=1}^{L}du_{\Lambda}\:\left\{ \prod_{\Lambda=1}^{L}\left[x\prod_{a=1}^{\Lambda-1}(1-u_{a})x^{\epsilon_{\lambda}}u_{\Lambda}^{\epsilon_{\Lambda}}\prod_{a=1}^{\Lambda-1}(1-u_{a})^{\epsilon_{\Lambda}}\right]\;x^{\alpha_{\mathrm{remn}}}\prod_{\Lambda=1}^{L}(1-u_{\Lambda})^{\alpha_{\mathrm{remn}}}\right\} 
\]
\[
=x^{\alpha_{\mathrm{remn}+\sum_{\Lambda}\tilde{\epsilon}_{\Lambda}}}\int\prod_{\Lambda=1}^{L}du_{\Lambda}\:\left\{ \prod_{\Lambda=1}^{L}\left[\prod_{a=1}^{\Lambda-1}(1-u_{a})u_{\Lambda}^{\epsilon_{\Lambda}}\prod_{a=1}^{\Lambda-1}(1-u_{a})^{\epsilon_{\Lambda}}(1-u_{\Lambda})^{\alpha_{\mathrm{remn}}}\right]\right\} 
\]
\end{comment}
\begin{align}
 & U_{i}^{+}\Big(\{r_{Nk}\},x\Big)=x^{\alpha_{\mathrm{remn}+\sum_{\Lambda}\tilde{\epsilon}_{\Lambda}}}\int\prod_{\Lambda=1}^{L}du_{\Lambda}\\
 & \quad\Bigg\{\prod_{\Lambda=1}^{L}\left[u_{\Lambda}^{\epsilon_{\Lambda}}\prod_{a=1}^{\Lambda-1}(1-u_{a})^{\tilde{\epsilon}_{\Lambda}}(1-u_{\Lambda})^{\alpha_{\mathrm{remn}}}\right]\Bigg\},\nonumber 
\end{align}
where I used $\tilde{\epsilon}_{\Lambda}=\epsilon_{\Lambda}+1$, for
convenience. One has  
\begin{align}
\prod_{\Lambda=1}^{L}\prod_{a=1}^{\Lambda-1}(1-u_{a})^{\tilde{\epsilon}_{\Lambda}} & =\prod_{a=1}^{L}\prod_{\Lambda=a+1}^{L}(1-u_{a})^{\tilde{\epsilon}_{\Lambda}}\\
 & =\prod_{\Lambda=1}^{L}\prod_{a=\Lambda+1}^{L}(1-u_{\Lambda})^{\tilde{\epsilon}_{a}}\\
 & =\prod_{\Lambda=1}^{L}(1-u_{\Lambda})^{\sum_{a=\Lambda+1}^{L}\tilde{\epsilon}_{a}},
\end{align}
which provides 
\begin{align}
 & U_{i}^{+}\Big(\{r_{Nk}\},x\Big)=x^{\alpha_{\mathrm{remn}+\sum_{\Lambda}\tilde{\epsilon}_{\Lambda}}}\int\prod_{\Lambda=1}^{L}du_{\Lambda}\\
 & \quad\Bigg\{\prod_{\Lambda=1}^{L}\left[u_{\Lambda}^{\epsilon_{\Lambda}}(1-u_{\Lambda})^{\sum_{a=\Lambda+1}^{L}\tilde{\epsilon}_{a}}(1-u_{\Lambda})^{\alpha_{\mathrm{remn}}}\right]\Bigg\}.\nonumber 
\end{align}
Defining
\begin{align}
 & \alpha=\alpha_{\mathrm{remn}}+\sum_{\Lambda=1}^{L}\tilde{\epsilon}_{\Lambda},\label{alpha}\\
 & \gamma_{\Lambda}=\alpha_{\mathrm{remn}}+\sum_{a=\Lambda+1}^{L}\tilde{\epsilon}_{a},\label{gamma}
\end{align}
one finds
\begin{equation}
U_{i}^{+}\Big(\{r_{Nk}\},x\Big)=x^{\alpha}\prod_{\Lambda=1}^{L}\int_{0}^{1}du_{\Lambda}u_{\Lambda}^{\epsilon_{\Lambda}}(1-u_{\Lambda})^{\gamma_{\Lambda}}.
\end{equation}
The integral can be done [$\int_{0}^{1}t^{x-1}(1-t)^{y-1}dt$ is the
Euler beta function], and one gets, with $x=x_{i}^{+}$,

\begin{equation}
U_{i}^{+}\Big(\{r_{Nk}\},x_{i}^{+}\Big)=(x_{i}^{+})^{\alpha}\prod_{\Lambda=1}^{L}\frac{\Gamma(1+\epsilon_{\Lambda})\Gamma(1+\gamma_{\Lambda})}{\Gamma(2+\epsilon_{\Lambda}+\gamma_{\Lambda})}.
\end{equation}
Using the relation $1+\epsilon_{\Lambda}+\gamma_{\Lambda}=\gamma_{\Lambda-1}$,
one gets
\begin{equation}
U_{i}^{+}\Big(\{r_{Nk}\},x_{i}^{+}\Big)=(x_{i}^{+})^{\alpha}\prod_{\Lambda=1}^{L}\frac{\Gamma(1+\epsilon_{\Lambda})\Gamma(1+\gamma_{\Lambda})}{\Gamma(1+\gamma_{\Lambda-1})}.\label{U-plus-i}
\end{equation}
In the following, the different factors in this expression will be
discussed. One has 
\begin{equation}
\prod_{\Lambda=1}^{L}=\prod_{\underset{\pi(k)=i}{k=1}}^{AB}\prod_{\lambda=1}^{l_{k}},
\end{equation}
which actually indicates the relation between the indices $\Lambda$
and the pair of indices $k$ and $\lambda$. So one gets 
\begin{align}
\prod_{\Lambda=1}^{L}\Gamma(1+\epsilon_{\Lambda}) & =\prod_{\underset{\pi(k)=i}{k=1}}^{AB}\prod_{\lambda=1}^{l_{k}}\Gamma(1+\epsilon_{\lambda k})\\
 & =\prod_{\underset{\pi(k)=i}{k=1}}^{AB}\prod_{N=1}^{4}\prod_{\lambda\in I_{Nk}}\Gamma(1\!+\!\beta_{N}),
\end{align}
where I split $\prod_{\lambda=1}^{l_{k}}$ into four products corresponding
to the four Pomeron types 1 - 4, and where I used Eq. (\ref{epsilon-lambda-k}).
So one finds
\begin{equation}
\prod_{\Lambda=1}^{L}\Gamma(1+\epsilon_{\Lambda})=\prod_{\underset{\pi(k)=i}{k=1}}^{AB}\prod_{N=1}^{4}\Gamma(1\!+\!\beta_{N})^{r_{Nk}}.\label{prod-gamma-1}
\end{equation}
Concerning $(x_{i}^{+})^{\alpha}$, I use Eq. (\ref{alpha}) and $\tilde{\epsilon}_{\Lambda}=\epsilon_{\Lambda}+1$,
and I define $\tilde{\beta}_{N}=\beta_{N}+1$, to get
\begin{align}
\alpha & =\alpha_{\mathrm{remn}}+\sum_{\Lambda=1}^{L}\tilde{\epsilon}_{\Lambda}\\
 & =\alpha_{\mathrm{remn}}+\sum_{\underset{\pi(k)=i}{k=1}}^{AB}\sum_{\lambda=1}^{l_{k}}\tilde{\epsilon}_{\lambda k}\\
 & =\alpha_{\mathrm{remn}}+\sum_{\underset{\pi(k)=i}{k=1}}^{AB}\sum_{N=1}^{4}\sum_{\lambda\in I_{Nk}}\tilde{\beta}_{N}\:\\
 & =\alpha_{\mathrm{remn}}+\sum_{\underset{\pi(k)=i}{k=1}}^{AB}\sum_{N=1}^{4}r_{Nk}\tilde{\beta}_{N}\:,
\end{align}
which gives
\begin{equation}
(x_{i}^{+})^{\alpha}=(x_{i}^{+})^{\alpha_{\mathrm{remn}}}\prod_{\underset{\pi(k)=i}{k=1}}^{AB}\prod_{N=1}^{4}(x_{i}^{+})^{r_{Nk}\tilde{\beta}_{N}}.\label{alpha-1}
\end{equation}
Furthermore, one has
\begin{align}
 & \prod_{\Lambda=1}^{L}\frac{\Gamma(1+\gamma_{\Lambda})}{\Gamma(1+\gamma_{\Lambda-1})}=\frac{\Gamma(1+\gamma_{L})}{\Gamma(1+\gamma_{0})}\label{ratio-gamma-1}\\
 & \qquad=\frac{\Gamma(1+\alpha_{\mathrm{remn}})}{\Gamma(1+\alpha_{\mathrm{remn}}+\sum_{\Lambda=1}^{L}\tilde{\epsilon}_{\Lambda})}\label{ratio-gamma-2}\\
 & \qquad=\frac{\Gamma(1+\alpha_{\mathrm{remn}})}{\Gamma(1+\alpha_{\mathrm{remn}}+\sum_{\underset{\pi(k)=i}{k=1}}^{AB}\sum_{N=1}^{4}r_{Nk}\tilde{\beta}_{N})}.\label{ratio-gamma-3}
\end{align}
Defining a function $g(z)$ as 
\begin{equation}
g(z)=\frac{\Gamma(1+\alpha_{\mathrm{remn}})}{\Gamma(1+\alpha_{\mathrm{remn}}+z)},
\end{equation}
one gets 
\begin{equation}
\prod_{\Lambda=1}^{L}\frac{\Gamma(1+\gamma_{\Lambda})}{\Gamma(1+\gamma_{\Lambda-1})}=g(\sum_{\underset{\pi(k)=i}{k=1}}^{AB}\sum_{N=1}^{4}r_{Nk}\tilde{\beta}_{N}).\label{ratio-gamma-4}
\end{equation}
Inserting Eqs. (\ref{prod-gamma-1}), (\ref{alpha-1}), and (\ref{ratio-gamma-4})
into Eq. (\ref{U-plus-i}), one gets%
\begin{comment}
\begin{equation}
U_{i}^{+}\Big(\{r_{Nk}\},x_{i}^{+}\Big)=(x_{i}^{+})^{\alpha}\prod_{\Lambda=1}^{L}\frac{\Gamma(1+\epsilon_{\Lambda})\Gamma(1+\gamma_{\Lambda})}{\Gamma(1+\gamma_{\Lambda-1})}.\label{U-plus-i-2}
\end{equation}
\end{comment}
\begin{align}
 & U_{i}^{+}\Big(\{r_{Nk}\},x_{i}^{+}\Big)=(x_{i}^{+})^{\alpha_{\mathrm{remn}}}\prod_{\underset{\pi(k)=i}{k=1}}^{AB}\prod_{N=1}^{4}(x_{i}^{+})^{r_{Nk}\tilde{\beta}_{N}}\nonumber \\
 & \prod_{\underset{\pi(k)=i}{k=1}}^{AB}\prod_{N=1}^{4}\Gamma(1\!+\!\beta_{N})^{r_{Nk}}g(\sum_{\underset{\pi(k)=i}{k=1}}^{AB}\sum_{N=1}^{4}r_{Nk}\tilde{\beta}_{N}).\label{U-plus-i-1}
\end{align}
{} A corresponding expression can be found for $U_{j}^{-}\Big(\{r_{Nk}\},x_{j}^{-}\Big)$.
The two expressions 
\begin{equation}
U^{+}\Big(\{r_{Nk}\},\{x_{i}^{+}\}\Big)=\prod_{i=1}^{A}U_{i}^{+}\Big(\{r_{Nk}\},x_{i}^{+}\Big),
\end{equation}
\begin{equation}
U^{-}\Big(\{r_{Nk}\},\{x_{j}^{-}\}\Big)=\prod_{j=1}^{B}U_{j}^{-}\Big(\{r_{Nk}\},x_{j}^{-}\Big),
\end{equation}
may be inserted into Eq. (\ref{W-AB-2}), using Eq. (\ref{U-plus-i-1})
and the corresponding $U_{j}^{-}$ expression, and one gets%
\begin{comment}
\begin{align*}
 & W_{AB}=\sum_{r_{1k}=0}^{\infty}\!\!\!\cdots\!\!\!\sum_{r_{4k}=0}^{\infty}\cdots\sum_{r_{1,AB}=0}^{\infty}\!\!\!\cdots\!\!\!\sum_{r_{4,AB}=0}^{\infty}\\
 & \prod_{k=1}^{AB}\frac{(-\alpha_{1})^{r_{1k}}}{r_{1k}!}\frac{(-\alpha_{2})^{r_{2k}}}{r_{2k}!}\frac{(-\alpha_{3})^{r_{3k}}}{r_{3k}!}\frac{(-\alpha_{4})^{r_{4k}}}{r_{4k}!}\\
 & \quad U^{+}\Big(\{r_{Nk}\},\{x_{i}^{+}\}\Big)\:U^{-}\Big(\{r_{Nk}\},\{x_{i}^{-}\}\Big)\,,
\end{align*}

\textemdash \textemdash \textemdash \textemdash \textendash{}

\begin{align*}
 & W_{AB}=\prod_{i=1}^{A}(x_{i}^{+})^{\alpha_{\mathrm{remn}}}\prod_{j=1}^{B}(x_{j}^{-})^{\alpha_{\mathrm{remn}}}\sum_{\{r_{Nk}\}}\\
 & \prod_{k=1}^{AB}\frac{(-\alpha_{1})^{r_{1k}}}{r_{1k}!}\frac{(-\alpha_{2})^{r_{2k}}}{r_{2k}!}\frac{(-\alpha_{3})^{r_{3k}}}{r_{3k}!}\frac{(-\alpha_{4})^{r_{4k}}}{r_{4k}!}\\
 & \prod_{i=1}^{A}\Bigg\{\prod_{\underset{\pi(k)=i}{k=1}}^{AB}\prod_{N=1}^{4}(x_{i}^{+})^{r_{Nk}\tilde{\beta}_{N}}\\
 & \prod_{\underset{\pi(k)=i}{k=1}}^{AB}\prod_{N=1}^{4}\Gamma(1\!+\!\beta_{N})^{r_{Nk}}g(\sum_{\underset{\pi(k)=i}{k=1}}^{AB}\sum_{N=1}^{4}r_{Nk}\tilde{\beta}_{N})\Bigg\}\\
 & \,\,target
\end{align*}
\end{comment}
\medskip{}

\begin{align}
 & W_{AB}=\label{W-AB-1}\\
 & \prod_{i=1}^{A}(x_{i}^{+})^{\alpha_{\mathrm{remn}}}\prod_{j=1}^{B}(x_{j}^{-})^{\alpha_{\mathrm{remn}}}\sum_{\{r_{Nk}\}}\Bigg\{\prod_{k=1}^{AB}\prod_{N=1}^{4}\frac{(-\alpha_{N})^{r_{Nk}}}{r_{Nk}!}\nonumber \\
 & \prod_{i=1}^{A}\Bigg[\prod_{\underset{\pi(k)=i}{k=1}}^{AB}\prod_{N=1}^{4}\left(\Gamma(\tilde{\beta}_{N})(x_{i}^{+})^{\tilde{\beta}_{N}}\right)^{r_{Nk}}g(\sum_{\underset{\pi(k)=i}{k=1}}^{AB}\sum_{N=1}^{4}r_{Nk}\tilde{\beta}_{N})\Bigg]\nonumber \\
 & \prod_{j=1}^{B}\Bigg[\prod_{\underset{\tau(k)=j}{k=1}}^{AB}\prod_{N=1}^{4}\left(\Gamma(\tilde{\beta}_{N})(x_{j}^{-})^{\tilde{\beta}_{N}}\right)^{r_{Nk}}g(\sum_{\underset{\tau(k)=j}{k=1}}^{AB}\sum_{N=1}^{4}r_{Nk}\tilde{\beta}_{N})\Bigg]\,\Bigg\},\nonumber 
\end{align}

\noindent
where $\sum_{\{r_{Nk}\}}$ means summing all the indices $r_{Nk}$,
with $1\le N\le4$ and with $1\le k\le AB$, from zero to infinity. I
use $\tilde{\beta}_{N}=\beta_{N}+1$. A similar formula (with
different notations) has been found in Ref. \cite{Drescher:2000ha}.%

\subsection{Doing the infinite sums}

\begin{comment}
GfunParK -> alpha beta betap for Gtilde 

atildg= alpD {*} exp(-b2 

/ delD ) /xminDf{*}{*}dble(epsG+gamb) 

{*}utgam2(btildgp(i,k)){*}utgam2(btildgpp(i,k))

{*}utgam1(1.+alplea){*}utgam1(1.+alplea)

/ utgam2( 1+alplea+btildgp ) / utgam2( 1+alplea+btildgpp ) 

\[
\frac{\Gamma(1+\alpha_{\mathrm{remn}})}{\Gamma(1+\alpha_{\mathrm{remn}}+\sum m_{n}b_{n})}=\prod(\frac{\Gamma(1+\alpha_{\mathrm{remn}})}{\Gamma(1+\alpha_{\mathrm{remn}}+b_{n})})^{m_{n}}
\]
\end{comment}

Equation (\ref{W-AB-1}) is the best one can do, without further assumptions.
At least the integrations could be done, and (although time consuming)
the infinite sums can be performed on a powerful computer (they converge).\\

However, a ``small'' modification would allow one to simplify the expression
enormously, and this modification concerns the function $g(z)=\Gamma(1+\alpha_{\mathrm{remn}})$
$/\Gamma(1+\alpha_{\mathrm{remn}}+z)$. The argument of $g$ is of
the form $\sum_{\lambda}\tilde{\beta}_{\lambda}$ with given coefficients
$\tilde{\beta}_{\lambda}$. If one would have\smallskip{}

\begin{equation}
g\left(\sum_{\lambda}\tilde{\beta}_{\lambda}\right)=c_{1}\prod_{\lambda}c_{2}\,g(c_{3}\,\tilde{\beta}_{\lambda}),\label{g-property}
\end{equation}

\noindent
with three parameters $c_{\mu}$, then one obtains
\begin{align}
 & g(\sum_{\underset{\pi(k)=i}{k=1}}^{AB}\sum_{N=1}^{4}r_{Nk}\tilde{\beta}_{N})=c_{1}\!\!\prod_{\underset{\pi(k)=i}{k=1}}^{AB}\prod_{N=1}^{4}\!\left(c_{2}\,g(c_{3}\,\tilde{\beta}_{N})\right)^{r_{Nk}},\\
 & g(\sum_{\underset{\tau(k)=j}{k=1}}^{AB}\sum_{N=1}^{4}r_{Nk}\tilde{\beta}_{N})=c_{1}\!\!\prod_{\underset{\tau(k)=j}{k=1}}^{AB}\prod_{N=1}^{4}\!\left(c_{2}\,g(c_{3}\,\tilde{\beta}_{N}\right)^{r_{Nk}},
\end{align}
and one finds for $W_{AB}$ the expression%
\begin{comment}
\begin{align}
 & W_{AB}=\label{W-AB-3}\\
 & \prod_{i=1}^{A}c_{1}(x_{i}^{+})^{\alpha_{\mathrm{remn}}}\prod_{j=1}^{B}c_{1}(x_{j}^{-})^{\alpha_{\mathrm{remn}}}\sum_{\{r_{Nk}\}}\Bigg\{\prod_{k=1}^{AB}\prod_{N=1}^{4}\frac{(-\alpha_{N})^{r_{Nk}}}{r_{Nk}!}\nonumber \\
 & \prod_{i=1}^{A}\Bigg[\prod_{\underset{\pi(k)=i}{k=1}}^{AB}\prod_{N=1}^{4}\left(\Gamma(\tilde{\beta}_{N})(x_{i}^{+})^{\tilde{\beta}_{N}}\right)^{r_{Nk}}\prod_{\underset{\pi(k)=i}{k=1}}^{AB}\prod_{N=1}^{4}\left(c_{2}\,g(c_{3}\,\tilde{\beta}_{N})\right)^{r_{Nk}}\Bigg]\nonumber \\
 & \prod_{j=1}^{B}\Bigg[\prod_{\underset{\tau(k)=j}{k=1}}^{AB}\prod_{N=1}^{4}\left(\Gamma(\tilde{\beta}_{N})(x_{j}^{-})^{\tilde{\beta}_{N}}\right)^{r_{Nk}}\prod_{\underset{\tau(k)=j}{k=1}}^{AB}\prod_{N=1}^{4}\left(c_{2}\,g(c_{3}\,\tilde{\beta}_{N}\right)^{r_{Nk}}\Bigg]\,\Bigg\},\nonumber 
\end{align}
\end{comment}
\begin{align}
 & W_{AB}=\label{W-AB-4}\\
 & \prod_{i=1}^{A}c_{1}(x_{i}^{+})^{\alpha_{\mathrm{remn}}}\prod_{j=1}^{B}c_{1}(x_{j}^{-})^{\alpha_{\mathrm{remn}}}\sum_{\{r_{Nk}\}}\prod_{k=1}^{AB}\prod_{N=1}^{4}\Bigg\{\frac{(-\alpha_{N})^{r_{Nk}}}{r_{Nk}!}\nonumber \\
 & \qquad\qquad\Bigg[\left(\Gamma(\tilde{\beta}_{N})(x_{\pi(k)}^{+})^{\tilde{\beta}_{N}}\right)^{r_{Nk}}\left(c_{2}\,g(c_{3}\,\tilde{\beta}_{N})\right)^{r_{Nk}}\Bigg]\nonumber \\
 & \qquad\qquad\Bigg[\left(\Gamma(\tilde{\beta}_{N})(x_{\tau(k)}^{-})^{\tilde{\beta}_{N}}\right)^{r_{Nk}}\left(c_{2}\,g(c_{3}\,\tilde{\beta}_{N}\right)^{r_{Nk}}\Bigg]\,\Bigg\}.\nonumber 
\end{align}
Defining 
\begin{equation}
D_{N}=\Gamma(\tilde{\beta}_{N})\,c_{2}\,g(c_{3}\,\tilde{\beta}_{N})=\frac{\Gamma(\tilde{\beta}_{N})\,c_{2}\,\Gamma(1+\alpha_{\mathrm{remn}})}{\Gamma(1+\alpha_{\mathrm{remn}}+c_{3}\,\tilde{\beta}_{N})},
\end{equation}
one gets
\begin{align}
 & W_{AB}=\prod_{i=1}^{A}c_{1}(x_{i}^{+})^{\alpha_{\mathrm{remn}}}\prod_{j=1}^{B}c_{1}(x_{j}^{-})^{\alpha_{\mathrm{remn}}}\label{W-AB-5}\\
 & \qquad\sum_{\{r_{Nk}\}}\prod_{k=1}^{AB}\prod_{N=1}^{4}\frac{1}{r_{Nk}!}\Bigg[-\alpha_{N}(x_{\pi(k)}^{+})^{\tilde{\beta}_{N}}(x_{\tau(k)}^{-})^{\tilde{\beta}_{N}}D_{N}^{\:2}\Bigg]^{r_{Nk}}.\nonumber 
\end{align}
Exchanging sum and product, one gets
\begin{align}
 & W_{AB}=\prod_{i=1}^{A}c_{1}(x_{i}^{+})^{\alpha_{\mathrm{remn}}}\prod_{j=1}^{B}c_{1}(x_{j}^{-})^{\alpha_{\mathrm{remn}}}\label{W-AB-6}\\
 & \qquad\prod_{k=1}^{AB}\prod_{N=1}^{4}\sum_{r_{Nk}}\frac{1}{r_{Nk}!}\Bigg[-\alpha_{N}(x_{\pi(k)}^{+})^{\tilde{\beta}_{N}}(x_{\tau(k)}^{-})^{\tilde{\beta}_{N}}D_{N}^{\:2}\Bigg]^{r_{Nk}},\nonumber 
\end{align}
where one recognizes the power series of the exponential function,
and so one gets finally\smallskip{}

\begin{align}
 & W_{AB}=\prod_{i=1}^{A}c_{1}(x_{i}^{+})^{\alpha_{\mathrm{remn}}}\prod_{j=1}^{B}c_{1}(x_{j}^{-})^{\alpha_{\mathrm{remn}}}\label{W-AB-7}\\
 & \qquad\qquad\prod_{k=1}^{AB}\exp\left(-\tilde{G}(x_{\pi(k)}^{+}x_{\tau(k)}^{-})\right),\nonumber 
\end{align}

\noindent
with 
\begin{equation}
\tilde{G}(x)=\sum_{N=1}^{4}\tilde{\alpha}_{N}x^{\tilde{\beta}_{N}}
\end{equation}
and with
\begin{align}
\tilde{\alpha}_{N} & =\alpha_{N}D_{N}^{\:2}=\alpha_{N}\left(\frac{\Gamma(\tilde{\beta}_{N})\,c_{2}\,\Gamma(1+\alpha_{\mathrm{remn}})}{\Gamma(1+\alpha_{\mathrm{remn}}+c_{3}\,\tilde{\beta}_{N})}\right)^{2},\\
\tilde{\beta}_{N} & =\beta_{N}+1\;.
\end{align}
So the property in Eq. (\ref{g-property}) for the function $g$ provides
infinite sums of the form of power series of the exponential
function, and the exponentials make sure that one always has $W_{AB}\ge0$.
\begin{comment}

\section*{END}
\end{comment}

\end{document}